\documentclass[11pt,a4paper]{article}
\pdfoutput=1
\usepackage{jheppub}
\usepackage{amsthm,amsbsy,amsfonts,mathrsfs,enumerate,float,wrapfig,amsmath}
\usepackage{subfigure}
\usepackage{longtable}
\usepackage{adjustbox} 
\usepackage {dashrule}
\allowdisplaybreaks[4]
\newcommand{\be}{\begin{equation}}
\newcommand{\ee}{\end{equation}}
\newcommand{\nn}{\nonumber}

\def\us#1_#2{\underset{#2}{#1}}
\def\os#1^#2{\overset{#2}{#1}}
\newcommand{\fsp}{\mathfrak{sp}}

\newcommand{\fg}{\mathfrak{g}}
\newcommand{\fsu}{\mathfrak{su}}
\newcommand{\fe}{\mathfrak{e}}
\newcommand{\fu}{\mathfrak{u}}
\newcommand{\ff}{\mathfrak{f}}

\usepackage{tikz}
\usetikzlibrary{calc}
\usetikzlibrary{decorations}
\pgfdeclaredecoration{ignore}{final}
{
\state{final}{}
}

 \pgfdeclaremetadecoration{middle}{initial}{
    \state{initial}[
        width={(\pgfmetadecoratedpathlength - \the\pgfdecorationsegmentlength)/2},
        next state=middle
    ]
    {\decoration{moveto}}

    \state{middle}[
        width={\the\pgfdecorationsegmentlength},
        next state=final
    ]
    {\decoration{curveto}}

    \state{final}
    {\decoration{ignore}}
}
\tikzset{middle segment/.style={decoration={middle},decorate, segment length=#1}}

\newcommand{\U}{\mathrm{U}}
\newcommand{\SU}{\mathrm{SU}}
\newcommand{\SO}{\mathrm{SO}}
\newcommand{\Sp}{\mathrm{Sp}}
\newcommand{\fso}{\mathfrak{so}}

\graphicspath{ {./tikzfigures/} }

\renewcommand{\emptyset}{\varnothing}
\renewcommand{\hat}{\widehat}
\renewcommand{\tilde}{\widetilde}

\usepackage[T1]{fontenc}

\title{6d/5d exceptional gauge theories from web diagrams}

\author[a]{Hirotaka Hayashi,}
\author[b,c]{Hee-Cheol Kim,}
\author[d]{Kantaro Ohmori}

\affiliation[a]{Department of Physics, School of Science, Tokai University, 4-1-1 Kitakaname, Hiratsuka-shi, Kanagawa 259-1292, Japan}
\affiliation[b]{Department of Physics, POSTECH, Pohang 790-784, Korea}
\affiliation[c]{Asia Pacific Center for Theoretical Physics, Postech, Pohang 37673, Korea}
\affiliation[d]{Simons Center for Geometry and Physics, SUNY, Stony Brook NY 11794, USA}

\emailAdd{h.hayashi@tokai.ac.jp}
\emailAdd{heecheol1@gmail.com}
\emailAdd{komori@scgp.stonybrook.edu}

\abstract{
We construct novel web diagrams with a trivalent or quadrivalent gluing for various 6d/5d theories from certain Higgsings of 6d conformal matter theories on a circle. The theories realized on the web diagrams include 5d Kaluza-Klein theories from circle compactifications of the 6d $G_2$ gauge theory with $4$ flavors, the 6d $F_4$ gauge theory with $3$ flavors, the 6d $E_6$ gauge theory with $4$ flavors and the 6d $E_7$ gauge theory with $3$ flavors. The Higgsings also give rise to 5d Kaluza-Klein theories from twisted compactifications of 6d theories including the 5d pure $\SU(3)$ gauge theory with the Chern-Simons level $9$ and the 5d pure $\SU(4)$ gauge theory with the Chern-Simons level $8$. We also compute the Nekrasov partition functions of the theories by applying the topological vertex formalism to the newly obtained web diagrams. }

\begin{document}
\preprint{
\begin{flushright}
\tt 
\end{flushright}
}

\maketitle


\section{Introduction}
5-brane brane web diagrams in Type IIB string theory \cite{Aharony:1997ju, Aharony:1997bh} are a useful tool to study various aspects of five-dimensional (5d) $\mathcal{N}=1$ supersymmetric field theories. One application of 5-brane web diagrams may be the computation of the Nekrasov partition functions of 5d theories realized on 5-brane webs. Since some 5-brane web diagrams are dual to Calabi-Yau threefolds in M-theory \cite{Leung:1997tw}, the Nekrasov partition functions of the 5d theories on the self-dual $\Omega$-background may be interepreted as the all genus topological string partition functions for the dual Calabi-Yau threefolds \cite{Gopakumar:1998ii, Gopakumar:1998jq}. When 5-brane web diagrams are dual to toric Calabi-Yau threefolds one can compute the Nekraov partition functions of the 5d theories \cite{Iqbal:2003ix, Iqbal:2003zz, Eguchi:2003sj, Hollowood:2003cv} by the topological vertex formalism \cite{Iqbal:2002we, Aganagic:2003db}. In terms of 5d gauge theories such cases correspond to gauge theories with SU-type gauge groups\footnote{It is also possible to realize certain non-Lagrangian theories such as the $E_0$ theory from 5-brane webs dual to toric Calabi-Yau threefolds.}.

On the other hand, a large class of 5d gauge theories with a superconformal fixed point has been classified from a field theoretic analysis \cite{Seiberg:1996bd, Intriligator:1997pq, Jefferson:2017ahm} and also from M-theory compactified on Calabi-Yau threefolds \cite{Morrison:1996xf, Douglas:1996xp, DelZotto:2017pti, Xie:2017pfl, Jefferson:2018irk, Bhardwaj:2018yhy, Bhardwaj:2018vuu, Apruzzi:2018nre, Apruzzi:2019vpe, Apruzzi:2019opn, Bhardwaj:2019fzv, Apruzzi:2019kgb, Apruzzi:2019enx, Bhardwaj:2019xeg, Bhardwaj:2020gyu, Bhardwaj:2020kim}. Then natural questions are twofolds. One is if we can realize other gauge theories from 5-brane webs. The other is if we can compute the Nekrasov partitioin functions of the 5d theories realized on the 5-brane webs using the topological vertex formalism. It is indeed possible to realize different gauge groups\footnote{In this paper we do not distinguish the global structure of gauge groups.} such as $\SO(N)$ and $\Sp(N)$ by introducing orientifolds in 5-brane web diagrams \cite{Brunner:1997gk, Bergman:2015dpa, Zafrir:2015ftn, Hayashi:2015vhy}. We can further realize an exceptional gauge group $G_2$ from 5-brane webs with an O5-plane \cite{Hayashi:2018bkd}. Recently 5-brane webs for $\SU(6)$ gauge theories with hypermultiplets in the rank-3 antisymmetric representation have been constructed in \cite{Ohmori:2018ona, Hayashi:2019yxj}. Interestingly it is also possible to compute the Nekrasov partition functions of most of the theories realized on the 5-brane webs by extending the topological vertex formalism. In \cite{Hayashi:2013qwa, Hayashi:2014wfa, Kim:2015jba, Hayashi:2015xla}, the formalism has been extended for 5-brane webs with configurations where some 5-branes jump over other 5-branes. Such configurations may appear, for example, in 5-brane webs with a resolved O7$^-$-plane. The topological vertex formalism for 5-brane webs with O5-planes has also been  developed in \cite{Kim:2017jqn, Hayashi:2020hhb,Kim:2021cua}.

A new type of web diagrams was introduced in \cite{Hayashi:2017jze}.  Each of the web diagram is given by gluing three or four toric diagrams. The gluing may be interpreted as gauging the diagonal part of flavor symmetries of three or four matter theories. We call the former type of the gluing trivalent gluing or trivalent gauging while we call the latter type of the gluing quadrivalent gluing or quadrivalent gauging. Using web diagrams with a trivalent gluing it is possible to construct the 5d pure $E_6, E_7, E_8$ gauge theories. Furthermore a prescription of applying the topological vertex to web diagrams with a trivalent gluing or quadrivalent gluing has also been developed in \cite{Hayashi:2017jze} and the Nekrasov partition functions of the 5d pure $E_6, E_7, E_8$ gauge theories have been computed.

Therefore including the webs with a trivalent gluing it was possible to realize exceptional gauge groups from web diagrams but one exception in the classification of simple Lie algebras is $F_4$. In this paper we fill in the last piece and provide web diagrams using the trivalent or quadrivalent gluing for all the exceptional gauge groups $G_2, F_4, E_6, E_7, E_8$. The web diagrams considered in \cite{Hayashi:2017jze} are made from a trivalent or quadrivalent gluing of toric diagrams and hence only simple-laced Lie algebras were realized. We extend the construction by considering trivalent or quadrivalent gluing of non-toric diagrams, which enables us to realize non-simply-laced Lie algebras. More conretely the web diagrams we construct include the ones for the 5d $G_2$ gauge theory with $4$ flavors, the 5d $F_4$ gauge theory with $3$ flavors, the 5d $E_6$ gauge theory with $4$ flavors and the 5d $E_7$ gauge theory with $3$ flavors (with mass parameters only for 2 flavors). Since the web diagrams are given by a trivalent or quadrivalent gluing it is possible to compute the Nekrasov partition functions of those theories and we give explicit expressions of the partition functions. 

Let us summarize our strategy to obtain the web diagrams. In order to obtain the web diagrams we make use of 5d gauge theory descriptions of six-dimensional (6d) conformal matter theories on a circle. A 6d conformal matter theory is realized on M5-branes probing an ADE singularity \cite{DelZotto:2014hpa}. We will call a theory realized on $k$ M5-branes probing a $G$-type singularity  $(G, G)_k$, which has the flavor symmetry $G \times G$. 
Through a chain of string dualities, 
the theory of type $(G, G)_k$ on a circle has a 5d gauge theory description given by a quiver theory whose quiver diagram is given by the affine $G$ Dynkin diagram. We call such a quiver theory as an affine $G$ Dynkin quiver theory \cite{DelZotto:2014hpa}. 
It is possible to realize these theories from web diagrams with a trivalent gluing for $G=E_6, E_7, E_8$ or with a quadrivalent gluing for $G=D_4$. The web diagrams show a part of the global symmetries explicitly from parallel external lines. Then one can perform Higgsings of these theories, which yields web diagrams made from the trivalent or quadrivalent gluing of non-toric diagrams. The low energy theories after the Higgsings may be determined by utilizing the analysis in \cite{Heckman:2016ssk}. One difference from the cases in \cite{Heckman:2016ssk} is that we consider Higgsings of 6d theories on a circle and hence the Higgsings may give rise to a 5d theory that is obtained from a circle compactification of a 6d theory with a twist. Note that we always keep the radius of the circle finite unless we take a 5d limit, and the Kaluza-Klein (KK) modes are not decoupled in the resulting Higgsed theories. The low energy theories which can be realized from the Higgsings of the web diagrams include circle compactifcations of the 6d $G_2$ gauge theory with $4$ flavors from $G=D_4, k=2$, the 6d $F_4$ gauge theory with $3$ flavors and the 6d $E_6$ gauge theory with $4$ flavors from $G=E_6, k=2$, the 6d $E_7$ gauge theory with $3$ flavors from $G=E_7, k=2$, and each theory possesses a tensor multiplet in 6d. Hence we can construct the web diagrams for the theories and their 5d limits yield the web diagrams for the 5d gauge theories with the exceptional gauge groups. Since the web diagrams are constructed by a trivalent or quadrivalent gluing, it is possible to compute the partition functions of the 6d theories on $T^2 \times \mathbb{R}^4$. 

Along the way of obtaining the web diagrams for the exceptional gauge theories, we also obtain web diagrams of various theories which arise from the Higgsings of $(G, G)_2$ on a circle. It is also possible to gauge one of the flavor symmetries of $(G, G)_k$ and we call the theory $(G, \underline{G})_k$ where $\underline{G}$ stands for the gauged symmetry group. We can similarly construct web diagrams which arise from Higgsings of $(G, \underline{G})_k$ on  a circle. For example a Higgsing of $(E_7, \underline{E_7})_1$ on a cirlce yields a circle compactification of the 6d $E_7$ gauge theory with a half-hypermultiplet in the fundamental representation in addition to a fundamental hypermultiplet and a tensor multiplet. Other examples of such theories are 5d $\SU(3)$ gauge theory with the Chern-Simons (CS) level $9$ from $G=D_4, k=1$ and the 5d $\SU(4)$ gauge theory with the CS level $8$ from $G = E_6, k=1$, each of which may be realized from a circle compactification of a 6d theory with a twist \cite{Jefferson:2018irk,Razamat:2018gro}. A 5-brane web diagram of the former theory has been constructed in \cite{Hayashi:2018lyv}. The partition functions of the theories on $S^1 \times \mathbb{R}^4$ has been recently computed in \cite{Hayashi:2020hhb, Kim:2021cua, Kim:2020hhh}. We give a quadrivalent or trivalent gluing realization by web diagrams of the theories and compute the Nekrasov partition functions of the theories using the web diagrams.  

The organization of this paper is as follows. In section \ref{sec:CMT}, we consider 6d theories which are described 
by $(G, G)_2$ with $G=D_4, E_6, E_7$. We then perfom Higgsings, which can be seen from the corresponding web diagrams, and identify the low energy theories for each Higgsing. In section \ref{sec:PF}, we compute the Nekrasov partition functions of gauge theories with exceptional gauge groups and also the Nekrasov partition functions of the marginal pure $\SU(3), \SU(4)$ gauge theories by utilizing the construction of the web diagrams in section \ref{sec:CMT}. We summarize our results and discuss future directions in section \ref{sec:concl}. For completeness we describe the relation between the Higgsings of web diagrams for the 
$(E_8, \underline{E_8})_1$ and the low energy theories in appendix \ref{sec:e8}. Appendix \ref{sec:top} summarizes the topological vertex formalism which we use in section \ref{sec:PF}.

\bigskip

\section{Brane webs and Higgs branches of conformal matter theories on a circle}
\label{sec:CMT}

In this section we study certain Higgsings of 6d theories on a circle that are given by gauging two minimal conformal matter theories\footnote{Some Higgsings may give rise to free hypermultiplets in the Higgsed theories but we will not mention their existence explicitly. }. More specifically the parent 6d theories we consider are 
$(D_4, D_4)_2$, $(E_6, E_6)_2$ and $(E_7, E_7)_2$. Circle compactifications of the 6d theories lead to 5d affine Dynkin quiver theories which can be realized by the trivalent or quadrivalent gluings of 5-brane web diagrams. Since we are interested in using the web diagrams to compute their partition functions, we will focus on Higgsings that can be manifestly realized from the diagrams. It turns out that such limited Higgsings yield web diagrams for 6d theories with an exceptional gauge group $G_2, F_4, E_6, E_7$ on a circle. In fact, we can also realize the web diagrams for the 5d pure $\SU(3)$ gauge theory with the CS level $9$ and the 5d pure $\SU(4)$ gauge theory with the CS level $8$. 

\subsection{Twisted compactifications of 6d theories}
\label{sec:twist}
The main focus of this paper is certain Higgsings of 6d theories, $(D_4, D_4)_2$, $(E_6, E_6)_2$ and $(E_7, E_7)_2$ on a circle. 
In fact it will turn out that the Higgsed theories may be divided into two classes. Some of the Higgsed theories are simple circle compactifications of 6d theories while the other Higgsed theories arise from circle compactificationsof 6d theories with twists. 
Here we first collect basic properties of circle compactifications of 6d theories with or withtout twists and set up the notation which we will use in the later sections. 

A circle compactification with or without a twist of a 6d $\mathcal{N}=(1,0)$ superconformal field theory (SCFT) may admit a 5d description. In the 5d description the KK modes arise as non-perturbative effects and the information of the KK modes are still contained in the 5d theory. Such 5d theories are sometimes dubbed as 5d KK theories. Let us first focus on 5d KK theories which are obtained from 6d theories on a circle. 5d $\mathcal{N}=1$ KK theories can be obtained from M-theory compactified on certain Calabi-Yau threefolds each of which is given by an elliptic fibration over a base $B$ \cite{Vafa:1996xn, Morrison:1996na, Morrison:1996pp, Bershadsky:1996nh}. The base $B$ is non-compact as we consider 5d KK theories with gravity decoupled. Then relevant physics is mainly characterized by a compact complex surface $S$ inside each Calabi-Yau threefold. The complex surface $S$ is given by a collection of Hirzebruch surfaces $S_{\alpha}$ with some points blown up \cite{Jefferson:2018irk, Bhardwaj:2019fzv}. The number of the Hirzebruch surfaces gives the dimension of the Coulomb branch moduli space of the corresponding 5d KK theory. 

In fact the compact complex surface $S$ may be also seen as a collection of $\mathbb{P}^1$'s on each of which we have an elliptic fiber. The $\mathbb{P}^1$'s are inside the base $B$ and the elliptic fiber may be degenerated on some of the base curves. 
Each component in the degenerated elliptic fiber on a $\mathbb{P}^1$ base $C_i$ corresponds to a collection of $\mathbb{P}^1$ fibers $f_{a, i}$ in Hirzebruch surfaces $S_{a, i}$ in $S$. 
The intersections between the $\mathbb{P}^1$ fibers $f_{a, i}$ with a fixed $i$ form a Dynkin diagram of an affine Lie algebra in the sense that the negative of the intersection numbers given by
\be\label{fScartan}
C_{ab}  =  - f_{a, i}\cdot S_{b, i}
\ee
becomes the Cartan matrix for the affine Lie algebra. We will use this viewpoint for specifying the geometry as well as the corresponding 5d KK theory. More specifically we will use the following notation, which obeys the one in \cite{Bhardwaj:2019fzv}. Let $\Omega^{ij}$ be the negative of the intersection number between the $C_i$ and the $C_j$ in $B$. Also suppose the elliptic fiber on the $C_i$ degenerates to $\mathbb{P}^1$ fibers whose intersection forms a Dynkin diagram of an affine Lie algebra $\mathfrak{g}_i^{(1)}$ with the Cartan matrix given by \eqref{fScartan}. Although $\Omega^{ij}$ with $j \neq i$ can be $0, -1, -2$, we will only encounter examples with $\Omega^{ij} = -1$. When $\Omega^{ij} = -1$ we write
\be\label{twonodes}
\os{\Omega^{ii}}^{\mathfrak{g}_i^{(1)}}-\os{\Omega^{jj}}^{\mathfrak{g}_j^{(1)}}
\ee
When the associated Lie algebra is trivial, we write $\mathfrak{g}_i^{(1)} = \fsp(0)^{(1)}$ when $\Omega^{ii}=1$ and $\mathfrak{g}_i^{(1)} = \fsu(1)^{(1)}$ when $\Omega^{ii} = 2$. In the case of $\fg_i = \fsp(n_i)$ without fundamental hypermultiplets, there can be two types depending on the 6d discrete theta angle. When the discrete theta angle is physically relevant we will write the discrete theta angle as the subscript of $\fsp(n_i)$. For $\fg_i = \fsp(0)$ the discrete theta is only physical when its adjacent node has an $\fsu(8)$ gauge algebra \cite{Mekareeya:2017jgc}.

The notation \eqref{twonodes} is also useful to read off the corresponding 6d uplift. In the 6d uplift, the number of the base $\mathbb{P}^1$'s is the number of tensor multiplets. $\mathfrak{g}_i$ gives the gauge algebra and it determines vector multiplets in the theory. Furthermore, the content of hypermultiplets is almost fixed by the anomaly cancellation once the intersection matrix $\Omega^{ij}$ is given\footnote{\label{ft:ambiguous}When we consider 6d theories with a tensor multiplet there are some cases where $\fg_i$ and $\Omega^{ii}$ are not enough to fix the matter content. For $\fg_i = \fsu(m) \; m\geq 3$ with $\Omega^{ii} = 1$ the matter content is either $(m+8)$ hypermultiplets in the fundamental representation and a hypermultiplet in the rank-2 antisymmetric representation or $(m-8)$ hypermultiplets in the fundamental representation and a hypermultiplet in the rank-2 symmetric representation. The latter case can be distinguished by denoting it by $\fsu(\hat{m})$.  For $\fg=\fsu(6)$ there is further a different case which has $15$ hypermutliplets in the fundamental representation and a half-hypermultiplet in the rank-3 antisymmetric representation, which can be differentiated by $\fsu(\tilde{6})$. Finally when $\fg = \fso(12)$ with $\Omega^{ii} = k\; (k=1, 2)$, we have either $(8-k)$ hypermultiplets in the vector representation and $(4-k)$ half-hypermultiplets in the spinor representation or $(8-k)$ hypermultiplets in the vector representation, $(3-k)$ half-hypermultiplets in the spinor representation and a half-hypermultiplet in the cospinor representation. The latter case may be distinguished by $\fso(\hat{12})$. }. There can be hypermultiplets in a representation $\mathcal{R}_i$ of $\mathfrak{g}_i$ and also hypermultiplets in a mixed representation $\mathcal{R}_{ij}$ of $\mathfrak{g}_i \oplus \mathfrak{g}_j$. When $\Omega^{ii} = 1, \mathfrak{g}_i = \fsp(n_i), \Omega^{jj} = k, \mathfrak{g}_j = \fso(7) \text{ or } \fso(8)$ and $\Omega^{ij} = -1$, there may be two possibilities for the mixed representation between $\fsp(n_i)$ and $ \fso(7) \text{ or } \fso(8)$. The two possibilities are $(\text{V}, \text{V})$ and $(\text{V}, \text{S})$. Here $\text{V}$ represents the vector representation and $\text{S}$ denotes the spinor representation. We will use \eqref{twonodes} for $(\text{V}, \text{V})$ and the case with $(\text{V}, \text{S})$ is written by\footnote{For $\fso(8)$ the physically relevant case is 
\be
\os{1}^{\mathfrak{sp}(n_i)}\hdashrule[0.5ex]{0.5cm}{1pt}{0.5mm}\os{k}^{\mathfrak{so}(8)}-\os{1}^{\mathfrak{sp}(n_j)},
\ee
due to the outer autmorphism of $\fso(8)$.
}
\be
\os{\Omega^{ii}}^{\mathfrak{g}_i^{(1)}}\hdashrule[0.5ex]{0.5cm}{1pt}{0.5mm}\os{\Omega^{jj}}^{\mathfrak{g}_j^{(1)}}
\ee
for the corresponding geometry and the 5d KK theory. 

Although the notation above will be enough to specify the theories in the examples we will consider later, we will also put flavor symmetry algebras in a square bracket below $\Omega^{ii}$. However there are some subtle cases. When we have $\fsu(n)\; (n \geq 3)$ algberas, then $\U(1)$ global symmetries associated to matter of the $\fsu(n)$'s may be broken by Adler-Bell-Jakiw (ABJ) anomalies. The remaining $\U(1)$ symmetry may not be localized on a base curve and we will not write Abelian symmetries associated to matter of $\fsu(n)$ algebras explicitly. In the case of one flavor we will write $N_f = 1$ below $\Omega^{ii}$. Another subtle case is 
\be\label{su2on2}
\us\os{2}^{\fsu(2)}_{\left[\fso(7)\right]}.
\ee
Although the naive flavor symmetry algebra expected  from the tensor branch effective field theory is $\fso(8)$, the geomtry indicates that only the $\fso(7)$ subalgebra is realized at the ultraviolet (UV) fixed point. This reduction of  the symmetry also nesseary for the consistensy with the dualities in lower dimensional theories \cite{Ohmori:2015pia}. When we have more than one factors of \eqref{su2on2}, determining the flavor symmetry could be complicated. Hence we will only write the number of flavors of $\fsu(2)$ in those cases. A prescription to determine flavor symmetries in the subtle cases has been presented in \cite{Apruzzi:2020eqi}.

We then consider geometries which give rise to 5d $\mathcal{N}=1$ KK theories coming from twisted compactifications of 6d theories. One type of twists arises from permutation of base $\mathbb{P}^1$ curves $C_i$. For a permutation $i \to \sigma(i)$, when the characterization of a 6d theory satisfies
\begin{align}
\mathfrak{g}_{\sigma(i)} &= \mathfrak{g}_i,\\
\Omega^{\sigma(i)\sigma(j)} &= \Omega^{ij},
\end{align}
and also $\mathcal{R}_{\sigma(i)} = \mathcal{R}_i, \mathcal{R}_{\sigma(i)\sigma(j)} = \mathcal{R}_{ij}$, we can consider a geometry that realizes a 5d KK theory from a circle compactification of the 6d theory with the twist. The number of base $\mathbb{P}^1$'s becomes the number of orbits under the permutation of $\sigma$. Let $I, J$ parameterize the orbits. Then the intersection matrix after the twist becomes
\be\label{twistedomega}
\Omega_{\sigma}^{IJ} = \sum_{j \in J}\Omega^{ij},
\ee
where $i$ denotes a node in the orbit $I$. Then the intersection form $\Omega^{IJ}_{\sigma}$ may become non-symmetric. The intersection form \eqref{twistedomega} may be interpreted as a Dirac paring between BPS strings and particles \cite{DelZotto:2020sop}. When $\Omega_{\sigma}^{IJ} \geq \Omega_{\sigma}^{JI} > 0$ we put the number $-\Omega_{\sigma}^{IJ}$ above the line between the node $I$ and the node $J$. In particular when $\Omega_{\sigma}^{IJ} > \Omega_{\sigma}^{JI}$, the line between the nodes is replaced with an arrow from $I$ to $J$. In the case when the diagonal entry changes by $\ell_I = \Omega^{ii} - \Omega_{\sigma}^{II}$ we can introduce $\ell_I$ edges for the node $I$ but we will not have this case in this paper. 

We can consider other type of twists which acts non-trivially on components of the elliptic fiber without exchanging base $\mathbb{P}^1$ curves. 
In order to see which twist is possible we here 
carefully analyze discrete components of a flavor symmetry acting on hypermultiplets in a gauge theory with eight real supercharges. We will divide the analysis by the type of representations of a gauge group. 
In this section we analyze only the classical subgroup; there can be a gauge-flavor mixed anomaly which reduces the flavor group.

\paragraph{Complex representation.}
We denote the scalars in the hypermultiplets by $\vec{Q} \in \mathcal{R}$ collectively. Here we assume the total representation $\mathcal{R}$ is the direct sum of $N_f$ copies of the same irreducible complex representation $R$ of the gauge group and $N_f$ copies of its complex conjugate: $\mathcal{R} = R^{\oplus N_f}\oplus \bar{R}^{\oplus N_f}$.
We write each component as $\vec Q = (Q_a^i,\tilde{Q}^{\bar{a}}_i)$ where $a$ and $\bar{a}$ are the gauge-representation indices and $i$ is the flavor index.
$\vec{Q}$ admits the antisymmetric gauge-invariant bilinear form:
\begin{equation}
    \langle \vec{Q},\vec{P}\rangle= \delta^a_{\bar{a}}(Q^i_a \tilde{P}^{\bar{a}}_i - P^i_a \tilde{Q}^{\bar{a}}_i). \label{eq:skewform}
\end{equation}
For the flavor symmetry to commute with the $R$-symmetry, it should preserve the bilinear form.
A unitary matrix $U\in \U(N_f)$ acting as $(Q^i,\tilde{Q}_i) \mapsto (U_{ij}Q^j,(U_{ij})^*\tilde{Q}_j)$ preserves \eqref{eq:skewform} and the kinetic term.\footnote{In 6d the $\U(1)$ part of $\U(N_f)$ can be explicitly broken by the ABJ-type anomaly. 
We do not take that effect into account here. Also, if the gauge group has a nontrivial center, the flavor group faithfully acting on the local operators is the quotient of the group $F$ in the main text by the center.}
However, the discrete component of the symmetry can come when we consider the charge conjugation on the gauge index: $a \leftrightarrow \bar{a}$. 
To preserve \eqref{eq:skewform}, the corresponding symmetry $M$ should act on the hypermultiplets as 
\begin{equation}
    M : (Q^i_a,\tilde{Q}_i^{\bar{a}}) \mapsto (- \tilde{Q}_i^{\bar{a}}, Q^i_a).
\end{equation}
Note that $M$ does not square to the identity, and rather $M^2 = - 1$,
which is a nontrivial element in $\U(N_f)$.
Formally, an element of the classical global symmetry group $F$ 
can be written as a pair $(U,M^a)$ with $a=0,1$, and the multiplication among them is
\begin{equation}
    (U_1,M^{a_1})\cdot (U_2,M^{a_2}) = ((-1)^{a_1a_2}U_1 (C^{a_1}\cdot U_2),M^{a_1+a_2 \, \text{mod 2}}),
\end{equation} 
where $C$ is the charge conjugation automorphism: $C\cdot U_2 = U_2^*$.
In other words, the global group $F$ is the extension of $\mathbb{Z}_2$ by $\U(N_f)$:
\begin{equation}
    1 \to \U(N_f) \to F \to \mathbb{Z}_2  \to 1.
\end{equation}
As $(U,1)$ and $(U,M)$ is not connected by a path in $F$, the $F$ has two disconnected components:
\begin{equation}
    \pi_0(F) = \mathbb{Z}_2.
\end{equation} 
When $N_f$ is even, we have an order 2 element $\tilde{M} = (\Omega_{N_f},M)$; $\tilde{M}^2 =1$. Here, the $\Omega_{N_f}$ is the symplectic identity matrix of size $N_f$:
\begin{equation}
    \Omega_{N_f} = \begin{pmatrix}
        0 & I_{\frac{N_f}{2}}\\
        -I_{\frac{N_f}{2}} & 0\\
    \end{pmatrix},
\end{equation}
where $I_N$ is the $N\times N$ identity matrix.
We can use $U$ and $\tilde{M}$ instead of $M$ to generate the whole group $F$, which means $F$ is isomorphic to the semidirect product $\U(N_f)\rtimes \mathbb{Z}_2$.
In other words when $N_f$ is even we can regard the matter content as $\frac{N_f}{2}$ half-hypermultiplets in $R$ and $\frac{N_f}{2}$ half-hypermultiplets in $\bar{R}$, and $\tilde{M}$ just swaps them.
When $N_f$ is odd we do not have such a splitting and $F$ is not isomorphic to the semidirect product.
In particular, when $N_f = 1$ the group $F$ is called the $\text{Pin}(2)^-$ group.

We 
are interested in circle compactifications twisted by an element $(T,M)$ disconnected from the identity. The global symmetry of the reduced theory is the commutant of the twist element in the global group $F$. If we conjugate $(T,M)$ by a general element $(U,M^a)$, we get 
\begin{equation}
    \begin{split}
    (U,M^a)\cdot(T,M)\cdot(U,M^a)^{-1} &= ((-1)^a U C^a \cdot T,M^{a+1 \text{ mod 2}} ) \cdot((-1)^a C^a\cdot U^\dag,M^a) \\
    &= (U (C^a \cdot T) U^\top ,M).
    \end{split}
\end{equation}
When $N_f$ is even and we take $T$ to be the symplectic identity $T = \Omega_{N_f}$, i.e.\ the twist element is $\tilde{M}$, the commutant is $\Sp\left(\frac{N_f}{2}\right)\times \mathbb{Z}_2$.
When $N_f$ is odd, we can take $T$ to be the following $N_f\times N_f$ matrix $\Omega'_{N_f}$: 
\begin{equation}
    \Omega_{N_f}' = \begin{pmatrix}
        0 & I_{\frac{N_f-1}{2}} & 0\\
        -I_{\frac{N_f-1}{2}} & 0 &0\\
        0&0&1
    \end{pmatrix}.
\end{equation}
The connecteced part of the remaining flavor group is $\Sp\left(\frac{N_f-1}{2}\right)$.
Another choice of the twist is $T = I_{N_f}$.\footnote{
    For a generic $T$, the connected part of the remaining flavor group is broken down to $\U(1)^{\left\lfloor\frac{N_f}{2}\right\rfloor}$. This breaking and also the enhancement to two different groups ($\Sp$ and $\SO$) can be easily be seen from the T-duality between the shift-orientifold and a pair of orientifolds with opposite charges when the system can be realized by branes in the string theory \cite{Keurentjes:2000bs}.
    When $N_f$ is odd, one of the branes is trapped at the orientifold in the T-dual frame cutting down the rank by a half.
} In this case, the remaining flavor symmetry is an $\mathrm{O}(N_f)$-extension of $\mathbb{Z}_2$.
When $N_f$ is even, this extension is trivial.
As the twist involves an outer automorphism of the gauge group, the twisted compactification also reduces the gauge group. Then a gauge algebra $\fg$ of a gauge group after the twisted compactification may be thought of as a twisted affine Lie algebra in the sense that the intersection matrix given by \eqref{fScartan} gives rise to the Cartan matrix of the twisted affine Lie algebra. As the twised Lie algebra comes from an order two outer automorphism of $\fg$ we will denote the twisted affine Lie algebra by $\fg^{(2)}$. 

We will encounter this type of twist in later sections and the examples we will see include
\be\label{sutwist}
\us\os{2}^{\fsu(n)}_{[\fsu(2n)]} \qquad \to \qquad \us\os{2}^{\fsu(n)^{(2)}}_{[\fsu(2n)^{(2)}]}, 
\ee
for $n \geq 3$ and\footnote{When $n=4$ the flavor symmetry algebra becomes $\fsu(2)^{(1)}$.}
\be\label{e6twist}
\us\os{n}^{\fe_6}_{[\fsu(6-n) \oplus \fu(1)]} \qquad \to \qquad \us\os{n}^{\fe_6^{(2)}}_{\left[\fsu\left(6-n\right)^{(2)}\right]}.
\ee
Note that the twist is possible even when $n$ is odd in \eqref{e6twist} \cite{Bhardwaj:2020kim}. The subalgebra without the affine node of $\fsu(n)^{(2)}$ is $\fsp\left(\left\lfloor\frac{n}{2}\right\rfloor\right)$, which is the algebra of the continuous part of the flavor symmetry obtained above.

\paragraph{Pseudo-real representation.}
Next, consider the case where the hypermultiplets are in a pseudo-real representation. To be precise, we assumes that the scalars $\vec{Q} = (Q_a^i)$ are in the direct sum $\mathcal{R}$ of $2N_f$ copies of an irreducible pseudo-real representation $R$: $\mathcal{R} = R^{\oplus 2N_f}$. Again, $a$ is the gauge index and $i$ is the flavor index.
Here, by convention, $N_f$ takes values in half-integers.
The 
supersymmetry with eight supercharges requires an antisymmetric gauge-invariant bilinear form. The pseudo-real representation $R$ is equipped with an invariant antisymmetric form $J^{ab}$, so we can just use it:
\begin{equation}
    \langle \vec{Q},\vec{P} \rangle = \delta_{ij}J^{ab}Q^i_a P^j_b.
\end{equation}
When $2N_f$ is even, the classical flavor group $F$ preserving this pairing and the kinetic term is $F = \mathrm{O}(2N_f,\mathbb{C})\cap \U(2N_f) = \mathrm{O}(2N_f)$. 
In particular, it has disconnected part:
\begin{equation}
    \pi_0(F) = \pi_0(\mathrm{O}(2N_f)) = \mathbb{Z}_2.
\end{equation}
When $2N_f$ is odd, then a $\mathbb{Z}_2$ element of determinant $-1$ in $\mathrm{O}(2N_f)$ is gauged and the flavor symmetry becomes $\SO(2N_f)$.

When $2N_f$ is even, if this disconnected component survives under quantum (ABJ-type) anomalies, we can do a twisted compactification using an element $T$ disconnected from the identity in $\mathrm{O}(2N_f)$. 
A choice with a maximal remaining flavor group is $ T = \mathrm{diag}(1,1,\cdots,1,-1)$, and the connected component of the remaining group is $\SO(2N_f-1)$.
This twist does not involve an outer automorphism of the gauge group, so the gauge group can remain the same after the compactificatioin. When $2N_f$ is odd, 
$\SO(2N_f)$ is connected and we do not have a non-trivial element disconnected from the identity. 

\paragraph{Strictly real representation}
Lastly, we consider the hypermultiplets in $N_f$ copies of a strictly real irreducible representation $R$ of a gauge group. 
The scalars in the multiplets value in the $2N_f$ copies of the complexification of R, and the paring is
\begin{equation}
    \langle \vec{Q},\vec{P}\rangle = \Omega_{ij} S^{ab} Q_a^i P_b^j,
\end{equation}
where $S^{ab}$ is the invariant symmetric bilinear form of $R$.
The flavor group preserving this pairing and the kinetic term is Sp$(N_f)$ and in particular it is connected.
Hence there is no non-trivial element $T$ which is disconnected from the identity. 

So far we have focused on gauge theories with hypermultiplets in a single representation. When we consider gauge theories with hypermultiplets in multiple repersentations 
the flavor symmetry group consists of multiple factors. Then it is possible to consider a twist which exchanges the same factors of the flavor symmetry. Let us consider cases where the exchange is induced from a non-trivial outer automorphism of a gauge algebra $\fg$. An outer autormorphism of a Lie algebra $\mathfrak{g}$ may be understood from an autormorphism of a Dynkin diagram of $\mathfrak{g}$. The relevant gauge algebras in 6d gauge theories are $\fg = \fso(8)$ and $\fso(12)$ since the spinor representation is not allowed when $\fg = \fso(2m) \; m \geq 7$ due to the anomaly cancellation condition. Both Lie algebras have an order two outer automorphism whereas $\fso(8)$ admits an order three outer automorhpism. 6d $\SO(8)$ and $\SO(12)$ gauge theories may admit two spinor repersentations with different chirality which are not related by complex conjugation. We will call one of them as the spinor representation and the other as the cospinor representation. 
The order two outer automorphism exchanges the spinor representation with the cospinor representation while the order three outer automorphism induces an exchange among the vector, spinor and cospinor representation of $\fso(8)$. Therefore the twist by the order two automorphism is possible when the number of (half-)hypermultiplets\footnote{The spinor and the cospinor representations are pseudo-real for $\fg = \fso(12)$. In this case half-hypermultiplets are allowed. On the other hand, the spinor and the cospinor representations of $\SO(8)$ are real and thus half-hypermultiplets are not allowed. } in the spinor represenation is equal to the number of (half-)hypermultiplets in the cospinor representation. The twist by the order three outer automorphism of $\fso(8)$ is possible when the numbers of hypermultiplets in the vector, spinor and cospinor representations are all the same. After the twisted compactifications, the gauge algebras become twisted affine Lie algebras. The twists by the order two outer automorphism of $\fso(8)$ and $\fso(12)$ yield $\fso(8)^{(2)}$ and $\fso(12)^{(2)}$ respectively and the twist by the order three outer automorphism $\fso(8)$ gives $\fso(8)^{(3)}$. 
The examples we will see later include 
\be\label{so8twist0}
\us\os{n}^{\fso(8)}_{\left[\fsp(4-n)\oplus \fsp(4-n) \oplus \fsp(4-n)\right]} \qquad \to \qquad \us\os{n}^{\fso(8)^{(2)}}_{\left[\fsp(4-n)^{(1)} \oplus \fsp(4-n)^{(1)}\right]}, \qquad (1 \leq n \leq 4), 
\ee
for the twist which involves the order two outer automorphism and 
\be\label{so8twist}
\us\os{n}^{\fso(8)}_{\left[\fsp(4-n)\oplus \fsp(4-n) \oplus \fsp(4-n)\right]} \qquad \to \qquad \us\os{n}^{\fso(8)^{(3)}}_{\left[\fsp(4-n)^{(1)}\right]}, \qquad (1 \leq n \leq 4), 
\ee
for the twist which involves the order three outer automorphism. 

For a 6d gauge theory with a single gauge group, one may consider a combination of a twist which exchanges different representations and a twist which acts on a single representation. Since the latter twist for a strictly real representation is trivial a non-trivial case may happen for a pseudo-real representation. Then the only possibility of the gauge algebra is $\fso(12)$. In order to have the combintation of the twists, the 6d theory needs to have even number of half-hypermultiplets in the spinor representation and the same number of half-hypermultiplets in the cospinor representation. However there is no 6d $\SO(12)$ gauge theory with such matter content which has a SCFT fixed point and we do not have an example of with the combination of the twists\footnote{There may be a 6d little string theory with $\fso(12)$ gauge algebra and eight hypermultiplets in the vector representation, two half-hypermultiplets in the spinor representation and two half-hypermultiplets in the cospinor representation \cite{Bhardwaj:2015oru}. In this case it may be possible to consider a combination of a twist which exchanges the spinor representation and the cospinor representation and a twist which makes use of an element disconnected from identify in $\mathrm{O}(2) \times \mathrm{O}(2)$ of the flavor symmetry.}.

In general, it is also possible to consider a twist which acts non-trivially on components of the elliptic fiber and the base $\mathbb{P}^1$ curves of a 6d theory has the corresponding discrete symmetry.  

After a twisted compactification a 5d KK theory is characterized with gauge algebras $\fg_I^{(q_I)}$ and $\Omega^{IJ}_{\sigma}$ of \eqref{twistedomega}. The corresponding Calabi-Yau threefolds in M-theory compactifications become genus one fibered Calabi-Yau threefolds \cite{Bhardwaj:2019fzv}. The gauge algebra $\fg_I^{(q_I)}$ represents how the torus fiber is degenerated and components $f_{A, I}$ of the torus fiber form a Dynkin diagram of $\fg_I^{(q_I)}$ where the Cartan matrix is given by \eqref{fScartan}. Namely each $f_{A, I}$ corresponds to a node in the Dynkin diagram of $\fg_I^{(q_I)}$. Then for each base curve it is possible to define a fiber 
\be\label{fiberI}
f_I = \sum_{A}d_Af_{A, I},
\ee
where $d_I$ is a mark associated to each node of the Dynkin diagaram of $\fg_I^{(q_I)}$. The gluing rule of the fiber \eqref{fiberI} on base curves next to each other is found to be \cite{Bhardwaj:2019fzv}
\be\label{gluingrule}
q_I\left(-\Omega_{\sigma}^{JI}\right)f_I \sim q_J\left(-\Omega_{\sigma}^{IJ}\right)f_J.
\ee

\subsection{$(D_4, D_4)$ conformal matter}
\label{sec:d4d4}
We start from Higgsings of the theory $(D_4, D_4)_2$ on a circle. 
The Higgsings have been studied in \cite{Kim:2019dqn} by using a 5-brane web with two O5-planes. Here we will redo the analysis by realizing the theories using the quadrivalent gluing of four 5-brane web diagrams. 

The basic building block is the minimal $(D_4, D_4)$ conformal matter. The minimal $(D_4, D_4)$ conformal matter theory is nothing but the E-string theory. The E-string theory compactified on a circle may be realized by the $\SU(2)$ gauging of four 5d $T_2$ theories. The $T_2$ theory is formally thought of as the $[\SU(2)] - \SU(1)$ theory where $[G]$ implies a flavor node with the flavor symmetry group $G$. Then gauging the diagonal part of the flavor symmetry groups of four copies of the $T_2$ theory yields an affine $D_4$ Dynkin quiver theory,
\be
\SU(1)  -{\underset{\text{\normalsize$\SU(1)$}}{\underset{\textstyle\vert}{\overset{\overset{\text{\normalsize$\SU(1)$}}{\textstyle\vert}}{\SU(2)}}}}- \SU(1).\label{min.affineD4}
\ee
Each $\SU(1)$ node may be interpreted as two flavors attached to the middle $\SU(2)$ node \cite{Bergman:2014kza, Hayashi:2014hfa}. Hence \eqref{min.affineD4} is equivalent to the $\SU(2)$ gauge theory with eight flavors, which has a 
UV completion as the E-string theory. The discrete theta angle of the $\SU(2)$ node is unphysical since it can be absorbed by the sign of the mass parameter of a hypermultiplet in the fundamental representation. 

\begin{figure}[t]
\centering
\includegraphics[width=6cm]{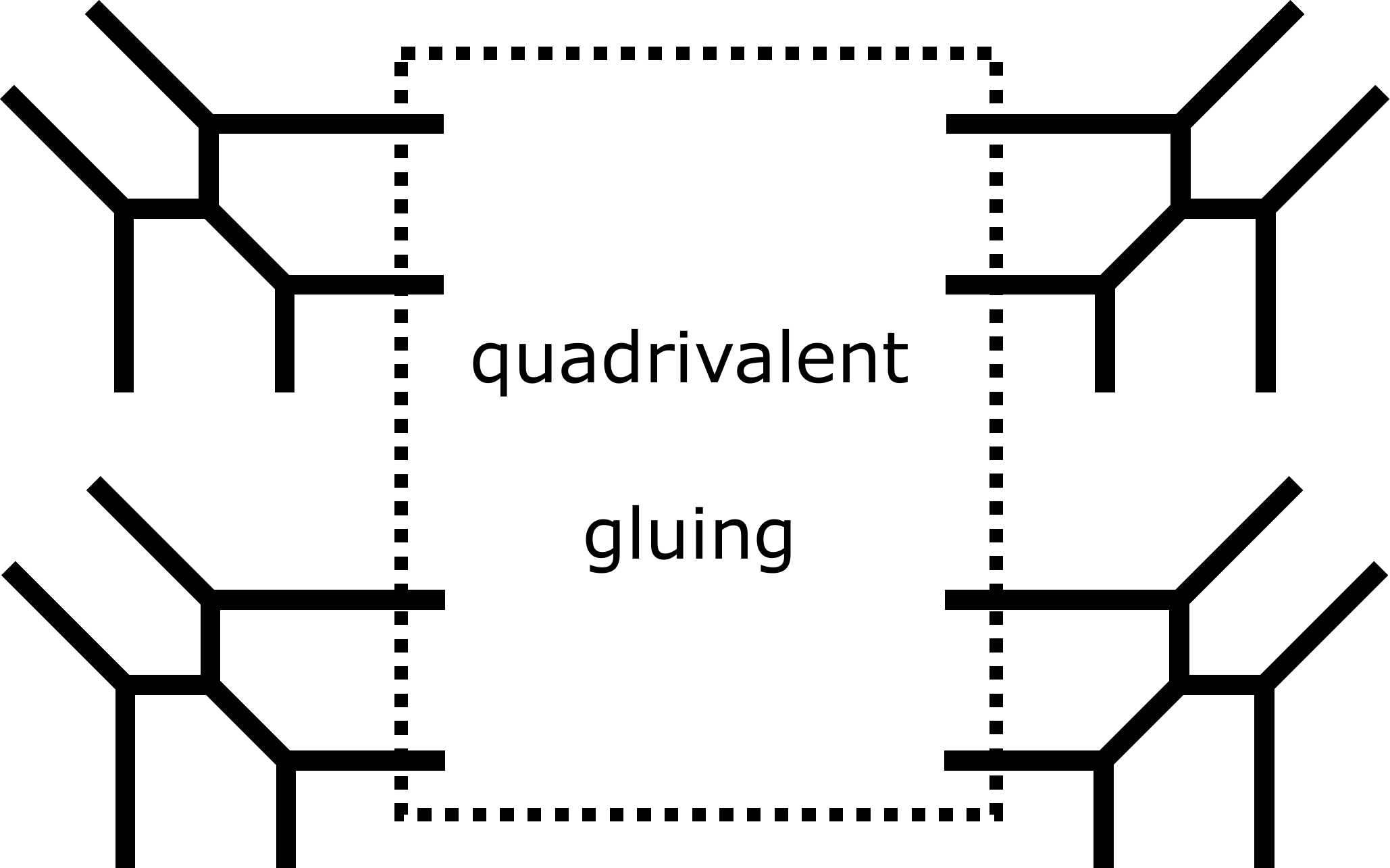}
\caption{
The web diagram for the minimal $(D_4, D_4)$ conformal matter theory on a circle. 
}
\label{fig:d4d4}
\end{figure}
The 5d affine $D_4$ Dynkin quiver theory \eqref{min.affineD4} is realized by the quadrivalent gluing of four 5-brane web diagrams each of which gives the $T_2$ theory. The web diagram is depicted in Figure \ref{fig:d4d4}. A $(p, q)$ 7-brane may be put on the end of an external $(p, q)$ 5-brane and the symmetry on the 7-branes implies a flavor symmetry of the web diagram \cite{DeWolfe:1999hj}. For example when we focus on the upper left diagram in Figure \ref{fig:d4d4}, we can put a $(1, -1)$ 7-brane on the end of each external $(1, -1)$ 5-brane. The two $(1, -1)$ 7-brane may be put on top of each other after a deformation, and hence it implies an $\SU(2)$ flavor symmetry. In total the manifest flavor symmetry from the diagram in Figure \ref{fig:d4d4} is $\SU(2)^8$ which is a subgroup of $\SO(8) \times \SO(8)$. More specifically the $\SU(2)^4$ flavor symmetry on the upper side of the diagram is a subgroup of one $\SO(8)$ and the other $\SU(2)^4$ flavor symmetry on the lower side of the diagram is a subgroup of the other $\SO(8)$. Therefore the eight external legs altogether on the upper side will be associated to one $\SO(8)$  and the eight external legs on the lower side will be associated to the other $\SO(8)$. 

On the other hand, 
the E-string theory can be also realized by an F-theory compactification on a non-compact elliptically fibered Calabi-Yau threefold. In terms of the notation introduced in section \ref{sec:twist} the geometry of the E-string is represented by
\be
\us\os1^{\fsp(0)^{(1)}}_{\left[\fe_8^{(1)}\right]}, \label{Estring}
\ee
where 
the algebra in the bracket below the self-intersection number represents the flavor algebra. 
Namely the diagram in Figure \ref{fig:d4d4} is a web diagram realization of the theory \eqref{Estring}. 

\begin{figure}[t]
\centering
\subfigure[]{\label{fig:d4d4d4}
\includegraphics[width=6cm]{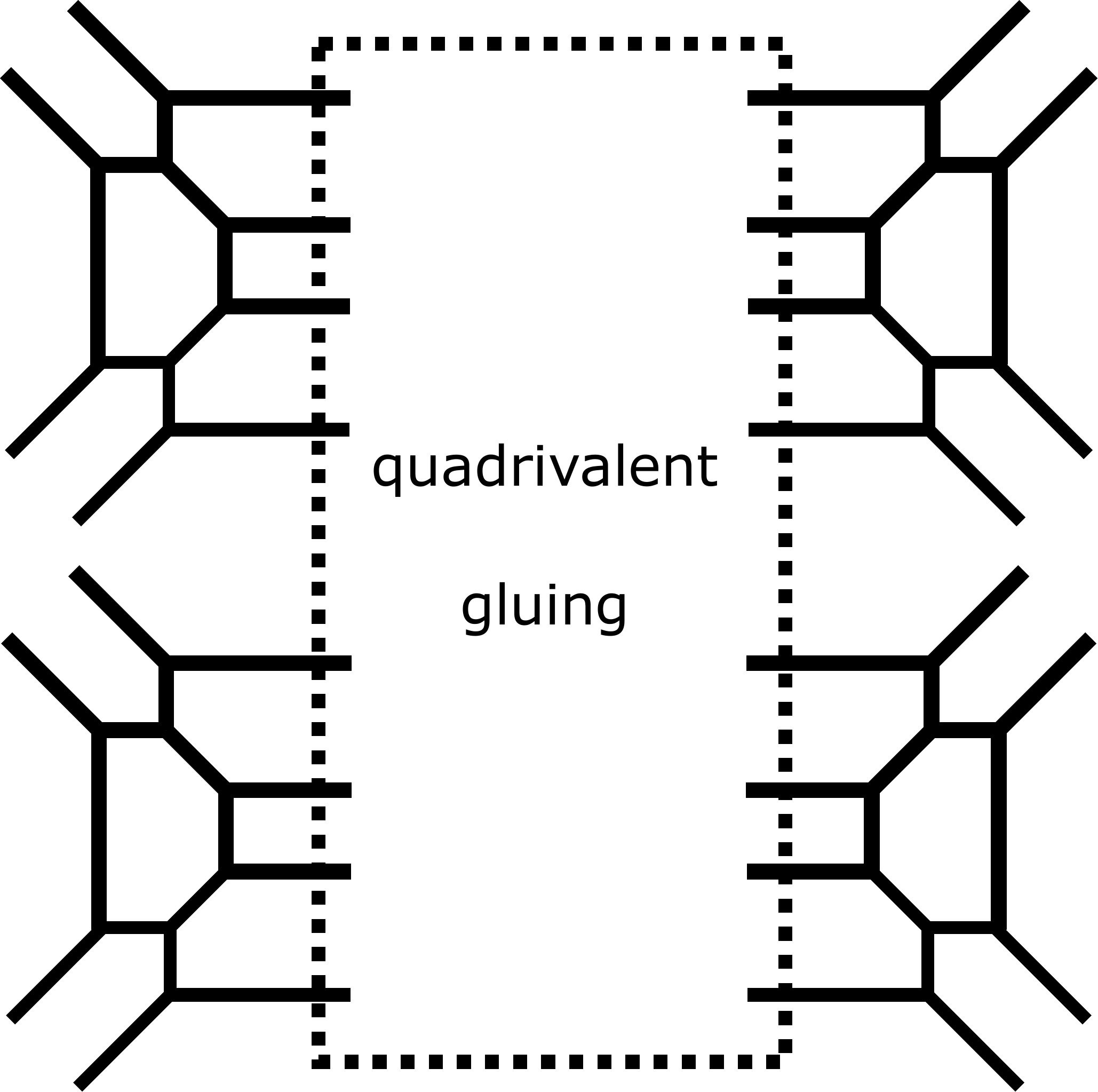}}
\hspace{0.5cm}
\subfigure[]{\label{fig:gd4d4}
\includegraphics[width=6cm]{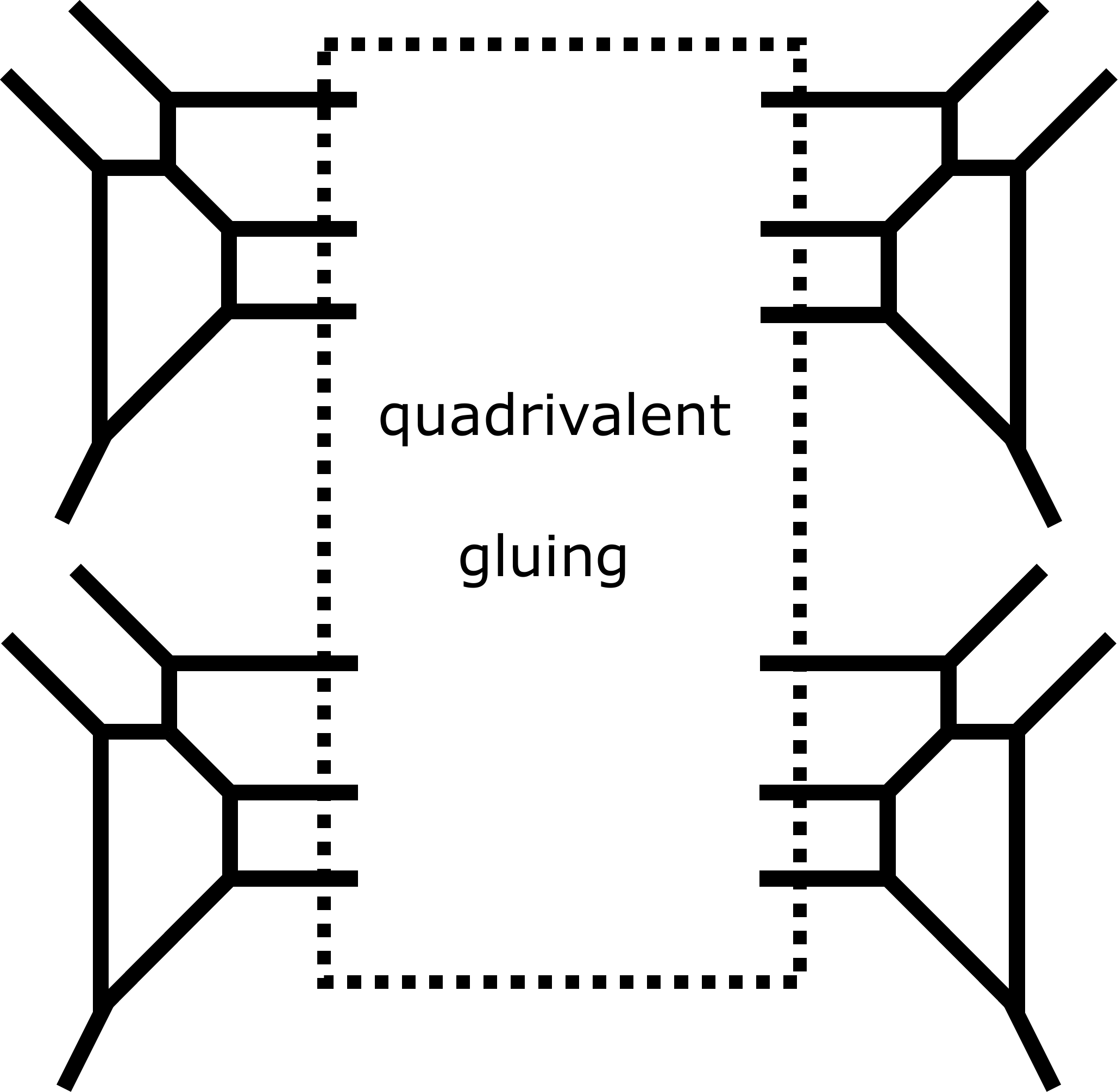}}
\caption{(a): The web diagram for the theory in \eqref{d4d4d4}, which gives a circle compactification of $(D_4, D_4)_2$. (b): The web diagram for the theory in \eqref{gd4d4}, which gives a circle compactification of $(D_4, \underline{D_4})_1$. }
\label{fig:gd4d4d4}
\end{figure}

Let us then consider combining two minimal $(D_4, D_4)$ conformal matter theories to obtain $(D_4, D_4)_2$. We take $\SO(8)$ subgroup of each $E_8$ flavor symmetry and gauge the diagonal subgroup of $\SO(8) \times \SO(8)$. The resutling theory then becomes $(D_4, D_4)_2$ and $(D_4, D_4)_2$ on a circle is described by
\be\label{d4d4d4}
\us\os1^{\fsp(0)^{(1)}}_{\left[\fso(8)^{(1)}\right]}-\os4^{\fso(8)^{(1)}}-\us\os1^{\fsp(0)^{(1)}}_{\left[\fso(8)^{(1)}\right]}.
\ee
The theory \eqref{d4d4d4} has a 5d gauge theory description and it is given by the following affine $D_4$ Dynkin quiver theory,
\be
\SU(2)_0  -{\underset{\text{\normalsize$\SU(2)_0$}}{\underset{\textstyle\vert}{\overset{\overset{\text{\normalsize$\SU(2)_0$}}{\textstyle\vert}}{\SU(4)_0}}}}- \SU(2)_0. \label{affineD4}
\ee
The subscript of $\SU(4)$ stands for the CS level and the subscript of $\SU(2)$ represents the discrete theta angle. Hereafter also a subscript of an $\SU(n)$ gauge node represents the CS level and a subscript of an $\Sp(n)$ gauge node stands for the discrete theta angle for simplicity. 
The $\SO(8)$ gauging may be also realized from the viewpoint of the web diagram by combining two copies of the diagram in Figure \ref{fig:d4d4}. Remember that the eight external legs on the lower side of the diagram are associated to one of the $\SO(8)$. Hence the $\SO(8)$ gauging is realized by connecting the eight external lines to another copy of the diagram. The diagram after the gauging is given in Figure \ref{fig:d4d4d4}. 
The quadrivalent gluing part is fixed by requiring that the gauging locally gives the $\SU(4)$ gauge group with the zero CS level. 

We are interested in Higgsings of the theory \eqref{d4d4d4} or \eqref{affineD4}. Since we will eventually use the web diagram for computing the partition functions of some theories, we focus on Higgsings which can be explicitly seen from the diagram. Since the $\SU(2)^8$ flavor symmetry is manifestly seen from the diagram in Figure \ref{fig:d4d4d4}, we consider Higgsings which break a part of or all the $\SU(2)^8$. Note that such a Higgsing is realized non-perturbatively as the Higgs branches appearing when instanton particles become massless in terms of the affine $D_4$ Dynkin quiver theory \eqref{affineD4}. In the following we will label the Higgsing by $[(a_1, a_2, a_3, a_4), (b_1, b_2, b_3, b_4)]$ where $a_i, b_i \; (i=1, 2, 3, 4)$ are either $0$ or $2$, and $0$ means no Higgsing and $2$ means the Higgsing which breaks one $\SU(2)$. The two brackets implies the two $\SU(2)^4$ each of which is a subgroup of $\SO(8)$.

\begin{figure}[t]
\centering
\subfigure[]{\label{fig:su2higgs}
\includegraphics[width=6cm]{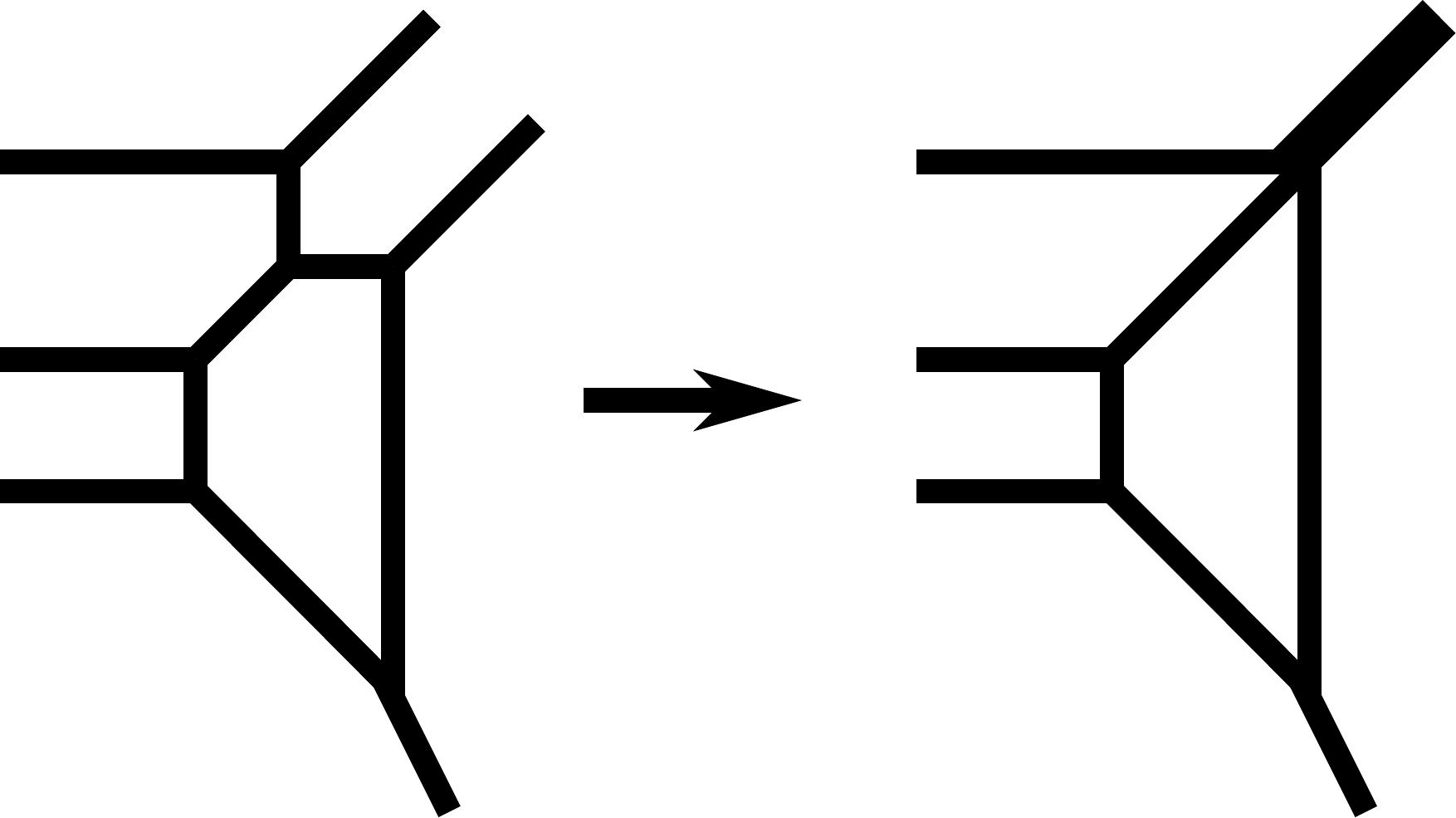}}
\hspace{0.5cm}
\subfigure[]{\label{fig:gd4d4su2higgs1}
\includegraphics[width=4cm]{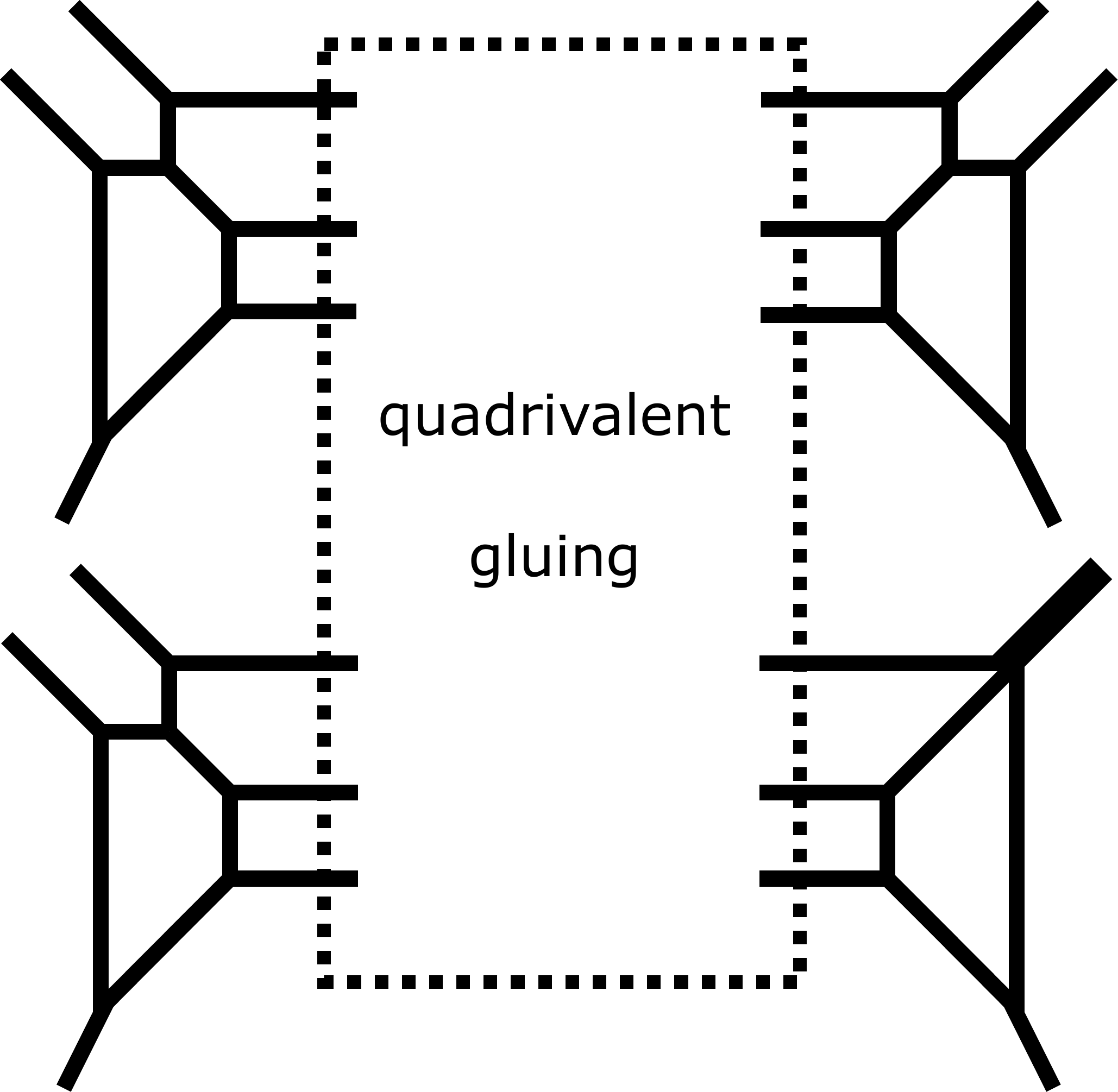}}
\caption{(a): The Higgsing which breaks an $\SU(2)$ for one of the four diagrams which are glued together in Figure \ref{fig:gd4d4}. (b): The web diagram of the theory which is realized after the Higgsing of the type $(2, 0, 0, 0)$. }
\label{fig:su2higgsex}
\end{figure}
Before studying the Higgsings labeled by $[(a_1, a_2, a_3, a_4), (b_1, b_2, b_3, b_4)]$  for the theory \eqref{d4d4d4} or \eqref{affineD4}, it is useful to consider 
Higgsings of the theory $(D_4, \underline{D_4})$ on a circle. The theory is described by 
\be\label{gd4d4}
\us\os1^{\fsp(0)^{(1)}}_{\left[\fso(8)^{(1)}\right]}-\os4^{\fso(8)^{(1)}},
\ee
and its web diagram is depicted in Figure \ref{fig:gd4d4}. The $\SU(2)^4$ flavor symmetry in $\SO(8)$ can be explicitly seen from the diagram and we consider Higgsings which break the $\SU(2)^4$. We will label the Higgsings by $(a_1, a_2, a_3, a_4)$ where $a_i\; (i=1, 2, 3, 4)$ are $2$ or $0$, which again means whether the Higgsing breaks an $\SU(2)$ or not. In terms of the web diagram, the $\SU(2)$ Higgsing is realized by putting two parallel external 5-branes on one 7-brane, which is depicted in Figure \ref{fig:su2higgs}. For example, the theory obtained after the Higgsing of the type $(2, 0, 0, 0)$ is realized by the web diagram where one of the four diagrams in Figure \ref{fig:gd4d4} is replaced with the diagram on the right in Figure \ref{fig:su2higgs}, which is illustrated in Figure \ref{fig:gd4d4su2higgs1}.  

Higgsings of a conformal matter theory are characterized by nilpotent orbits of a Lie algebra associated to the flavor symmetry of the theory and the correspondence between the nilpotent orbits and the resulting theories at low energies after the Higgsings have been determined in \cite{Heckman:2016ssk}. 
Hence 
we will relate the Higgsing label $(a_1, a_2, a_3, a_4)$ with the nilpotent orbits of $\fso(8)$ to determine the theory obtained after applying the Higgsing $(a_1, a_2, a_3, a_4)$ to the theory $(D_4, \underline{D_4})$ on a circle. 
One difference is that the analysis in \cite{Heckman:2016ssk} has considered Higgsings of 6d theories without a circle compactification but here we consider Higgsings of the 6d theory on a circle. Therefore the Higgsings may give rise to a theory that is obtained by a twisted compactification of a 6d theory, which has been also noted in \cite{Kim:2019dqn}. 

\begin{figure}[t]
\centering
\includegraphics[width=6cm]{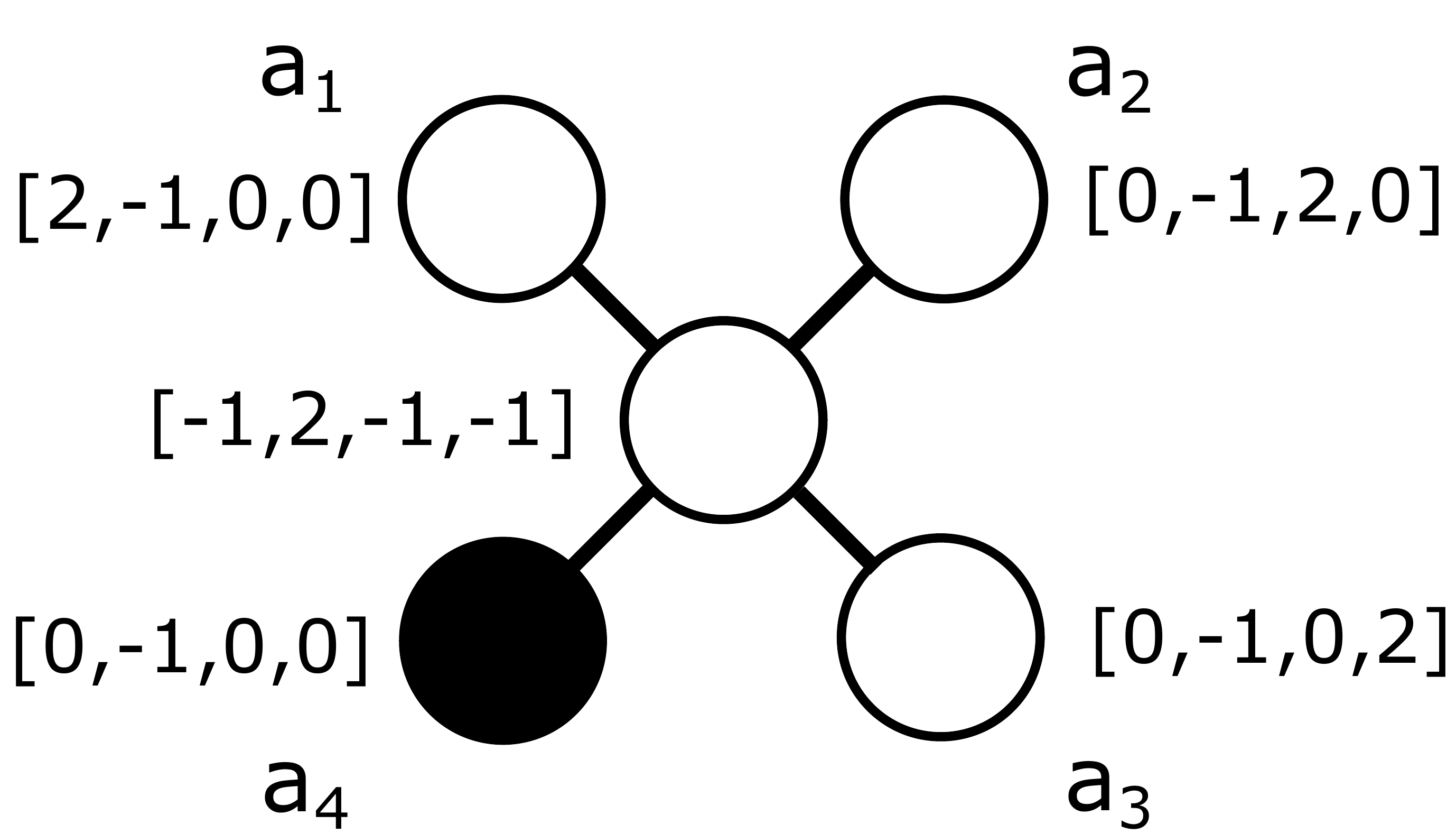}
\caption{The affine $D_4$ Dynkin diagram with the affine node given in black. The four digits in a bracket indicate the Dynkin label for each node. $a_i$ next to a circle implies that this node is Higgsed when $a_i = 2$.  
}
\label{fig:affined4dynkin}
\end{figure}
Let us then relate the Higgsings labeled by $(a_1, a_2, a_3, a_4)$ with the nilpotent orbits of $\fso(8)$. The breaking labeled by $(a_1, a_2, a_3, a_4)$ of the $\SU(2)^4$ flavor symmetry is achieved by giving a vacuum expectation value (vev) to a nilpotent element in $\fso(8)$. The four $\SU(2)$'s are associated to the four nodes except for the middle node in the affine $D_4$ Dynkin diagram and the Higgsing of $a_i = 2$ breaks an $\SU(2)$ associated to one of the four nodes. We assign $a_i\; (i=1, 2, 3, 4)$ to the four nodes in the affine $D_4$ Dynkin diagram as in Figure \ref{fig:affined4dynkin}. From the assignment in Figure \ref{fig:affined4dynkin} it is possible to find an explicit nilpotent element corresponding to the Higgsing labeled by $(a_1, a_2, a_3, a_4)$. In this case the orbit of the nilpotent element is characterized by a certain partition of $8$ and the end result of the correspondence is\footnote{In this paper, generally, we do the nilpotent Higgsing using the element $e_{\mathfrak{h}} = \sum_{\alpha} \rho(E_{\alpha})$ where $\alpha$ runs through the simple roots of a subalgebra $\mathfrak{h}$ and $\rho$ is the embedding into the ambient algebra. To identify the nilpotent orbit containing $e_{\mathfrak{h}}$, we computed the corresponding weighted Dynkin diagram, explained in, e.g.\ \cite{Chacaltana:2012zy}, helped by the \texttt{Mathematica} package \texttt{LieART} \cite{Feger:2012bs,Feger:2019tvk}. From the weighted Dynkin diagram one can read off the corresponding partition when the ambient algebra is classical, or the Bala-Cater (B-C) label when the ambient algebra is exceptional.} 
\begin{equation}\label{su24so8corr}
\begin{split}
(0, 0, 0, 0) &\leftrightarrow [1, 1, 1, 1, 1, 1, 1, 1],\cr
(2, 0, 0, 0), (0, 2, 0, 0), (0, 0, 2, 0), (0, 0, 0, 2) &\leftrightarrow [2, 2, 1, 1, 1, 1],\cr
(2, 2, 0, 0), (0, 0, 2, 2) &\leftrightarrow [2, 2, 2, 2]_I,\cr
(2, 0, 2, 0), (0, 2, 0, 2) &\leftrightarrow [2, 2, 2, 2]_{II},\cr
(2,0, 0, 2), (0, 2, 2, 0) &\leftrightarrow [3, 1, 1, 1, 1, 1],\cr
(2, 2, 2, 0), (2, 2, 0, 2), (2, 0, 2, 2), (0, 2, 2, 2) &\leftrightarrow [3, 2, 2, 1],\cr
(2, 2, 2, 2) &\leftrightarrow [3, 3, 1, 1].
\end{split}
\end{equation}
Since the four nodes associated to $a_1, a_2, a_3, a_4$ are equivalent to each other, all the Higgsings with the labels of two $2$'s and two $0$'s lead to the same theory although they are associated to three different nilpotent orbits. 

Using the correspondence \eqref{su24so8corr} it is possible to relate the Higgsings labeled by $(a_1, a_2, a_3, a_4)$ with the theories after the Higgsings from the result in \cite{Heckman:2016ssk}. However, there is one more step to identify the Higgsed theory after the Higgsing $(a_1, a_2, a_3, a_4)$ since the Higgsings considered in \cite{Heckman:2016ssk} are Higgsings of 6d theories without a circle compactification. The Higgsings in our cases may lead to circle compactifications of the Higgsed 6d theories with twists. Indeed some of the Higgsings necesarily involves a Higgsing associated to an affine node of a Dynkin diagram. In thoses cases we expect that the resulting theories are 5d KK theories obtained by twisted compactifications of 6d theories if the original 6d theories on $S^1$ admit a non-trivial twist. For practical computations, we will take the following strategy to determine if the Higgsed theory is that from a twisted compactification of a 6d theory. If the numbers of Coulomb branch moduli and mass parameters of the 5d KK theory obtained by a circle compactification of the 6d Higgsed theory in \cite{Heckman:2016ssk} agree with the numbers of Coulomb branch moduli and mass parameters read from the web diagrams, then we identify the Higgsed theory with a circle compactification of the 6d Higgsed theory in \cite{Heckman:2016ssk} without a twist . If the numbers do not match with each other, then we identify the Higgsed theory with a twisted compactification of the 6d theory in \cite{Heckman:2016ssk}. We can determine the twist from a discrete symmetry of the Higgsed theories and also the matching of the numbers of Coulomb branch moduli and mass parameters after the twist.

Using this strategy it is possible to determine the relation between the Higgsing $(a_1, a_2, a_3, a_4)$ and the Higgsed theories and the result is summarized in Table \ref{tb:d4_1}. The Higgsing label in Table \ref{tb:d4_1} is in the order $a_1 \geq a_2 \geq a_3 \geq a_4$ since the other orders give the same theory. 
\begin{center}
\begin{longtable}{c|c|c|c}
\caption{\label{tb:d4_1} Higgsings associated to $\SU(2)^4$ of \eqref{gd4d4}. }\\
$\text{Higgsing}$&$\text{partition}$&$\text{twist}$&$\text{theory}$
\\[3 pt]
\hline
\rule[-10pt]{0pt}{30pt}
$(0, 0, 0, 0)$ & $[1^8]$ &$1$ & $\us\os1^{\fsp(0)^{(1)}}_{\left[\fso(8)^{(1)}\right]}-\os4^{\fso(8)^{(1)}}$
\\[10 pt]
\hline
\rule[-10pt]{0pt}{30pt}
$(2, 0, 0, 0)$ & $[2^2,1^4]$&$1$ & $\us\os3^{\fso(8)^{(1)}}_{\left[\fsp(1)^{(1)} \oplus \fsp(1)^{(1)} \oplus \fsp(1)^{(1)}\right]}$
\\[10 pt]
\hline
\rule[-10pt]{0pt}{30pt}
$(2, 2, 0, 0)$ & $[2^4]_I, [3, 1^5], [2^4]_{II}$&$1$ & $\us\os3^{\fso(7)^{(1)}}_{\left[\fsp(2)^{(1)}\right]}$
\\[10 pt]
\hline
\rule[-10pt]{0pt}{30pt}
$(2, 2, 2, 0)$ & $[3, 2^2, 1]$&$1$ & $\us\os3^{\fg_2^{(1)}}_{\left[\fsp(1)^{(1)}\right]}$
\\[10 pt]
\hline
\rule[-10pt]{0pt}{30pt}
$(2, 2, 2, 2)$ & $[3^2, 1^2]$ &$Z_2$ & $\os3^{\fsu(3)^{(2)}}$
\\
\hline
\end{longtable}
\end{center}
\begin{figure}[t]
\centering
\subfigure[]{\label{fig:puresu3k9v2}
\includegraphics[width=6cm]{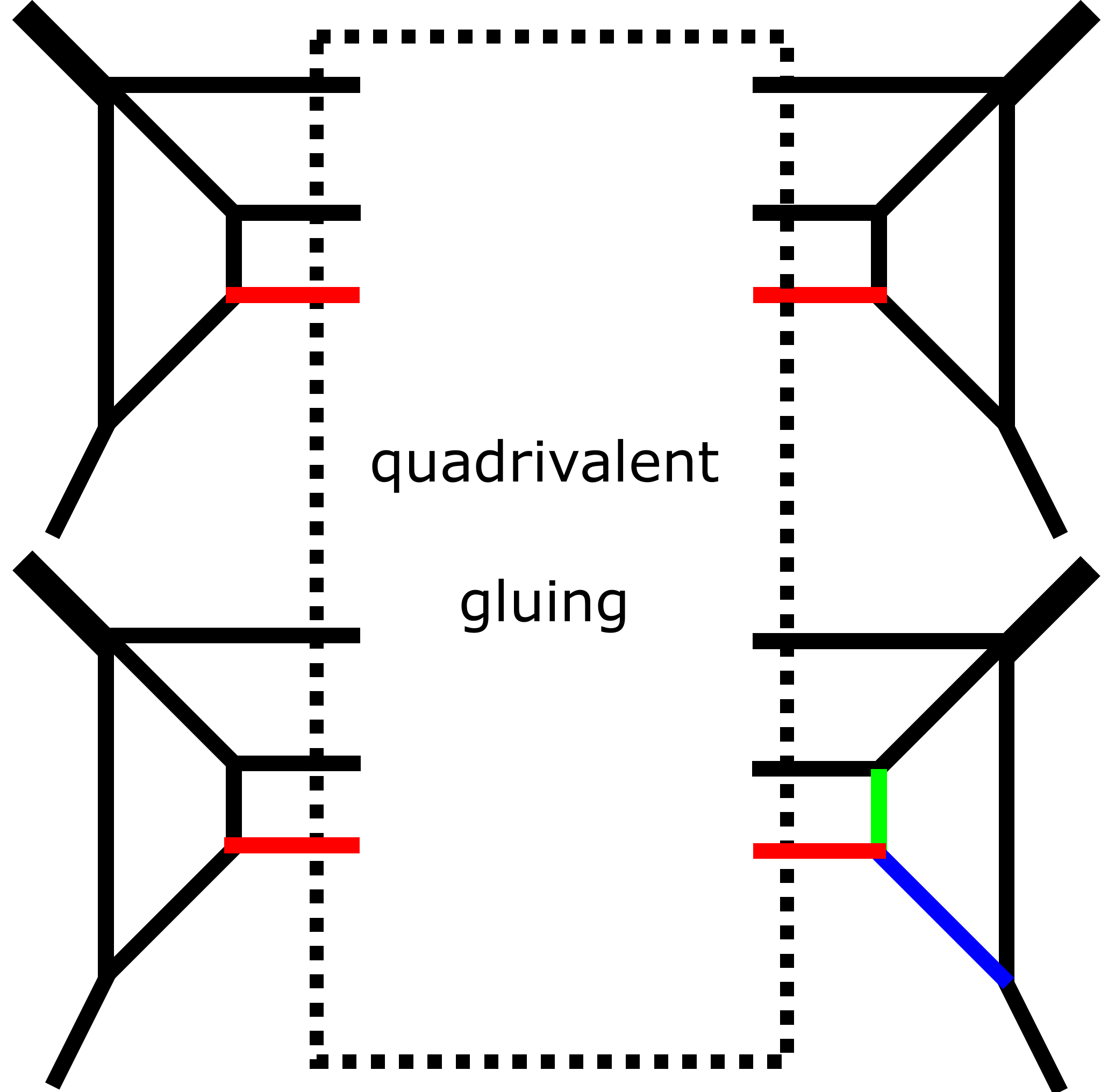}}
\hspace{1cm}
\subfigure[]{\label{fig:twistedsu3}
\includegraphics[width=4cm]{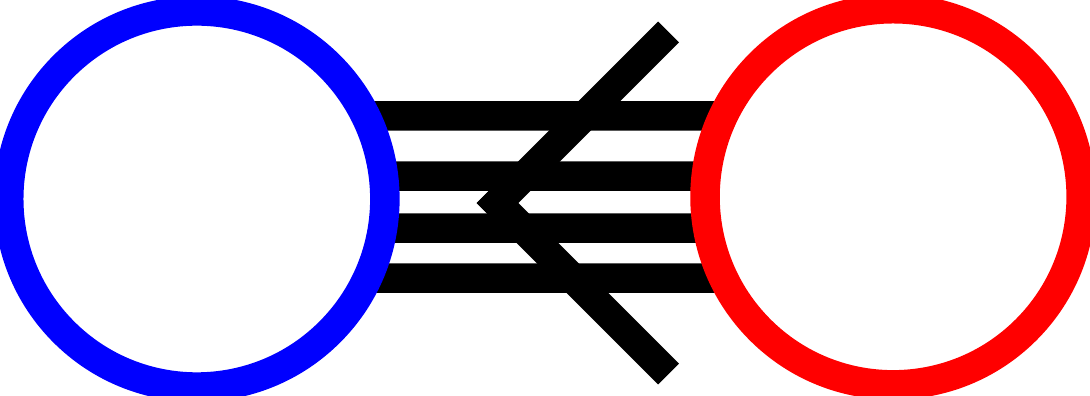}}
\caption{(a): The web diagram of the theory obtained after the Higgsing $(2, 2, 2, 2)$. (b): The Dynkin diagram of the twisted affine Lie algebra $\fsu(3)^{(2)}$. 
The lines in red and blue in Figure \ref{fig:puresu3k9v2} correspond to the nodes in red and blue in this figure respectively. }
\label{fig:twistedsu3web}
\end{figure}
We can see in Table \ref{tb:d4_1} that a twisted compactification of a 6d theory already appears in this example. Let us closely look at the Higgsing $(2, 2, 2, 2)$, which is in the last row in Table \ref{tb:d4_1}. The web diagram for the $(2, 2, 2, 2)$ Higgsing is depicted in Figure \ref{fig:puresu3k9v2}. The Higgsing necessarily involves a Higgsing which breaks an $\SU(2)$ associated to the affine node of the affine $D_4$ Dynkin diagram. In this case, we expect that the resulting theory is given by a twisted compactification of the 6d theory. In terms of the relation \eqref{su24so8corr} the Higgsings $(2, 2, 2, 2)$ corresponds to the partition $[3, 3, 1, 1]$. The Higgsing giving a vev of the nilpotent orbit labeled by $[3, 3, 1, 1]$ of the 6d theory $(D_4, \underline{D_4})$ gives rise to \cite{Heckman:2016ssk},
\be
\os3^{\fsu(3)}. \label{puresu3}
\ee
A circle compactification without a twist of the theory \eqref{puresu3} yields a 5d KK theory with $3$ Coulomb branch moduli and one mass parameter. On the other hand the number of Coulomb branch moduli from the web diagram in Figure \ref{fig:puresu3k9v2} is $2$ and it has one mass parameter. Namely we have a mismatch of the number of the Coulomb branch moduli, which also suggests that the Higgsing $(2, 2, 2, 2)$ generates a theory obtained by a twisted compactification of a 6d theory. 
To see which twist is possible, note that the 6d pure $\SU(3)$ gauge theory has a discrete symmetry given by the charge conjugation. Hence it is possible to consider the twisted compactification of the theory \eqref{puresu3} which yields 
\be
\os3^{\fsu(3)^{(2)}}. \label{puresu3twist}
\ee
The twist formally corresponds to $N_f = 0$ of the complex representation discussed in section \ref{sec:twist}. 
The 5d KK theory described by\eqref{puresu3twist} has $2$ Coulomb branch moduli and one mass parameter and the numbers agree with those from the web diagram in Figure \ref{fig:puresu3k9v2}. 
Therefore we argue that the web diagram in Figure \ref{fig:puresu3k9v2} corresponds to the geometry \eqref{puresu3twist}. 
It is also possible to see the appearance of the twisted affine Lie algebra $\fsu(3)^{(2)}$ from the web diagram in Figure \ref{fig:twistedsu3web}. By using the intersection numbers given by \eqref{fScartan}, 
the intersection matrix between the fiber class in red and that in blue 
in Figure \ref{fig:puresu3k9v2} becomes the Cartan matrix of the twisted affine Lie algebra $\fsu(3)^{(2)}$ and the fibers form the Dynkin diagram of $\fsu(3)^{(2)}$, which is depicted in Figure \ref{fig:twistedsu3}.

It has been also argued that the theory \eqref{puresu3twist} gives rise to the 5d $\SU(3)$ gauge theory with the CS level $9$ \cite{Jefferson:2018irk} and hence the diagram in Figure \ref{fig:puresu3k9v2} gives a realization of the 5d pure $\SU(3)$ gauge theory with the CS level $9$ using the web diagram with the quadrivalent gluing. W-bosons corresponding to simple roots of $\SU(3)$ are associated to the fiber class in green 
and that in blue in Figure \ref{fig:puresu3k9v2}. 

\begin{figure}[t]
\centering
\includegraphics[width=14cm]{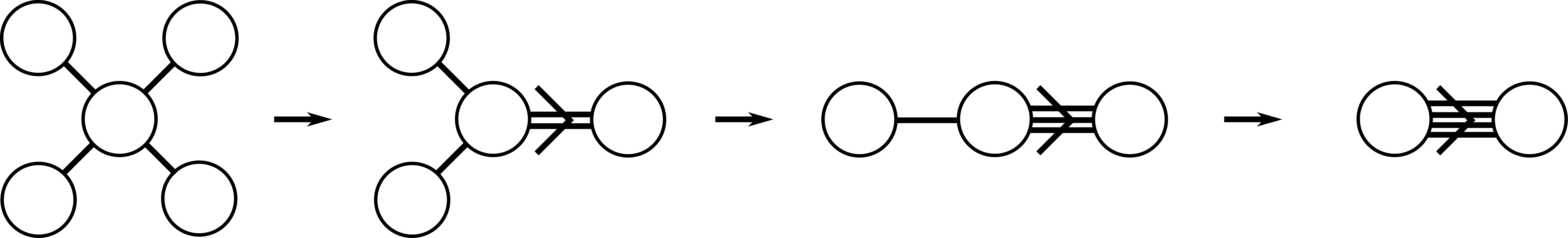}
\caption{The change of the Dynkin diagram in the process of the successive Higgsings of $\SU(2)$.}
\label{fig:d4dynkinchain}
\end{figure}
One can also clearly see a pattern of the change of the Dynkin diagrams by the successive Higgsings associated to the $\SU(2)$. For the each $\SU(2)$ Higgsing the algebra changes as $\fso(8)^{(1)} \to \fso(7)^{(1)} \to \fg_2^{(1)} \to \fsu(3)^{(2)}$ statring fom $(2,0,0,0)$. The change of the Dynkin diagram is depicted in Figure \ref{fig:d4dynkinchain}. Namely the effect of each $\SU(2)$ Higgsing in terms of the Dynkin diagram is that it folds the diagram by identifying two nodes with one line added between the middle node and the identified node. The appearance of $\fsu(3)^{(2)}$ is natural from this respect. 

\begin{figure}[t]
\centering
\subfigure[]{\label{fig:g2on3}
\includegraphics[width=6cm]{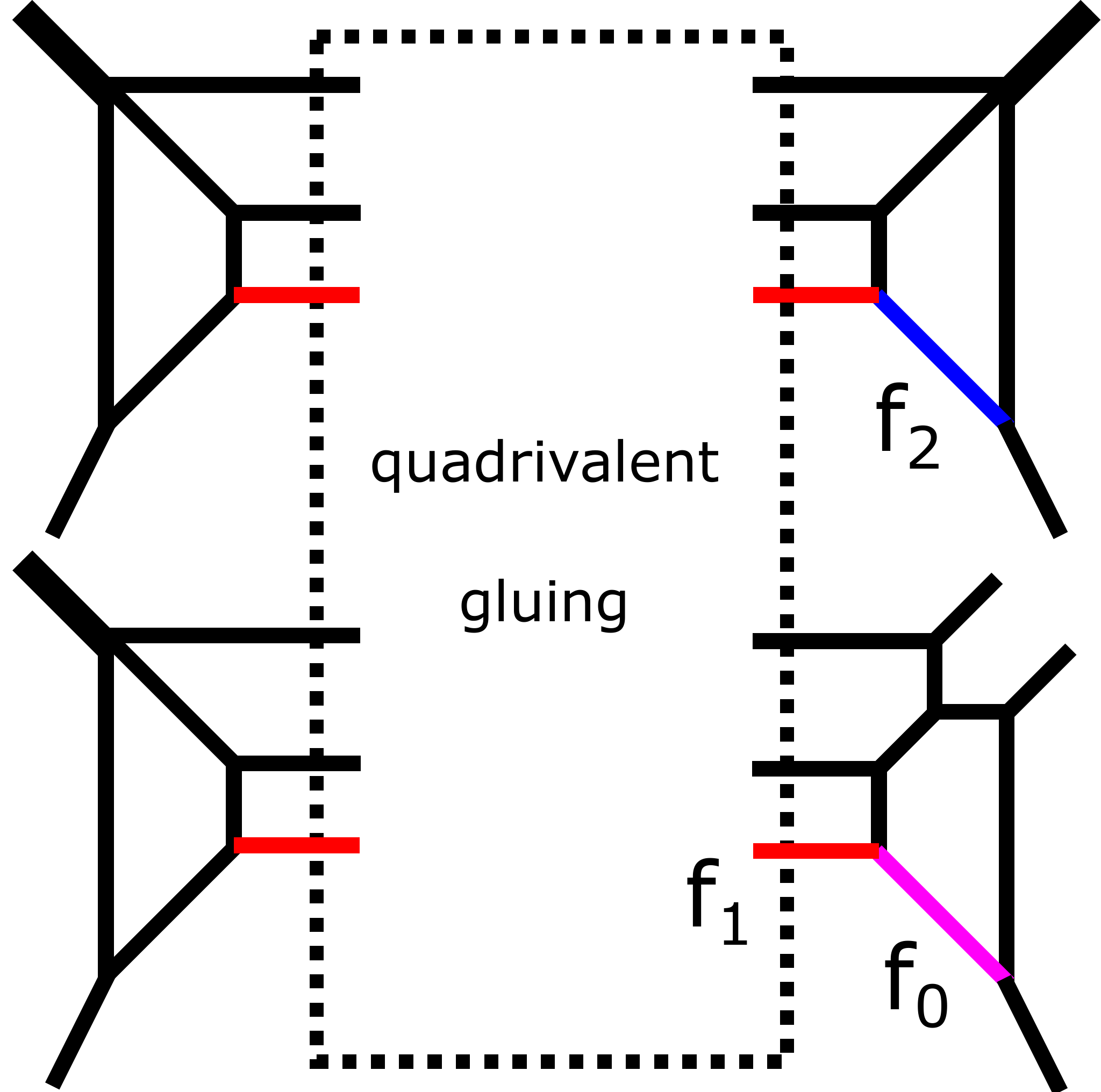}}
\hspace{1cm}
\subfigure[]{\label{fig:g2fiber}
\includegraphics[width=4cm]{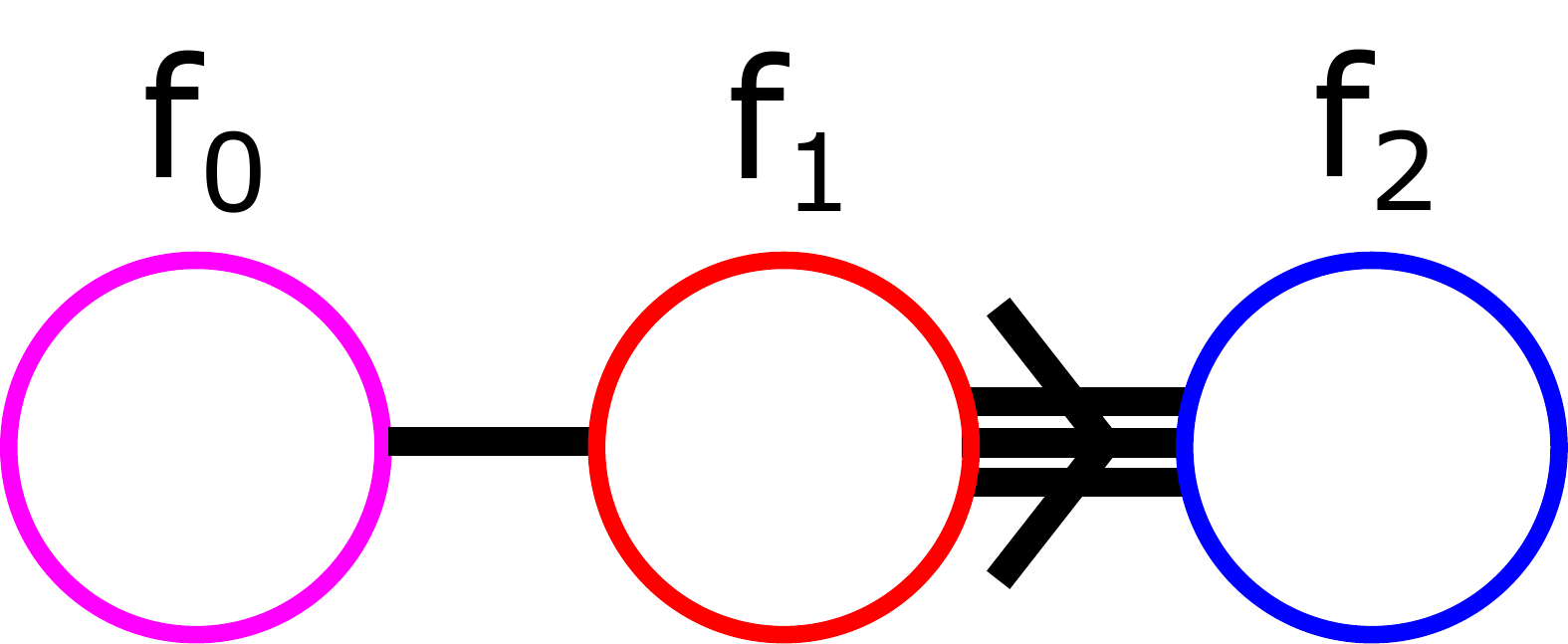}}
\caption{(a): The web diagram of the theory obtained after the Higgsing $(2, 2, 2, 0)$. (b): The Dynkin diagram of the affine Lie algebra $\fg_2^{(1)}$. 
The lines in red, blue and magenta in Figure \ref{fig:g2on3} correspond to the nodes in red, blue and magenta in this figure respectively. }
\label{fig:g2web}
\end{figure}

It is also illustrative to see how the torus fiber changes in the process $(2, 2, 2, 0) \to (2, 2, 2, 2)$. The web diagram after the Higgsing $(2, 2, 2, 0)$ is depicted in Figure \ref{fig:g2on3}. 
The intersection among the fiber classes $f_0, f_1, f_2$ form the affine $G_2$ Dynkin diagram as in Figure \ref{fig:g2fiber}. In this case the elliptic fiber is given by the combination \eqref{fiberI}, which becomes 
\be\label{g2t2}
f_{\fg_2^{(1)}} = f_0 + 2f_1 + 3f_2,
\ee 
and an M2-brane which is wrapped on the curve \eqref{g2t2} corresponds to a KK mode with unit momentum in the 5d KK theory. The mass of the particle is $\frac{1}{R_{\text{6d}}}$ where $R_{\text{6d}}$ is the radius of the $S^1$ compactification of the 6d theory which is 6d $G_2$ gauge theory with one flavor and a tensor multiplet. By performing the Higgsing which breaks the remaining $\SU(2)$ flavor symmetry we obtain the web diagram in Figure \ref{fig:puresu3k9v2} where $f_0$ is identified with $f_2$. Then the combination \eqref{g2t2} can be written as
\be\label{fKKg2}
f_{KK} = f_0 + 2f_1 + 3f_2 = 2f_1 + 4f_2.
\ee
On the other hand the fiber \eqref{fiberI} for the twisted affine Lie algebra $\fsu(3)^{(2)}$ is given by 
\be\label{fsu32}
f_{\fsu(3)^{(2)}} = f_1 + 2f_2.
\ee
Combining \eqref{fKKg2} with \eqref{fsu32} gives
\be
f_{KK} = 2f_{\fsu(3)^{(2)}}. 
\ee
Then an M2-brane wrapped on $f_{\fsu(3)^{(2)}}$ corresponds to a fractional KK mode with mass $\frac{1}{2R_{\text{6d}}}$.

\begin{figure}[t]
\centering
\includegraphics[width=6cm]{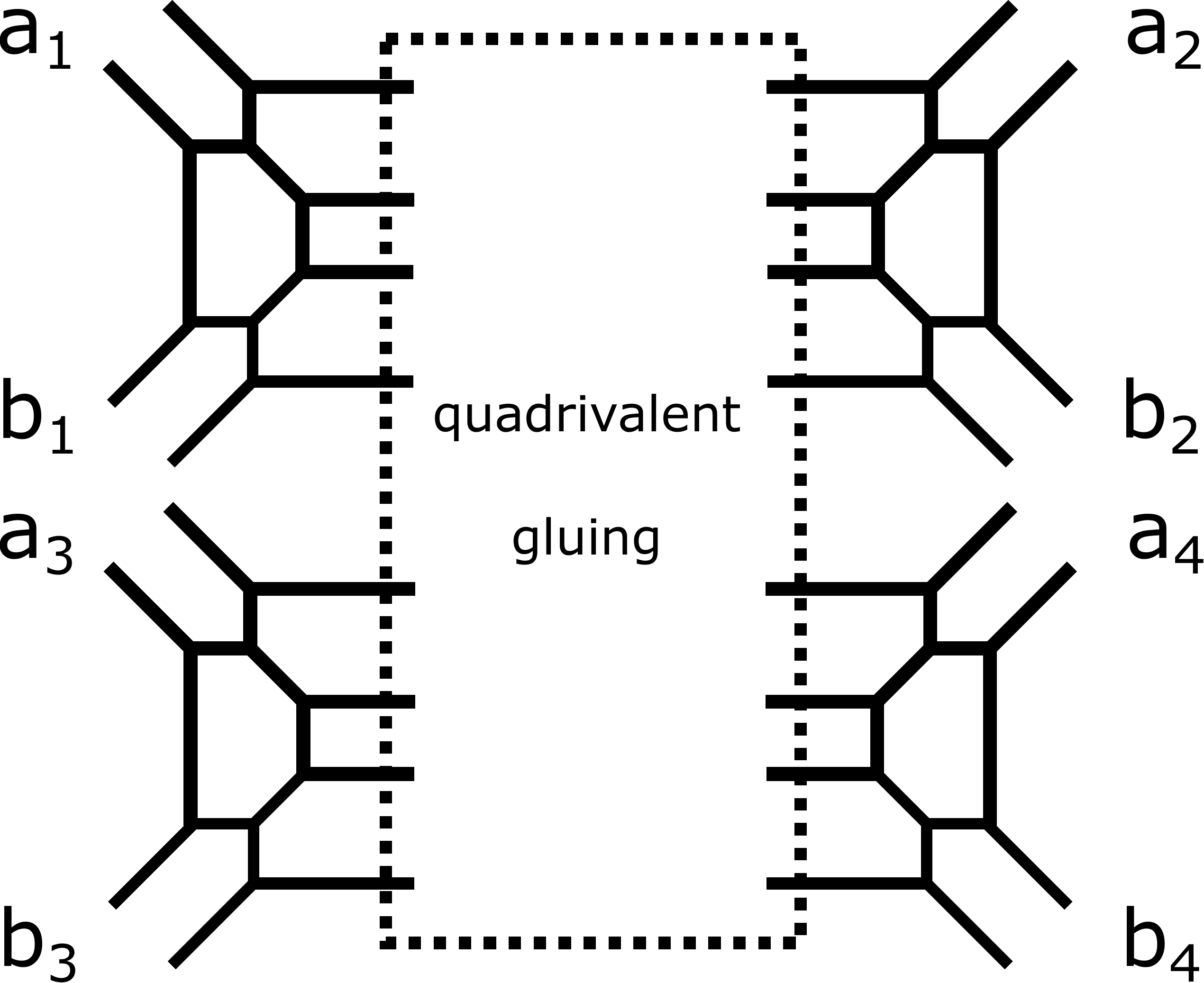}
\caption{The labeling of $[(a_1, a_2, a_3, a_4), (b_1, b_2, b_3, b_4)]$ for each $\SU(2)$ in the web diagram in Figure \ref{fig:d4d4d4}
}
\label{fig:d4d4d4higgs}
\end{figure}
We then move on to the Higgsings labeled by $[(a_1, a_2, a_3, a_4), (b_1, b_2, b_3. b_4)]$ of the theory \eqref{d4d4d4}. The notation here is that when $a_i$ or $b_i$ is $2$ then we perform the Higgsing for the parallel external lines next to $a_i$ or $b_i$ in Figure \ref{fig:d4d4d4higgs}. For obtaining the theory after the Higgsing labeled by $[(a_1, a_2, a_3, a_4), (b_1, b_2, b_3. b_4)]$,
we can combine the theory which arises from $(a_1, a_2, a_3, a_4)$ with the theory from $(b_1, b_2, b_3, b_4)$ when either of the theories still has the $\fso(8)^{(1)}$ algebra. More specifically, suppose we consider the Higgsing $(a_1, a_2, a_3, a_4)$ or $(b_1, b_2, b_3, b_4)$ which gives
\be
\os{n_2}^{\fg_2}-\os{\left(4 - n_1\right)}^{\fso(8)^{(1)}},
\ee
and the other Higgsing which gives 
\be
\os{\left(4 - m_1\right)}^{\fg'_1}-\os{m_2}^{\fg'_2},
\ee
when we apply the Higgsing to the theory \eqref{gd4d4}. Then the rule of the combining the two theories is given by
\begin{equation}\label{d4combine}
\begin{split}
&\left[\os{n_2}^{\fg_2}-\os{\left(4 - n_1\right)}^{\fso(8)^{(1)}}\right] + \left[\os{\left(4 - m_1\right)}^{\fg'_1}-\os{m_2}^{\fg'_2}\right]\cr
\to&\qquad \os{n_2}^{\fg_2}-\os{\left(4-n_1-m_1\right)}^{\fg'_1}-\os{m_2}^{\fg'_2}.
\end{split}
\end{equation}
The flavor algebra can be determined from the matter content which satsifies the anomaly cacellation as in \cite{Heckman:2015bfa} with a possible twist.

\begin{figure}[t]
\centering
\includegraphics[width=12cm]{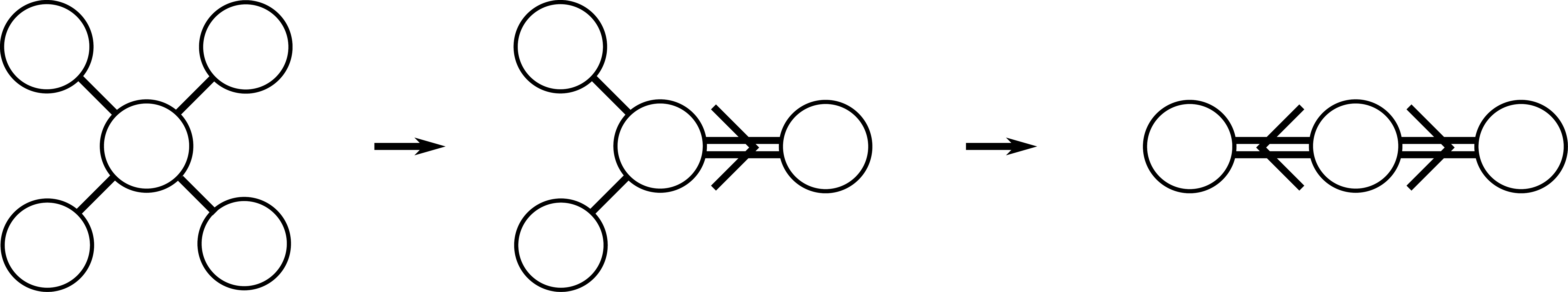}
\caption{The change of the Dynkin diagram after the Higgsings of $(2, 2, 0, 0)$ and then the Higgsing of $(0, 0, 2, 2)$.}
\label{fig:twistedaffinesu4}
\end{figure}

When both the $(a_1, a_2, a_3, a_4)$ and $(b_1, b_2, b_3, b_4)$ cases Higgs the $\fso(8)^{(1)}$ algebra in each theory we need to be careful of combining the two theories. For example let us consider the cases where two of $a_i$ and two of $b_i$ are $2$ and the others are zero. All such cases have one base curve with the self-intersection number $-2$. Hence we can specify the theory by determining the algebra on the curve. The algebra can be determined from the Higgsed web diagrams by computing the Cartan matrix of the algebra using \eqref{fScartan}. 
There are three possible Higgsings and the correspondence between the Higgsings and the resulting theories is
\begin{align}
[(2, 2, 0, 0), (2, 2, 0, 0)] \quad&\leftrightarrow \quad \us\os2^{\fso(7)^{(1)}}_{\left[\fsp(1)^{(1)} \oplus \fsp(4)^{(1)}\right]},\label{22toso7}\\
[(2, 2, 0, 0), (2, 0, 2, 0)] \quad &\leftrightarrow \quad \us\os2^{\fg_2^{(1)}}_{\left[\fsp(4)^{(1)}\right]}, \label{22tog2}\\
[(2, 2, 0, 0), (0, 0, 2, 2)] \quad &\leftrightarrow \quad \us\os2^{\fsu(4)^{(2)}}_{\left[\fsu(8)^{(2)}\right]}, \label{22tosu4}
\end{align}
where we fixed $(a_1, a_2, a_3, a_4) = (2, 2, 0, 0)$ since the other choices give the same result after the appropriate rearrangement of $b_i$'s. Also for \eqref{22tog2}, $(b_1, b_2, b_3, b_4) = (2, 0, 0, 2), (0, 2, 2, 0)$ and $(0, 2, 0, 2)$ yield the same result as \eqref{22tog2}. Note that the Higgsing \eqref{22toso7} does not change the number of the Coulomb branch moduli compared to $[(2, 2, 0, 0), (2, 0, 0, 0)]$, In other words the change of $[(2, 2, 0, 0), (2, 0, 0, 0)]$ into $[(2, 2, 0, 0), (2, 2, 0, 0)]$ is not a Higgsing but simply tuning a mass parameter. 
Note that the twisted affine Lie algbera $\fsu(4)^{(2)}$ appears in \eqref{22tosu4}, which comes from the twist of \eqref{sutwist}. It is also illustrative to see the change in terms of the Dynkin diagram for the Higgsing \eqref{22tosu4}, which is depicted in Figure \ref{fig:twistedaffinesu4}. 

Next we consider the cases where three of $a_i$'s and two of $b_i$'s are two and the others are zero. The correspondence between the Higgsings and the resulting theories is 
\begin{align}
[(2, 2, 2, 0), (2, 2, 0, 0)], [(2, 2, 2, 0), (2, 0, 2, 0)], [(2, 2, 2, 0), (0, 2, 2, 0)]  &\leftrightarrow \us\os2^{\fg_2^{(1)}}_{\left[\fsp(4)^{(1)}\right]}\label{222g2},\\
[(2, 2, 2, 0), (2, 0, 0, 2)], [(2, 2, 2, 0), (0, 2, 0, 2)], [(2, 2, 2, 0), (0, 0, 2, 2)] &\leftrightarrow  \us\os2^{\fsu(3)^{(2)}}_{\left[\fsu(6)^{(2)}\right]}. \label{222su3}
\end{align}
We again fixed $(a_1, a_2, a_3, a_4) = (2, 2, 2, 0)$ without loss of generality. In the theory of \eqref{222g2} one mass parameter is turned off compared to the general case. 

The other cases where we have more than five $2$'s in $[(a_1, a_2, a_3, a_4), (b_1, b_2, b_3, b_4)]$ do not lead to new theories. The case of $[(2, 2, 2, 2, 0), (2, 2, 2, 0)]$ gives rise to 
\be
\us\os2^{\fg_2^{(1)}}_{\left[\fsp(4)^{(1)}\right]},
\ee
with two mass parameters turned off. The others yields
\be
\us\os2^{\fsu(3)^{(2)}}_{\left[\fsu(6)^{(2)}\right]}, 
\ee
and the number of mass parameters which are turned off is the number of $2$'s in $[(a_1, a_2, a_3, a_4), (b_1, b_2, b_3, b_4)]$ minus $5$. 


The result of the relation between the Higgsings of $[(a_1, a_2, a_3, a_4), (b_1, b_2, b_3, b_4)]$ and the resulting theories is summarize in Table \ref{tb:d4_2}. 
\begin{center}
\begin{longtable}{c|c|c}
\caption{ \label{tb:d4_2} Higgsings associated to $\SU(2)^4 \times \SU(2)^4$ of \eqref{d4d4d4}. AS represents a hypermultiplet in the rank-2 antisymmetric representation of a gauge group which is connected to $[\text{AS}]$.}\\
$\text{Higgsing}$&$\text{theory}$&$\text{A 5d description}$
\\[3 pt]
\hline
\rule[-10pt]{0pt}{30pt}
$\left[(0, 0, 0, 0), (0, 0, 0, 0)\right]$&$\quad\us\os1^{\fsp(0)^{(1)}}_{\left[\fso(8)^{(1)}\right]}-\os4^{\fso(8)^{(1)}}-\us\os1^{\fsp(0)^{(1)}}_{\left[\fso(8)^{(1)}\right]}\quad$&$\SU(2)_0  -{\underset{\text{\normalsize$\SU(2)_0$}}{\underset{\textstyle\vert}{\overset{\overset{\text{\normalsize$\SU(2)_0$}}{\textstyle\vert}}{\SU(4)_0}}}}- \SU(2)_0$
\\[10 pt]
\hline
\rule[-10pt]{0pt}{30pt}
$\left[(2, 0, 0, 0), (0, 0, 0, 0)\right]$ &$\quad\us\os3^{\fso(8)^{(1)}}_{\left[\fsp(1)^{(1)} \oplus \fsp(1)^{(1)} \oplus \fsp(1)^{(1)}\right]}-\us\os1^{\fsp(0)^{(1)}}_{\left[\fso(8)^{(1)}\right]}\quad$&$\SU(2)_0  -{\underset{\text{\normalsize$\SU(2)_0$}}{\underset{\textstyle\vert}{\overset{\overset{\text{\normalsize$\SU(2)_0$}}{\textstyle\vert}}{\SU(4)_{\pm 1}}}}}- \left[\text{AS}\right]$
\\[10 pt]
\hline
\rule[-10pt]{0pt}{30pt}
$\left[(2, 2, 0, 0), (0, 0, 0, 0)\right]$ &$\quad\us\os3^{\fso(7)^{(1)}}_{\left[\fsp(2)^{(1)}\right]}-\us\os1^{\fsp(0)^{(1)}}_{\left[\fso(9)^{(1)}\right]}\quad$&$\SU(2)_0  -{\overset{\overset{\text{\normalsize$\SU(2)_0$}}{\textstyle\vert}}{\SU(4)_{\pm 2}}}- \left[2\text{AS}\right]$
\\[10 pt]
\hline
\rule[-10pt]{0pt}{30pt}
$\left[(2, 0, 0, 0), (2, 0, 0, 0)_4\right] $&$\quad \us\os2^{\fso(8)^{(1)}}_{\left[\fsp(2)^{(1)} \oplus \fsp(2)^{(1)} \oplus \fsp(2)^{(1)}\right]} \quad$&$
\begin{array}{c}
\SU(2)_0  -{\overset{\overset{\text{\normalsize$\SU(2)_0$}}{\textstyle\vert}}{\SU(4)_{0}}}- \left[2\text{AS}\right]\\[5 pt]
\SU(2)_0  -{\overset{\overset{\text{\normalsize$\SU(2)_0$}}{\textstyle\vert}}{\Sp(2)_0}}- \SU(2)_0
\end{array}$
\\[10 pt]
\hline
\rule[-10pt]{0pt}{30pt}
$\left[(2, 2, 2, 0), (0, 0, 0, 0)\right]$ &$\quad\us\os3^{\fg_2^{(1)}}_{\left[\fsp(1)^{(1)}\right]}-\us\os1^{\fsp(0)^{(1)}}_{\left[\ff_4^{(1)}\right]}\quad$&$\SU(2)_0  -\SU(4)_{\pm 3}- \left[3\text{AS}\right]$
\\[10 pt]
\hline
\rule[-10pt]{0pt}{30pt}
$
\begin{array}{c}
\left[(2, 2, 0, 0), (2, 0, 0, 0)_4\right]\\
\left[(2, 2, 0, 0), (2, 2, 0, 0)\right]'
\end{array}$ &$\quad \us\os2^{\fso(7)^{(1)}}_{\left[\fsp(1)^{(1)} \oplus \fsp(4)^{(1)}\right]} \quad$&$
\begin{array}{c}
\SU(2)_0  -\SU(4)_{\pm 1}- \left[3\text{AS}\right]\\[5 pt]
\SU(2)_0  -{\overset{\overset{\text{\normalsize$\SU(2)_0$}}{\textstyle\vert}}{\Sp(2)_0}}- \left[\text{AS}\right]
\end{array}$
\\[10 pt]
\hline
\rule[-10pt]{0pt}{30pt}
$\left[(2, 2, 2, 2), (0, 0, 0, 0)\right]$ &$\quad\os3^{\fsu(3)^{(2)}}-\us\os1^{\fsp(0)^{(1)}}_{\left[\fe_6^{(2)}\right]}\quad$&$\SU(4)_{\pm 4}- \left[4\text{AS}\right]$
\\[10 pt]
\hline
\rule[-10pt]{0pt}{30pt}
$
\begin{array}{c}
\left[(2, 2, 2, 0), (2, 0, 0, 0)\right]\\
\left[(2, 2, 0, 0), (2, 0, 2, 0)_{2,2}\right]\\
\left[(2, 2, 2, 0), (2, 2, 0, 0)_{3,1}\right]'\\
\left[(2, 2, 2, 0), (2, 2, 2, 0)\right]''
\end{array}
$&$\quad \us\os2^{\fg_2^{(1)}}_{\left[\fsp(4)^{(1)}\right]} \quad$&$
\begin{array}{c}
\SU(4)_{\pm 2}- \left[4\text{AS}\right]\\[5 pt]
\SU(2)_0 - \Sp(2)_0 -\left[2\text{AS}\right]
\end{array}$
\\[10 pt]
\hline
\rule[-10pt]{0pt}{30pt}
$\left[(2, 2, 0, 0), (0, 0, 2, 2)\right]$ &$\quad\us\os2^{\fsu(4)^{(2)}}_{\left[\fsu(8)^{(2)}\right]}\quad$&$\SU(4)_{0}- \left[4\text{AS}\right]$
\\[10 pt]
\hline
\rule[-10pt]{0pt}{30pt}
$\begin{array}{c}
\left[(2, 2, 2, 2), (2, 0, 0, 0)_4\right]\\
\left[(2, 2, 2, 0), (2, 0, 0, 2)_{3,1}\right]\\
\left[(2, 2, 2, 0), (2, 2, 0, 2)_{3,1}\right]'\\
\left[(2, 2, 2, 2), (2, 2, 0, 0)_4\right]'\\
\left[(2, 2, 2, 2), (2, 2, 2, 0)_4\right]''\\
\left[(2, 2, 2, 2), (2, 2, 2, 2)\right]'''
\end{array}$ &$\quad \us\os2^{\fsu(3)^{(2)}}_{\left[\fsu(6)^{(2)}\right]} \quad$&$\Sp(2)_0- \left[3\text{AS}\right]$
\\[10 pt]
\hline
\end{longtable}
\end{center}
In Table \ref{tb:d4_2} we fixed $a_1 \geq a_2 \geq a_3 \geq a_4$. The subscript $(n, m)\; (n+m=4)$ for $(b_1, b_2, b_3, b_4)$ means that the permutation of the first $n$ entries and the permutation of the next $m$ entries give the same theory. For the permutation of all the four entries, we simiply wrote $4$ for the subscript. The number of prime marks implies the number of mass parameters which are turned off compared to the general cases.


In fact the theories listed in Table \ref{tb:d4_2} have a 5d gauge theory description. The original theory \eqref{d4d4d4} has a 5d gauge theory description \eqref{affineD4}. Higgsing one $\SU(2)$ yields a hypermultiplet in the rank-2 antisymmetric represetation of $\SU(4)$ and also change the CS level by $\pm 1$ \cite{Bergman:2015dpa}. In our convention a Higgsing of $a_i=2$ increases the CS level by one and a Higgsing of $b_i = 2$ decreases the CS level by one. The rank-2 antisymmetric representation of $\SU(4)$ is a real representation and the flavor symmetry is $\Sp(N_f)$ for $N_f$ hypermultiplets in the rank-2 antisymmetric representation of $\SU(4)$. For example we consider a Higgsing $a_i = 2$ and we have a rank-2 antisymmetric hypermutliplet, the $\SU(2)$ associated to $b_i$ is the flavor symmetry associated to the rank-2 antisymmetric hypermultiplet. Then the Higgsing $b_i=2$ changes the gauge group $\SU(4)$ into $\Sp(2)$. Similarly the quiver $\Sp(2)_0 - \SU(2)_0$ has an $\SU(2)^2$ flavor symmetry and Higgsing one $\SU(2)$ gives a hypermultipet in the rank-2 antisymmetric representation of $\Sp(2)$. 
In this way, one can find a 5d gauge theory description for each theory labeled by $[(a_1, a_2, a_3, a_4), (b_1, b_2, b_3, b_4)]$. The result is summarized in the third column in Table \ref{tb:d4_2}. The negative sign of the CS level for each theory can be obtained by switching $(a_1, a_2, a_3, a_4)$ with $(b_1, b_2, b_3, b_4)$. From this perspective the number of the prime marks means the number of massless rank-2 antisymmetric hypermultiplets of $\Sp(2)$. Some of the 5d theories in Table \eqref{tb:d4_2} have a single gauge group and the classification of UV complete 5d gauge theories with a simple gauge group has been done in \cite{Bhardwaj:2020gyu}. The result in Table \ref{tb:d4_2} is consistent with the result in \cite{Bhardwaj:2020gyu}.

\subsection{$(E_6, E_6)$ conformal matter}
\label{sec:e6e6}

Next we consdier the theory $(E_6, E_6)_2$ which is obtained by gauging two minimal $(E_6, E_6)$ conformal matter theories and study Higgsings of the theory $(E_6, E_6)_2$ on a circle. For that we can basically repeat the process of what we have done for Higgsing the theory \eqref{d4d4d4} in section \ref{sec:d4d4}. 

\begin{figure}[t]
\centering
\includegraphics[width=6cm]{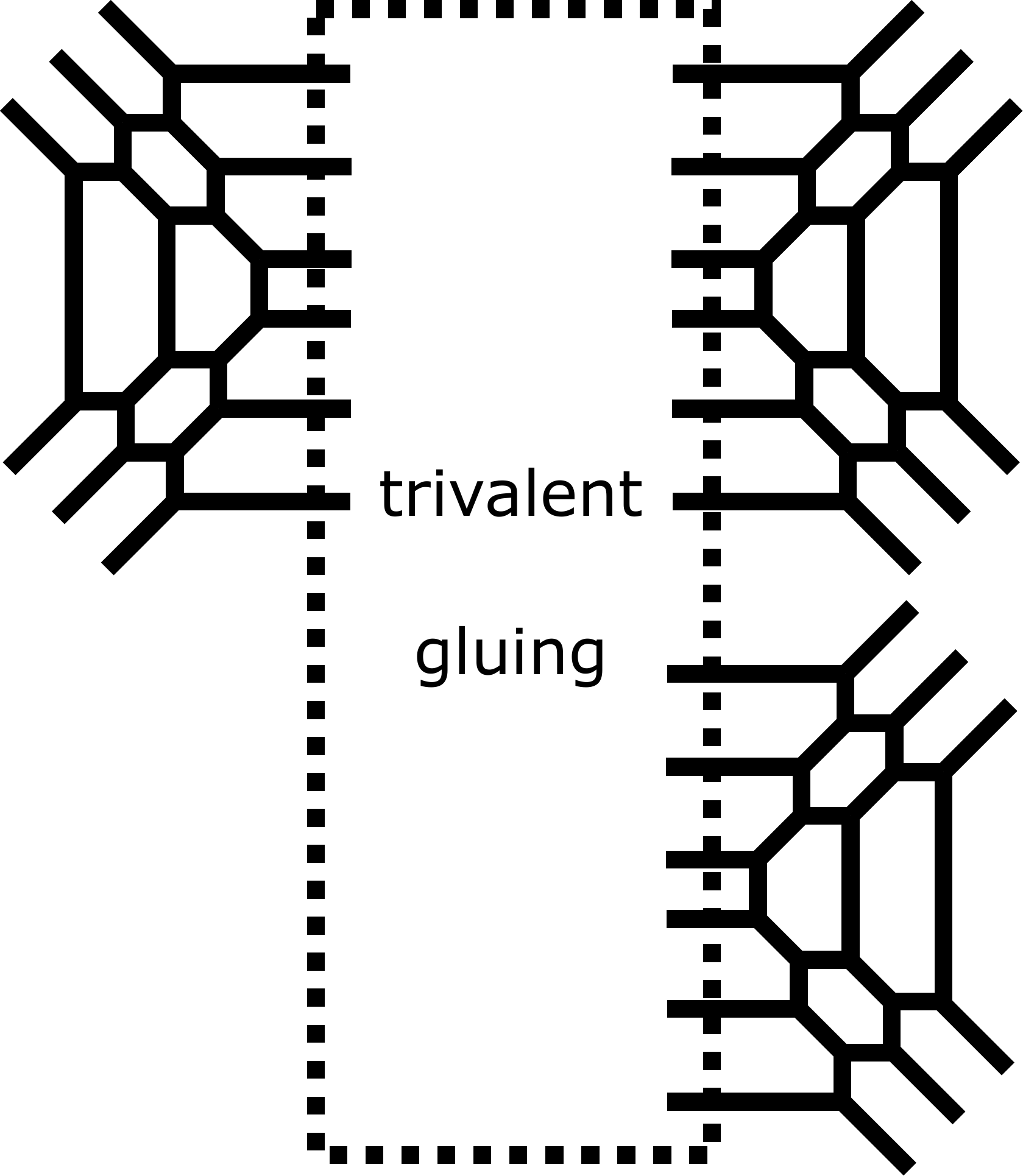}
\caption{
The web diagram for the theory \eqref{e6e6e6}, which is $(E_6, E_6)_2$ on $S^1$. 
}
\label{fig:e6e6e6}
\end{figure}

The basic building block is the minimal $(E_6, E_6)$ conformal matter theory or $(E_6, E_6)_1$. 
Using the notation in section \ref{sec:twist}, the theory $(E_6, E_6)_1$ on a circle is described by
\be
\us\os1^{\fsp(0)^{(1)}}_{\left[\fe_6^{(1)}\right]}-\os3^{\fsu(3)^{(1)}}-\us\os1^{\fsp(0)^{(1)}}_{\left[\fe_6^{(1)}\right]}.\label{e6e6}
\ee
It also has a 5d gauge theory description and it is given by the following affine $E_6$ Dynkin quiver theory,
\be
\SU(1)  - \SU(2) -{\overset{\text{\normalsize$\SU(1)$}}{\overset{\textstyle\vert}{\overset{\overset{\text{\normalsize$\SU(2)$}}{\textstyle\vert}}{\SU(3)_0}}}}- \SU(2) - \SU(1).\label{min.affineE6}
\ee
Each $\SU(1)$ gauge node may be understood are two flavors attached to the adjacent $\SU(2)$ gauge node and the discrete theta of the $\SU(2)$ gauge nodes are not physical. 
This theory can be realized by the trivalent $\SU(3)$ gauging of three copies of the 5d $T_3$ theory.
\begin{figure}[t]
\centering
\includegraphics[width=6cm]{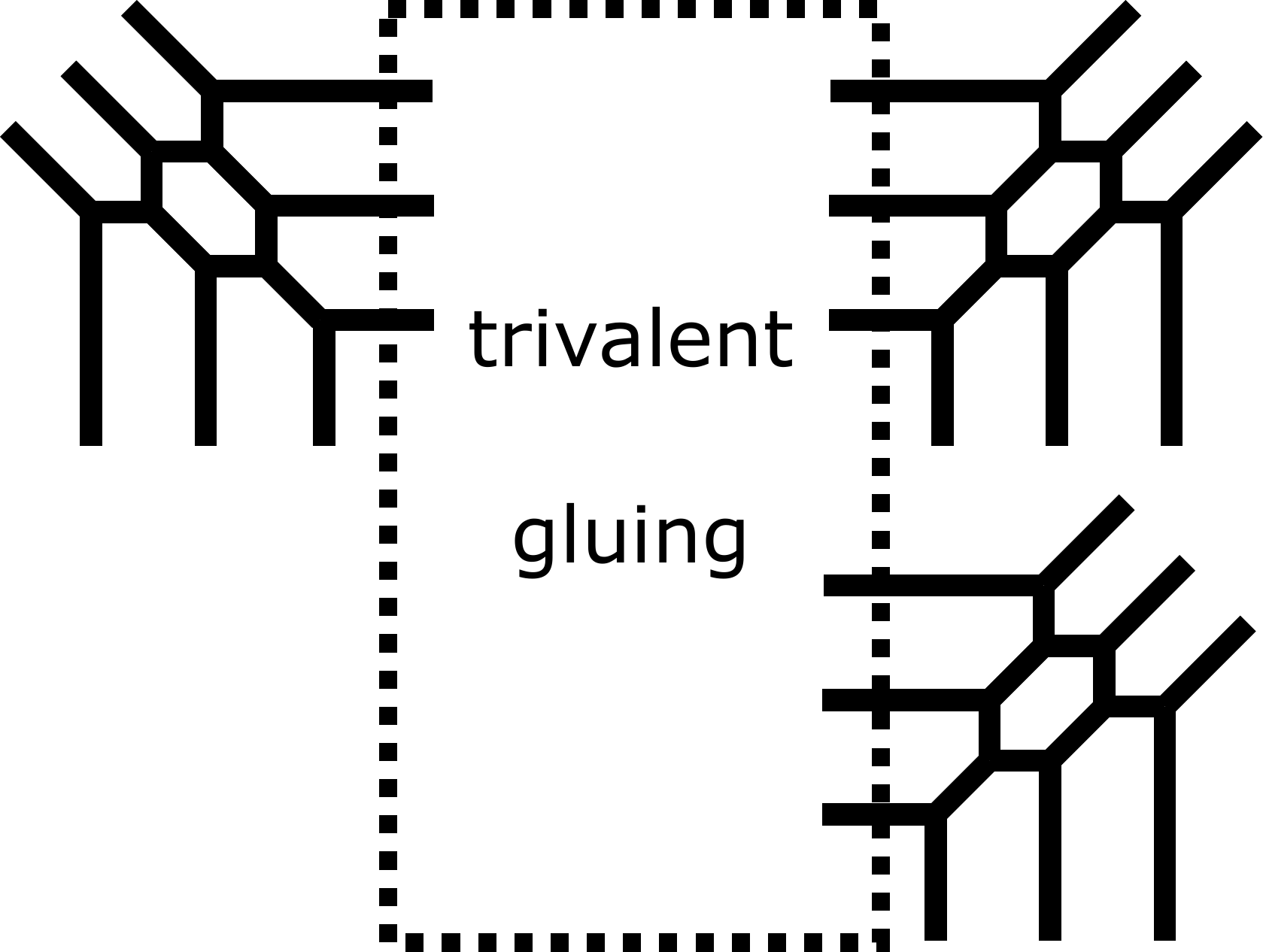}
\caption{The web diagram for the minimal $(E_6, E_6)$ conformal matter theory on a circle. The trivalent gluing is done so that the central $\SU(3)$ gauge node has the zero CS level. }
\label{fig:e6e6}
\end{figure}
The web diagram is depicted in Figure \ref{fig:e6e6}. The web diagram shows $\SU(3)^3 \times \SU(3)^3$ flavor symmetry explicitly which is a subgroup of $E_6 \times E_6$. To obtain the theory of $(E_6, E_6)_2$, we connect the two copies of the minimal $(E_6, E_6)$ conformal matter theories by gauging the diagonal subgroup of $E_6 \times E_6$ each of which comes from each minimal $(E_6, E_6)$ conformal matter theory. Then the resulting theory, which is $(E_6, E_6)_2$, on a circle is given by 
\be
\us\os1^{\fsp(0)^{(1)}}_{\left[\fe_6^{(1)}\right]}-\os3^{\fsu(3)^{(1)}}-\os1^{\fsp(0)^{(1)}}-\os6^{\fe_6^{(1)}}-\os1^{\fsp(0)^{(1)}}-\os3^{\fsu(3)^{(1)}}-\us\os1^{\fsp(0)^{(1)}}_{\left[\fe_6^{(1)}\right]}, \label{e6e6e6}
\ee
and its 5d gauge theory description is 
\be
\SU(2)_0  - \SU(4)_0 -{\overset{\text{\normalsize$\SU(2)_0$}}{\overset{\textstyle\vert}{\overset{\overset{\text{\normalsize$\SU(4)_0$}}{\textstyle\vert}}{\SU(6)_0}}}}- \SU(4)_0 - \SU(2)_0.\label{e6quiver}
\ee
The web diagram of the theory \eqref{e6e6e6} or equivalently \eqref{e6quiver} is obtained by applying the same gauging to the two copies of the diagram in Figure \ref{fig:e6e6} and it is depicted in Figure \ref{fig:e6e6e6}. 
The trivalent gauging part is fixed by requiring that the $\SU(6)$ gauge node has the CS level zero. 
The web diagram shows the $\SU(3)^3 \times \SU(3)^3$ flavor symmetry explicitly which is a subgroup of $E_6 \times E_6$.

\begin{figure}[t]
\centering
\subfigure[]{\label{fig:ge6e6}
\includegraphics[width=4cm]{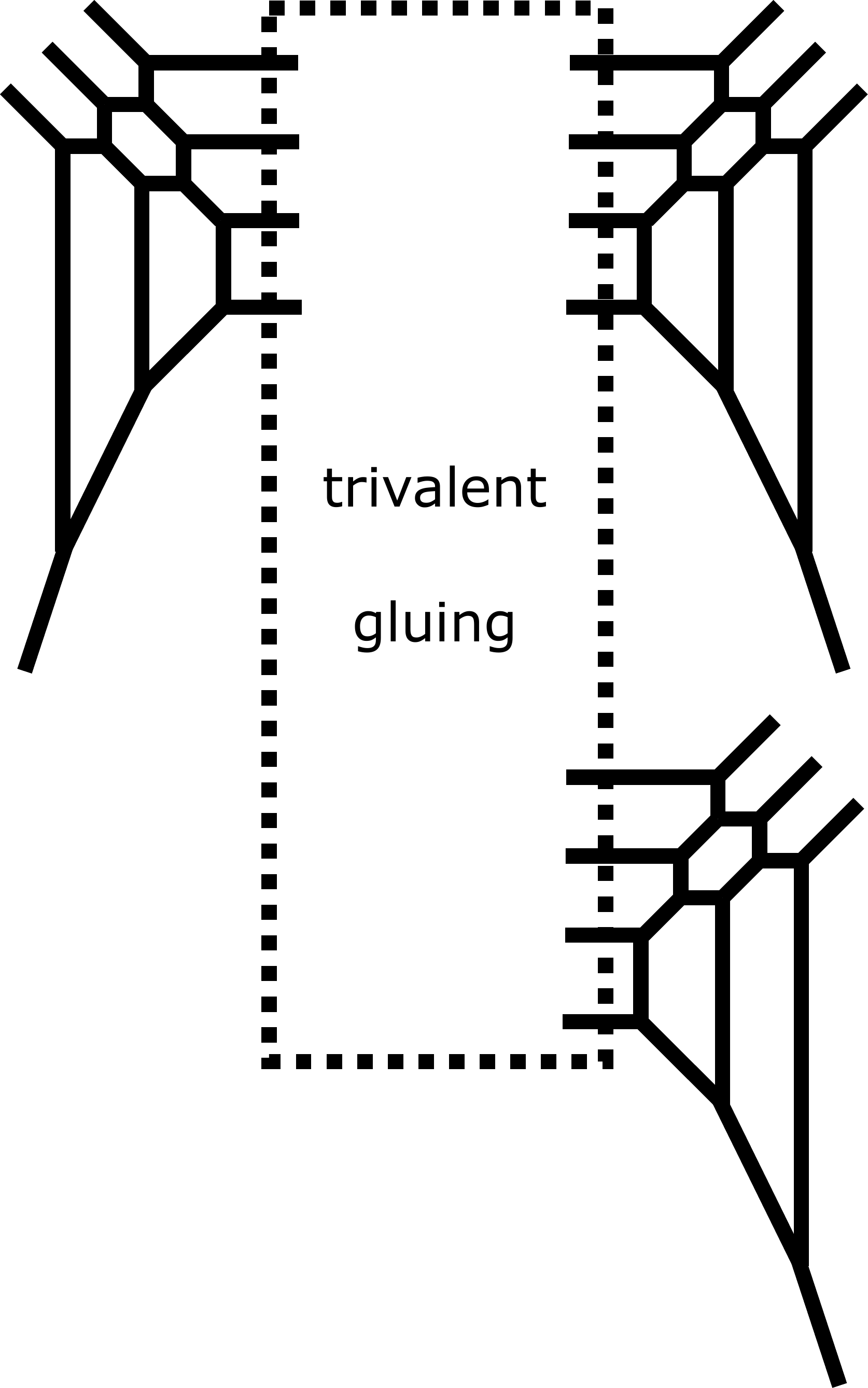}}
\hspace{1cm}
\subfigure[]{\label{fig:affinee6dynkin}
\includegraphics[width=6cm]{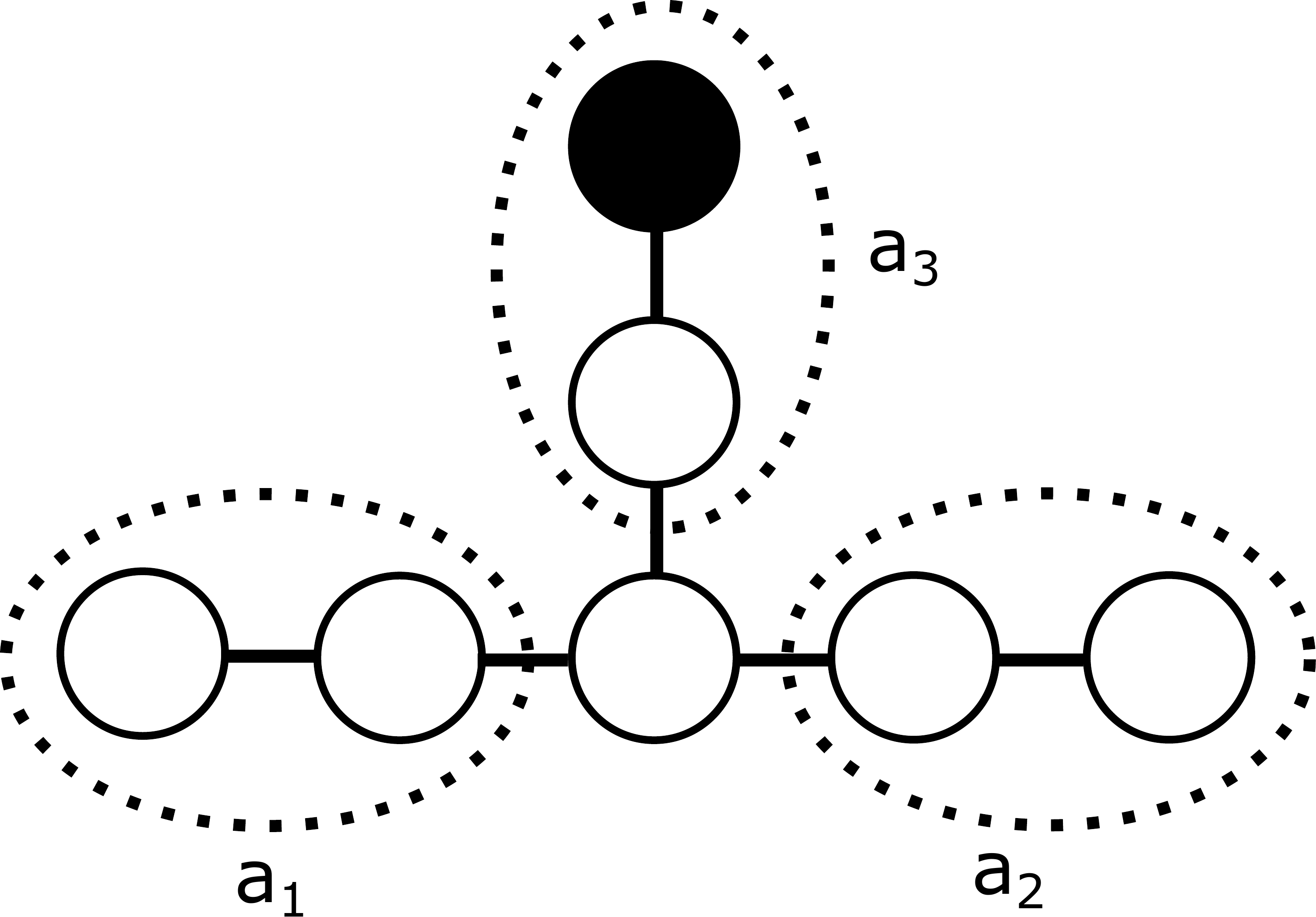}}
\caption{(a): The web diagram for the theory \eqref{ge6e6}, which is $(E_6, \underline{E_6})$ on $S^1$. (b): The affine $E_6$ Dynkin diagram with the affine node given in black. $a_i$ next to a dotted circle implies that one node in the dotted circle is Higgsed when $a_i = 2$ and two nodes in the dotted circle are Higgsed when $a_i = 3$.}
\label{fig:ge6e6e6}
\end{figure}

We are interested in Higgsings of the theory \eqref{e6e6e6}. Since we will make use of web diagrams for computing the parititon functions later, we focus on the Higgsings breaking $\SU(3)^3 \times \SU(3)^3$, which can be realized by the web diagram with several external 5-branes put on a 7-brane. Before studying such Higgsings we first consider the theory $(E_6, \underline{E_6})$ which is obtained by gauging one of the $E_6$ flavor symmetries of the minimal $(E_6, E_6)$ conformal matter theory and study Higgsings of the theory $(E_6, \underline{E_6})$ on a circle. 
Namely the theory we first consider is given by
\be
\us\os1^{\fsp(0)^{(1)}}_{\left[\fe_6^{(1)}\right]}-\os3^{\fsu(3)^{(1)}}-\os1^{\fsp(0)^{(1)}}-\os6^{\fe_6^{(1)}}, \label{ge6e6}
\ee
and it is realized on the web diagram in Figure \ref{fig:ge6e6}. The web diagram in Figure \ref{fig:ge6e6} shows the $\SU(3)^3$ flavor symmetry explicitly and we consider Higgsings which break the $\SU(3)^3$ flavor symmetry. The Higgsings can be labeled by $(a_1, a_2, a_3)$ where $a_i\; (i=1, 2, 3)$ are either $0, 2$ or $3$ ansd $a_i = n$ means the Higgsings which breaks $\SU(n)$ inside an $\SU(3)$. 

For identifying the theories after the Higgsings we take the same strategy which we have done in section \ref{sec:d4d4}. 
In the current cases, the Higgsings are characterized by nilpotent orbits of $\fe_6$. Since it is not associated to a classical group the nilpotent orbits are not any more labeled by parititons but we can still classify them by the B-C labels or weighted Dynkin diagrams. We associate $a_1, a_2$ and $a_3$ to nodes of the affine $E_6$ Dynkin diagram as in Figure \ref{fig:affinee6dynkin}.
Then we can compute the weighted Dynkin diagrams corresponding to the Higgsings $(a_1, a_2, a_3)$. The weighted Dynkin diagrams are related to the B-C labels and  
hence we can utilize the relation between the B-C labels and the Higgsed theories obtained in \cite{Heckman:2016ssk}. 
When a Higgsing breaks a symmetry associated to the affine node of the affine $E_6$ Dynkin diagram, then the Higgsing may give rise to a 5d KK theory from a twisted compactification of a 6d theory. 
The twist may be inferred from the matching between the numbers of Coulomb branch moduli and the mass parameters of the 5d KK theory from the twisted compactification and those from the web diagram. The result is summarized in Table \ref{tb:e6_1}. 
\begin{center}
\begin{longtable}{c|c|c|c}
\caption{\label{tb:e6_1} Higgsings associated to $\SU(3)^3$ of \eqref{ge6e6}. }\\
$\text{Higgsing}$&$\text{B-C Label}$&$\text{twist}$&$\text{theory}$
\\[3 pt]
\hline
\rule[-10pt]{0pt}{30pt}
$(0, 0, 0)$ &$ 0 $ &$1$ &$\quad \us\os1^{\fsp(0)^{(1)}}_{\left[\fe_6^{(1)}\right]}-\os3^{\fsu(3)^{(1)}}-\os1^{\fsp(0)^{(1)}}-\os6^{\fe_6^{(1)}}$
\\[10 pt]
\hline
\rule[-10pt]{0pt}{30pt}
$(2, 0, 0)$ &$ A_1$&$1$ &$ \us\os2^{\fsu(3)^{(1)}}_{\left[\fsu(6)^{(1)}\right]}-\os1^{\fsp(0)^{(1)}}-\os6^{\fe_6^{(1)}}$
\\[10 pt]
\hline
\rule[-10pt]{0pt}{30pt}
$(2, 2, 0)$ & $2A_1$&$1$ &$ \us\os2^{\fsu(2)^{(1)}}_{\left[\fso(7)^{(1)}\right]}-\us\os1^{\fsp(0)^{(1)}}_{\left[\fu(1)^{(1)}\right]}-\os6^{\fe_6^{(1)}}$
\\[10 pt]
\hline
\rule[-10pt]{0pt}{30pt}
$(2, 2, 2) $& $3A_1(ns)$&$1$ & $\us\os2^{\fsu(1)^{(1)}}_{\left[\fsu(2)^{(1)}\right]}-\us\os1^{\fsp(0)^{(1)}}_{\left[\fsu(3)^{(1)}\right]}-\os6^{\fe_6^{(1)}}$
\\[10 pt]
\hline
\rule[-10pt]{0pt}{30pt}
$(3, 0, 0) $& $A_2$&$1$ & $\us\os1^{\fsp(0)^{(1)}}_{\left[\fsu(3)^{(1)}\right]}-\os6^{\fe_6^{(1)}}-\us\os1^{\fsp(0)^{(1)}}_{\left[\fsu(3)^{(1)}\right]}$
\\[10 pt]
\hline
\rule[-10pt]{0pt}{30pt}
$(3, 2, 0) $& $A_2 + A_1$&$1$ &$ \us\os1^{\fsp(0)^{(1)}}_{\left[\fsu(3)^{(1)}\right]}-\us\os5^{\fe_6^{(1)}}_{\left[\fu(1)^{(1)}\right]}$
\\[10 pt]
\hline
\rule[-10pt]{0pt}{30pt}
$(3, 3, 0)$ &$ 2A_2 $&$1$& $\us\os1^{\fsp(0)^{(1)}}_{\left[\fg_2^{(1)}\right]}-\os5^{\ff_4^{(1)}}$
\\[10 pt]
\hline
\rule[-10pt]{0pt}{30pt}
$(3, 2, 2) $&$ A_2 + 2A_1 $&$1$&$ \us\os4^{\fe_6^{(1)}}_{\left[\fsu(2)^{(1)}\oplus\fu(1)^{(1)}\right]}$
\\[10 pt]
\hline
\rule[-10pt]{0pt}{30pt}
$(3, 3, 2)$ & $2A_2 + A_1$&$1$ & $\us\os4^{\ff_4^{(1)}}_{\left[\fsp(1)^{(1)}\right]}$
\\[10 pt]
\hline
\rule[-10pt]{0pt}{30pt}
$(3, 3, 3)$ &$ D_4(a_1)$&$Z_3$&$ \os4^{\fso(8)^{(3)}}$
\\[10 pt]
\hline
\end{longtable}
\end{center}

\begin{figure}[t]
\centering
\subfigure[]{\label{fig:puresu4k8v2}
\includegraphics[width=5cm]{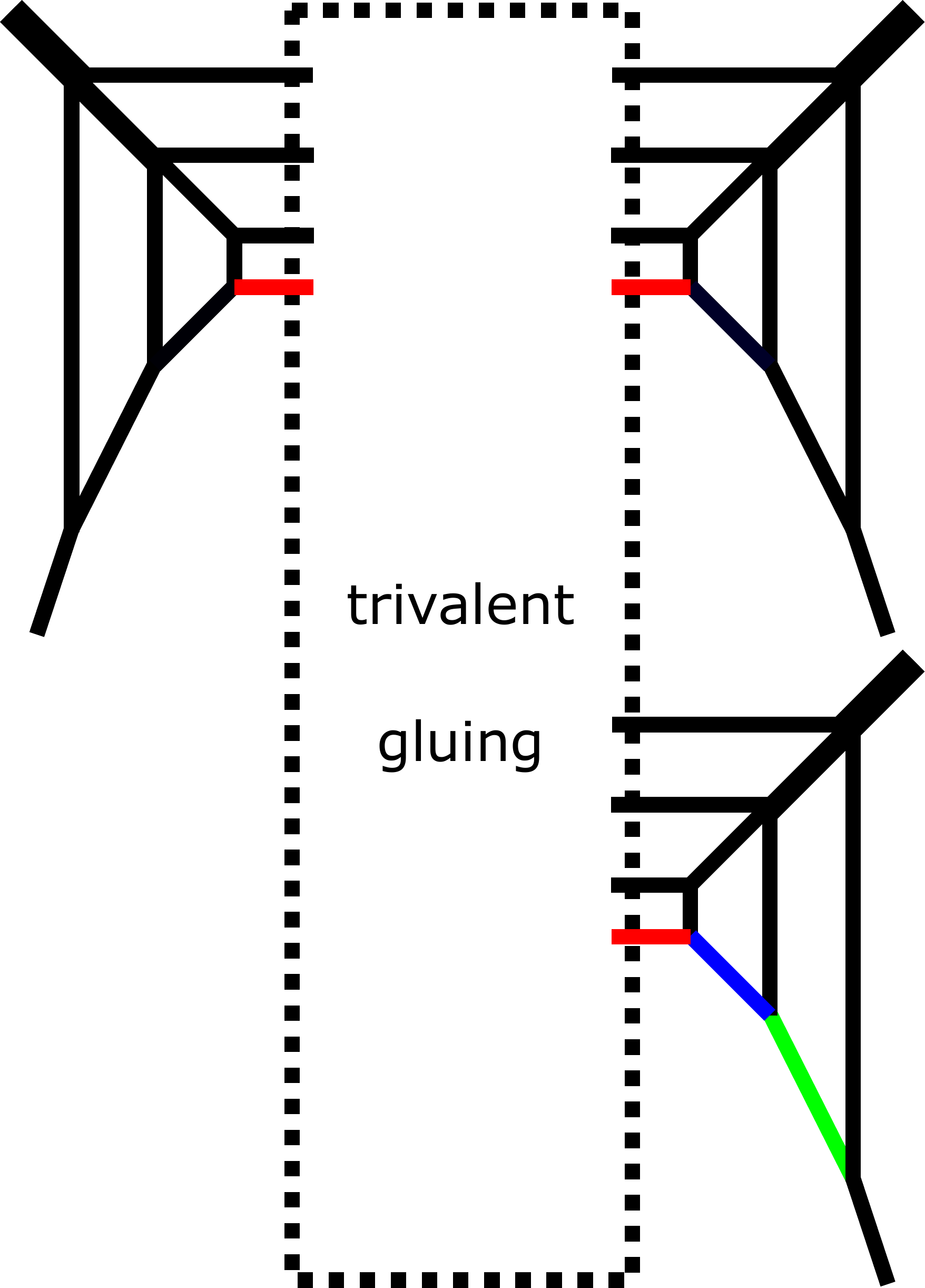}}
\hspace{1cm}
\subfigure[]{\label{fig:twistedd4}
\includegraphics[width=6cm]{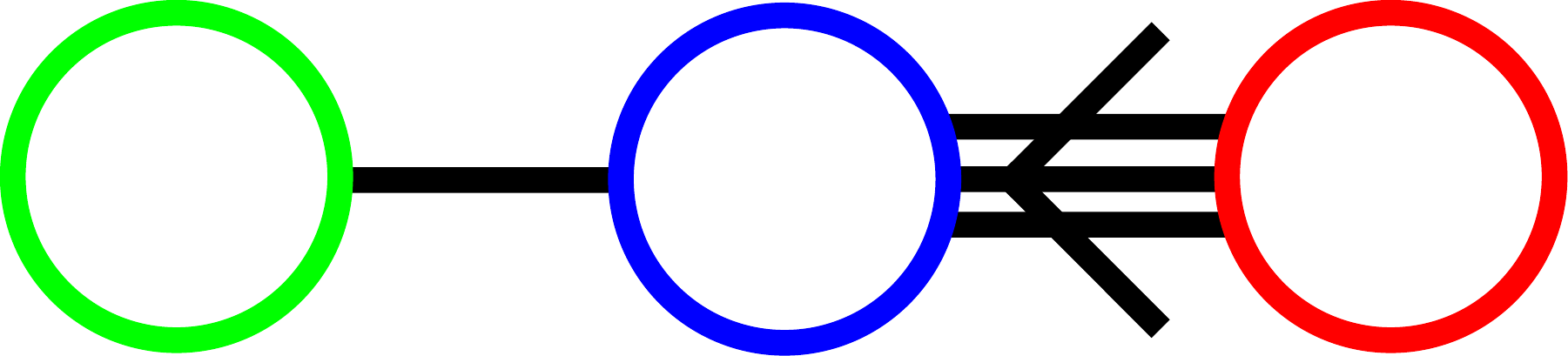}}
\caption{(a): The web diagram of the theory obtained after the Higgsing $(3, 3, 3)$. (b): The Dynkin diagram of the twisted affine Lie algebra $\fso(8)^{(3)}$ which is formed by the colored lines in Figure \ref{fig:puresu4k8v2}. The lines in red, blue and green in Figure \ref{fig:puresu4k8v2} correspond to the nodes in red, blue and green in this figure respectively. }
\label{fig:twistedd4web}
\end{figure}
Let us look at the case $(3, 3, 3)$. The web diagram after applying the Higgsing $(3, 3, 3)$ is depicted in Figure \ref{fig:puresu4k8v2}. In this case the Higgsing involves the affine node of the affine $E_6$ Dynkin diagram in Figure \ref{fig:affinee6dynkin}. Hence we expect that the Higgsed theory is a twisted compactification of a 6d theory. In 6d, the Higgsing associated to the B-C label $D_4(a_1)$ yields \cite{Heckman:2016ssk}
\be
\os4^{\fso(8)}.\label{pureso8}
\ee
Indeed, this theory has a discrete $S_3$ symmetry related to the outer automorphism of $\fso(8)$ and we can consider a compactification of the theory on a circle with a twist given by $Z_2$ or $Z_3$. From \eqref{so8twist0} and \eqref{so8twist} with $n=4$, the $Z_2$ twist gives 
\be
\os4^{\fso(8)^{(2)}},\label{pureso8twist0}
\ee
while the $Z_3$ twist yields
\be
\os4^{\fso(8)^{(3)}}.\label{pureso8twist}
\ee
The 5d KK theory of \eqref{pureso8twist0} has $4$ Coulomb branch moduli and the 5d KK theory of \eqref{pureso8twist} contains $3$ Coulomb branch moduli. On the other hand the web diagram in Figure \ref{fig:puresu4k8v2} has $3$ Coulomb branch moduli. This implies that the $(3, 3, 3)$ Higgsing gives \eqref{pureso8twist} which is a circle compactification of \eqref{pureso8} with the $Z_3$ twist. This theory is listed in the last row of Table \ref{tb:e6_1}. 

We can also determine the algebra on the $(-4)$-curve from the web diagram after the Higgsing labeled by $(3, 3, 3)$, which is depicted in Figure \ref{fig:puresu4k8v2}. By using the intersection \eqref{fScartan} it is possible to see that the intersection among the three fiber classes in red, blue and green in Figure \ref{fig:puresu4k8v2} forms the Dynkin diagram of the twisted affine Lie algebra $\fso(8)^{(3)}$ as in Figure \ref{fig:twistedd4}. 
It has been also argued that the theory \eqref{pureso8twist} gives rise to the 5d $\SU(4)$ gauge theory with the CS level $8$ \cite{Razamat:2018gro}. Hence the diagram in Figure \ref{fig:puresu4k8v2} realizes the 5d pure $\SU(4)$ gauge theory with the CS level $8$ and we will compute the partition function of the theory using this diagram in section \ref{sec:puremarginal}. 

\begin{figure}[t]
\centering
\subfigure[]{\label{fig:f4on4}
\includegraphics[width=5cm]{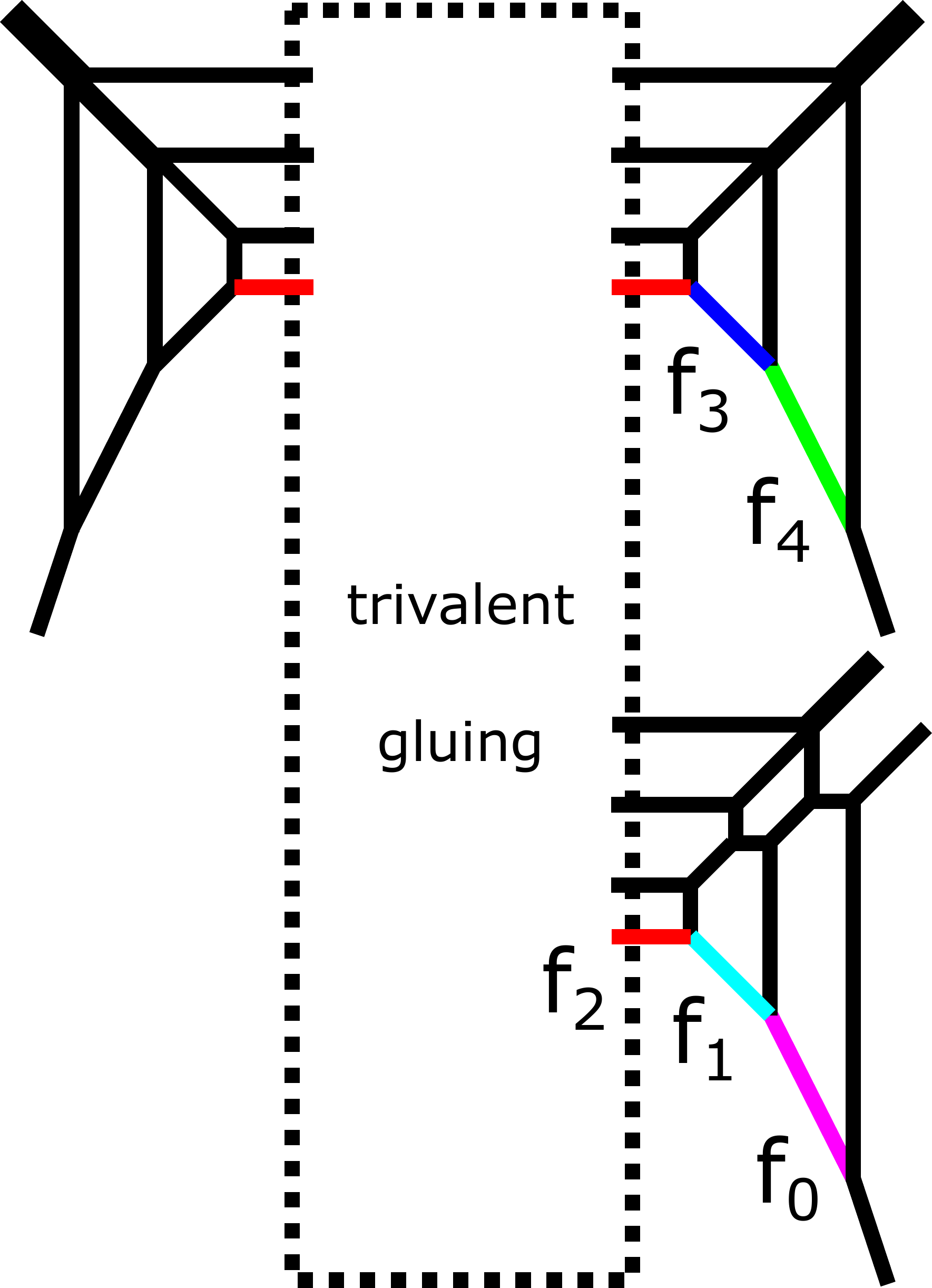}}
\hspace{1cm}
\subfigure[]{\label{fig:f4fiber}
\includegraphics[width=6cm]{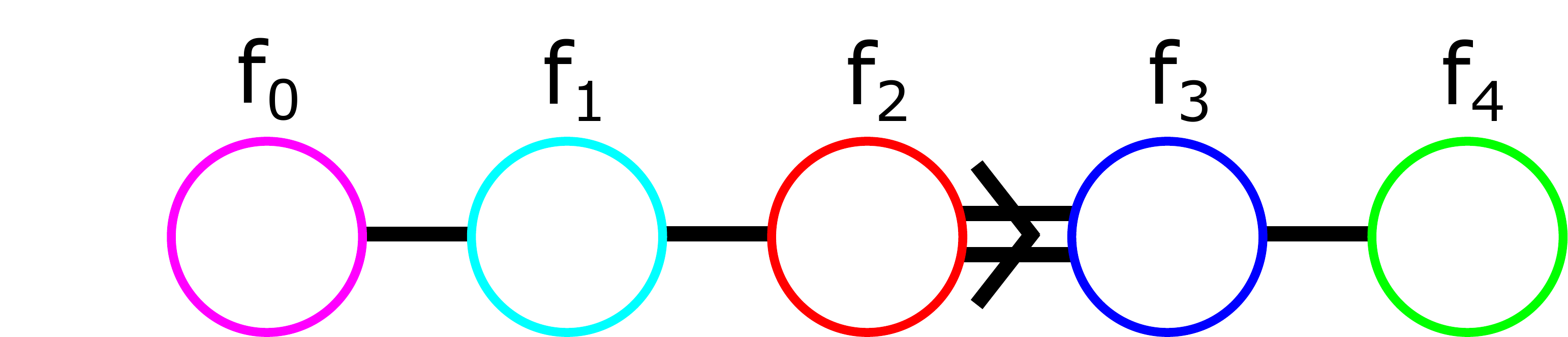}}
\caption{(a): The web diagram of the theory obtained after the Higgsing $(3, 3, 2)$. (b): The Dynkin diagram of the affine Lie algebra $\ff_4^{(1)}$ which is formed by the colored lines in Figure \ref{fig:puresu4k8v2}. The lines in magenta, cyan, red, blue and green in Figure \ref{fig:puresu4k8v2} correspond to the nodes in magenta, cyan, red, blue and green in this figure respectively. }
\label{fig:f4web}
\end{figure}

Let us also see the change of fibers in the process $(3, 3, 2) \to (3, 3, 3)$. The web diagram after the Higgsing $(3, 3, 2)$ is depicted in Figure \ref{fig:f4on4}. 
The fiber classes $f_0, f_1, f_2, f_3, f_4$ form the Dynkin diagram of $\ff_4^{(1)}$ as in Figure \ref{fig:f4fiber}. The the elliptic fiber of \eqref{fiberI} is given by
\be\label{f4t2}
f_{KK} = f_{\ff_4^{(1)}} = f_0 + 2f_1 + 3f_2 + 4f_3 + 2f_4.  
\ee
An M2-brane wrapped on \eqref{f4t2} gives a KK mode with mass $\frac{1}{R_{\text{6d}}}$. 
When we further apply the further Higgsing to the diagram in Figure \ref{fig:f4on4}, the diagram becomes the one in Figure \ref{fig:puresu4k8v2} with the identification of fibers $f_1 = f_3, f_0 = f_4$. Then \eqref{f4t2} is 
\be
f_{KK} = f_4 + 2f_3 + 3f_2 + 4f_3 + 2f_4 = 3f_{\fso(8)^{(3)}},
\ee
where
\be\label{fso83}
f_{\fso(8)^{(3)}} = f_2 + 2f_3 + f_4,
\ee
which is the fiber \eqref{fiberI} for the twisted affine Lie algebra $\fso(8)^{(3)}$. Then an M2-brane which is wrapped on the curve \eqref{fso83} gives a fractional KK mode with mass $\frac{1}{3R_{\text{6d}}}$.

\begin{figure}[t]
\centering
\includegraphics[width=5cm]{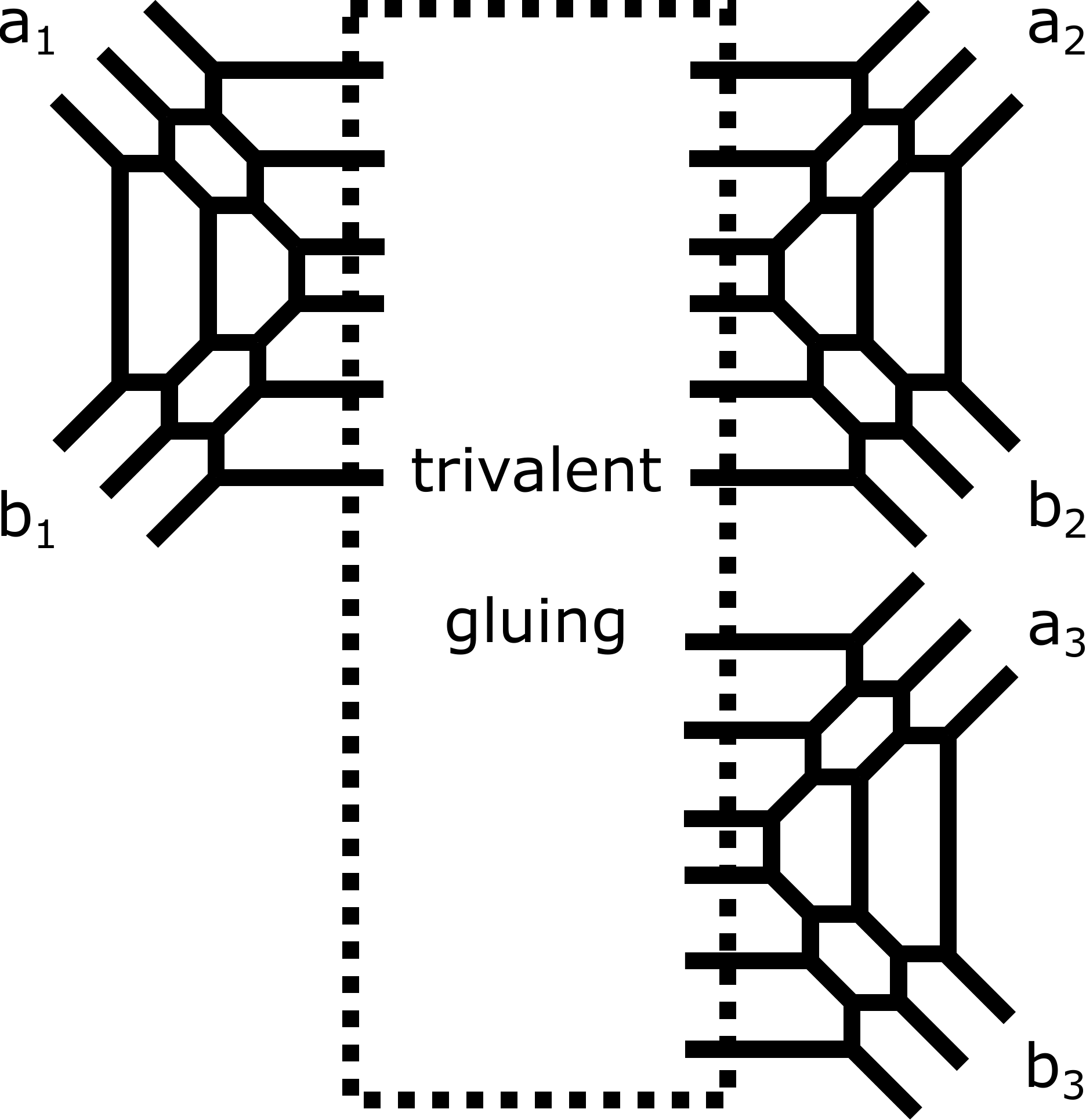}
\caption{The labeling of $[(a_1, a_2, a_3), (b_1, b_2, b_3)]$ for each $\SU(3)$ in the web diagram in Figure \ref{fig:e6e6e6}
}
\label{fig:e6e6e6higgs}
\end{figure}
We then consider Higgsings of the theory \eqref{e6e6e6}. Since we focus on Higgsings which break the $\SU(3)^3 \times \SU(3)^3$ flavor symmetry which can be explicitly seen from the diagram in Figure \ref{fig:e6e6e6}, the Higgsings can be labeled by $[(a_1, a_2, a_3), (b_1, b_2, b_3)]$ where $a_i, b_i\; (i=1, 2, 3)$ are either $0, 2$ or $3$. We associate $a_i, b_i\;(i=1, 2, 3)$ to the web diagram as in Figure \ref{fig:e6e6e6higgs} and  $a_i = n$ or $b_i = n$ means the Higgsing which breaks $\SU(n)$ inside the $\SU(3)$ corresponding to $a_i$ or $b_i$ in Figure \ref{fig:e6e6e6higgs}. When we apply a Higgsing of $(a_1, a_2, a_3)$ or $(b_1, b_2, b_3)$ to the theory \eqref{ge6e6}, then the Higgsed theory is the one listed in Table \ref{tb:e6_1}. When we consider a Higgsing $[(a_1, a_2, a_3), (b_1, b_2, b_3)]$ where either of the Higgsing $(a_1, a_2, a_3)$ or $(b_1, b_2, b_3)$ preserves the $\fe_6^{(1)}$ algebra, then the theory after the Higgsing $[(a_1, a_2, a_3), (b_1, b_2, b_3)]$ is given by combining the theory from $(a_1, a_2, a_3)$ and the one from $(b_1, b_2, b_3)$ by the rule,
\begin{equation}\label{e6combine}
\begin{split}
&\left[\cdots \os{n_3}^{\fg_3}-\os{n_2}^{\fg_2}-\os{\left(6-n_1\right)}^{\fe_6^{(1)}}\right] + \left[\os{\left(6-m_1\right)}^{\fg'_1}-\os{m_2}^{\fg'_2}-\os{m_3}^{\fg'_3}\cdots \right]\cr
&\to \cdots \os{n_3}^{\fg_3}-\os{n_2}^{\fg_2}-\os{\left(6-n_1-m_1\right)}^{\fg'_1}-\os{m_2}^{\fg'_2}-\os{m_3}^{\fg'_3},\cdots.
\end{split}
\end{equation}

When both $(a_1, a_2, a_3)$ and $(b_1, b_2, b_3)$ Higgs the $\fe_6^{(1)}$ algebra then we work out the corresponding Higgsed web diagrams explicitly to determine the algebra on each curve. The result of such cases is summarized in Table \ref{tb:e6_2}. Table \ref{tb:e6_2} also includes single node cases which we will use for computing the paritition functions later. 
\begin{center}
\begin{longtable}{c|c|c}
\caption{\label{tb:e6_2} Higgsings associated to $\SU(3)^3 \times \SU(3)^3$ of \eqref{e6e6e6}. The table includes the cases where both $(a_1, a_2, a_3)$ and $(b_1, b_2, b_3)$ Higgs the $\fe_6^{(1)}$ algebra in \eqref{e6e6e6}. The other cases which is not listed in this table can be obtained by the rule \eqref{e6combine}.}\\
$\text{Higgsing}$&$\text{Theory}$&$\text{A 5d description}$
\\[3 pt]
\hline
\rule[-10pt]{0pt}{30pt}
$
\begin{array}{c}
\left[(3, 3, 0), (3, 2, 0)_3\right]\\
\left[(3, 3, 0), (3, 3, 0)\right]'
\end{array}
$&$\quad\us\os1^{\fsp(0)^{(1)}}_{\left[\fg_2^{(1)}\right]}-\us\os4^{\ff_4^{(1)}}_{\left[\fsp(1)^{(1)}\right]}-\us\os1^{\fsp(0)^{(1)}}_{\left[\fg_2^{(1)}\right]}\quad$&$
\begin{array}{c}
\SU(2)_0 - \SU(4)_0 -\Sp(3)_0 - \left[\text{AS}\right]\\[5 pt]
\SU(2)_0  -{\overset{\overset{\text{\normalsize$\SU(2)_0$}}{\textstyle\vert}}{\SU(6)_{\pm 1}}}- \left[\text{AS}\right]\\[5 pt]
\left[\text{AS}\right] - \SU(4)_{\pm 1} - \Sp(3)_{\pi} - \SU(2)_0
\end{array}$
\\[10 pt]
\hline
\rule[-10pt]{0pt}{30pt}
$\left[(3, 3, 0), (3, 0, 3)_{2,1}\right]$&$\quad\us\os1^{\fsp(0)^{(1)}}_{\left[\fso(8)^{(3)}\right]}-\os4^{\fso(8)^{(3)}}-\us\os1^{\fsp(0)^{(1)}}_{\left[\fso(8)^{(3)}\right]}\quad$&$\SU(2)_0 -\Sp(3)_0  - \SU(2)_0$
\\[10 pt]
\hline
\rule[-10pt]{0pt}{30pt}
$
\begin{array}{c}
\left[(3, 3, 0), (3, 2, 2)_{3}\right]\\
\left[(3, 3, 0), (3, 3, 2)\right]'
\end{array}
$&$\quad\us\os3^{\ff_4^{(1)}}_{\left[\fsp(2)^{(1)}\right]}-\us\os1^{\fsp(0)^{(1)}}_{\left[\fg_2^{(1)}\right]}\quad$&$
\begin{array}{c}
[\text{AS}] - \SU(4)_{\pm 1} -\Sp(3)_0 - \left[\text{AS}\right]\\[5 pt]
\SU(2)_0 - \SU(6)_{\pm 1} - \left[2\text{AS}\right]
\end{array}$
\\[10 pt]
\hline
\rule[-10pt]{0pt}{30pt}
$
\begin{array}{c}
\left[(3, 3, 0), (3, 2, 3)_{2,1}\right]\\
\left[(3, 3, 3), (3, 2, 0)_3\right]\\
\left[(3, 3, 3), (3, 3, 0)\right]'
\end{array}
$&$\quad\us\os3^{\fso(8)^{(3)}}_{\left[\fsp(1)^{(1)}\right]}-\us\os1^{\fsp(0)^{(1)}}_{\left[\fso(8)^{(3)}\right]}\quad$&$[\text{AS}]  -\Sp(3)_0 - \SU(2)_0$
\\[10 pt]
\hline
\rule[-10pt]{0pt}{30pt}
$\left[(3, 2, 2), (3, 2, 2)_3\right]$&$\quad\us\os2^{\fe_6^{(1)}}_{\left[\fsu(4)^{(1)} \oplus \fu(1)^{(1)}\right]}\quad$&$
\begin{array}{c}
\Sp(2)_0  - \SU(6)_0- \left[2\text{AS}\right]\\[5 pt]
\Sp(2)_0 - \Sp(3)_0 - \Sp(2)_0
\end{array}$
\\[10 pt]
\hline
\rule[-10pt]{0pt}{30pt}
$
\begin{array}{c}
\left[(3, 3, 2), (3, 2, 2)_3\right]\\
\left[(3, 3, 2), (3, 3, 2)_{2, 1}\right]'
\end{array}
$&$\quad\us\os2^{\ff_4^{(1)}}_{\left[\fsp(3)^{(1)}\right]}\quad$&$
\begin{array}{c}
\SU(6)_{\pm 1} - \left[3\text{AS}\right]\\[5 pt]
\Sp(2)_0 - \Sp(3)_0 - \left[\text{AS}\right]
\end{array}$
\\[10 pt]
\hline
\rule[-10pt]{0pt}{30pt}
$
\begin{array}{c}
\left[(3, 3, 2), (3, 2, 3)_{2,1}\right]\\
\left[(3, 3, 3), (3, 3, 2)\right]'\\
\left[(3, 3, 3), (3, 3, 3)\right]''
\end{array}$&$\quad\us\os2^{\fso(8)^{(3)}}_{\left[\fsp(2)^{(1)} \right]}\quad$&$ \Sp(3)_0 - \left[2\text{AS}\right]$
\\[10 pt]
\hline
\end{longtable}
\end{center}
For $(a_1, a_2, a_3)$, we fixed $a_1 \geq a_2 \geq a_3$ and the other notation also follows from the one in Table \ref{tb:d4_2}.


\begin{figure}[t]
\centering
\includegraphics[width=10cm]{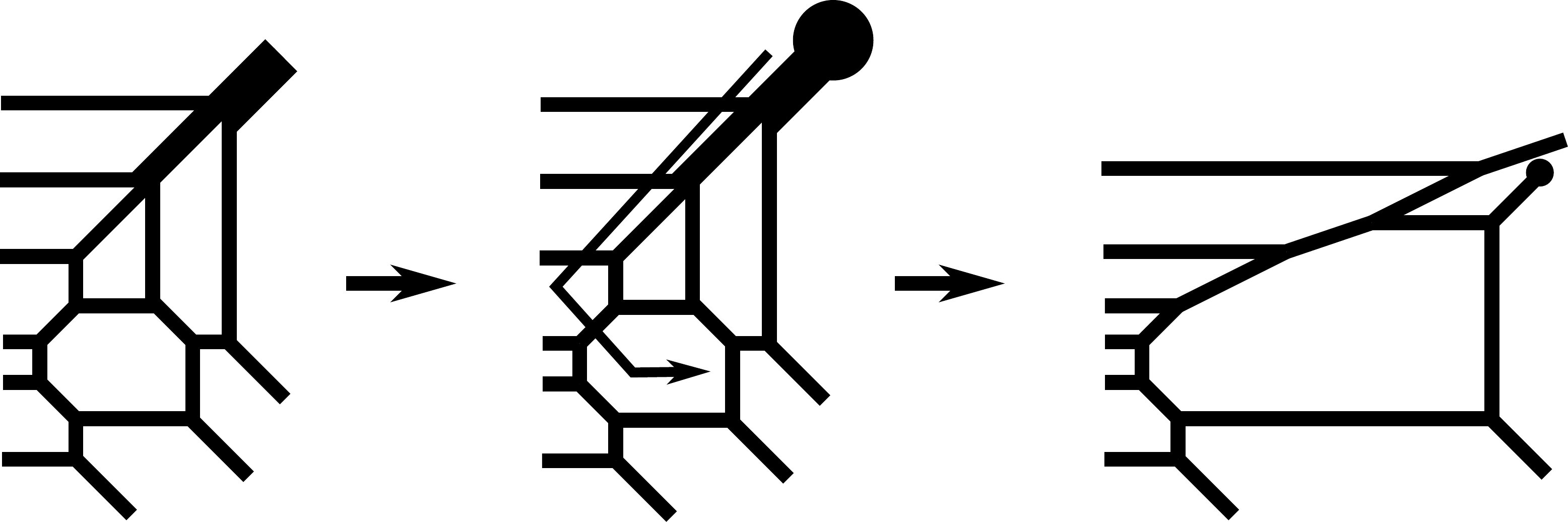}
\caption{A deformation of the diagram after the Higgsings which breaks $\SU(3)$ to a diagram of the $\SU(6)_1 - \SU(2)_0$ theory.
}
\label{fig:30higgs}
\end{figure}
Applying the Higgsings to the theory \eqref{e6quiver}, which is a gauge theory description of \eqref{e6e6e6}, we can obtain 5d gauge theory descriptions of the theories after the Higgsings. We can focus on one tail $\SU(6)_0 - \SU(4)_0 - \SU(2)_0$ of the quiver \eqref{e6quiver} and see how it changes by a Higgsing $(a_i, b_i)$ associated to the tail. An $\SU(2)$ Higgsing is the same as the one we considered in section \ref{sec:d4d4}. 
Namely $(a_i, b_i) = (2, 0), (0, 2)$ leads to $\SU(6)_0 - \SU(4)_{\pm 1} - \left[\text{AS}\right]$ and $(a_i, b_i) = (2, 2)$ yields $\SU(6)_0 - \Sp(2)_0$. 
For the Higgsing $(a_i, b_i) = (3, 0)$, which is given in the diagram on the left in Figure \ref{fig:30higgs}, moving a 7-brane yields the diagram on the right in Figure \ref{fig:30higgs}. The web diagram realizes the $\SU(6)_1 - \SU(2)_0$ theory. Similarly the Higgsing $(a_i, b_i) = (0, 3)$ gives rise to $\SU(6)_{-1} - \SU(2)_0$. Furthremore the Higgsing $(a_i, b_i) = (3, 2), (2, 3)$ give rise to $\SU(6)_1 - [\text{AS}]$ and $\SU(6)_{-1} - [\text{AS}]$ respectively \cite{Bergman:2015dpa}. Finally the Higgsing given by $(a_i, b_i) = (3, 3)$ yields the $\Sp(3)$ gauge theory with the zero discrete theta angle. Namely we have the following Higgsing chain, 
\begin{align}
\begin{tikzpicture}
\node at (0, 0) {$\SU(6)_0 - \SU(4)_0 - \SU(2)_0$};
\node at (-3, -0.75) {$a_i = 2, b_i = 0\; \swarrow$};
\node at (3, -0.75) {$\searrow\;a_i = 0, b_i = 2$};
\node at (-3, -1.5) {$\SU(6)_0 - \SU(4)_1 - \left[\text{AS}\right]$};
\node at (3, -1.5) {$\SU(6)_0 - \SU(4)_{-1} - \left[\text{AS}\right]$};
\node at (-6, -2.25) {$a_i = 3, b_i = 0\; \downarrow$};
\node at (6, -2.25) {$\downarrow\;a_i = 0, b_i = 3$};
\node at (0, -2.25) {$\searrow\;a_i = 2, b_i = 2\;\swarrow$};
\node at (-4.5, -3) {$\SU(6)_1 - \SU(2)_0$};
\node at (4.5, -3) {$\SU(6)_{-1} - \SU(2)_0$};
\node at (0, -3) {$\SU(6)_0 - \Sp(2)_0$};
\node at (-2.5, -3.75) {$\searrow\;a_i=3, b_i = 2\;\swarrow$};
\node at (2.5, -3.75) {$\searrow\;a_i = 2, b_i = 3\;\swarrow$};
\node at (-2.5, -4.5) {$\SU(6)_1 - \left[\text{AS}\right]$};
\node at (2.5, -4.5) {$\SU(6)_{-1} - \left[\text{AS}\right]$};
\node at (0, -5.25) {$\searrow\;a_i=3, b_i = 3\;\swarrow$};
\node at (0, -6) {$\Sp(3)_0$};
\end{tikzpicture}\label{33higgs}
\end{align}
We can also use 
\be\label{sp3v1}
\SU(6)_{k} - \SU(4)_0 - \SU(2)_0 \qquad  \substack{(3, 3)\;\text{Higgs}\\\longrightarrow}  \qquad \Sp(3)_{\theta = |k|\pi \; (\text{mod}\;2\pi)},
\ee
and 
\be\label{sp3v2}
[\text{AS}]-\SU(6)_{k} - \SU(4)_0 - \SU(2)_0 \qquad  \substack{(3, 3)\;\text{Higgs}\\\longrightarrow}  \qquad [\text{AS}] - \Sp(3)_{\theta = |k+1|\pi \; (\text{mod}\;2\pi)},
\ee
for some small values of $k$. 
Then we can obtain a 5d gauge theory description for each theory given by the Higgsing labeled by $[(a_1, a_2, a_3), (b_1, b_2, b_3)]$ from the Higgsing chain in \eqref{33higgs} as well as \eqref{sp3v1} and \eqref{sp3v2}. For the theories listed in Table \ref{tb:e6_2}, their 5d gauge theory descriptions are displayed in the third column in Table \ref{tb:e6_2}.

\subsection{$(E_7, E_7)$ conformal matter}
\label{sec:e7e7}

\begin{figure}[t]
\centering
\includegraphics[width=6cm]{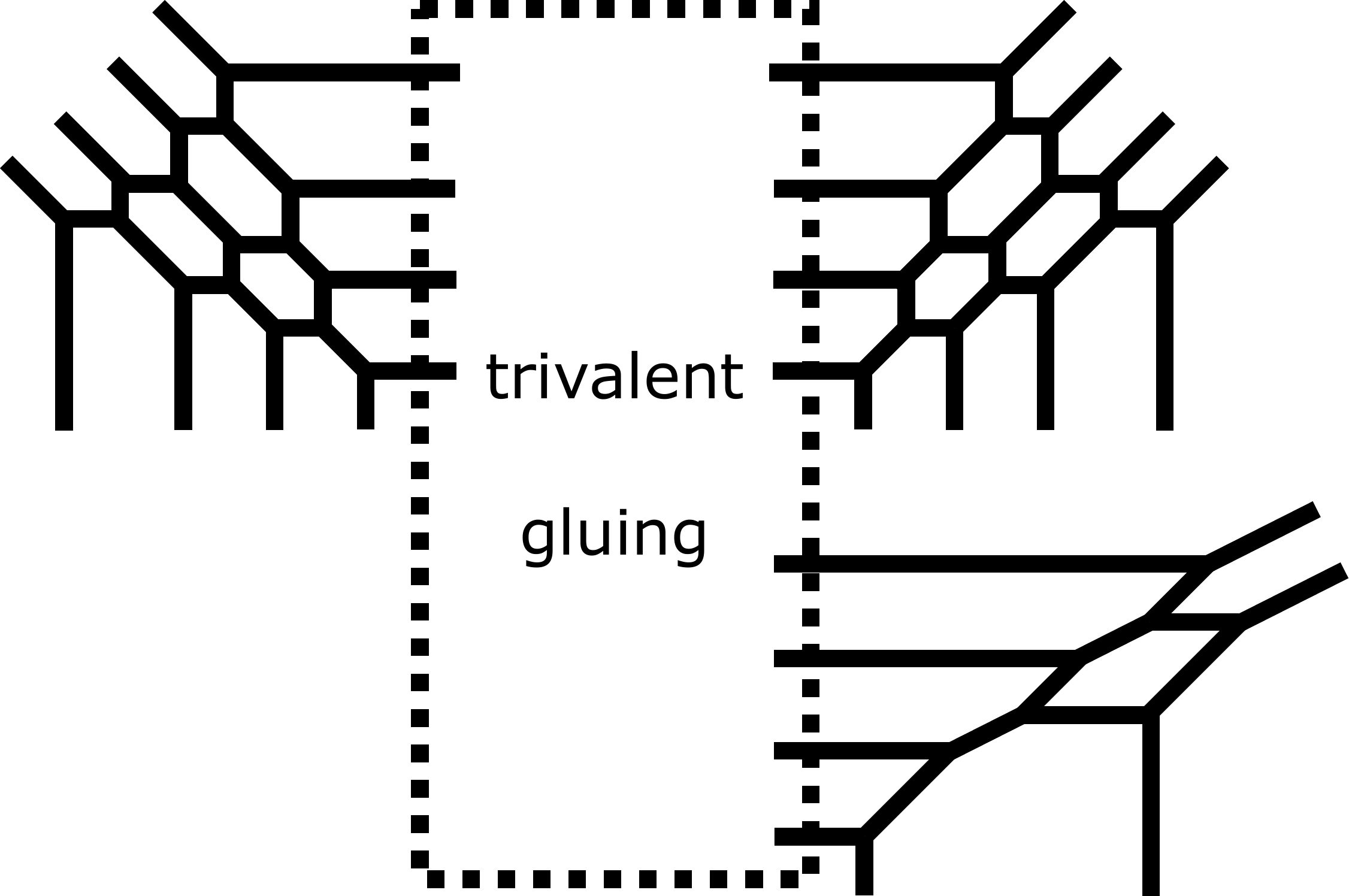}
\caption{The web diagram for the minimal $(E_7, E_7)$ conformal matter theory on a circle. The trivalent gluing is done so that the central $\SU(4)$ gauge node has the zero CS level. }
\label{fig:e7e7}
\end{figure}
\begin{figure}[t]
\centering
\includegraphics[width=6cm]{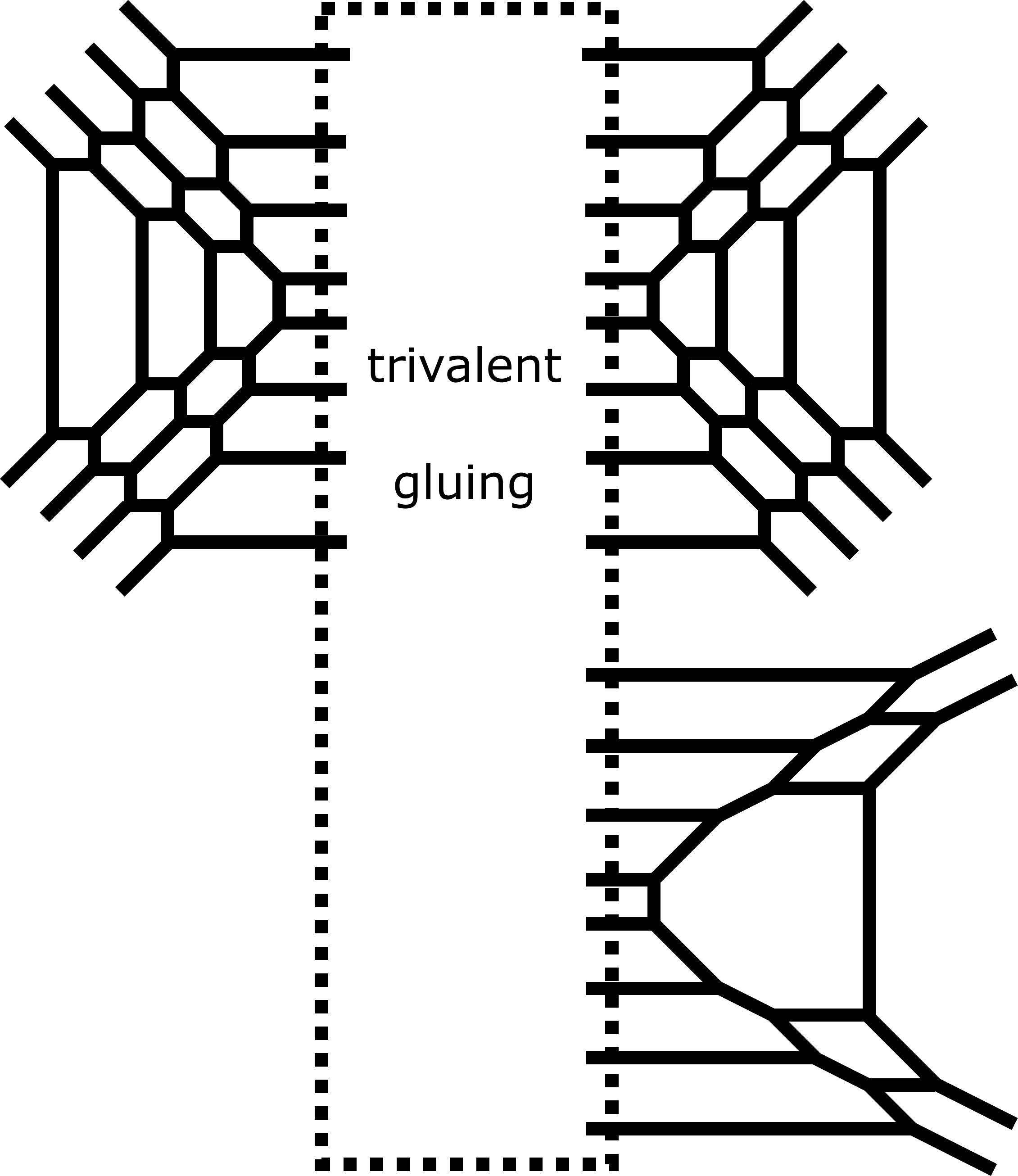}
\caption{The web diagram for the theory \eqref{e7e7e7} or \eqref{e7quiver}, which is $(E_7, E_7)_2$ on $S^1$. }
\label{fig:e7e7e7}
\end{figure}

Lastly we consdier the theory $(E_7, E_7)_2$ which is obtained by gauging two minimal $(E_7, E_7)$ conformal matter theories and study Higgsings of the theory $(E_7, E_7)_2$ on a circle. The basic building block is the minimal $(E_7, E_7)$ conformal matter theory and the theory $(E_7, E_7)_1$ on $S^1$ is given by\footnote{Although a dotted line is used between a $(-2)$-curve with $\fsu(2)^{(-1)}$ and the $(-3)$-curve with $\fso(7)^{(1)}$ in \eqref{e7e7}, we do not have an ambiguity in this case since there is no matter in the vector representation for the theory characterized by $\fso(7)^{(1)}$ on the $(-3)$-curve.}
\begin{align}
\us\os1^{\fsp(0)^{(1)}}_{\left[\fe_7^{(1)}\right]}-\os2^{\fsu(2)^{(1)}}\hdashrule[0.5ex]{0.5cm}{1pt}{0.5mm}\os3^{\fso(7)^{(1)}}\hdashrule[0.5ex]{0.5cm}{1pt}{0.5mm}\os2^{\fsu(2)^{(1)}}-\us\os1^{\fsp(0)^{(1)}}_{\left[\fe_7^{(1)}\right]}.\label{e7e7}
\end{align} 
The theory $(E_7, E_7)_1$ on a circle has a 5d gauge theory description, which is the following affine $E_7$ Dynkin quiver theory, 
\be
\SU(1)  - \SU(2) - \SU(3) -{\overset{\overset{\text{\normalsize$\SU(2)_0$}}{\textstyle\vert}}{\SU(4)_0}}- \SU(3)_0 - \SU(2) - \SU(1).\label{min.affineE7}
\ee
Again one $\SU(1)$ node represents two flavors attached to the adjacent $\SU(2)$ gauge node and the discrete theta angle of such $\SU(2)$ gauge node is unphysical. 
The 5d theory \eqref{min.affineE7} may be realized on a web diagram with a trivalent $\SU(4)$ gauging. The web diagram is depicted in Figure \ref{fig:e7e7}. The two upper diagrams are the web diagrams for the $T_4$ theory. From the diagram, the external legs extending in the upper direction and the lower direction explicitly show two copies of an $\SU(4) \times \SU(4) \times \SU(2)$ flavor symmetry, which is a subgroup of $E_7 \times E_7$. We can again construct the theory $(E_7, E_7)_2$ 
by combining two minimal $(E_7, E_7)$ conformal matter theories. The theory $(E_7, E_7)_2$ on a circle is 
\begin{align}
\us\os1^{\fsp(0)^{(1)}}_{\left[\fe_7^{(1)}\right]}-\os2^{\fsu(2)^{(1)}}\hdashrule[0.5ex]{0.5cm}{1pt}{0.5mm}\os3^{\fso(7)^{(1)}}\hdashrule[0.5ex]{0.5cm}{1pt}{0.5mm}\os2^{\fsu(2)^{(1)}}-\os1^{\fsp(0)^{(1)}}-\os8^{\fe_7^{(1)}}-\os1^{\fsp(0)^{(1)}}-\os2^{\fsu(2)^{(1)}}\hdashrule[0.5ex]{0.5cm}{1pt}{0.5mm}\os3^{\fso(7)^{(1)}}\hdashrule[0.5ex]{0.5cm}{1pt}{0.5mm}\os2^{\fsu(2)^{(1)}}\us\os1^{\fsp(0)^{(1)}}_{\left[\fe_7^{(1)}\right]}.\label{e7e7e7}
\end{align} 
This theory \eqref{e7e7e7} also has a 5d gauge theory description given by 
\be
\SU(2)_0  - \SU(4)_0 - \SU(6)_0 -{\overset{\overset{\text{\normalsize$\SU(4)_0$}}{\textstyle\vert}}{\SU(8)_0}}- \SU(6)_0 - \SU(4)_0 - \SU(2)_0,\label{e7quiver}
\ee
and the web diagram realizing \eqref{e7quiver} is depicted in Figure \ref{fig:e7e7e7}. The trivalent $\SU(8)$ gauging is fixed by requiring that the $\SU(8)$ gauge theory has no CS level.
The web diagram in Figure \ref{fig:e7e7e7} shows the $\left(\SU(4)^2 \times \SU(2)\right) \times \left(\SU(4)^2 \times \SU(2)\right)$ flavor symmetry explicitly.

\begin{figure}[t]
\centering
\subfigure[]{\label{fig:ge7e7}
\includegraphics[width=4cm]{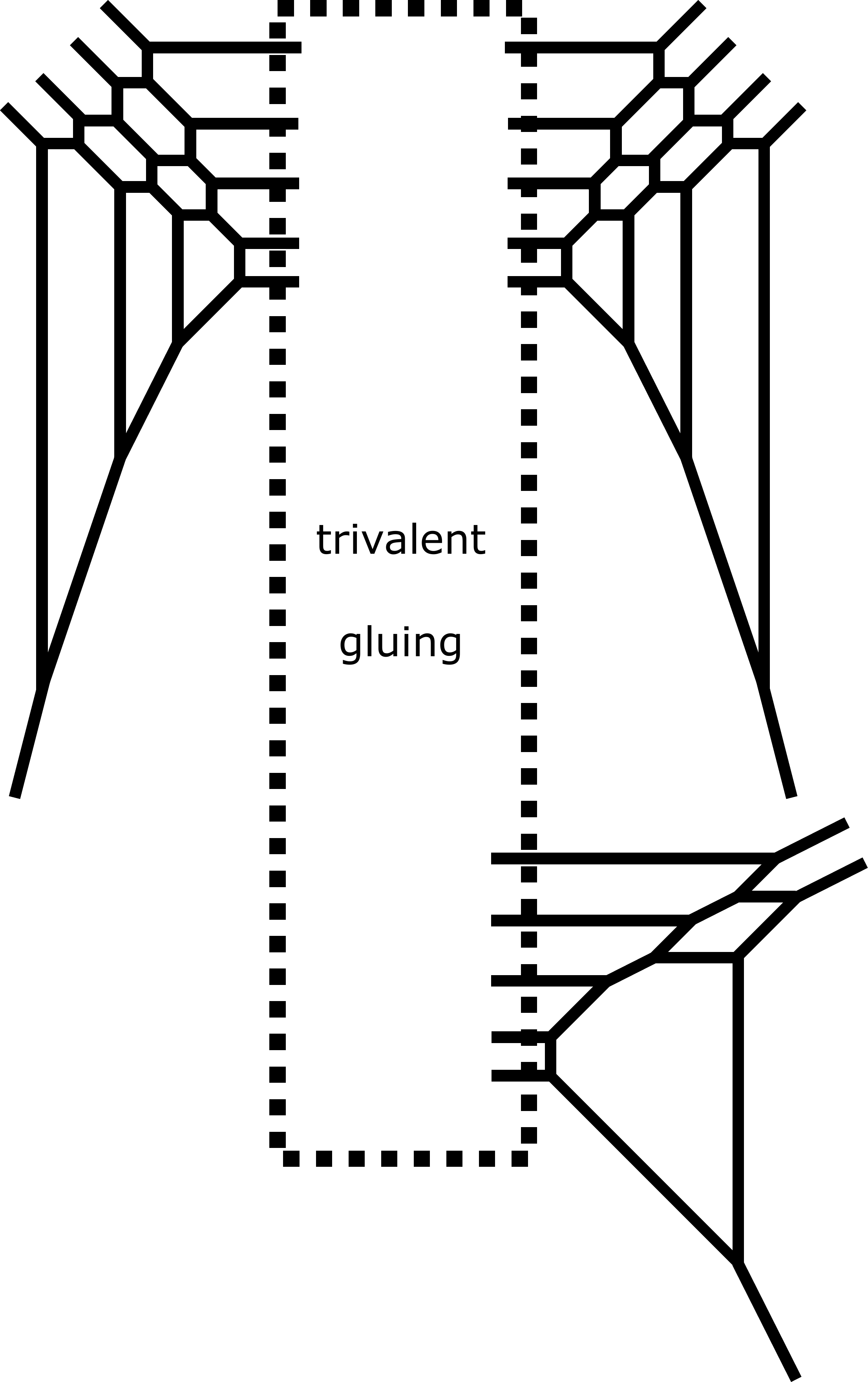}}
\hspace{1cm}
\subfigure[]{\label{fig:affinee7dynkin}
\includegraphics[width=7cm]{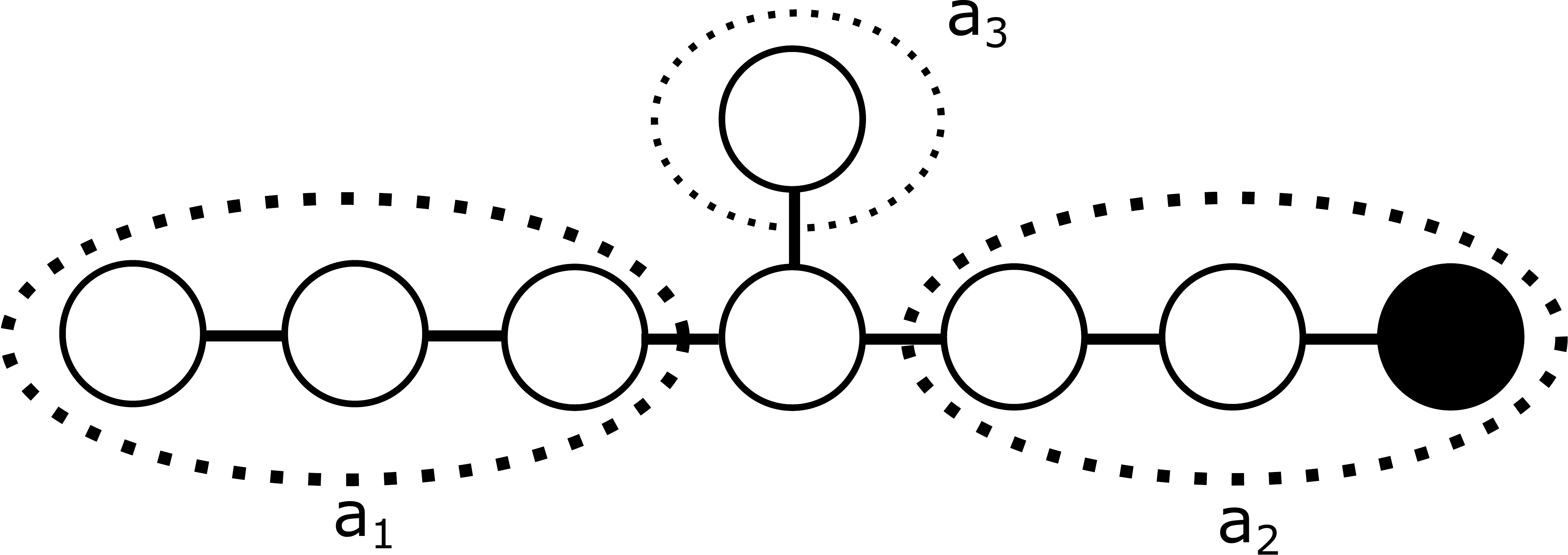}}
\caption{(a): The web diagram for the theory \eqref{ge7e7}, which is $(E_7, \underline{E_7})$ on $S^1$. (b): The affine $E_7$ Dynkin diagram with the affine node given in black. $a_i$ next to a dotted circle implies that $k$ consecutive nodes in the dotted line are Higgsed when $a_i = k+1$. When $a_1 = (2, 2)$ ($a_2 = (2, 2)$) then this implies a Higgsing of the two nodes except for the middle one in the three nodes circled by the dotted line next to $a_1$ ($a_2$). }
\label{fig:ge7e7e7}
\end{figure}

We are interested in the theories which are obtained after Higgsing \eqref{e7e7e7}. Before studying Higgsings of the theory \eqref{e7e7e7}, it is again useful to study Higgsings of the theory $(E_7, \underline{E_7})$ on $S^1$ given by,
\be
\us\os1^{\fsp(0)^{(1)}}_{\left[\fe_7^{(1)}\right]}-\os2^{\fsu(2)^{(1)}}\hdashrule[0.5ex]{0.5cm}{1pt}{0.5mm}\os3^{\fso(7)^{(1)}}\hdashrule[0.5ex]{0.5cm}{1pt}{0.5mm}\os2^{\fsu(2)^{(1)}}-\os1^{\fsp(0)^{(1)}}-\os8^{\fe_7^{(1)}}. \label{ge7e7}
\ee
The web diagram of this theory is depicted in Figure \ref{fig:ge7e7}. 
The flavor symmetry associated to the lower external legs are gauged and we are left with an $E_7$ flavor symmetry out of which an $\SU(4) \times \SU(4) \times \SU(2)$ flavor symmetry can be explicitly seen from the diagram. We label Higgsings which break the $\SU(4) \times \SU(4) \times \SU(2)$ flavor symmetry by $(a_1, a_2, a_3)$. $a_1, a_2, a_3$ are associated to the $\SU(4), \SU(4), \SU(2)$ of the $\SU(4) \times \SU(2) \times \SU(2)$ flavor symmetry, and $a_1, a_2$ are either $0, 2, 3, 4$ or $(2, 2)$ and $a_3$ is either $0$ or $2$. $a_i = n$ means an $\SU(n)$ part is broken and $a_1 =(2, 2)$ or $a_2 =(2, 2)$ corresponds to the Higgsing which breaks an $\SU(2) \times \SU(2)$ inside an $\SU(4)$. 

We can identify the theories after a Higgsing labeled by $(a_1, a_2, a_3)$ by making use of the result in \cite{Heckman:2016ssk} using the same stragety in section \ref{sec:d4d4} and \ref{sec:e6e6}. 
We first associate $(a_1, a_2, a_3)$ to the nodes in the affine $E_7$ Dynkin diagram as in Figure \ref{fig:affinee7dynkin} and 
then compute the weighted Dynkin diagram 
corresponding to the Higgsing $(a_1, a_2, a_3)$. Then using the relation between the weighted Dynkin diagrams and the B-C labels we can identify the B-C labels of the Higgsings. In \cite{Heckman:2016ssk}, various Higgsings of a theory obtained by connecting $(E_7, E_7)$ minimal conformal matter theories have been determined and they are classified by nilpotent orbits of $\fe_7$ or the B-C labels. 
From the relation between the Higgsing $(a_1, a_2, a_3)$ and the B-C label, we can associate the Higgsing $(a_1, a_2, a_3)$ to the corresponding Higgsed theory obtained in \cite{Heckman:2016ssk}. Again some of the Higgsings 
break a symmetry corresponding to the affine node of the affine $E_7$ Dynkin diagram and such a case may correspond to a twisted compactifications of a 6d theory when the original 6d theory has a discrete symmetry discussed in section \ref{sec:twist}. 
The twist may be inferred from the matching between the numbers of Coulomb branch moduli and the mass parameters of the 5d KK theory from the twisted compactification and those from the web diagram. 
We summarize the result in Table \ref{tb:e7_1}. 
\begin{center}
\begin{longtable}{c|c|c|c}
\caption{ \label{tb:e7_1} Higgsings associated to $\SU(4)^2\times \SU(2)$ of \eqref{ge7e7}. $N_f = \frac{1}{2}$ implies a half-hypermultiplet in the fundamental representation. }\\
$\text{Higgsing}$&$\text{B-C Label}$&$\text{twist}$&$\text{theory}$
\\[3 pt]
\hline
\rule[-10pt]{0pt}{30pt}
$(0, 0, 0)$ &$ 0$&$1$ &$\us\os1^{\fsp(0)^{(1)}}_{\left[\fe_7^{(1)}\right]}-\os2^{\fsu(2)^{(1)}}\hdashrule[0.5ex]{0.5cm}{1pt}{0.5mm}\os3^{\fso(7)^{(1)}}\hdashrule[0.5ex]{0.5cm}{1pt}{0.5mm}\os2^{\fsu(2)^{(1)}}-\os1^{\fsp(0)^{(1)}}-\os8^{\fe_7^{(1)}}$
\\[10 pt]
\hline
\rule[-10pt]{0pt}{30pt}
$(2, 0, 0), (0,0, 2)$ &$ A_1$&$1$ &$\us\os1^{\fsp(1)^{(1)}}_{\left[\fso(12)^{(1)}\right]}\hdashrule[0.5ex]{0.5cm}{1pt}{0.5mm}\os3^{\fso(7)^{(1)}}\hdashrule[0.5ex]{0.5cm}{1pt}{0.5mm}\os2^{\fsu(2)^{(1)}}-\os1^{\fsp(0)^{(1)}}-\os8^{\fe_7^{(1)}}$
\\[10 pt]
\hline
\rule[-10pt]{0pt}{30pt}
$(2, 2, 0), ((2,2), 0, 0), (0, 2, 2)$ &$ 2A_1$&$1$ &$ \us\os1^{\fsp(0)^{(1)}}_{\left[\fso(9)^{(1)}\right]}\hdashrule[0.5ex]{0.5cm}{1pt}{0.5mm}\us\os3^{\fso(7)^{(1)}}_{\left[\fsp(1)^{(1)}\right]}\hdashrule[0.5ex]{0.5cm}{1pt}{0.5mm}\os2^{\fsu(2)^{(1)}}-\os1^{\fsp(0)^{(1)}}-\os8^{\fe_7^{(1)}}$
\\[10 pt]
\hline
\rule[-10pt]{0pt}{30pt}
$(2, 2, 2), ((2,2), 2, 0)$ & $\left(3A_1\right)' $&$1$&$\us\os2^{\fso(7)^{(1)}}_{\left[\fsp(1)^{(1)} \oplus \fsp(3)^{(1)}\right]}\hdashrule[0.5ex]{0.5cm}{1pt}{0.5mm}\os2^{\fsu(2)^{(1)}}-\os1^{\fsp(0)^{(1)}}-\os8^{\fe_7^{(1)}}$
\\[10 pt]
\hline
\rule[-10pt]{0pt}{30pt}
$((2,2), 0, 2)$ & $\left(3A_1\right)''$&$1$ &$\us\os3^{\fg_2^{(1)}}_{\left[\ff_4^{(1)}\right]}-\os2^{\fsu(2)^{(1)}}-\os1^{\fsp(0)^{(1)}}-\os8^{\fe_7^{(1)}}$
\\[10 pt]
\hline
\rule[-10pt]{0pt}{30pt}
$((2,2), 2, 2) $& $4A_1$ &$1$&$ \us\os2^{\fg_2^{(1)}}_{\left[\fsp(3)^{(1)}\right]}-\os2^{\fsu(2)^{(1)}}-\os1^{\fsp(0)^{(1)}}-\os8^{\fe_7^{(1)}}$
\\[10 pt]
\hline
\rule[-10pt]{0pt}{30pt}
$(3,0,0) $&$ A_2 $&$1$&$ \us\os2^{\fsu(4)^{(1)}}_{\left[\fsu(6)^{(1)}\right]}-\os2^{\fsu(2)^{(1)}}-\os1^{\fsp(0)^{(1)}}-\os8^{\fe_7^{(1)}}$
\\[10 pt]
\hline
\rule[-10pt]{0pt}{30pt}
$((2,2),(2,2),0) $&$ A_2$&$Z_2$&$ \us\os2^{\fsu(4)^{(2)}}_{\left[\fsu(6)^{(2)}\right]}-\os2^{\fsu(2)^{(1)}}-\os1^{\fsp(0)^{(1)}}-\os8^{\fe_7^{(1)}}$
\\[10 pt]
\hline
\rule[-10pt]{0pt}{30pt}
$(3, 2, 0), (3, 0, 2) $&$ A_2 + A_1 $&$1$ &$
\begin{array}{c}
\us\os2^{\fsu(3)^{(1)}}_{\left[\fsu(4)^{(1)}\right]}-\us\os2^{\fsu(2)^{(1)}}_{\left[N_f = 1\right]}-\os1^{\fsp(0)^{(1)}}-\os8^{\fe_7^{(1)}}\\[10pt]
\text{flavor:}\quad \fsu(4)^{(1)} \oplus \fu(1)^{(1)}
\end{array}$
\\[10 pt]
\hline
\rule[-10pt]{0pt}{30pt}
$((2,2), (2,2), 2) $&$A_2 + A_1$&$Z_2$ &$ \us\os2^{\fsu(3)^{(2)}}_{\left[\fsu(4)^{(2)}\right]}-\os2^{\fsu(2)^{(1)}}-\os1^{\fsp(0)^{(1)}}-\os8^{\fe_7^{(1)}}$
\\[10 pt]
\hline
\rule[-10pt]{0pt}{30pt}
$((2, 2), 3, 0), (3, 2, 2)$ & $A_2 + 2A_1  $&$1$&$
\begin{array}{c}
\us\os2^{\fsu(2)^{(1)}}_{\left[N_f=2\right]}-\us\os2^{\fsu(2)^{(1)}}_{\left[N_f=2\right]}-\os1^{\fsp(0)^{(1)}}-\os8^{\fe_7^{(1)}}\\[10pt]
\text{flavor:}\quad \fsu(2)^{(1)}\oplus \fsu(2)^{(1)} \oplus \fsu(2)^{(1)}
\end{array}$
\\[10 pt]
\hline
\rule[-10pt]{0pt}{30pt}
$(3, 3, 0) $&$ 2A_2 $&$1$&$\us\os2^{\fsu(2)^{(1)}}_{\left[\fg_2^{(1)}\right]}-\os2^{\fsu(1)^{(1)}}-\us\os1^{\fsp(0)^{(1)}}_{\left[\fsu(2)^{(1)}\right]}-\os8^{\fe_7^{(1)}}$
\\[10 pt]
\hline
\rule[-10pt]{0pt}{30pt}
$((2,2), 3, 2) $&$ A_2 + 3A_1 $&$1$&$\os2^{\fsu(1)^{(1)}}-\us\os2^{\fsu(2)^{(1)}}_{\left[\fg_2^{(1)}\right]}-\us\os1^{\fsp(0)^{(1)}}_{\left[\fsu(2)^{(1)}\right]}-\os8^{\fe_7^{(1)}}$
\\[10 pt]
\hline
\rule[-10pt]{0pt}{30pt}
$(3, 3, 2)$ &$ 2A_2 + A_1$&$1$ &$\us\os2^{\fsu(1)^{(1)}}_{\left[\fsu(2)^{(1)}\right]}-\os2^{\fsu(1)^{(1)}}-\us\os1^{\fsp(0)^{(1)}}_{\left[\fsu(2)^{(1)}\right]}-\os8^{\fe_7^{(1)}}$
\\[10 pt]
\hline
\rule[-10pt]{0pt}{30pt}
$(4, 0, 0) $&$ A_3 $&$1$&$\us\os2^{\fsu(2)^{(1)}}_{\left[\fso(7)^{(1)}\right]}-\os1^{\fsp(0)^{(1)}}-
\os8^{\fe_7^{(1)}}-\us\os1^{\fsp(0)^{(1)}}_{\left[\fsu(2)^{(1)}\right]}$
\\[10 pt]
\hline
\rule[-10pt]{0pt}{30pt}
$(4, 2, 0) $&$ \left(A_3 + A_1\right)' $&$1$&$\us\os2^{\fsu(1)^{(1)}}_{\left[\fsu(2)^{(1)}\right]}-\us\os1^{\fsp(0)^{(1)}}_{\left[\fsu(2)^{(1)}\right]}-
\os8^{\fe_7^{(1)}}-\us\os1^{\fsp(0)^{(1)}}_{\left[\fsu(2)^{(1)}\right]}$
\\[10 pt]
\hline
\rule[-10pt]{0pt}{30pt}
$(4, 0, 2) $&$ \left(A_3 + A_1\right)'' $&$1$&$\us\os2^{\fsu(2)^{(1)}}_{\left[\fso(7)^{(1)}\right]}-\os1^{\fsp(0)^{(1)}}-\us\os7^{\fe_7^{(1)}}_{\left[N_f = \frac{1}{2}\right]}$
\\[10 pt]
\hline
\rule[-10pt]{0pt}{30pt}
$(4, 2, 2) $&$ A_3 + 2A_1 $&$1$&$\us\os2^{\fsu(1)^{(1)}}_{\left[\fsu(2)^{(1)}\right]}-\us\os1^{\fsp(0)^{(1)}}_{\left[\fsu(2)^{(1)}\right]}-\us\os7^{\fe_7^{(1)}}_{\left[N_f  = \frac{1}{2}\right]}$
\\[10 pt]
\hline
\rule[-10pt]{0pt}{30pt}
$(4, (2,2), 0) $&$ D_4(a_1)$&$Z_2$&$\us\os1^{\fsp(0)^{(1)}}_{\left[\fsu(2)^{(1)}\right]}-
\os8^{\fe_7^{(1)}}\os\rightarrow^{2}\us\os1^{\fsp(0)^{(1)}}_{\left[\fsu(2)^{(1)}\right]}$
\\[10 pt]
\hline
\rule[-10pt]{0pt}{30pt}
$(4, (2,2), 2)$ &$ D_4(a_1) + A_1$&$Z_2$&$\us\os1^{\fsp(0)^{(1)}}_{\left[\fsu(2)^{(1)}\right]}\os\leftarrow^2\us\os7^{\fe_7^{(1)}}_{\left[N_f = \frac{1}{2}\right]}$
\\[10 pt]
\hline
\rule[-10pt]{0pt}{30pt}
$(4, 3, 0) $&$ A_3 + A_2 $&$1$&$\us\os1^{\fsp(0)^{(1)}}_{\left[\fsu(2)^{(1)}\right]}-\us\os6^{\fe_7^{(1)}}_{\left[\fso(2)^{(1)}\right]}$
\\[10 pt]
\hline
\rule[-10pt]{0pt}{30pt}
$(4, 3, 2) $&$ A_3 + A_2 + A_1 $&$1$&$\us\os5^{\fe_7^{(1)}}_{\left[\fso(3)^{(1)}\right]}$
\\[10 pt]
\hline
\rule[-10pt]{0pt}{30pt}
$(4, 4, 0) $&$ A_4$&$Z_2$&$\us\os1^{\fsp(0)^{(1)}}_{\left[\fsu(3)^{(2)}\right]}-\os6^{\fe_6^{(2)}}$
\\[10 pt]
\hline
\rule[-10pt]{0pt}{30pt}
$(4, 4, 2)$ &$ A_4 + A_1$ &$Z_2$&$\os5^{\fe_6^{(2)}}$
\\[10 pt]
\hline
\end{longtable}
\end{center}

\begin{figure}[t]
\centering
\subfigure[]{\label{fig:twistede6on2}
\includegraphics[width=5cm]{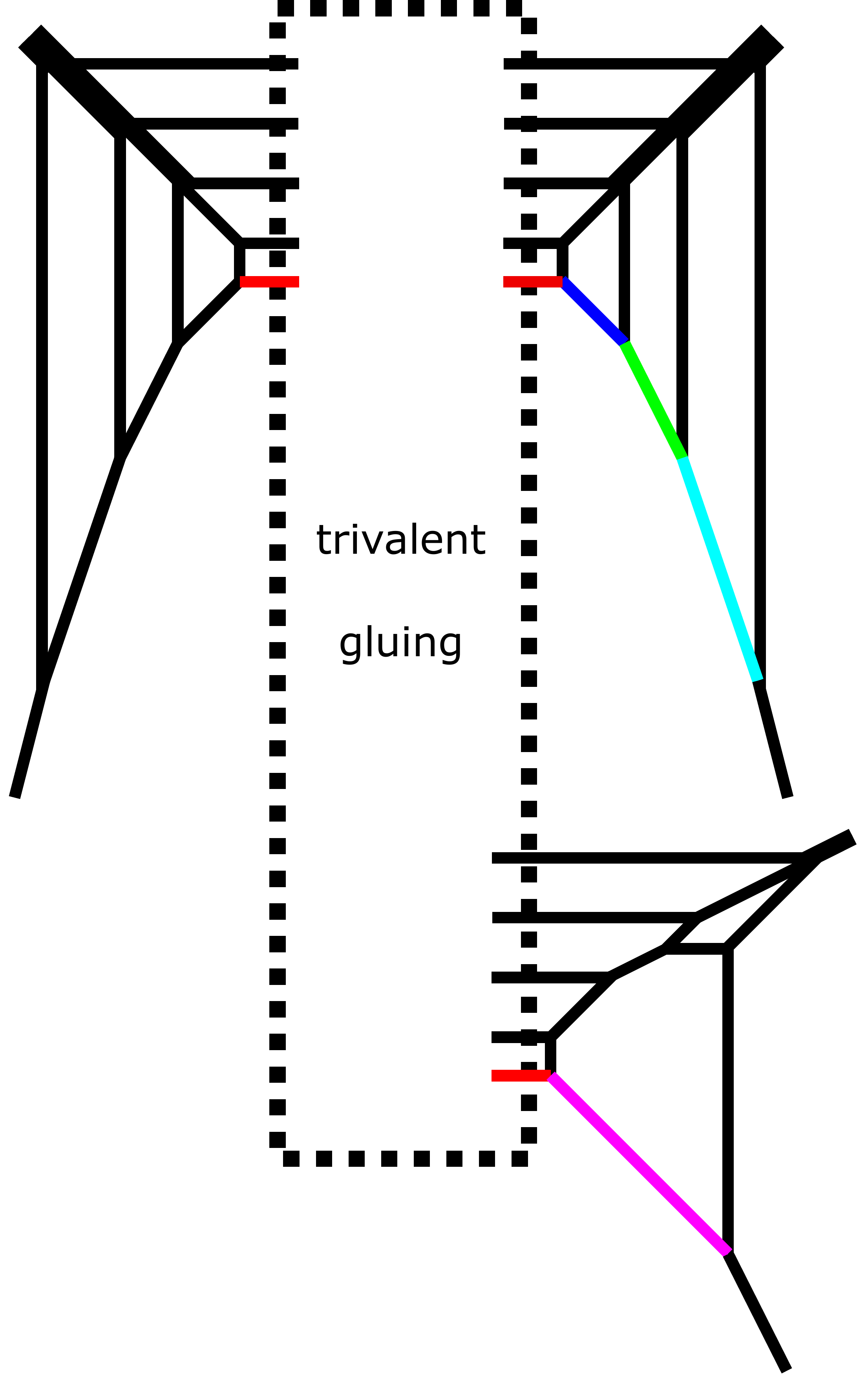}}
\hspace{1cm}
\subfigure[]{\label{fig:twistede6}
\includegraphics[width=6cm]{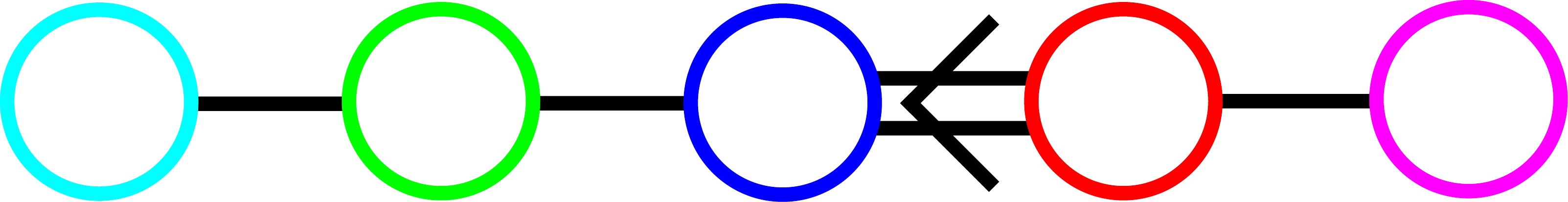}}
\caption{(a): The web diagram of the theory obtained after the Higgsing $(4, 4, 2)$. (b): The Dynkin diagram of the twisted affine Lie algebra $\fe_6^{(2)}$ which is formed by the colored lines in Figure \ref{fig:twistede6on2}. Each line in non-black color in Figure \ref{fig:puresu4k8v2} corresponds to the node in the same color in this figure. }
\label{fig:twistede6web}
\end{figure}
For the Higgsing $(4, 4, 0)$ or $(4, 4, 2)$ one can see that the twisted $\fe_6^{(2)}$ algebra is realized from the corresponding web diagram. For example, the web diagram for the case of $(4, 4, 2)$ is depicted in Figure \ref{fig:twistede6on2}. 
By using the intersection \eqref{fScartan},  the intersection among the fiber classes drawn as lines in non-black color in Figure \ref{fig:twistede6on2} yields the Dynkin diagram of the twisted affine Lie algebra of $\fe_6^{(2)}$. The explicit relation between the fiber classes and the nodes of the Dynkin diagram is also depicted in Figure \ref{fig:twistede6}. 
In the $(4,4,2)$ case, the 6d theory is the $E_6$ gauge theory with a single fundamental hypermultiplet and a tensor multiplet. 
The analysis in section \ref{sec:twist} shows that this theory admits a non-trivial twist, which breaks the flavor group down to a discrete group. This indicates that the reduced theory does not have any matter hypermultiplet, which is the case of \eqref{e6twist} with $n=5$. The reference \cite{Bhardwaj:2020kim} has also proposed the same twist and the analysis here is consistent with it.

\begin{figure}[t]
\centering
\includegraphics[width=6cm]{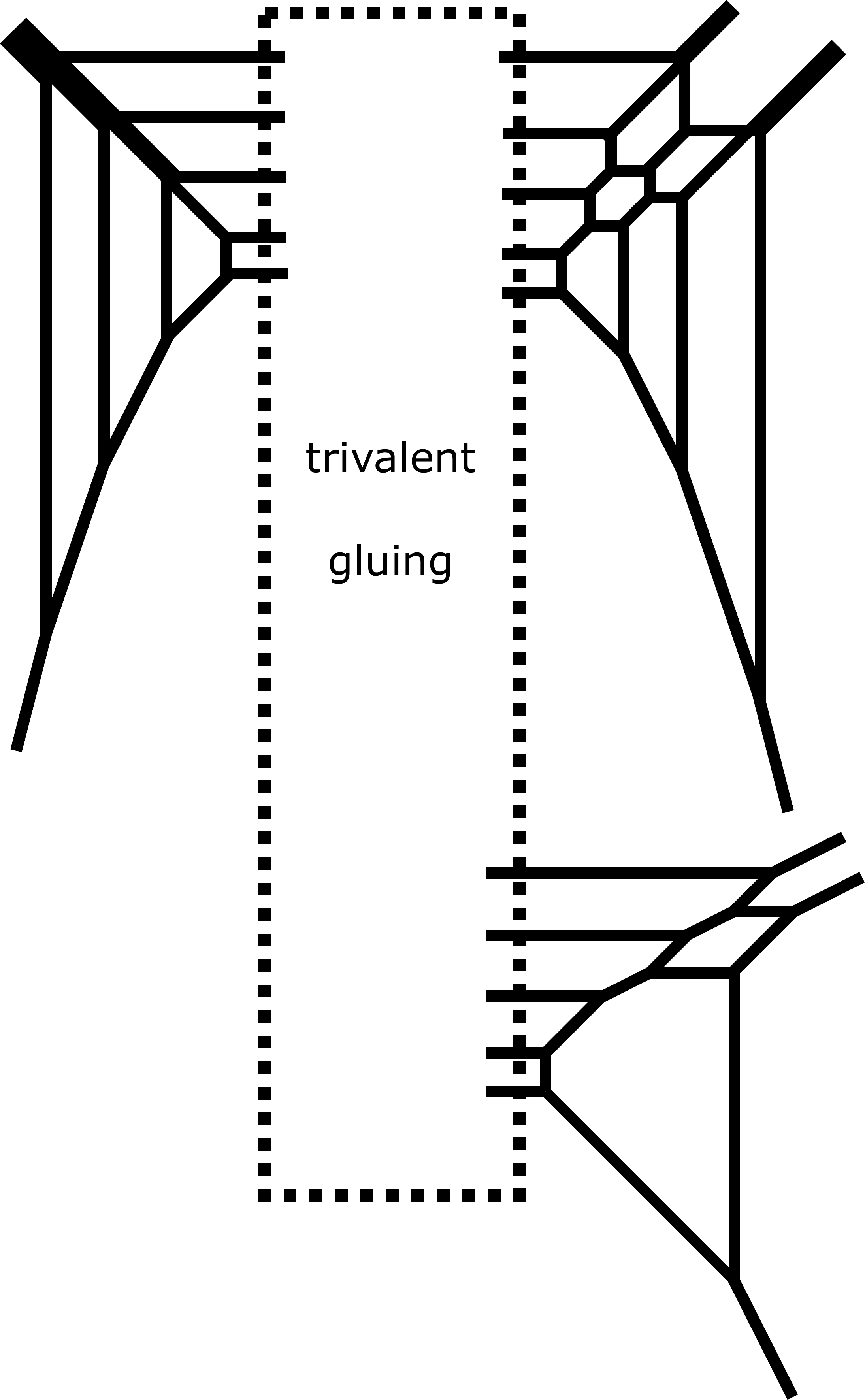}
\caption{The web diagram of the theory obtained after the Higgsing $(4, (2, 2), 0)$.
}
\label{fig:twiste7}
\end{figure}

It turns out that some of the Higgsings of the theory $(E_7, \underline{E_7})$ on $S^1$ correspond to twisted compactifications where the twist exchanges tensor multiplets of the parent 6d theory. For example, let us consider the Higgsing $(4, (2, 2), 0)$. The Higgsing necessarily involves the affine node of the affine $E_7$ Dynkin diagram in Figure \ref{fig:affinee7dynkin} and we expect that the Higgsing gives a 5d KK theory which is obtained by a twisted compactification of a 6d theory. In 6d, the Higgsing associated to the B-C label $D_4(a_1)$ of the theory $(E_7, \underline{E_7})$ gives \cite{Heckman:2016ssk}
\be\label{e7e7D4a1}
\us\os1^{\fsp(0)^{(1)}}_{\left[\fsu(2)^{(1)}\right]}-\us\os8^{\fe_7^{(1)}}_{\us|_{\text{\normalsize$\us\os1^{\fsp(0)^{(1)}}_{\left[\fsu(2)^{(1)}\right]}$}}} -\us\os1^{\fsp(0)^{(1)}}_{\left[\fsu(2)^{(1)}\right]}
\ee
The 5d KK theory from \eqref{e7e7D4a1} has $11$ Coulomb branch moduli and $4$ mass parametesr. On the other hand the web diagram of the theory after the Higgsing $(4, (2, 2), 0)$ is depicted in Figure \ref{fig:twiste7}.
The 5d theory from the web diagram has $10$ Coulomb branch moduli and $3$ mass parameters and hence there is a mismatch of the number of parameters. We can see that the theory \eqref{e7e7D4a1} has a symmetry which exchanges some of the E-string theories attached to the $(-8)$-curve. In order to generate a 5d KK theory with $10$ Coulomb branch moduli let us consider a twist which exchanges two of the E-string theories in \eqref{e7e7D4a1}. For concreteness we label the nodes as 
\be\label{e7e7D4a1v2}
\us\os{1_1}^{\fsp(0)^{(1)}}_{\left[\fsu(2)^{(1)}\right]}-\us\os{8_2}^{\fe_7^{(1)}}_{\us|_{\text{\normalsize$\us\os{1_4}^{\fsp(0)^{(1)}}_{\left[\fsu(2)^{(1)}\right]}$}}} -\us\os{1_3}^{\fsp(0)^{(1)}}_{\left[\fsu(2)^{(1)}\right]},
\ee
where the subscripts of $\Omega^{ii}$ represent the label. The intersection matrix between the base curves is given by
\begin{align}
\Omega^{D_4(a_1)} = \left(
\begin{array}{cccc}
1 & -1 & 0 & 0\\
-1 & 8 & -1 & -1\\
0 & -1 & 1 & 0\\
0 & -1 & 0 & 1
\end{array}
\right).
\end{align}
Then we consider a $Z_2$ twist which exchanges the third node with the fourth node. The matrix after the twist computed by \eqref{twistedomega} becomes
\begin{align}\label{twistedomegae7}
\Omega_{\sigma}^{D_4(a_1)} = \left(
\begin{array}{ccc}
1 & -1 & 0 \\
-1 & 8 & -2\\
0 & -1 & 1
\end{array}
\right).
\end{align}
From \eqref{twistedomegae7} the 5d KK theory after the twist can be represented by
\be\label{e7D4a1twisted}
\us\os{1_1}^{\fsp(0)^{(1)}}_{\left[\fsu(2)^{(1)}\right]}-
\os{8_2}^{\fe_7^{(1)}}\os\rightarrow^{2}\us\os{1_3}^{\fsp(0)^{(1)}}_{\left[\fsu(2)^{(1)}\right]}.
\ee
The 5d KK theory from \eqref{e7D4a1twisted} has $10$ Coulomb branch moduli and $3$ mass parameters. which agree with the numbers from the web diagram in Figure \ref{fig:twiste7}. Therefore we claim that the $(4, (2, 2), 0)$ Higgsing gives rise to the 5d KK theory of \eqref{e7D4a1twisted}.

\begin{figure}[t]
\centering
\includegraphics[width=8cm]{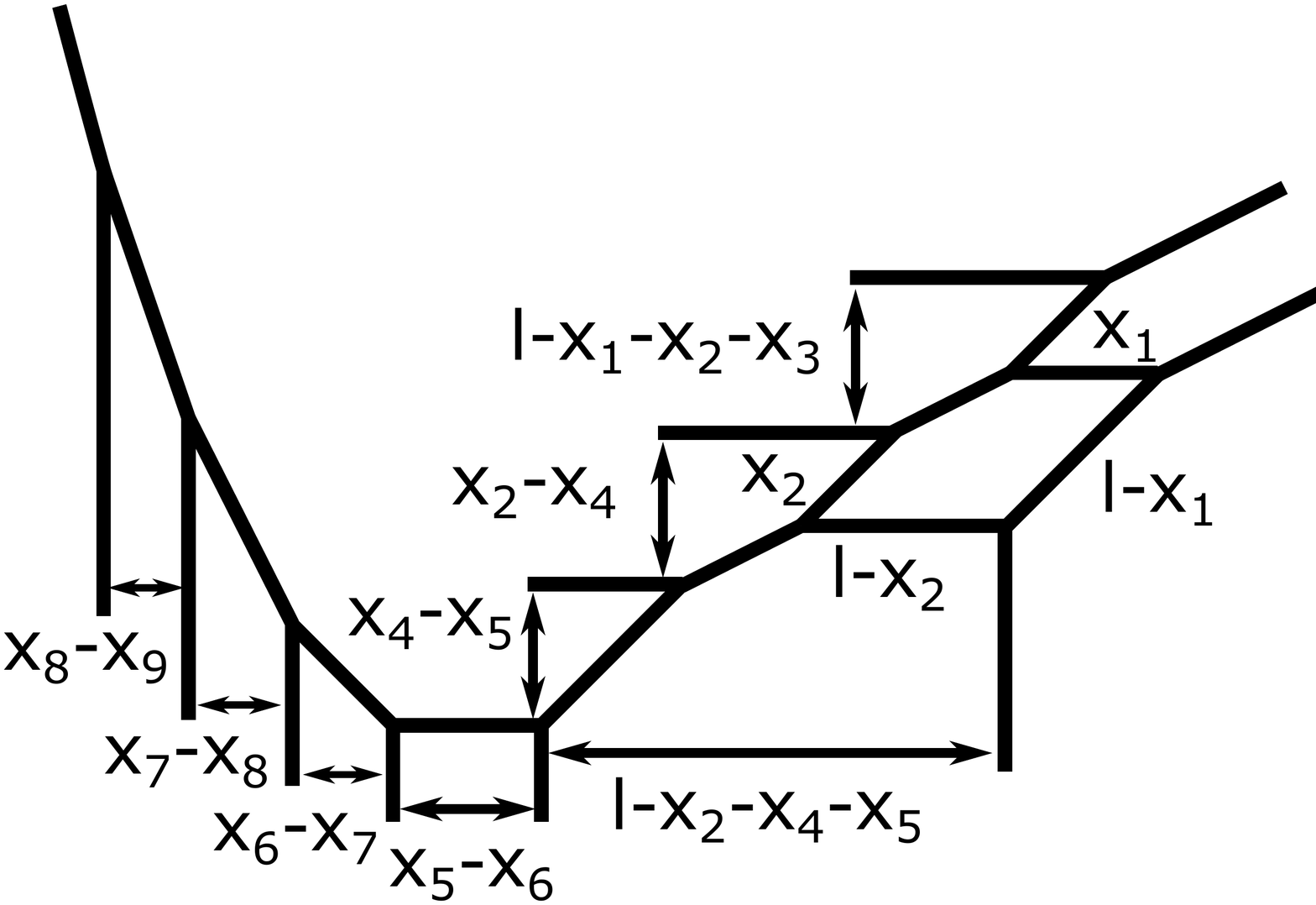}
\caption{The web diagram of dP$_9$ which corresponds to the E-string on the left  in \eqref{e7D4a1twisted}. $l$ is the hyperplane class of $\mathbb{P}^2$ and $x_i\; (i=1, \cdots, 9)$ are the exceptional curve classes. A curve class next to a double arrow implies that the length of the double arrow is related to the volume of the curve class. }
\label{fig:lestring}
\end{figure}

In fact we can see the web diagram in Figure \ref{fig:twiste7} is consistent with \eqref{twistedomegae7}. The Coulomb branch moduli for the left and right E-strings in \eqref{e7D4a1twisted} come from the faces in the upper-right and lower-right diagrams in Figure \ref{fig:twiste7} respectively. After performing flop transitions, we can extract the diagram corresponding to the left E-string and it is drawn in Figure \ref{fig:lestring}. 
The geometry of the E-string is characterized by a rational elliptic surface or del Pezzo $9$ surface (dP$_9$) and curves in dP$_9$ are spanned by the hyperplane class $l$ of $\mathbb{P}^2$ and exceptional curve classes $x_i\; (i=1, \cdots, 9)$. The genus one curve is then given by 
\be
f_{\fsp(0)^{(1)}_1} = 3l - \sum_{i=1}^9x_i,
\ee
where the subspcript of the $\fsp(0)^{(1)}$ implies the first node in \eqref{e7D4a1twisted}. From the embedding of the diagram in Figure \ref{fig:lestring} into the diagram in Figure \ref{fig:twiste7}, we can see that the eight $(-2)$-curve classes next to the double arrows in Figure \ref{fig:lestring} are the fibers which form the affine $E_7$ Dynkin diagram for the $\fe_7^{(1)}$ gauge algebra in the center of \eqref{e7D4a1twisted}. Then the genus one fiber on the $(-8)$-curve in \eqref{e7D4a1twisted} is given by
\begin{equation}
\begin{split}
f_{\fe_7^{(1)}} &= (l-x_1 -x_2 - x_3) + 2(x_2 - x_4) + 3(x_4 - x_5) + 4(x_5 - x_6) + 3(x_6 - x_7) \cr
&\hspace{5cm}+ 2(x_7 - x_8) + (x_8 - x_9) + 2(l-x_2 - x_4 - x_5)\cr
&=3l - \sum_{i=1}^9x_i.
\end{split}
\end{equation}
Hence the torus fiber on the left $(-1)$-curve and that on the center $(-8)$-curve in \eqref{e7D4a1twisted} is glued as 
\be\label{le7fiber}
f_{\fsp(0)_1^{(1)}} \sim  f_{\fe_7^{(1)}}. 
\ee

\begin{figure}[t]
\centering
\includegraphics[width=8cm]{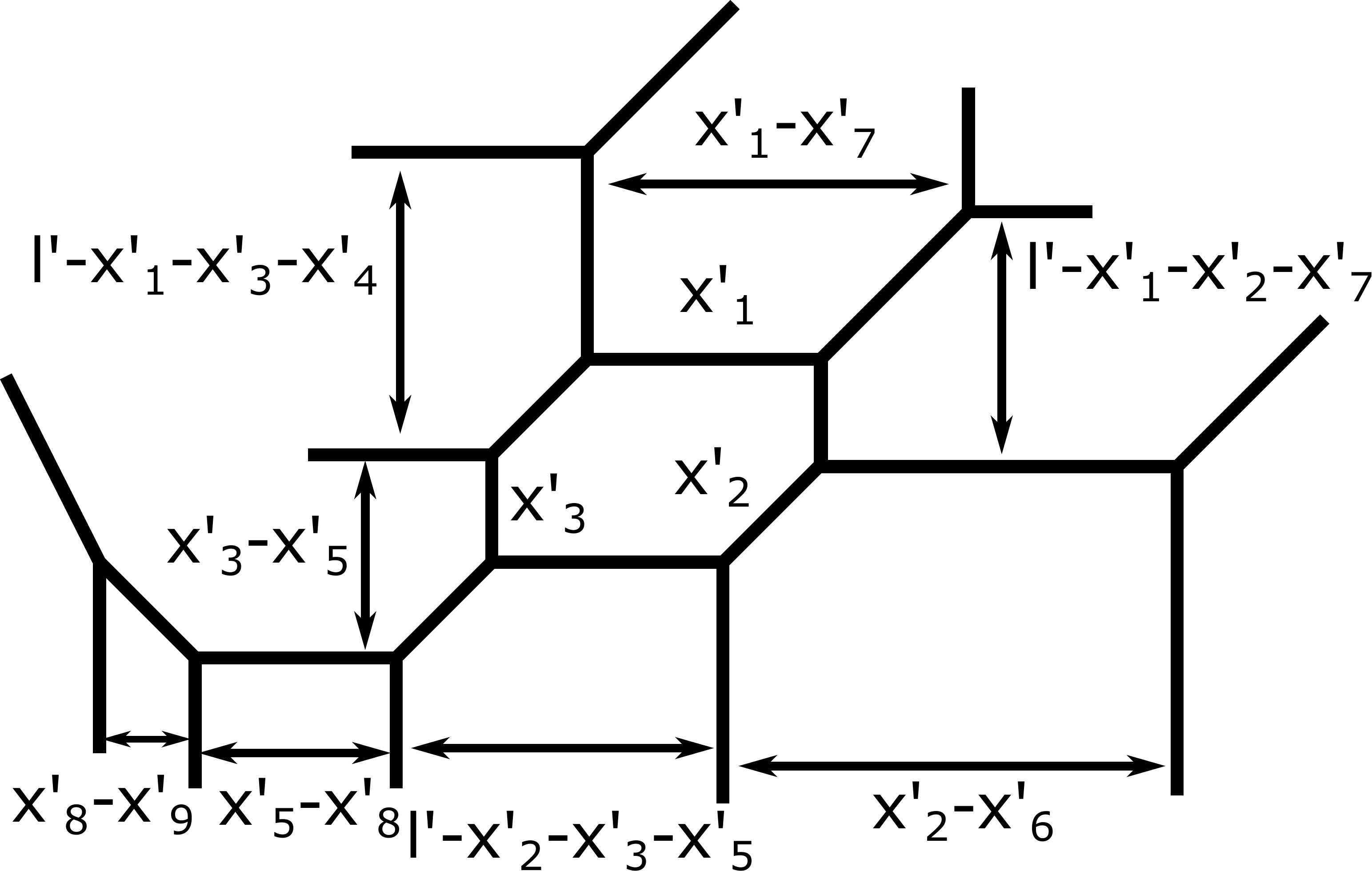}
\caption{The web diagram of dP$_9$ which corresponds to the E-string on the right in \eqref{e7D4a1twisted}. The notation for the curve classes is the same as the one in Figure \ref{fig:lestring}. The curve classes here are represented with a prime mark. }
\label{fig:restring}
\end{figure}

We then consider the E-string on the right of \eqref{e7D4a1twisted}. The web diagram corresponding to the right E-string in Figure \ref{fig:twiste7} is depicted in Figure \ref{fig:restring}. 
The torus fiber class of the dP$_9$ is again given by 
\be
f_{\fsp(0)_3^{(1)}} = 3l' - \sum_{i=1}^9x'_i,
\ee
where the subspcript of the $\fsp(0)^{(1)}$ represents the third node in \eqref{e7D4a1twisted}. The eight $(-2)$-curve classes next to the double arrows in Figure \ref{fig:restring} are fiber classes which form the affine $E_7$ Dynkin diagram for the $\fe_7^{(1)}$ gauge algebra in \eqref{e7D4a1twisted}. In terms of the curve classes in Figure \ref{fig:restring} the torus fiber on the $(-8)$-curve is given by 
\begin{equation}
\begin{split}
f_{\fe_7^{(1)}} &= (l'-x'_1 - x'_2 - x'_7) + 2(x'_2 - x'_6) + 3(l' - x'_2 -x'_3 - x'_5) + 4(x'_5 - x'_8)\cr
&\hspace{3cm} + 3(x'_3 - x'_5) + 2(l' - x'_1 - x'_3 - x'_4) + (x'_1 - x'_7) + 2(x'_8 - x'_9)\cr
&=2\left(3l' - \sum_{i=1}^9x'_i\right).
\end{split}
\end{equation}
Therefore the torus fiber on the right $(-1)$-curve and that on the center $(-8)$-curve in \eqref{e7D4a1twisted} is glued by 
\be\label{e7rfiber}
f_{\fe_7^{(1)}} \sim 2f_{\fsp(0)_3^{(1)}}. 
\ee
Therefore, from \eqref{le7fiber} and \eqref{e7rfiber}, the torus fibers are glued by 
\be\label{le7rfiber}
f_{\fsp(0)_1^{(1)}} \sim  f_{\fe_7^{(1)}} \sim 2f_{\fsp(0)_3^{(1)}}. 
\ee

On the other hand the gluing rule \eqref{gluingrule} implies 
\begin{align}
\left(-\Omega_{\sigma\;21}^{D_4(a_1)}\right)f_{\fsp(0)_1^{(1)}} \sim  \left(-\Omega_{\sigma\;12}^{D_4(a_1)}\right)f_{\fe_7^{(1)}}\label{gluinge7v1}\\
\left(-\Omega_{\sigma\;32}^{D_4(a_1)}\right)f_{\fe_7^{(1)}} \sim  \left(-\Omega_{\sigma\;23}^{D_4(a_1)}\right)f_{\fsp(0)_3^{(1)}}\label{gluinge7v3}
\end{align}
The gluing relations \eqref{gluinge7v1}, \eqref{gluinge7v3} with \eqref{twistedomegae7} are completely consistent with \eqref{le7rfiber} and this gives a support for the claim that the Higgsing $(4, (2, 2), 0)$ gives rise to the 5d KK theory from \eqref{e7D4a1twisted}.


\begin{figure}[t]
\centering
\includegraphics[width=6cm]{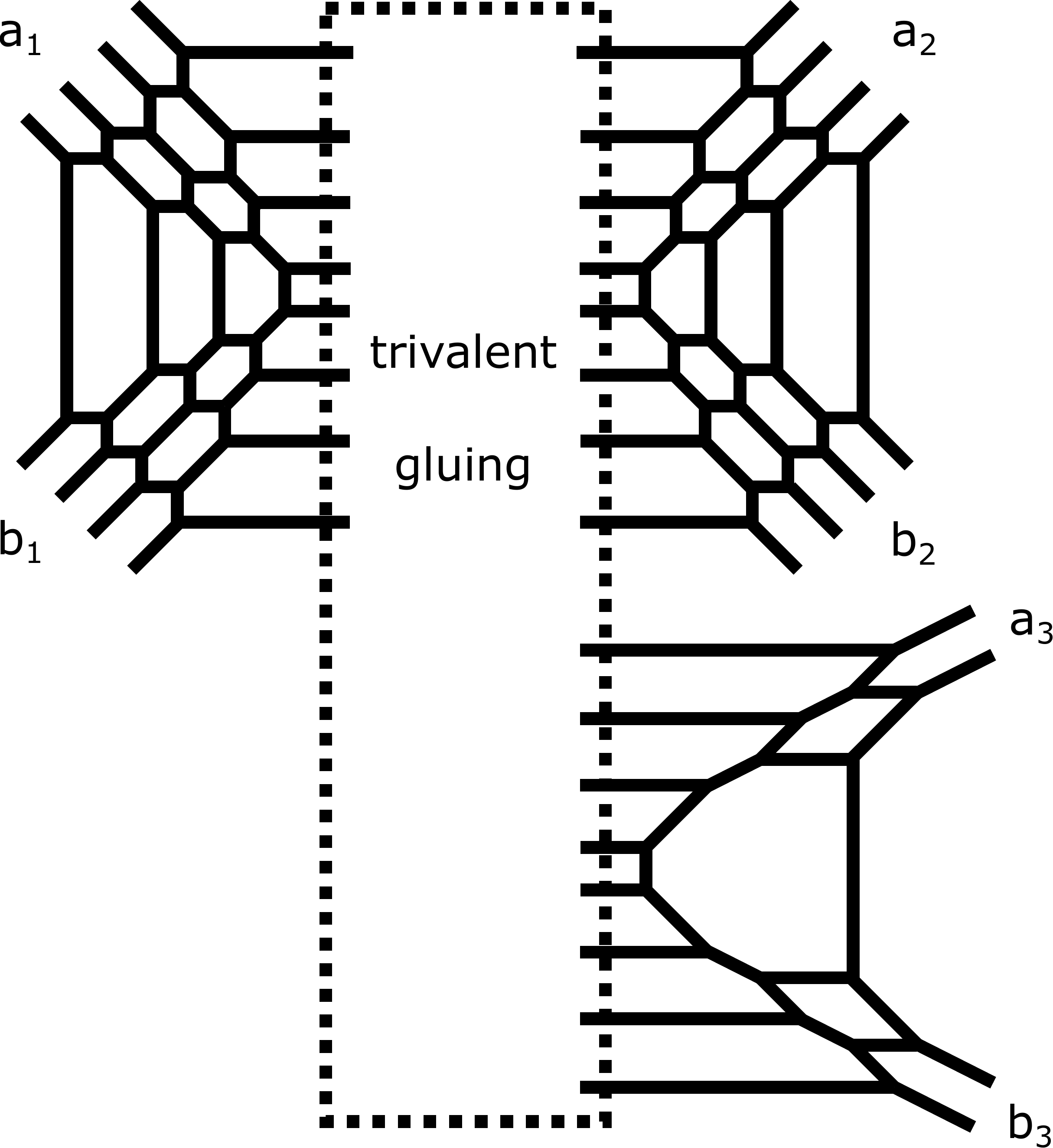}
\caption{The labeling of $[(a_1, a_2, a_3), (b_1, b_2, b_3)]$ for each $(\SU(4)^2 \times \SU(2)) \times (\SU(4)^2 \times \SU(2))$ in the web diagram in Figure \ref{fig:e7e7e7}
}
\label{fig:e7e7e7higgs}
\end{figure}
Let us then consider Higgsings of the theory \eqref{e6e6e6}. We can label the Higgsings by $[(a_1, a_2, a_3), (b_1, b_2, b_3)]$ associated to a breaking of the $[(\SU(4)^2 \times \SU(2)) \times (\SU(4)^2 \times \SU(2))]$ flavor symmetry 
which can be seen from the parallel external legs in Figure \ref{fig:e7e7e7higgs}. We will use the same notation which we used in Table \ref{tb:e7_1} for $(a_1, a_2, a_3)$ and similarly for $(b_1, b_2, b_3)$. For most of the cases we can make use of the Higgsings in Table \ref{tb:e7_1}. When we apply a Higgsing $(a_1, a_2, a_3)$ or $(b_1, b_2, b_3)$ to \eqref{ge7e7}, the resulting theory is the one given in Table \ref{tb:e7_1}. If either of the Higgsed theories contains the $\fe_7^{(1)}$ algebra, the theory associated to the Higgsing $[(a_1, a_2, a_3), (b_1, b_2, b_3)]$ is given by combining the theory from $(a_1, a_2, a_3)$ with the theory from $(b_1, b_2, b_3)$ by the following rule,
\begin{align}
&\left[\cdots \os{n_3}^{\fg_3}-\os{n_2}^{\fg_2}-\os{\left(8-n_1\right)}^{\fe_7^{(1)}}\right] + \left[\os{\left(8-m_1\right)}^{\fg'_1}-\os{m_2}^{\fg'_2}-\os{m_3}^{\fg'_3}\cdots \right]\cr
&\to \cdots \os{n_3}^{\fg_3}-\os{n_2}^{\fg_2}-\os{\left(8-n_1-m_1\right)}^{\fg'_1}-\os{m_2}^{\fg'_2}-\os{m_3}^{\fg'_3}\cdots.\label{e7combine}
\end{align}
Since the resulting theories are obtained straightforwardly from this rule, we will not display such cases explicitly. 

When both Higgsings change the $\fe_7^{(1)}$ algebra, we work out each case by appyling the Higgsing to the diagram in Figure \ref{fig:e7e7e7higgs}. In this example there are only three such cases, $[(4, 4, 0), (4, 4, 0)], [(4, 4, 2), (4, 4, 0)], [(4, 4, 2), (4, 4, 2)]$. For all the three cases the algebra on the glued curve turns out to be $\fe_6^{(2)}$ from the web diagrams and the result is summarized in Table \ref{tb:e7_2}. The order of $a_1, a_2, a_3$ is fixed as $a_1 \geq a_2 \geq a_3$ without loss of generality and we use the same notation as that we used in Table \ref{tb:d4_2}. Table \ref{tb:e7_2} also contains a single node theory with the $\fe_7^{(1)}$ algebra which we will use in the computation of the partition function in section \ref{sec:e7}. 
\begin{center}
\begin{longtable}{c|c|c}
\caption{\label{tb:e7_2} Higgsings associated to $\left(\SU(4)^2\times \SU(2)\right) \times \left(\SU(4)^2\times\SU(2)\right)$ of \eqref{e7e7e7}. The table includes the cases where both $(a_1, a_2, a_3)$ and $(b_1, b_2, b_3)$ Higgs the $\fe_7^{(1)}$ algebra. The other cases which are not listed in this table can be obtained by the rule \eqref{e7combine}.}\\
$\text{Higgsing}$&$\text{Theory}$&$\text{A 5d description}$
\\[3 pt]
\hline
\rule[-10pt]{0pt}{30pt}
$
\begin{array}{c}
\left[(4, 4, 0), (4, 3, 0)_{2,1}\right]\\
\left[(4, 4, 0), (4, 4, 0)\right]'
\end{array}
$&$\quad\us\os1^{\fsp(0)^{(1)}}_{\left[\fsu(3)^{(2)}\right]}-\us\os4^{\fe_6^{(2)}}_{\left[\fsu(2)^{(1)}\right]}-\us\os1^{\fsp(0)^{(1)}}_{\left[\fsu(3)^{(2)}\right]}\quad$&$\SU(4)_0 - \Sp(4)_0 - \left[\text{AS}\right]$
\\[10 pt]
\hline
\rule[-10pt]{0pt}{30pt}
$\left[(4, 3, 2), (4, 3, 2)_{2,1}\right]'$&$\quad\us\os2^{\fe_7^{(1)}}_{\left[\fso(6)^{(1)}\right]}\quad$&$
\begin{array}{c}
\left(\text{rank 1}\right)  - \SU(8)_0- \left[2\text{AS}\right]\\[5 pt]
\left(\text{rank 1}\right)  - \Sp(4)_{0} - \Sp(3)_0
\end{array}$
\\[10 pt]
\hline
\rule[-10pt]{0pt}{30pt}
$
\begin{array}{c}
\left[(4, 3, 2), (4, 4, 0)\right]\\
\left[(4, 4, 2), (4, 4, 0)\right]'
\end{array}$&$\quad\us\os3^{\fe_6^{(2)}}_{\left[\fsu(3)^{(2)}\right]}-\us\os1^{\fsp(0)^{(1)}}_{\left[\fsu(3)^{(2)}\right]}\quad$&$\left(\text{rank 2}\right)- \Sp(4)_{\pi} - \left[\text{AS}\right]$
\\[10 pt]
\hline
\rule[-10pt]{0pt}{30pt}
$
\begin{array}{c}
\left[(4, 4, 2), (4, 3, 2)_{2,1}\right]'\\
\left[(4, 4, 2), (4, 4, 2)\right]''
\end{array}$&$\quad\us\os2^{\fe_6^{(2)}}_{\left[\fsu(4)^{(2)}\right]}\quad$&$\left(\text{rank 1}\right)  - \Sp(4)_0- \left[\text{AS}\right]$
\\[10 pt]
\hline
\end{longtable}
\end{center}

As in Table \ref{tb:d4_2} or Table \ref{tb:e6_2}, the number of the prime marks in the left column in Table \ref{tb:e7_2} implies the number of mass parameters which are turned off. Such cases also appear in Table \ref{tb:d4_2} and Table \ref{tb:e6_2} and it has an interpretation of turning off mass parameters of the rank-2 antisymmetric hypermultiplets of an $\Sp$ gauge group in a 5d description. However the case of $[(4, 3, 2), (4, 3, 2)_{2,1}]'$ is different from those cases. When we apply the Higging $(4, 3, 2)$ to \eqref{ge7e7}, the resulting theory is the 5d KK theory from a circle compactification of the 6d $E_7$ gauge theory with a hypermultiplet and a half-hypermultiplet in the fundamental representation of $E_7$ as well as a tensor multiplet. 
Hence the Higgsing $[(4, 3, 2), (4, 3, 2)_{2,1}]'$ yields a 5d KK theory from a circle compactification of the 6d $E_7$ gauge theory with two hypermutliplets and two half-hypermutliplets in the fundamental representation of $E_7$ in addition to a tensor multiplet. 
Then the web diagram corresponding to the Higgsing $[(4, 3, 2), (4, 3, 2)]$ has three mass parameters; two are mass parameters for the fundamental hypermultiplets and one is the radius of the circle. In general we may turn on a mass parameter for two half-hypermutliplets and they behave as a massive hypermultiplet. However the diagram for the Higgsing $[(4, 3, 2), (4, 3, 2)_{2,1}]'$ does not admit such a mass parameter and the prime mark of $[(4, 3, 2), (4, 3, 2)]'$ implies the restriction of turning on the mass parameter in terms of the web diagram. 
Similarly the theories obtained by further Higgsings from the diagram of $[(4, 3, 2), (4, 3, 2)_{2,1}]'$ also have at least one less number of mass parameters compared to a generic case. For example the Higgsing $[(4,4,2), (4,3,2)_{2,1}]$ yields the $\fe_6^{(2)}$ on a $(-2)$-curve and the flavor algebra is $\fsu(4)^{(2)}$. From the web diagram it has two mass parameters although the $\fsu(4)^{(2)}$ algebra in general implies three mass parameters. The theory labeled by $[(4, 4, 2), (4, 4, 2)]$ further tunes one more mass parameter of the same theory.

\begin{figure}[t]
\centering
\includegraphics[width=6cm]{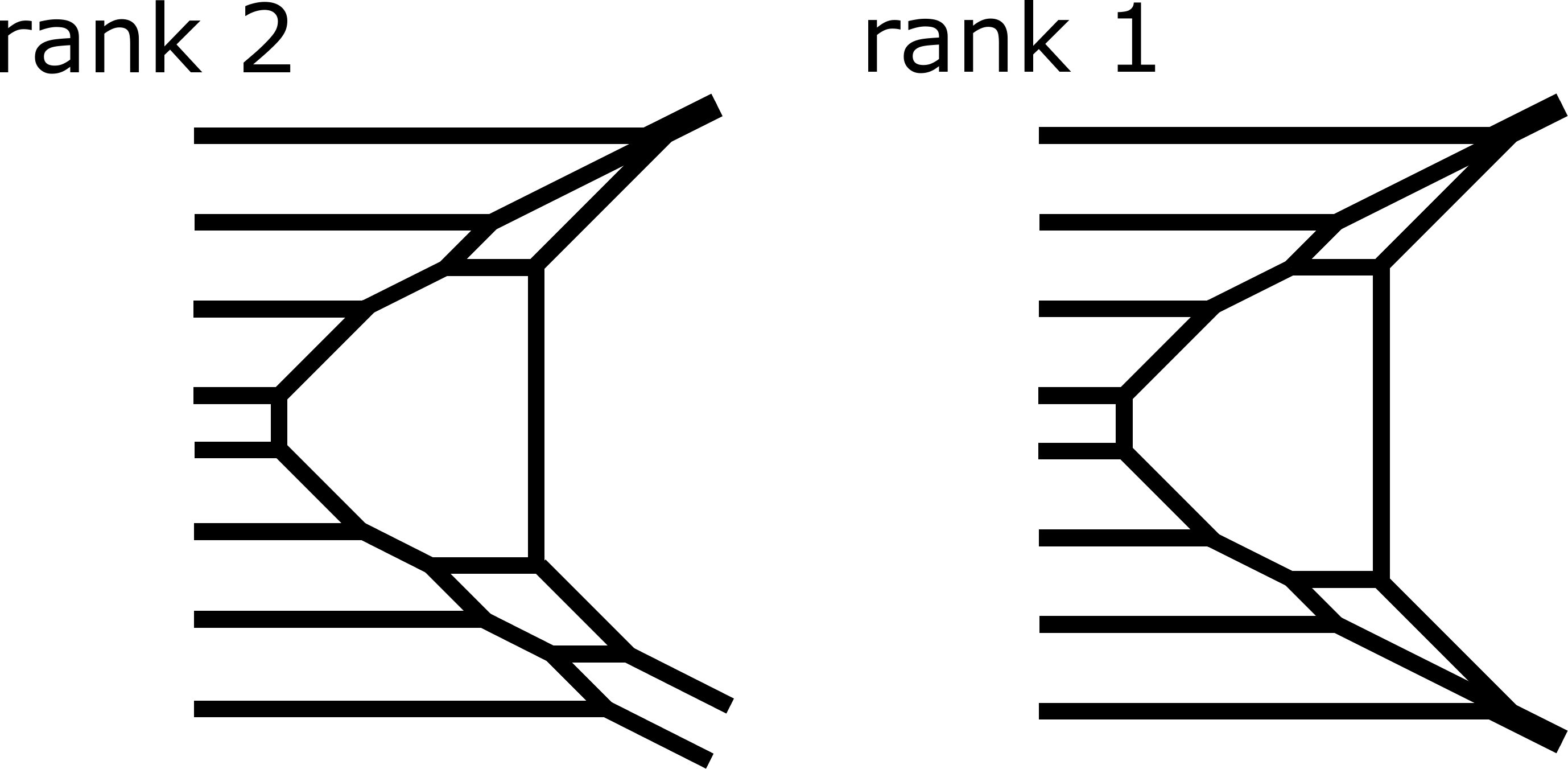}
\caption{The web diagrams for the theory denoted by $(\text{rank 1})$ and the theory denoted by $(\text{rank 2})$ in Table \ref{tb:e7_2}.
}
\label{fig:nonlag}
\end{figure}
We can also consider 5d gauge theory descriptions after the Higging $[(a_1, a_2, a_3), (b_1, b_2, b_3)]$. 
Examples of 5d gauge theory descriptions
are written in the third column in Table \ref{tb:e7_2}. The original theory \eqref{e7quiver} has two long quiver tails given by $\SU(8)_0 - \SU(6)_0 - \SU(4)_0 - \SU(2)_0$. The theory exhibits an $\SU(4) \times \SU(4)$ flavor symmetry, 
The $(4, 3)$ and $(3, 4)$ Higgsings yield the $\SU(8)$ gauge theory with a hypermultiplet in the rank-2 antisymmetric representation with the CS level $+ 1$ and $-1$ respectively \cite{Bergman:2015dpa}. This Higgsing gives the presence of the antisymmetric hypermultiplets in Table \ref{tb:e7_2}. Furthermore we have 
\be\label{sp4v1}
\SU(8)_{k} - \SU(6)_0 - \SU(4)_0 - \SU(2)_0 \qquad  \substack{(4, 4)\;\text{Higgs}\\\longrightarrow}  \qquad \Sp(4)_{\theta = |k+1|\pi \; (\text{mod}\;2\pi)},
\ee
\be\label{sp4v2}
[\text{AS}]-\SU(8)_{k} - \SU(6)_0 - \SU(4)_0 - \SU(2)_0 \qquad  \substack{(4, 4)\;\text{Higgs}\\\longrightarrow}  \qquad [\text{AS}] - \Sp(4)_{\theta = |k+1|\pi \; (\text{mod}\;2\pi)},
\ee
for some small values of $k$
$(\text{rank 1})$ and $(\text{rank 2})$ are non-Lagrangian theories and the rank indicates the dimension of the Coulomb branch moduli space. The web diagram of the theories are depicted in Figure \ref{fig:nonlag}. The theories show an explicit $\SU(8)$ flavor symmetry. For the last and the second to the last case in Table \ref{tb:e7_2}, an $\Sp(4)$ subgroup of the $\SU(8)$ is gauged. 


\bigskip

\section{Partition functions of 6d/5d exceptional gauge theories}
\label{sec:PF}
In this section, we compute the partition functions of some 6d theories on $T^2 \times \mathbb{R}^4$, which are obtained in section \ref{sec:CMT}, by the topological vertex formalism using the trivalent/quadrivalent gluing prescription proposed in \cite{Hayashi:2017jze}. We will also take a 5d limit of the partition functions and compare their perturbative part with the known results. Let us also stress that we can compute the partition functions for all the 6d theories considered in section \ref{sec:CMT}. We here select some interesting examples with exceptional gauge groups and compute their partition functions. 

\subsection{6d/5d $G_2$ gauge theory with matter}
\label{sec:g2}

We begin with the computation of the partition functioins for the 6d $G_2$ gauge theory with four flavors and a tensor multiplet on $T^2 \times \mathbb{R}^4$. We will also take a 5d limit to obtain the partition function of the 5d $G_2$ gauge theory with four flavors on $S^1 \times \mathbb{R}^4$.

\begin{figure}[t]
\centering
\includegraphics[width=7cm]{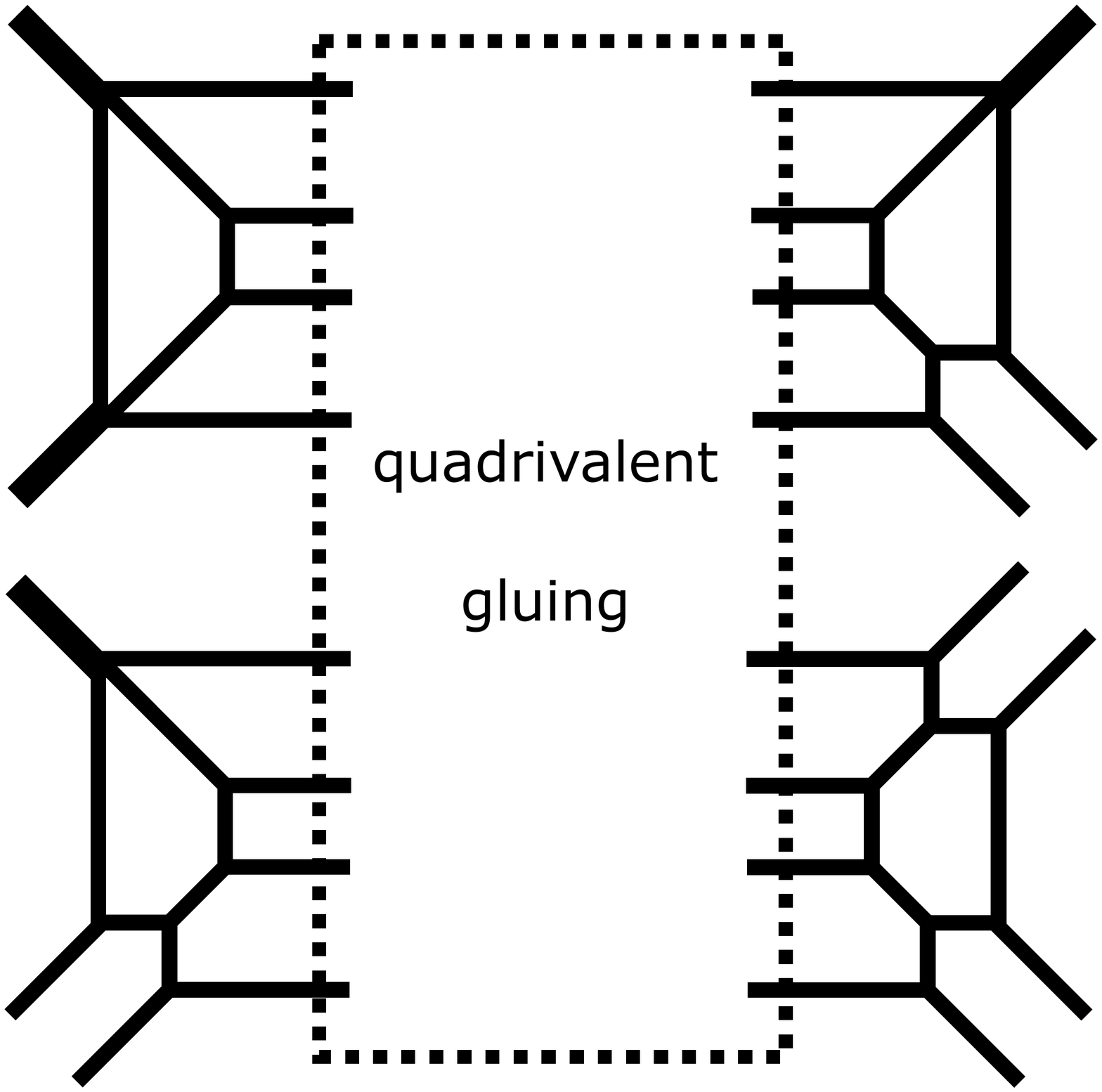}
\caption{(a): The web diagram for 
$\fg_2^{(1)}$ on $(-2)$-curve. 
}
\label{fig:g2on2}
\end{figure}

\paragraph{$\fg_2^{(1)}$ on $(-2)$-curve.}
The 6d $G_2$ gauge theory with four flavors and a tensor multiplet compactified on a circle arises on the $\left[(2,2,2,0), (2,0,0,0)\right]$ Higgs branch of the 
6d theory $(D_4, D_4)_2$ on $S^1$ as in Table \ref{tb:d4_2}. The original 
theory $(D_4, D_4)_2$ on $S^1$ is the affine $D_4$ Dynkin quiver theory \eqref{affineD4} and it is realized on the web diagram in Figure \ref{fig:d4d4d4}. Then a web diagram corresponding to the geometry $\fg_2^{(1)}$ on $(-2)$-curve, which yields the 6d $G_2$ gauge theory with four flavors and a tensor multiplet compactified on $S^1$, can be obtained by applying the $\left[(2,2,2,0), (2,0,0,0)\right]$ Higgsing to the diagram in Figure \ref{fig:d4d4d4}. From the diagram in Figure \ref{fig:d4d4d4} we can explicitly see the flavor symmetry $\SU(2)^4 \times \SU(2)^4$ from parallel external legs in the diagram. Then the $\left[(2,2,2,0), (2,0,0,0)\right]$ Higgsing breaks three $\SU(2)$'s in one $\SU(2)^4$ and another $\SU(2)$ in the other $\SU(2)^4$. The Higgsing which breaks one $\SU(2)$ is carried out by binding two parallel external 5-branes on one 7-brane \cite{Benini:2009gi}. Then the web diagram after applying the $\left[(2,2,2,0), (2,0,0,0)\right]$ Higgsing yields the diagram in Figure \ref{fig:g2on2}. 
The intersection between the fiber classes of the three faces 
forms the affine $G_2$ Dynkin diagram, which gives rise to the affine $G_2$ Lie algebra, in a similar matter to Figure \ref{fig:g2web}. 

\begin{figure}[t]
\centering
\includegraphics[width=8cm]{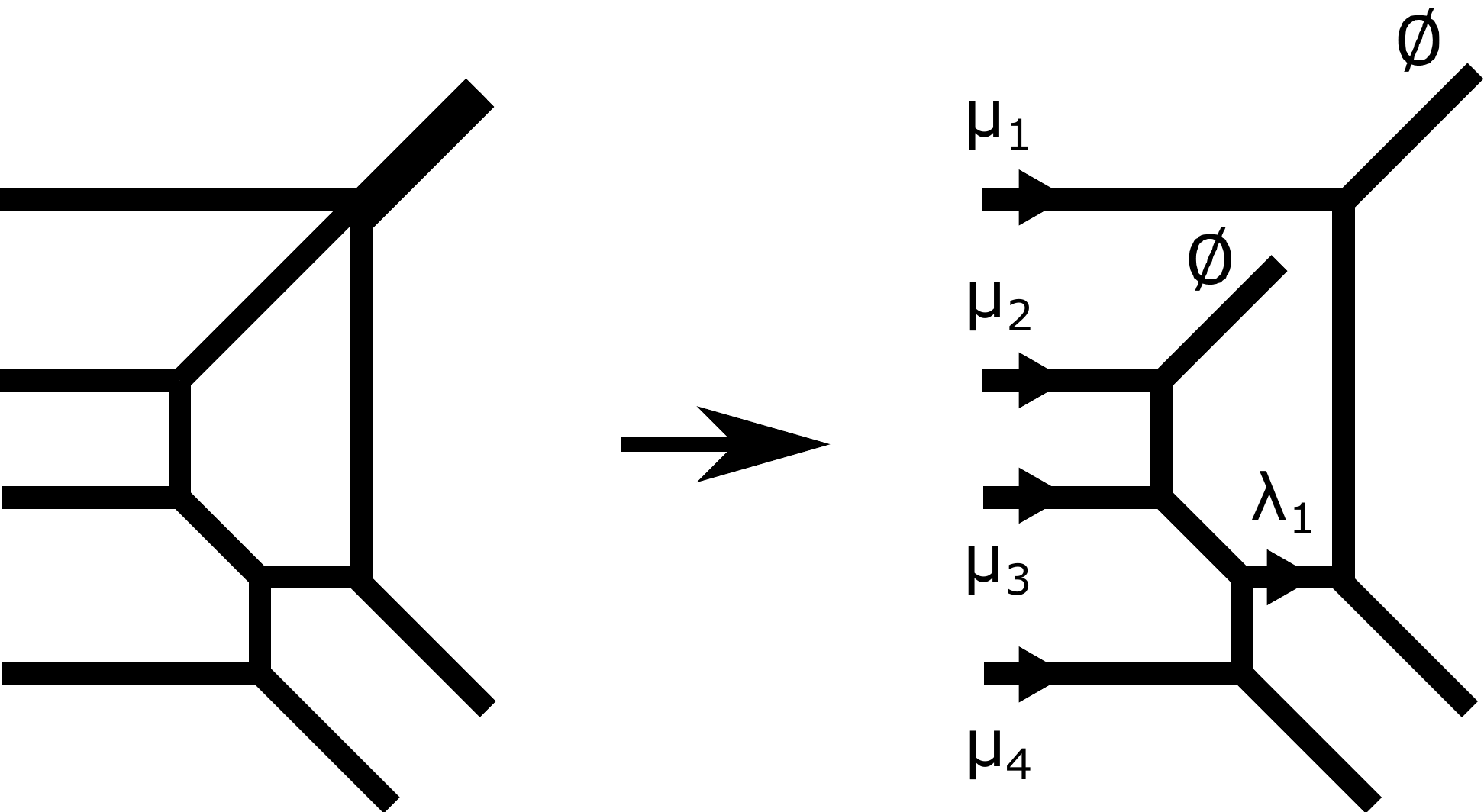}
\caption{The diagram on the right-hand side shows the assignment of Young diagrams for applying the topological vertex to the diagram on the left-hand side. }
\label{fig:nontoric}
\end{figure}

Since the circle compactification of the 6d $G_2$ gauge theory with four flavors and a tensor multiplet is realized on the web diagram in Figure \ref{fig:g2on2} with the quadrivalent gluing, we can apply the topological vertex \cite{Iqbal:2002we, Aganagic:2003db} with the gluing prescription developed in \cite{Hayashi:2017jze} to the diagram for computing the partition function of the 6d theory compactified on $T^2 \times \mathbb{R}^4$. Ref.~\cite{Hayashi:2017jze} considered examples which are made from a trivalent or quadrivalent gluing of toric diagrams. The examples include 6d/5d pure $\SO(8), E_6, E_7, E_8$ gauge theories, namely gauge theories with simply-laced gauge groups. Here we generalize the computations into a trivalent or quadrivalent gluing of non-toric diagrams, which will give rise to cases of non-simply-laced gauge groups as well as gauge theories with exceptional gauge groups with matter. Although each piece we glue is a non-toric diagram it is still possible to apply the topological vertex to a non-toric diagram when it is given by a Higgsing of a toric diagram. As briefly reviewed in section \ref{sec:vertex}, the end result in the unrefined cases is that we can simply apply the topological vertex in a usual way with trivial Young diagrams assigned for all the external legs including 5-branes put on a single 7-brane \cite{Hayashi:2013qwa, Hayashi:2014wfa, Kim:2015jba, Hayashi:2015xla}. For example for applying the topological vertex to the diagram on the left-hand side in Figure \ref{fig:nontoric}, which is one of the four diagrams which we glue in Figure \ref{fig:g2on2}, we can use the diagram on the right-hand side  in Figure \ref{fig:nontoric}.

\begin{figure}[t]
\centering
\subfigure[]{\label{fig:g2para1}
\includegraphics[width=4.5cm]{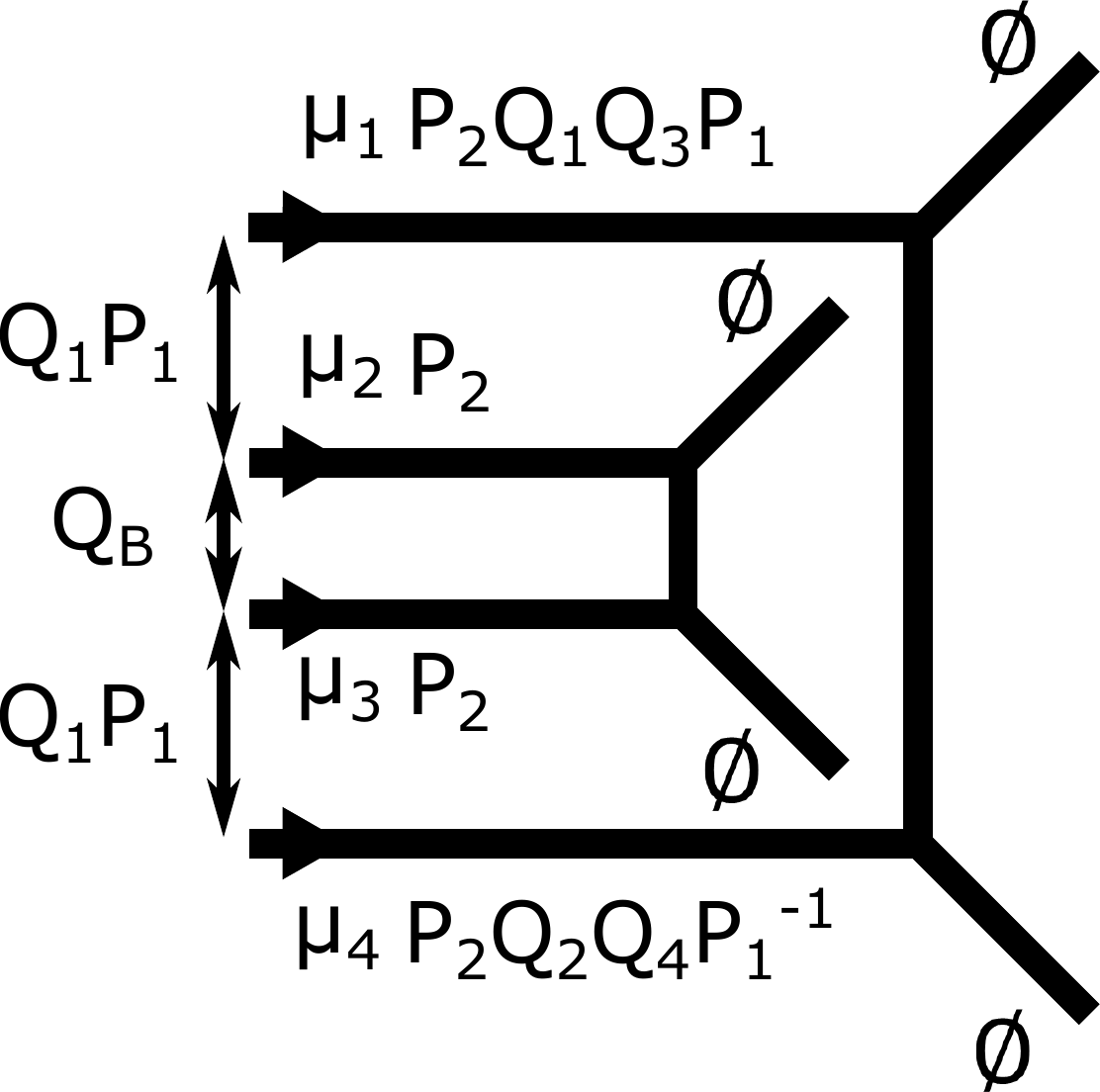}}
\hspace{1cm}
\subfigure[]{\label{fig:g2para2}
\includegraphics[width=4.5cm]{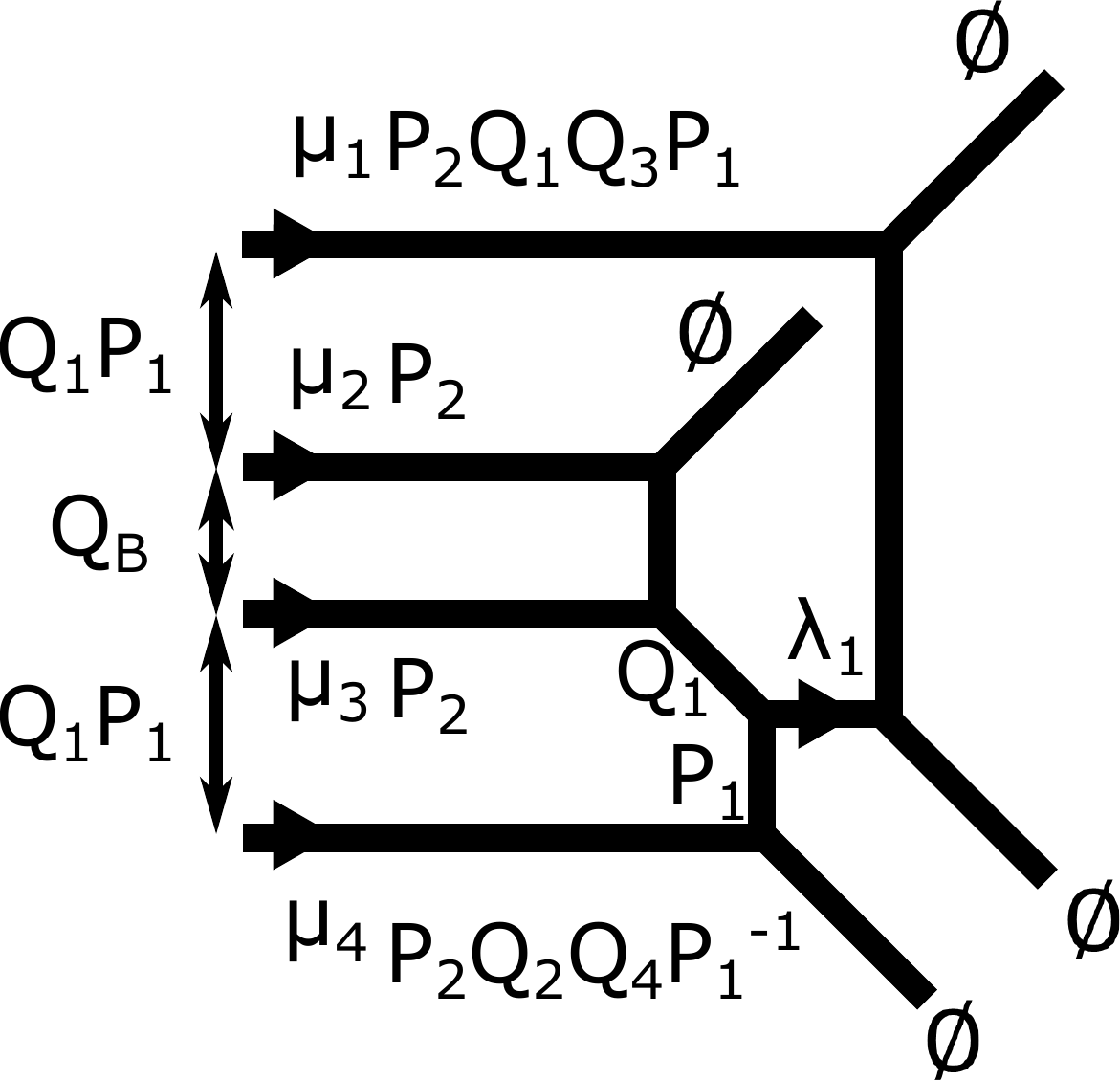}}
\subfigure[]{\label{fig:g2para3}
\includegraphics[width=4.5cm]{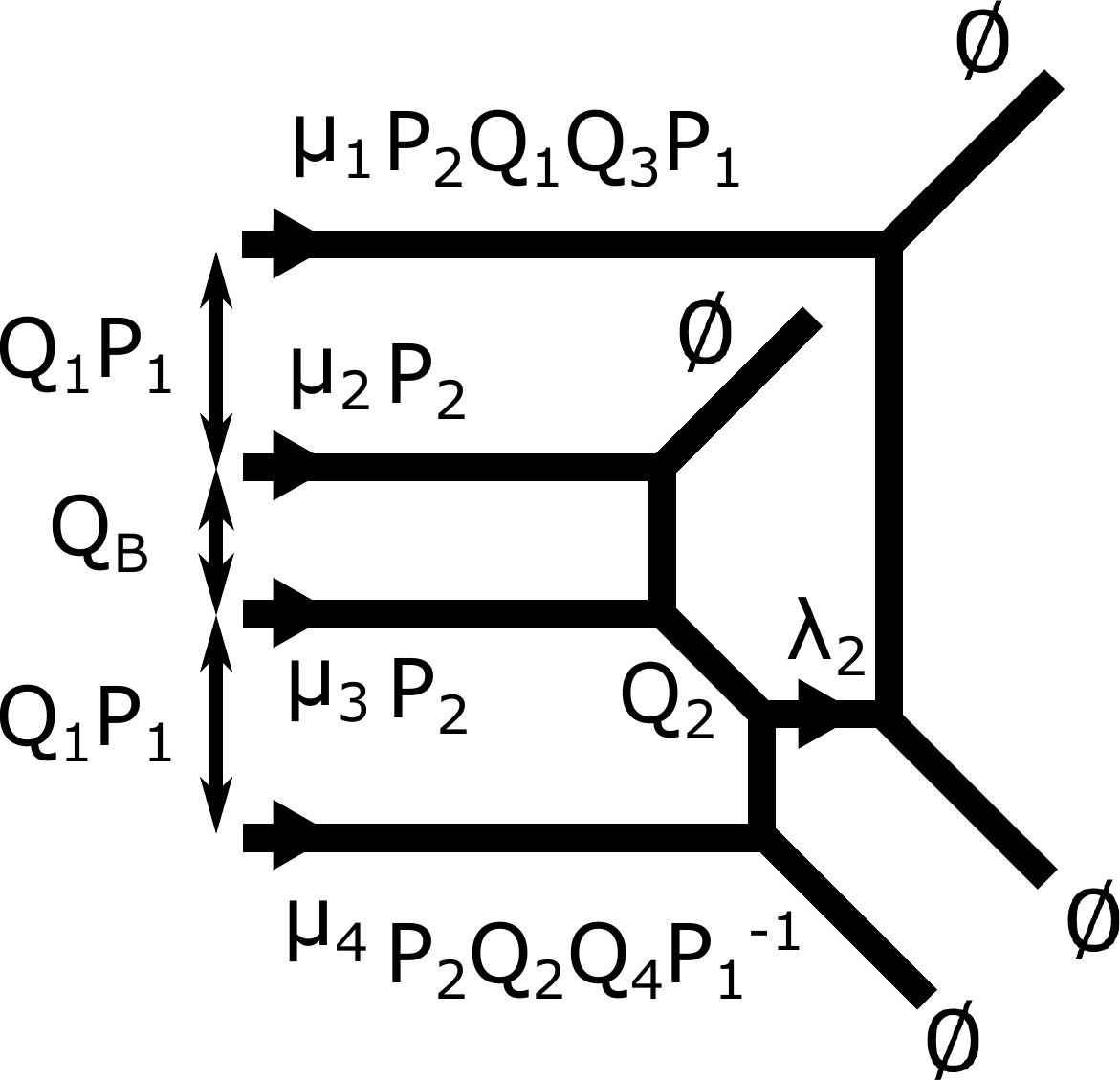}}
\hspace{1cm}
\subfigure[]{\label{fig:g2para4}
\includegraphics[width=4.5cm]{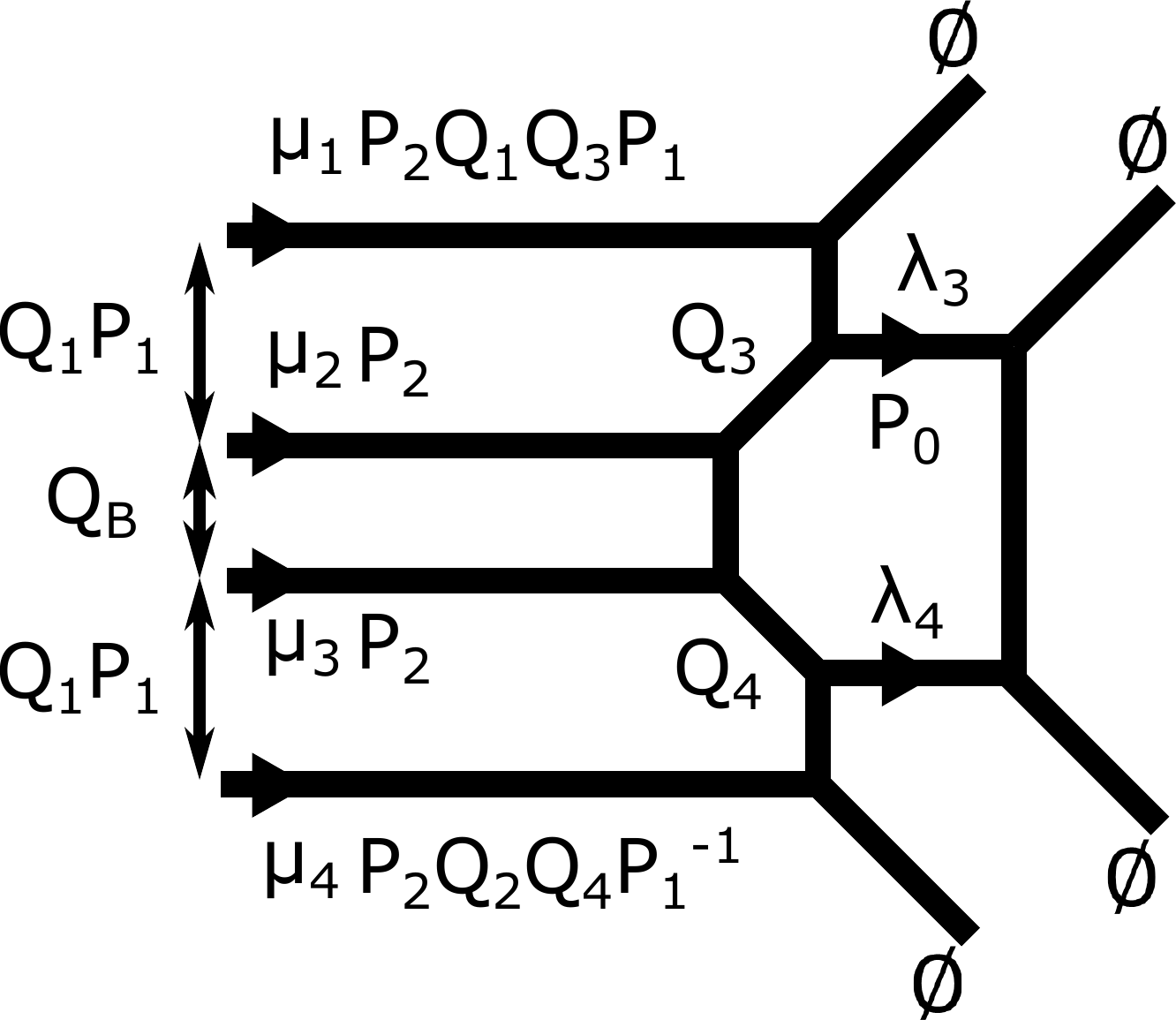}}
\caption{Assignment of Young diagrams ($\mu_1, \mu_2, \mu_3, \mu_4, \lambda_1, \lambda_2, \lambda_3, \lambda_3$) and K\"ahler parameters ($Q_1, Q_2, Q_3, Q_3, P_1, P_2, Q_B$) for each of the four diagrams in Figure \ref{fig:g2on2}.}
\label{fig:g2para}
\end{figure}

\begin{figure}[t]
\centering
\includegraphics[width=4cm]{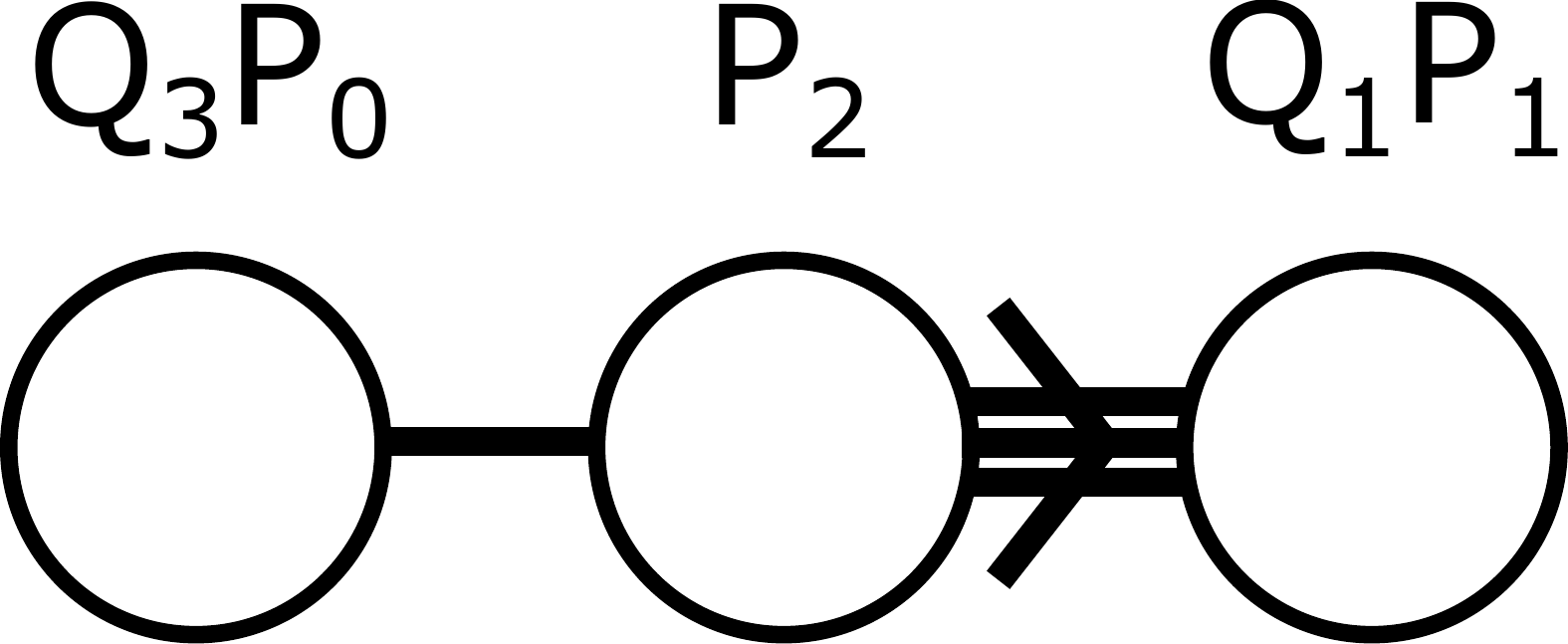}
\caption{The K\"ahler parameters of the fibers which form the affine $G_2$ Dynkin diagram.}
\label{fig:g2dynkin}
\end{figure}

Let us compute the partition function for the 6d $G_2$ gauge theory with four flavors and a tensor multiplet on $T^2 \times \mathbb{R}^4$ using the topological vertex with the quadrivalent gluing prescription. For that we first apply the topological vertex to each of the four diagrams in Figure \ref{fig:g2on2}. Namely we apply the topological vertex to the diagrams in Figure \ref{fig:g2para1} - Figure \ref{fig:g2para4}. 
We also choose a parameterization as shown in Figure \ref{fig:g2para}. Each parameter represents a K\"ahler parameter in the dual geometry and it is related to the lengh of the corresponding line\footnote{More precisely a parameter $Q$ in the figures is related to length $\ell$ of a line by $Q = e^{-\ell}$.}. Note also that we need to assign non-trivial Young diagrams for gluing lines. The K\"ahler parameters for the fiber classes which form the affine $G_2$ Dynkin diagram is given in Figure \ref{fig:g2dynkin}. 

\begin{figure}[t]
\centering
\subfigure[]{\label{fig:halfsu4}
\includegraphics[width=5cm]{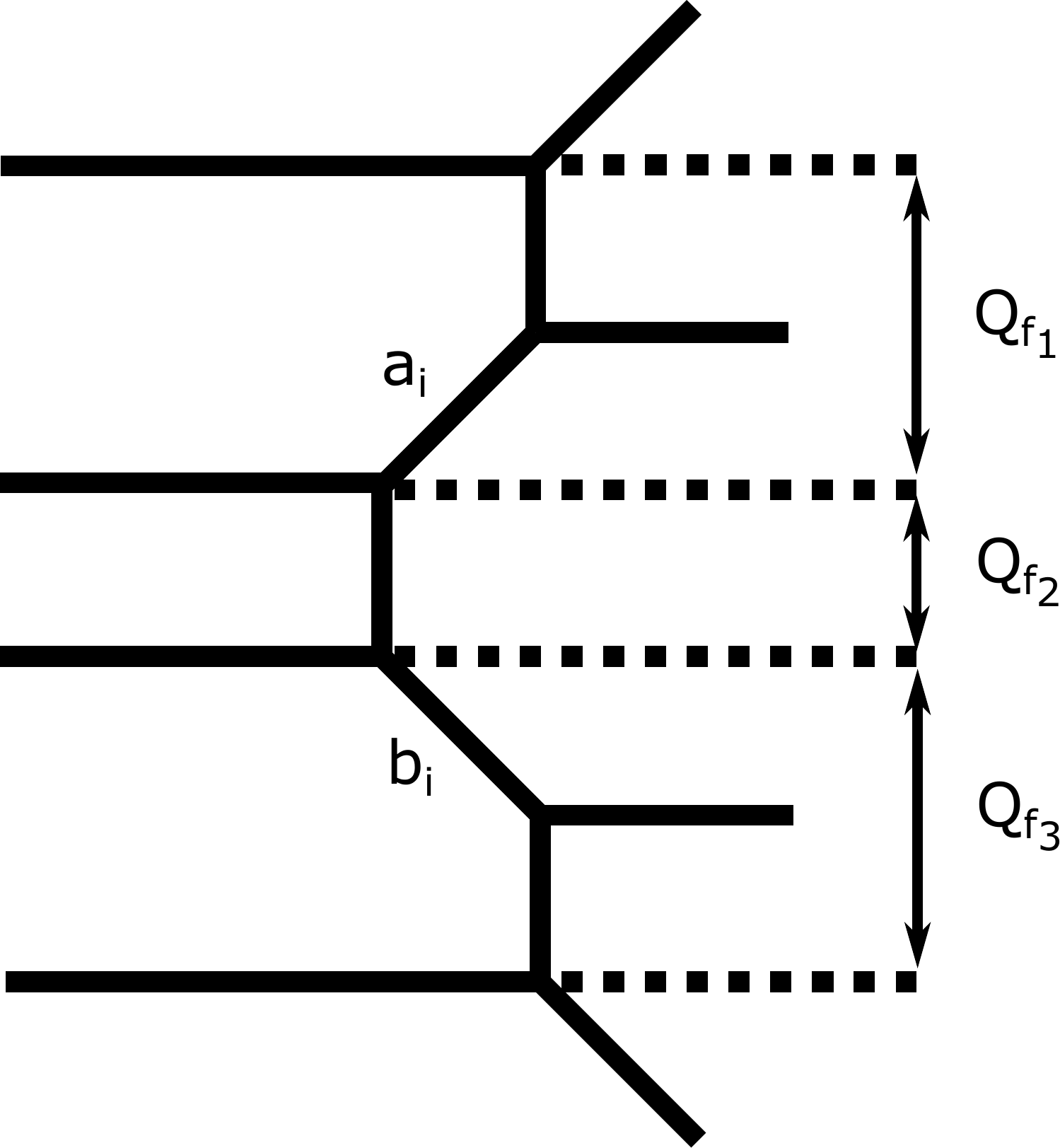}}
\hspace{1cm}
\subfigure[]{\label{fig:su4}
\includegraphics[width=7cm]{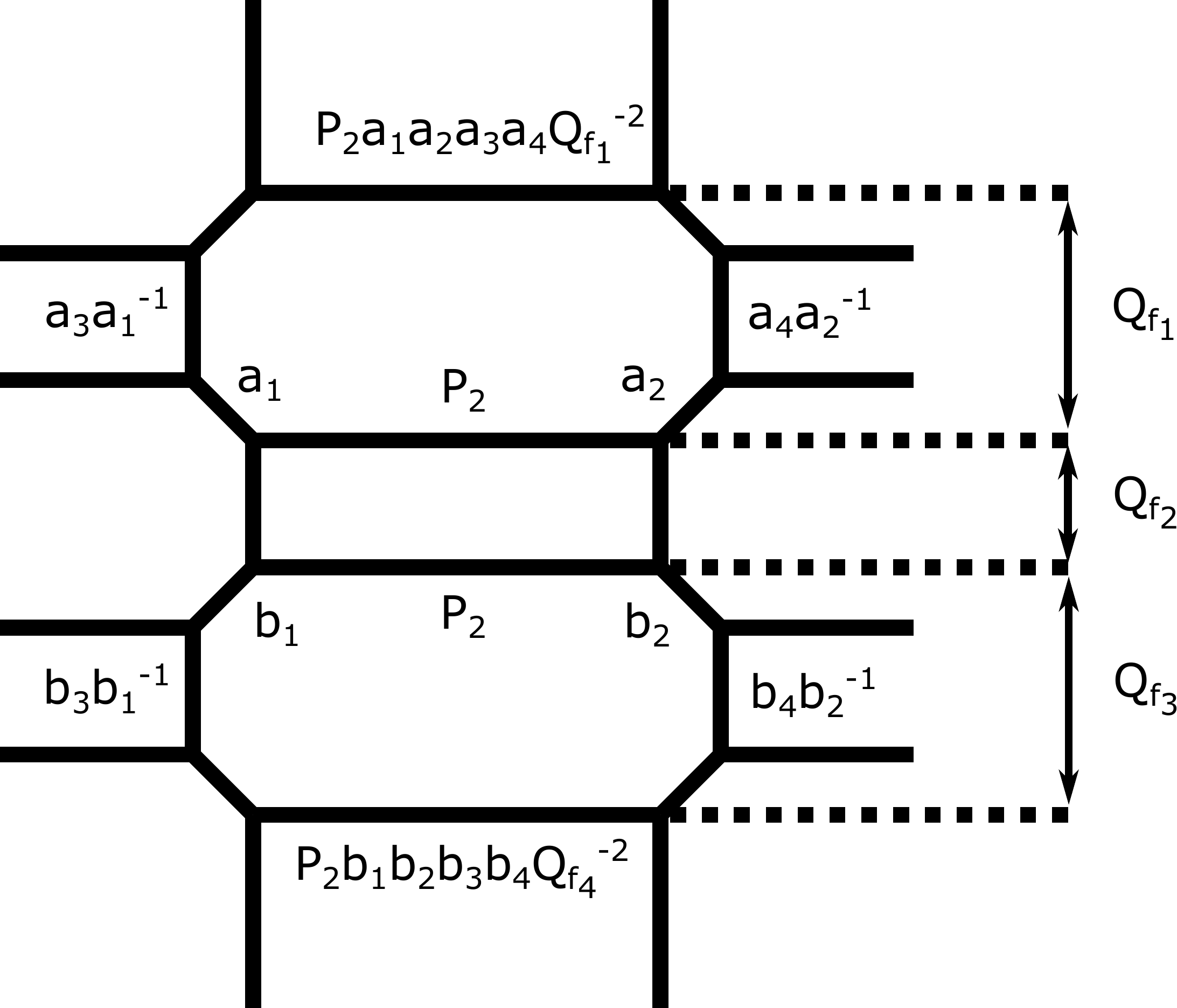}}
\caption{(a): 
A web diagram which describes a hypermultiplet in the bifundamental representation between the $\SU(4)$ and an $\SU(2)$ in the affine $D_4$ Dynkin quiver theory \eqref{affineD4}. (b): A web diagram which describes the $\SU(4)_0$ gauge theory with $8$ flavors.}
\label{fig:su4fig}
\end{figure}

In Figure \ref{fig:g2para}, we also write the parameters associated to the gluing lines. We assigned the K\"ahler parameter $P_2$ for the length of the gluing line which is the second or third from the top. The length of the top and the bottom gluing lines can be determined from the 
local structure around the quadrivalent gluing \cite{Hayashi:2019fsa}. 
To determine the K\"ahler parameters, we first write down a diagram which shows local structure of the gluing. Although we cannot depict the whole diagram on a plane, it is possible to write down a local diagram focusing on the quadrivalent gluing on a plane. A 5d gauge theory description of the original 6d theory $(D_4, D_4)_2$ on $S^1$ before the Higgsing $\left[(2,2,2,0), (2,0,0,0)\right]$ was the affine $D_4$ Dynkin quiver theory given in \eqref{affineD4}. The affine $D_4$ Dynkin quiver theory consists of the central $\SU(4)$ gauge node and four tails with $\SU(2)$ gauge groups. The web diagram representing a hypermultiplet in the bifundamental representation between the $\SU(4)$ and one of the four $\SU(2)$'s is depicted in Figure \ref{fig:halfsu4} where $i=1, 2, 3, 4$ for the four tails. The $\SU(4)$ gauge theory part in the affine $D_4$ Dynkin quiver theory is obtained by gluing four copies of the diagram in Figure \ref{fig:halfsu4} and hence it is locally given by the $\SU(4)$ gauge theory with $2 \times 4 = 8$ flavors with the zero Chern-Simons level. A web diagram which describes the $\SU(4)$ gauge theory is depicted in Figure \ref{fig:su4}. 
For one $\SU(2)$ Higgsing associated with upper external legs in Figure \ref{fig:d4d4d4} induces tuning $a_i =0$ for an $i$. On the other hand
one $\SU(2)$ Higgsing associated with lower external legs in Figure \ref{fig:d4d4d4} induces tuning $b_i =0$ for an $i$. Then performing the $\left[(2,2,2,0), (2,0,0,0)\right]$ Higgsing is realized by the tuning 
\begin{align}
a_1 = a_2 = a_3 = Q_{f_1}, \qquad b_1 = Q_{f_3}.
\end{align}
After the tuning the remaining parameters are related to the parameters in Figure \ref{fig:g2para} by
\begin{align}\label{SU4CB}
Q_{f_1} = Q_{f_3} = Q_1P_1, \quad Q_{f_2} = Q_B, 
\end{align}
and 
\begin{align}
a_4= Q_3, \quad b_2 = Q_1, \quad b_3 = Q_2, \quad b_4 = Q_4.
\end{align}
Then the K\"ahler parameters for the top and the bottom horizontal lines in Figure \ref{fig:su4} become
\begin{align}
P_2a_1a_2a_3a_4Q_{f_1}^{-2} &= P_2Q_1Q_3P_1,\\
P_2b_1b_2b_3b_4Q_{f_3}^{-2} &= P_2Q_2Q_4P_1^{-1},
\end{align}
which are exactly the K\"ahler parameters for the top and the bottom gluing lines written in Figure \ref{fig:g2para}. 

In terms of the computation the only remaining part which we still need to determine is the framing factors for the gluing lines. Comparing the web diagram before the Higgsing in Figure \ref{fig:d4d4d4} and the one after the Higgsing in Figure \ref{fig:g2on2}, the $(p, q)$-charge of the 5-branes attached to the gluing lines does not change. This implies that the framing factors also do not change by the Higgsing. The framing factors before the Higgsing can be read off from the local diagram around the central $\SU(4)$ gauge node given in FIgure \ref{fig:su4}. Therefore we will use the framing factors for the four gluing lines depicted in Figure \ref{fig:su4} for the framing factors of the four gluing lines in Figure \ref{fig:g2on2}.

We now have all the ingredients for computing the partition function of the theory realized by the web diagram in Figure \ref{fig:g2on2}. First we apply the toplogical vertex to the four diagrams in Figure \ref{fig:g2para}. Some formulae of the topological vertex formalism are summarized in appendix \ref{sec:top}. The contribution from each diagram is given by
\begin{align}
&Z^{\fg_2, \text{top}}_{1, \{\mu_i\}}\left(Q_B, \{Q_a\}, \{P_b\}\right)\cr
&\quad= q^{\frac{1}{2}\sum_{i=1}^4||\mu_i^t||^2}\left(\prod_{i=1}^4\tilde{Z}_{\mu_i^t}(q)\right)\mathcal{I}_{\mu_1, \mu_4}^-\left(Q_BQ_1^2P_1^2\right)\mathcal{I}_{\mu_2, \mu_3}^-\left(Q_B\right),\label{g2top1}\\
&Z^{\fg_2, \text{top}}_{2, \{\mu_i\},}\left(Q_B, \{Q_a\}, \{P_b\}\right)\cr
&\quad= q^{\frac{1}{2}\sum_{i=1}^4||\mu_i^t||^2}\left(\prod_{i=1}^4\tilde{Z}_{\mu_i^t}(q)\right)\mathcal{I}_{\mu_2, \mu_3}^-\left(Q_B\right)\mathcal{I}_{\mu_2, \mu_4}^-\left(Q_BQ_1P_1\right)\mathcal{I}_{\mu_3, \mu_4}^-\left(Q_1P_1\right)\cr
&\quad\sum_{\lambda_1}\left(-P_1\right)^{|\lambda_1|}q^{\frac{1}{2}\left(||\lambda_1||^2 + ||\lambda_1^t||^2\right)}\left(\tilde{Z}_{\lambda_1}(q)\tilde{Z}_{\lambda_1^t}(q)\right)\mathcal{I}_{\mu_1, \lambda_1}^-\left(Q_BQ_1^2P_1\right)\cr
&\quad\mathcal{I}_{\mu_2, \lambda_1}^+\left(Q_BQ_1\right)\mathcal{I}_{\mu_3, \lambda_1}^+\left(Q_1\right)\mathcal{I}_{\lambda_1, \mu_4}^+\left(P_1\right),\label{g2top2}\\
&Z^{\fg_2, \text{top}}_{3, \{\mu_i\}}\left(Q_B, \{Q_a\}, \{P_b\}\right)\cr
&\quad= q^{\frac{1}{2}\sum_{i=1}^4||\mu_i^t||^2}\left(\prod_{i=1}^4\tilde{Z}_{\mu_i^t}(q)\right)\mathcal{I}_{\mu_2, \mu_3}^-\left(Q_B\right)\mathcal{I}_{\mu_2, \mu_4}^-\left(Q_BQ_1P_1\right)\mathcal{I}_{\mu_3, \mu_4}^-\left(Q_1P_1\right)\cr
&\quad\sum_{\lambda_2}\left(-Q_1Q_2^{-1}P_1\right)^{|\lambda_2|}q^{\frac{1}{2}\left(||\lambda_2||^2 + ||\lambda_2^t||^2\right)}\left(\tilde{Z}_{\lambda_2}(q)\tilde{Z}_{\lambda_2^t}(q)\right)\mathcal{I}_{\mu_1, \lambda_2}^-\left(Q_BQ_1Q_2P_1\right)\cr
&\quad\mathcal{I}_{\mu_2, \lambda_2}^+\left(Q_BQ_2\right)\mathcal{I}_{\mu_3, \lambda_2}^+\left(Q_2\right)\mathcal{I}_{\lambda_2, \mu_4}^+\left(Q_1Q_2^{-1}P_1\right),\label{g2top3}\\
&Z^{\fg_2, \text{top}}_{4, \{\mu_i\}}\left(Q_B, \{Q_a\}, \{P_b\}\right)\cr
&\quad= q^{\frac{1}{2}\sum_{i=1}^4||\mu_i^t||^2}\left(\prod_{i=1}^4\tilde{Z}_{\mu_i^t}(q)\right)\mathcal{I}_{\mu_1, \mu_2}^-\left(Q_1P_1\right)\mathcal{I}_{\mu_1, \mu_3}^-\left(Q_BQ_1P_1\right)\mathcal{I}_{\mu_1, \mu_4}^-\left(Q_BQ_1^2P_1^2\right)\cr
&\quad\mathcal{I}_{\mu_2, \mu_3}^-\left(Q_B\right)\mathcal{I}_{\mu_2, \mu_4}^-\left(Q_BQ_1P_1\right)\mathcal{I}_{\mu_3, \mu_4}^-\left(Q_1P_1\right)\cr
&\quad\sum_{\lambda_3, \lambda_4}\left(-P_0\right)^{|\lambda_3|}\left(-P_0Q_3Q_4^{-1}\right)^{|\lambda_4|}q^{\frac{1}{2}\sum_{i=3,4}\left(||\lambda_i||^2 + ||\lambda_i^t||^2\right)}\left(\prod_{i=3,4}\tilde{Z}_{\lambda_i}(q)\tilde{Z}_{\lambda_i^t}(q)\right)\cr
&\quad\mathcal{I}_{\mu_1, \lambda_3}^+\left(Q_1Q_3^{-1}P_1\right)\mathcal{I}_{\mu_1, \lambda_4}^+\left(Q_BQ_1Q_4P_1\right)\mathcal{I}_{\lambda_3, \mu_2}^+\left(Q_3\right)\mathcal{I}_{\lambda_3, \mu_3}^+\left(Q_BQ_3\right)\mathcal{I}_{\lambda_3, \lambda_4}^-\left(Q_BQ_3Q_4\right)^2\cr
&\quad\mathcal{I}_{\lambda_3, \mu_4}^+\left(Q_BQ_3Q_1P_1\right)\mathcal{I}_{\mu_2, \lambda_4}^+\left(Q_BQ_4\right)\mathcal{I}_{\mu_3, \lambda_4}^+\left(Q_4\right)\mathcal{I}_{\lambda_4, \mu_4}^+\left(Q_1Q_4^{-1}P_1\right),\label{g2top4}
\end{align}
where we defined
\begin{align}
\mathcal{I}^{\pm}_{\mu, \nu}(Q) = \prod_{i,j=1}^{\infty}\left(1 - Qq^{i + j - 1 - \mu_i - \nu^t_j}\right)^{\pm 1}.
\end{align}
The notation of the form $\{A_b\}$ implies $A_1, A_2, \cdots$ collectively and we will use this notation in the later expressions also. 
When we simply sum over the Young diagrams $\mu_i, \; (i=1, 2, 3, 4)$ for the product of the four factors \eqref{g2top1} - \eqref{g2top4} with the appropriate K\"ahler parameters and the framing factors for the gluing legs, it does not lead to the correct result. One of the points of the quadrivalent gluing prescription is that we need to divide each of the four functions \eqref{g2top1} - \eqref{g2top4} roughly by the square root of the $\SU(4)$ vector mutliplet contribution before the summation of the Young diagrams. More precisely we divide each of them by 
\begin{align}
Z_{\{\mu_i\}}^{\text{SU(4)$_R$}}\left(Q_B, \{Q_a\}, \{P_b\}\right) &=q^{\frac{1}{2}\sum_{i=1}^4||\mu_i^t||^2}\left(\prod_{i=1}^4\tilde{Z}_{\mu_i^t}(q)\right)\prod_{1\leq i < j \leq 4}\mathcal{I}^-_{\mu_i, \mu_j}\left(Q_{f_i}Q_{f_{i+1}}\cdots Q_{f_{j-1}}\right),\label{SU4R}
\end{align}
where $Q_{f_i}$'s are given by \eqref{SU4CB}. Since we divide \eqref{g2top1} - \eqref{g2top4} by \eqref{SU4R}, we need to introduce $\SU(4)$ vector mutiplet contribution when we glue the four functions. The final result of the partition function is then given by
\begin{equation}\label{part.6dg2}
\begin{split}
&Z_{\fg_2^{(1)}, (-2)}^{\text{6d}} \cr
&= \sum_{\{\mu_i\}}(-P_2Q_1P_1Q_3)^{|\mu_1|}(-P_2)^{|\mu_2|+|\mu_3|}(-P_2Q_2Q_4P_1^{-1})^{|\mu_4|}\cr
&\hspace{1cm}Z_{\{\mu_i\}}^{\text{SU(4)$_L$}}\left(Q_B, \{Q_a\}, \{P_b\}\right) Z_{\{\mu_i\}}^{\text{SU(4)$_R$}}\left(Q_B, \{Q_a\}, \{P_b\}\right) f_{\mu_1}(q)^{-1}f_{\mu_2}(q)f_{\mu_3}(q)^{-1}f_{\mu_4}(q)\cr
&\hspace{1cm}Z^{\fg_2}_{1, \{\mu_i\}}\left(Q_B, \{Q_a\}, \{P_b\}\right)Z^{\fg_2}_{2, \{\mu_i\}}\left(Q_B, \{Q_a\}, \{P_b\}\right)Z^{\fg_2}_{3, \{\mu_i\}}\left(Q_B, \{Q_a\}, \{P_b\}\right)\cr
&\hspace{1cm}Z^{\fg_2}_{4, \{\mu_i\}}\left(Q_B, \{Q_a\}, \{P_b\}\right),
\end{split}
\end{equation}
where we defined
\begin{align}
Z^{\fg_2}_{j, \{\mu_i\}}\left(Q_B, \{Q_a\}, \{P_b\}\right):=\frac{Z^{\fg_2, \text{top}}_{j, \{\mu_i\}}\left(Q_B, \{Q_a\}, \{P_b\}\right)}{Z_{\{\mu_i\}}^{\text{SU(4)$_R$}}\left(Q_B, \{Q_a\}, \{P_b\}\right) },
\end{align}
for $j=1, 2, 3, 4$ and 
\begin{align}
Z_{\{\mu_i\}}^{\text{SU(4)$_L$}}\left(Q_B, \{Q_a\}, \{P_b\}\right) =q^{\frac{1}{2}\sum_{i=1}^4\left(||\mu_i||^2 - ||\mu_i^t||^2\right)}Z_{\{\mu_i\}}^{\text{SU(4)$_R$}}\left(Q_B, \{Q_a\}, \{P_b\}\right). \label{SU4L}
\end{align}

In general, 6d (and also 5d) partition functions computed from the topological vertex or the localization method may contain an extra factor \cite{Bergman:2013ala, Hayashi:2013qwa, Bao:2013pwa, Bergman:2013aca, Hwang:2014uwa} which is characterized by a factor indepedent from Coulomb branch moduli of 5d theories in the partition functions. We will use this nomenclature throughout this paper. The extra factor can contain contributions which are not in the BPS spectrum of 5d KK theories and we will focus on the part with the extra factor removed in the later comparison. The partition function \eqref{part.6dg2} may contain an extra factor and we consider factoring out the extra factor from \eqref{part.6dg2}. For that we reparameterize the length of lines in Figure \ref{fig:g2para} in terms of Coulomb branch moduli and mass parameters in a 5d description. 

First let us introduce mass parameters. 
In a web diagram, length between parallel external legs does not depend on Coulomb branch moduli and hence we can assign a mass parameter to the length between parallel external legs. 
From the web diagram in Figure \ref{fig:g2para}, each of Figure \ref{fig:g2para2}, Figure \ref{fig:g2para3}, Figure \ref{fig:g2para4} contain two parallel external legs and we can parameterize the lengths between them by mass parameters $M'_1, M'_2, M'_3, M'_4$ as
\begin{align}
P_1 = M'_1, \quad Q_1P_1Q_2^{-1} = M'_2, \quad P_0Q_1P_1Q_3^{-1} = M'_3, \quad P_0Q_1P_1Q_4^{-1} = M'_4. 
\end{align}
The web diagram in Figure \ref{fig:su4} implies that the length of the top gluing line in Figure \ref{fig:g2para} also characterizes the length between parallel external legs and we can assign 
\be
P_2Q_1Q_3P_1 = M'_0. 
\ee

On the other hand it is possible to introduce the Coulomb branch moduli by utilizing the intersection number between a complex surface and a complex curve in the dual geometry. Namely the dependence of the Coulomb branch moduli $A'_i\; (i=1, \cdots, \text{rank}(G))$ for the K\"ahler parameter $Q_C$ 
of a complex curve $C$ may be assigned by 
\be\label{QCB}
Q_C \propto A'^{n^C_i}_i,
\ee
where $n^C_i$ is the intersection number $n^C_i = -(C \cdot S_i)$ with a complex compact surface $S_i$ which corresponds to a compact face in a web diagram. In the current example we have three faces including the face whose fiber class gives the affine node in the affine $G_2$ Dynkin diagram in Figure \ref{fig:g2dynkin}. When we consider a geometry which consists of an elliptic fiber whose degeneration is given by an affine Lie algebra $\mathfrak{g}^{(1)}$ over a rational curve, the class of the elliptic fiber is given by \eqref{fiberI} and its K\"ahler parameter is obtained by
\be\label{elliptic}
Q_{\tau} = \prod_{a=0}^{n_d-1}F_{a}^{d_i},
\ee
where $n_d$ is the number of nodes for the affine Dynkin diagram for $\mathfrak{g}^{(1)}$, $F_i$ is the K\"ahler parameter for the fiber class of each surface and $d_i$ is the mark of the corresponding node of the affine Dynkin diagram. In the case of the web diagram in Figure \ref{fig:g2on2} for the geometry of $\mathfrak{g}_2^{(1)}$ on $(-2)$-curve, it is given by
\be\label{elliptic.g2}
Q_{\tau} = (Q_3P_0)(P_2)^2(Q_1P_1)^3 = M'^2_0M'_3.
\ee

With this parameterization we can determine an extra factor from the partition function \eqref{part.6dg2} by identifying a factor which only depends on $M'_i\; (i=0, 1, 2, 3, 4)$. Denoting such a factor by $Z_{\text{extra}}^{\fg_2^{(1)},(-2)}(M'_0, M'_1, M'_2, M'_3, M'_4)$ the partition function \eqref{part.6dg2} can be written as
\be\label{hatpart.6dg2}
Z_{\fg_{2}^{(1)},(-2)}^{\text{6d}} = \hat{Z}_{\fg_{2}^{(1)},(-2)}^{\text{6d}}\left(\{A'_a\}, \{M'_i\}\right)Z_{\text{extra}}^{\fg_{2}^{(1)},(-2)}\left(M'_0, M'_1, M'_2, M'_3, M'_4\right).
\ee 
We argue that the partition function $\hat{Z}_{\fg_{2}^{(1)},(-2)}^{\text{6d}}\left(\{A'_a\}, \{M'_i\}\right)$ in \eqref{part.6dg2} yields the partition function of the 6d $G_2$ gauge theory with four flavors and a tensor multiplet on $T^2 \times \mathbb{R}^4$ with an extra factor removed. Note that the extra factor part $Z_{\text{extra}}^{\fg_{2}^{(1)},(-2)}\left(M'_0, M'_1, M'_2, M'_3, M'_4\right)$ in \eqref{hatpart.6dg2} is not trivial as the diagram in Figure \ref{fig:g2on2} has parallel external lines. 

\paragraph{5d $G_2$ gauge theory with $4$ flavors.} 

It is possible to take a 5d limit to the web diagram in Figure \ref{fig:g2on2} and obtain a web diagram for the 5d $G_2$ gauge theory with four flavors. Note that the web diagram in Figure \ref{fig:g2on2} realizes the 6d $G_2$ gauge theory with four flavors and a tensor multiplet compactified on a circle $S^1$. Let the radius of the circle $R_{\text{6d}}$. When we consider the dual geometry in M-theory then the K\"ahler parameter of the elliptic class, $Q_{\tau}$, is roughly related to the radius by
\be\label{QR6d}
Q_{\tau} \sim e^{-\frac{1}{R_{\text{6d}}}} 
\ee
from the duality between M-theory and F-theory. Then the 5d limit $R_{\text{6d}} \to 0$ of the 6d theory on $S^1$ amounts to $Q_{\tau} \to 0$. In the current case, the limit can be realized by taking $P_0 \to 0$ with the other K$\ddot{\text{a}}$hler parameters $P_a, Q_b, Q_B \;(a=1, 2, b=1, 2, 3, 4)$ fixed. In terms of the affine $G_2$ Dynkin diagram in Figure \ref{fig:g2dynkin} formed by the fiber classes of the three faces, the 5d limit decouples the fiber class associated to the affine node. Hence the intersection form between the remaining fiber classes yields the $G_2$ Dynkin diagram, leading to the $G_2$ Lie algebra for the 5d $G_2$ gauge group. 
\begin{figure}[t]
\centering
\includegraphics[width=5cm]{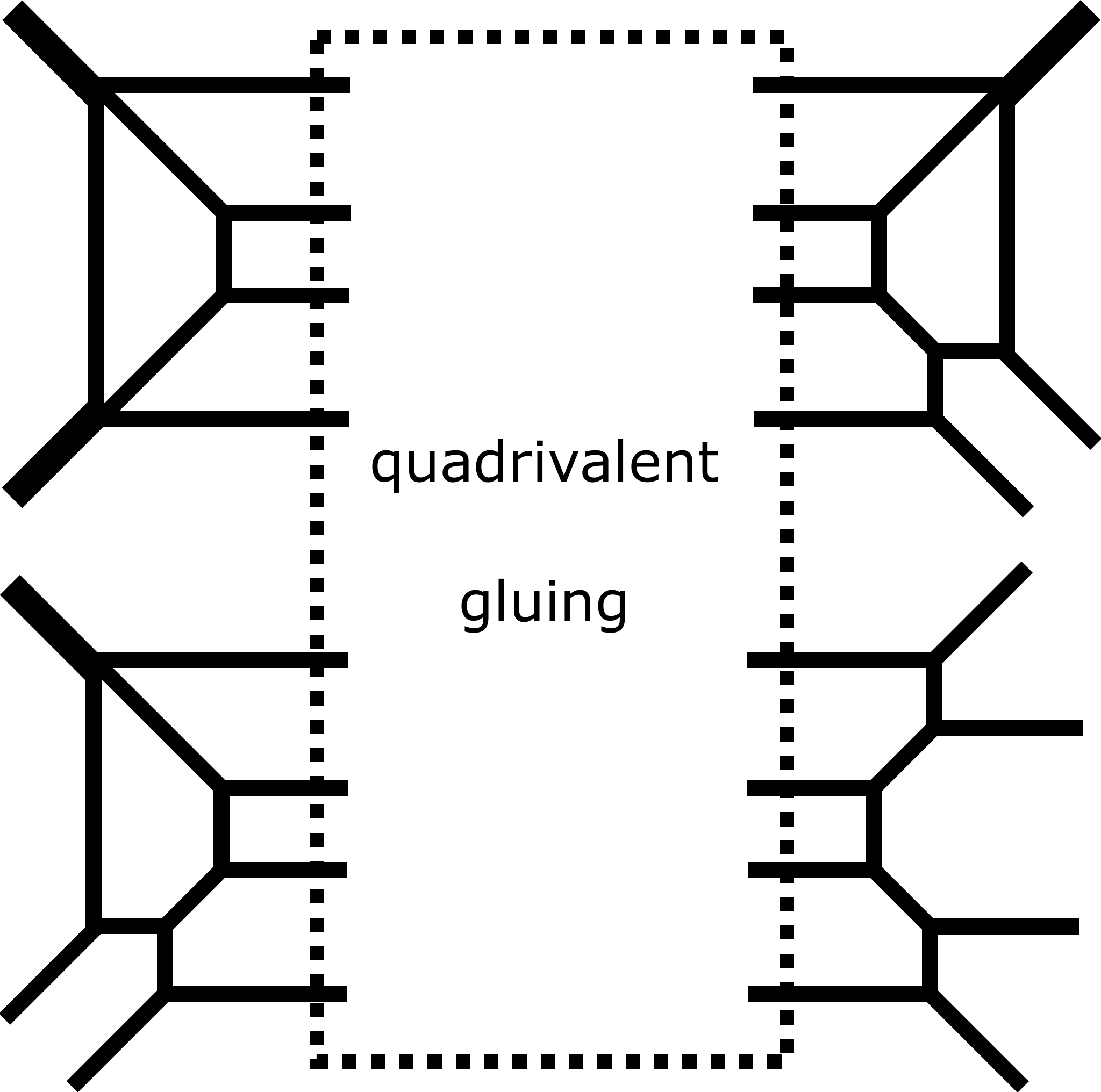}
\caption{
The web diagram for 5d $G_2$ gaue theory with $4$ flavors. }
\label{fig:g2w4f}
\end{figure}
The diagram after applying the limit $P_0 \to 0$ to the web diagram in Figure \ref{fig:g2on2} is depicted in Figure \ref{fig:g2w4f}. Since it corresponds to the 5d limit of the 6d $G_2$ gauge theory with four flavors and a tensor multiplet on $S^1$, the web diagram in Figure \ref{fig:g2w4f} realizes the 5d $G_2$ gauge theory with four flavors.  

It is then straightforward to compute the partition function of the 5d $G_2$ gauge theory with four flavors on $S^1 \times \mathbb{R}^4$ by applying the 5d limit to the partition function \eqref{part.6dg2}. 
The 5d partition function after the limit $P_0 \to 0$ may also contain an extra factor and it can be extracted by writing the partition function in terms of the gauge theory parameters of the 5d $G_2$ gauge theory with four flavors. The parameterization of the Coulomb branch moduli can be done in the same way as the parameterization of the 6d case given by \eqref{QCB}. Note that the fiber class for the affine node of the affine $G_2$ Dynkin diagram  is decoupled and the fiber classes form the $G_2$ Dynkin diagram. Then for a line corresponding to a state of a root or a weight with the Dynkin label $[l_1, l_2]$ of the $G_2$ Lie algebra, the Coulomb branch moduli dependence becomes $A_1^{l_1}A_2^{l_2}$ since the intesrection number \eqref{QCB} becomes the Dynkin label of the Lie algebra. 
In particular the K\"ahler parameters $P_2, Q_1P_1$ for the fiber classes in Figure \ref{fig:g2dynkin} only depend on the Coulomb branch moduli and we parameterize them by
\begin{align}\label{g2.CB}
Q_1P_1 = A_1^2A_2^{-1}, \qquad P_2 = A_1^{-3}A_2^2. 
\end{align}
Similarly, when fiber classes of surfaces form a Dynkin diagram of a Lie algebra $\mathfrak{g}$, we will parameterize the K\"ahler parameter for a fiber class $f$ with the Dynkin label $[l_1, l_2, \cdots l_{\text{rank}(\mathfrak{g})}]$ by
\be\label{QfAl}
Q_f = \prod_{i=1}^{\text{rank}(\mathfrak{g})}A_i^{l_i},
\ee
in the later computations of 5d partition functions. 

Since the diagram in Figure \ref{fig:g2w4f} contains four flavors there are four mass parameters for the flavors. A part of the contribution from the four flavors corresponding to a weight in the fundamental representation comes from a string with the length $Q_1, Q_2, Q_3$ or $Q_4$ and hence the four parameters depend on the four mass parameters. The Coulomb branch moduli dependence of the four K\"ahler parameters can be read off from the intersection numbers between the line and the faces as the intersection numbers correspond to the Dynkin labels. The Dynkin labels of the weights corresponding to the four parameters is 
\begin{align}
Q_1\;:\;[2, -1], \qquad Q_2\;:\;[2, -1], \qquad Q_3\;:\;[1, -1] \qquad Q_4\;:\;[1, -1].
\end{align}
Hence we introduce four mass parameters $M_1, M_2, M_3, M_4$ by
\begin{align}\label{g2.matter}
Q_1 = A_1^2A_2^{-1}M_1^{-1}, \quad Q_2 = A_1^2A_2^{-1}M_2^{-1}, \quad Q_3 = A_1A_2^{-1}M_3^{-1}, \quad Q_4 = A_1A_2^{-1}M_4^{-1}. 
\end{align}

There is one more mass parameter associated to the instanton fugacity. A way to determine the dependence on the instanton fugacity is the comparison with the effective prepotential on a Coulomb branch. For a 5d gauge theory with a gauge group $G$ and matter in the representation $R_f$, the effective prepotential on a Coulomb branch is given by \cite{Seiberg:1996bd, Morrison:1996xf, Intriligator:1997pq}
\be\label{prepotential}
\mathcal{F}(a) = \frac{1}{2}m_0h_{ij}a_ia_j + \frac{\kappa}{6}d_{ijk}a_ia_ja_k + \frac{1}{12}\left(\sum_{r \in\text{roots}}\left|r\cdot a\right|^3 - \sum_f\sum_{w\in R_f}\left|w\cdot a + m_f\right|^3\right),
\ee
where $a_i\;(i=1, 2, \cdots, \text{rank}(G))$ are the Coulomb branch moduli, $m_0$ is the inverse of the squared gauge coupling, $\kappa$ is the classical CS level, $m_f$ is the mass parameter for the matter labeled by $f$. $r$ is a root of the Lie algebra of $G$, $w$ is a weight of the representation $R_f$. Also we defined $h_{ij} = \text{Tr}(T_iT_j)$ and $d_{ijk}=\frac{1}{2}\text{Tr}\left(T_i\{T_j,T_k\}\right)$ where $T_i$'s are the Cartan generators. The derivative $\frac{\partial F}{\partial a_i}$ yields the tension of a monopole string. From a 5-brane web diagram a monopole string is given by a D3-brane filling a face and the tension is the area of the face. 

Let us then apply \eqref{prepotential} to the case of the 5d $G_2$ gauge theory with four flavors realized on the web diagram in Figure \ref{fig:g2w4f}. The parameters in \eqref{prepotential} are related to the K\"ahler parameters by $A_b = e^{-a_b}, M_i = e^{-m_i}, \; (b=1, 2, i=1, 2, 3, 4)$. The phase from the root part is determined by the choice \eqref{g2.CB}. The phase from the matter part is also determined from \eqref{g2.matter} from the fact that the effective mass is positive if it is given by a linear combination with positive coefficients of the lengths of lines in the web diagram in Figure \ref{fig:g2w4f}. Then the monopole string tension from the derivative by $a_2$ becomes
\be\label{g2area}
\frac{\partial F(a)}{\partial a_2} = (-3a_1 + 2a_2)(-2a_1 + 2a_2 + m_0 + m_3 + m_4).
\ee
From the relation between the tension of a monopole string and the area of a face \eqref{g2area} should be equal to the area of the central face in Figure \ref{fig:g2w4f}. From the diagram in Figure \ref{fig:su4}, the middle face is still locally described by $\mathbb{F}_0$ after the Higgsing and its area is simply given by $q_B(-3a_1 + 2a_2)$ with $Q_B = e^{-q_b}$. Hence we identify 
\be
Q_B = A_1^{-2}A_2^{2}M_3M_4M_0,
\ee
where $M_0 = e^{-m_0}$. Note that we have
\be\label{g2.m0}
Q_BQ_3Q_4 = M_0.
\ee
The relation \eqref{g2.m0} means that the length between parallel horizontal external lines turns out to be $m_0$.  

As mentioned before, the partition function of the 5d $G_2$ gauge theory with four flavors is obtained by applying the limit $P_0 \to 0$ to \eqref{part.6dg2} up to an extra factor. Namely we have 
\begin{align}\label{part.5dg2}
Z^{\text{5d}}_{\fg_2+ 4\text{F}} &= Z^{\text{6d}}_{\fg_2^{(1)}, (-2)}\Big|_{P_0 = 0}\cr
&=\hat{Z}^{\text{5d}}_{\fg_2+ 4\text{F}}\left(\{A_b\}, \{M_i\}\right)Z_{\text{extra}}^{\fg_2+ 4\text{F}}\left(M_0, M_1, M_2, M_3, M_4\right),
\end{align}
and we argue $\hat{Z}^{\text{5d}}_{\text{$\fg_2$+$4$F}}\left(\{A_b\}, \{M_i\}\right)$ yields the partition function of the 5d $G_2$ gauge theory with four flavors on $S^1 \times \mathbb{R}^4$ except for an extra factor which is independent of the Coulomb branch moduli .

In order to check the validity of the result \eqref{part.5dg2}, we compare the perturbative part of \eqref{part.5dg2} with the universal formula for the perturbative part of the 5d Nekrasov partition function of the 5d $G_2$ gauge theory with four flavors. The perturbative part also depends on a phase of the dual Calabi-Yau threefold and we use the phase corresponding to the diagram in Figure \ref{fig:g2w4f}, which is the same phase as the one which we used in the evaluation of \eqref{prepotential}. Then the perturbative part of the partition function of the 5d $G_2$ gauge theory with four flavors becomes
\begin{align}\label{5dg2pert}
Z^{\text{5d pert}}_{\mathfrak{g}_2+4\text{F}} = Z^{\mathfrak{g}_2}_{\text{cartan}}Z^{\mathfrak{g}_2}_{\text{roots}}Z^{\mathfrak{g}_2}_{\text{flavor 1}}Z^{\mathfrak{g}_2}_{\text{flavor 2}}Z^{\mathfrak{g}_2}_{\text{flavor 3}}Z^{\mathfrak{g}_2}_{\text{flavor 4}},
\end{align}
where each factor is given by
\begin{align}
Z^{\mathfrak{g}_2}_{\text{cartan}} &= \text{PE}\left[\frac{2q}{(1-q)^2}\right],\\
Z^{\mathfrak{g}_2}_{\text{roots}} &= \text{PE}\left[\frac{2q}{(1-q)^2}\left(\frac{A_1^3}{A_2}+\frac{A_1^2}{A_2}+A_1+A_2+\frac{A_2}{A_1}+\frac{A_2^2}{A_1^3}\right)\right],\\
Z^{\mathfrak{g}_2}_{\text{flavor 1}} &= \text{PE}\left[-\frac{q}{(1-q)^2}\left(\frac{A_1^2 M_1}{A_2}+\frac{A_1^2}{A_2 M_1}+A_1 M_1+\frac{A_1}{M_1}+\frac{A_2
   M_1}{A_1}+\frac{A_2}{A_1 M_1}+M_1\right)\right],\\
Z^{\mathfrak{g}_2}_{\text{flavor 2}} &= \text{PE}\left[-\frac{q}{(1-q)^2}\left(\frac{A_1^2 M_2}{A_2}+\frac{A_1^2}{A_2 M_2}+A_1 M_2+\frac{A_1}{M_2}+\frac{A_2M_2}{A_1}+\frac{A_2}{A_1 M_2}+M_2\right)\right],\\
Z^{\mathfrak{g}_2}_{\text{flavor 3}} &= \text{PE}\left[-\frac{q}{(1-q)^2}\left(\frac{A_1^2}{A_2 M_3}+A_1 M_3+\frac{A_1}{A_2 M_3}+\frac{A_1}{M_3}+\frac{A_2}{A_1M_3}+\frac{A_2}{A_1^2 M_3}+\frac{1}{M_3}\right)\right],\\
Z^{\mathfrak{g}_2}_{\text{flavor 4}} &= \text{PE}\left[-\frac{q}{(1-q)^2}\left(\frac{A_1^2}{A_2 M_4}+A_1 M_4+\frac{A_1}{A_2 M_4}+\frac{A_1}{M_4}+\frac{A_2}{A_1
   M_4}+\frac{A_2}{A_1^2 M_4}+\frac{1}{M_4}\right)\right].
\end{align}

On the other hand, the perturabtive part of \eqref{part.5dg2} can be extracted by taking the limit $M_0 \to 0$ or $Q_B \to 0$. The diagram in Figure \ref{fig:g2w4f} splits into a upper half part and a lower half part. The partition function from the upper half diagram gives 
\begin{align}\label{g2upper}
Z_{\text{upper}}^{\mathfrak{g}_2 +4\text{F}}&=\text{PE}\left[\frac{q}{(1-q)^2}\left\{\left(\frac{A_1^3}{A_2} - \frac{A_1^2}{A_2} - A_1 + A_2 - \frac{A_2}{A_1} + \frac{A_2^2}{A_1^3}\right) - \left(\frac{A_1^2}{A_2M_3} + A_1M_3 + \frac{A_1}{M_3} \right.\right.\right.\cr
&\qquad\left.\left.\left. +\frac{A_1}{A_2M_3}+ \frac{A_2}{A_1M_3} + \frac{A_2}{A_1^2M_3}\right) + \left(\frac{2}{M_3} + \frac{1}{M_3^2}\right)\right\}\right],
\end{align}
when we consider the summation until the order $P_2^5Q_3^6R_3^6$ where $R_3 := Q_1P_1Q_3^{-1}$. The partition function from the lower half diagram is given by
\begin{equation}\label{g2lower}
\begin{split}
&Z_{\text{lower}}^{\mathfrak{g}_2 +4\text{F}}\cr
&=\text{PE}\left[\frac{q}{(1-q)^2}\left\{\left(\frac{3A_1^2}{A_2} + 3A_1  + \frac{3A_2}{A_1} + \frac{A_2^2}{A_1^3} + \frac{A_1^3}{A_2} + A_2\right)\right.\right.\\
&- \left(\frac{A_1^2M_1}{A_2} + A_1M_1 + \frac{A_1}{M_1}  +\frac{A_2M_1}{A_1}+ \frac{A_2}{A_1M_1} + \frac{A_1^2}{A_2M_1}\right)\cr
&-\left(\frac{A_1^2M_2}{A_2} + A_1M_2 + \frac{A_1}{M_2}  +\frac{A_2M_2}{A_1}+ \frac{A_2}{A_1M_2} + \frac{A_1^2}{A_2M_2}\right)\cr
&-\left(\frac{A_1^2}{A_2M_4} + A_1M_4 + \frac{A_1}{M_4}+ \frac{A_1}{A_2M_4} + \frac{A_2}{A_1M_4} + \frac{A_2}{A_1^2M_4}\right)\cr
&\left.\left.+ \left(-2M_1 + M_1^2 -2M_2 + M_2^2 - \frac{2}{M_4} + \frac{1}{M_4^2} + \frac{M_1M_2}{M_4} + \frac{M_2}{M_1M_4} + \frac{M_1}{M_2M_4} +\frac{1}{M_1M_2M_4}\right)\right\}\right],
\end{split}
\end{equation}
when we consider the summation until the order $P_1^3P_2^2Q_1^3Q_2^2Q_4^2R_1^2R_2^3R_4^3$ where $R_1:=P_1^{-1}$, $R_2:=Q_1Q_2^{-1}P_1$, $ R_4:=Q_1Q_4^{-1}P_1$. Indeed we can see that 
\be
Z_{\text{upper}}^{\mathfrak{g}_2 +4\text{F}}Z_{\text{lower}}^{\mathfrak{g}_2 +4\text{F}}=Z^{\mathfrak{g}_2}_{\text{roots}}\left(\prod_{i=1}^4Z'^{\mathfrak{g}_2}_{\text{flavor $i$}}\right)Z_{\text{extra, pert}}^{\fg_2 + 4\text{F}},
\ee
where $Z'^{\mathfrak{g}_2}_{\text{flavor $i$}}$ is $Z^{\mathfrak{g}_2}_{\text{flavor $i$}}$ with the Coulomb branch independent part removed and 
\begin{align}
Z_{\text{extra, pert}}^{\fg_2 + 4\text{F}} &=\text{PE}\left[-2M_1 + M_1^2 -2M_2  + M_2^2 + \frac{2}{M_3} + \frac{1}{M_3^2}- \frac{2}{M_4} + \frac{1}{M_4^2}\right.\cr
&\hspace{4cm}\left.  + \frac{M_1M_2}{M_4} + \frac{M_2}{M_1M_4} + \frac{M_1}{M_2M_4} +\frac{1}{M_1M_2M_4}\right].
\end{align}
Hence \eqref{part.5dg2} in the limit $Q_B \to 0$ reproduced the terms carrying gauge charges in the perturbative part  of the Nekrasov partition function.

\subsection{6d/5d $F_4$ gauge theory with matter}
\label{sec:f4}
The next example is the partition function of the 6d $F_4$ gauge theory with three flavors and a tensor multiplet on $T^2 \times \mathbb{R}^4$. In fact the partition function may be also obtained from the partition function of the $E_6$ gauge theory with four flavors and a tensor multiplet, which will be discussed in section \ref{sec:e6}, by tuning parameters which realize the Higgsing from $E_6$ to $F_4$. However we will compute the partition function of the $F_4$ gauge theory independently here 
in order to demonstrate the direct application of the topological vertex to $F_4$ web diagrams. 

\begin{figure}[t]
\centering
\subfigure[]{\label{fig:f4on2}
\includegraphics[width=6cm]{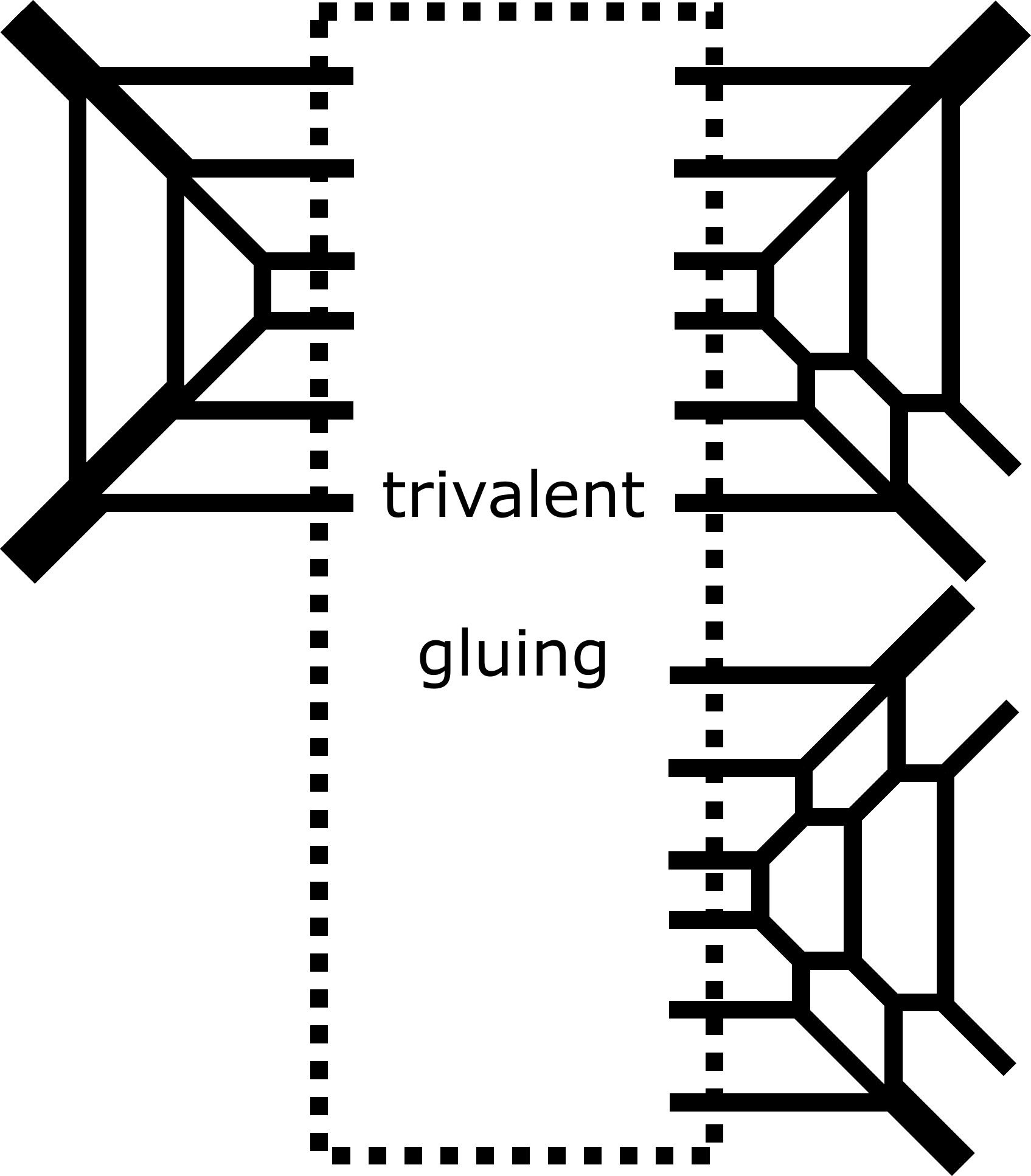}}
\hspace{1cm}
\subfigure[]{\label{fig:f4w3f}
\includegraphics[width=6cm]{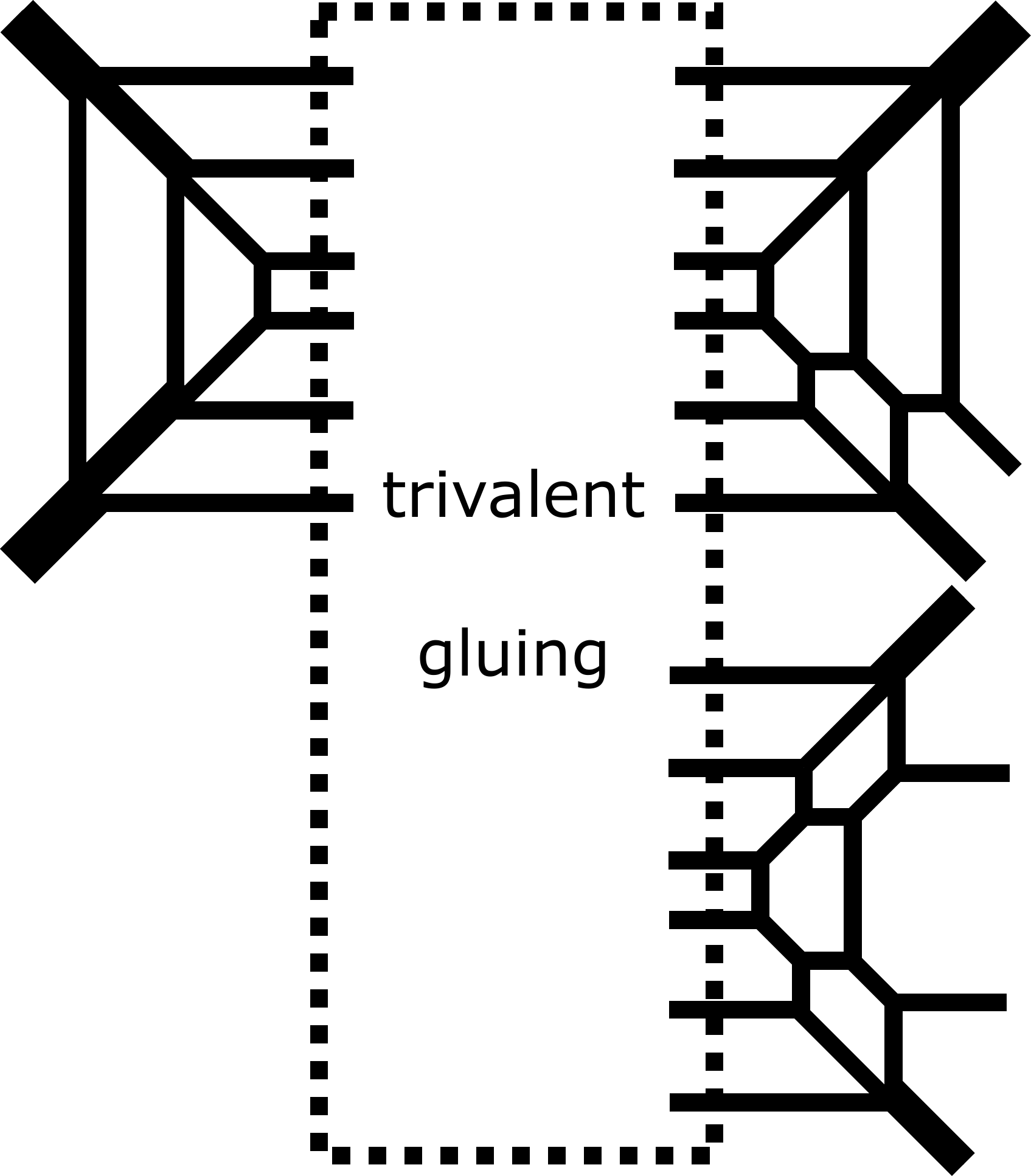}}
\caption{(a). The  web diagram for the geometry $\ff_4^{(1)}$ on $(-2)$-curve. (b). The web diagram for 5d $F_4$ gaue theory with $3$ flavors. }
\label{fig:f4}
\end{figure}

\paragraph{$\ff_4^{(1)}$ on $(-2)$-curve.}
The 6d $F_4$ gauge theory with three flavors and a tensor multiplet compactified on a circle arises as a low energy theory on the $\left[(3,3,2), (3,2,2)\right]$ Higgs branch of the 6d theory $(E_6, E_6)_2$ on $S^1$ 
as in Table \ref{tb:e6_2}. 
A 5d gauge theory description of the original 6d theory $(E_6, E_6)_2$ is the affine $E_6$ Dynkin quiver theory in 
\eqref{e6quiver}. The affine $E_6$ Dynkin quiver theory can be realized on a web diagram with a trivalent gluing and the diagram has been already depicted 
in Figure \ref{fig:e6e6e6}. Then the web diagram corresponding to the geometry $\ff_4^{(1)}$ on $(-2)$-curve, which yields the 6d $F_4$ gauge theory with three flavors and a tensor multiplet compactified on $S^1$, can be obtained by performing the $\left[(3, 3, 2), (3, 2, 2)\right]$ Higgsing to the web diagram in Figure \ref{fig:e6e6e6}. The web diagram after the Higgsing is depicted in Figure \ref{fig:f4on2}. 
The fiber classes of the five faces form the affine $F_4$ Dynkin diagram in a similar manner to Figure \ref{fig:f4web} as expected. As with the $G_2$ case, we can consider the 5d limit of the web diagram in Figure \ref{fig:f4on2}. The 5d limit was achieved by decoupling the fiber class associated to the affine node. We can also take a decoupling limit for the fiber class corresponding to the affine node of the affine $F_4$ Dynkin diagram, leading to the $F_4$ Dynkin diagram, and the resulting web diagram is drawn in Figure \ref{fig:f4w3f}. The decoupling 
realizes the 5d limit and the web diagram in Figure \ref{fig:f4w3f} yields the 5d $F_4$ gauge theory with $3$ hypermultiplets in the fundamental representation. 

\begin{figure}[t]
\centering
\subfigure[]{\label{fig:f4para1}
\includegraphics[width=4cm]{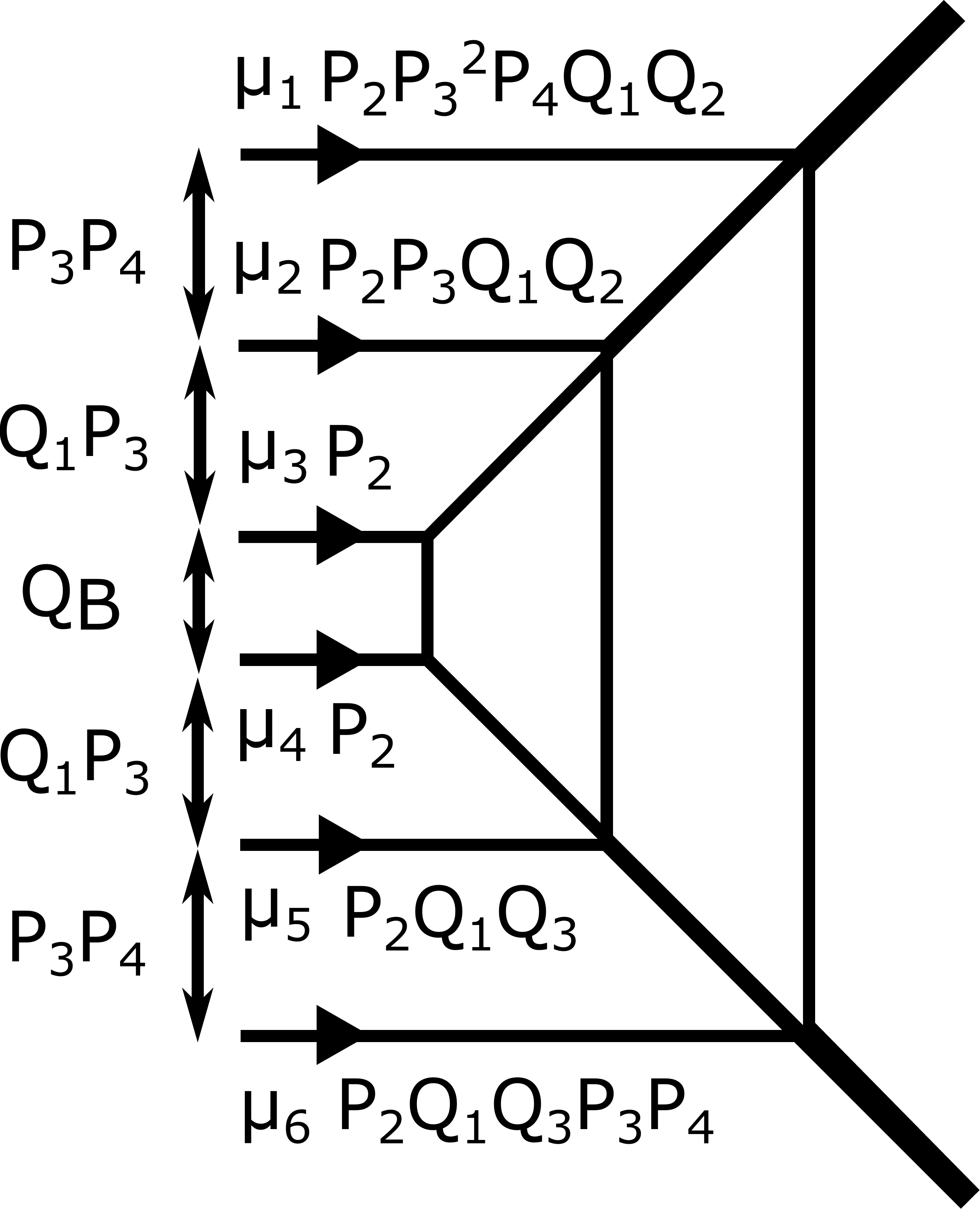}}
\hspace{1cm}
\subfigure[]{\label{fig:f4para2}
\includegraphics[width=4cm]{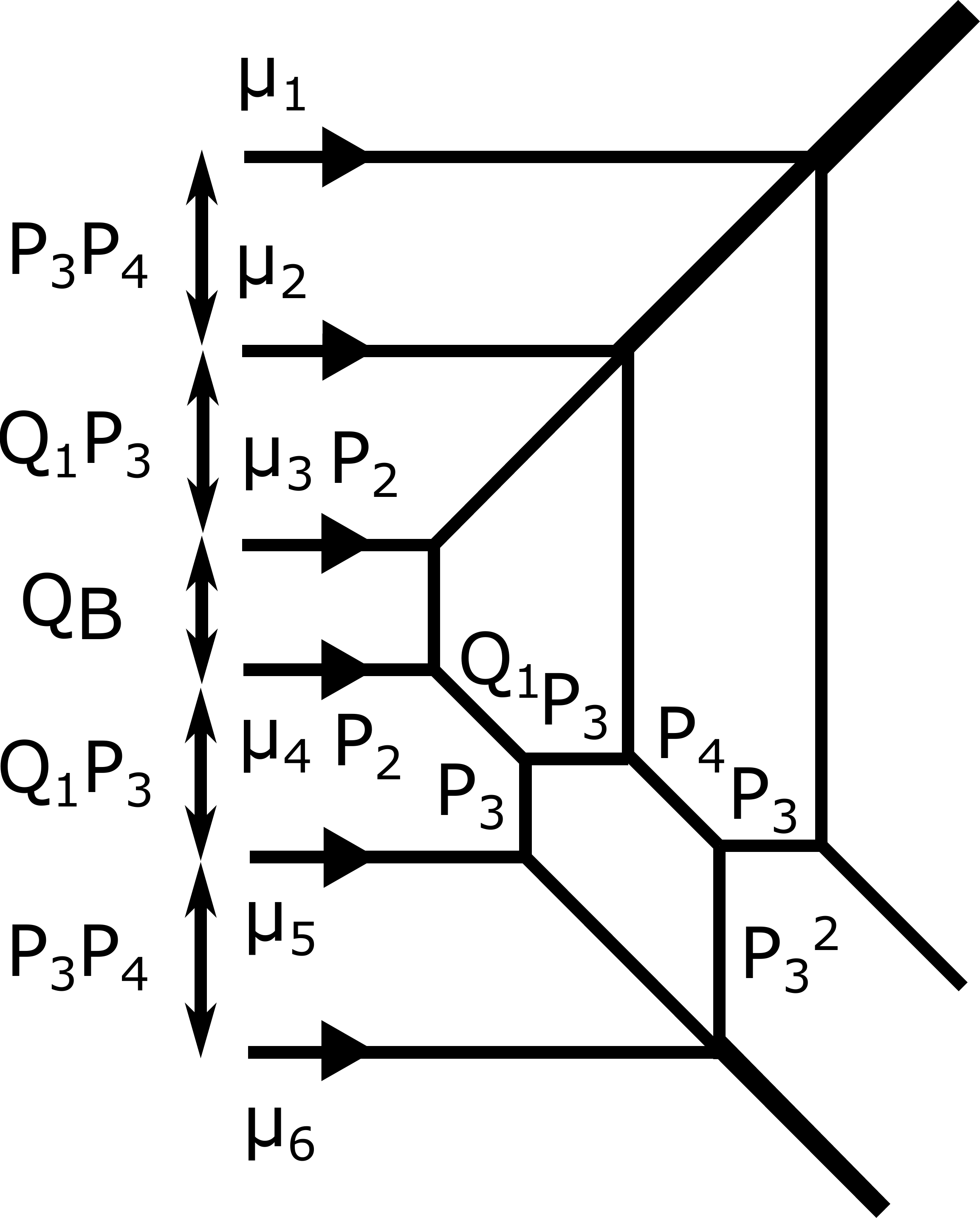}}
\hspace{1cm}
\subfigure[]{\label{fig:f4para3}
\includegraphics[width=4cm]{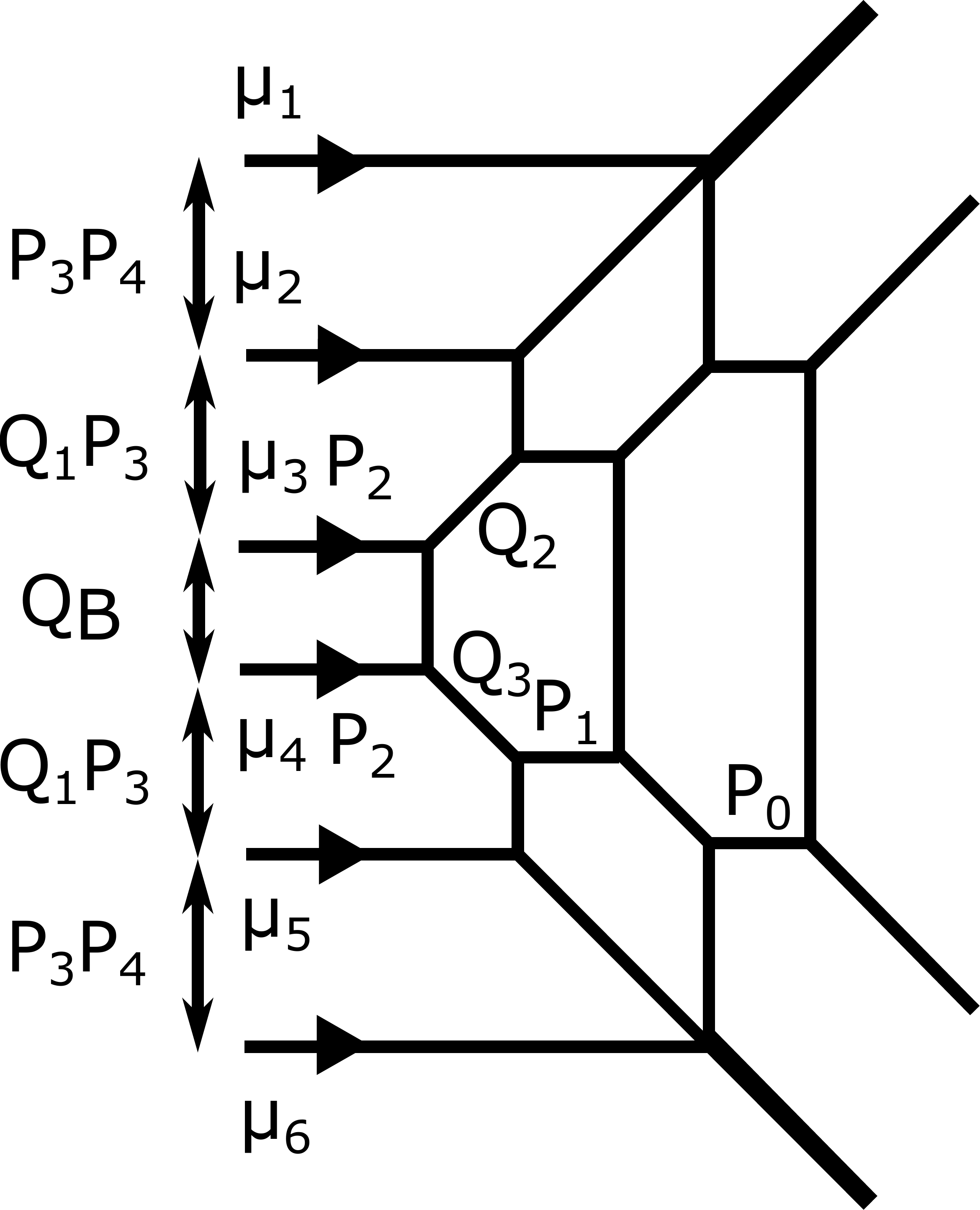}}
\caption{Parameterization of the web diagram for the geometry $\ff_4^{(1)}$ on $(-2)$-curve in Figure \ref{fig:f4on2}.}
\label{fig:f4para}
\end{figure}

\begin{figure}[t]
\centering
\includegraphics[width=6cm]{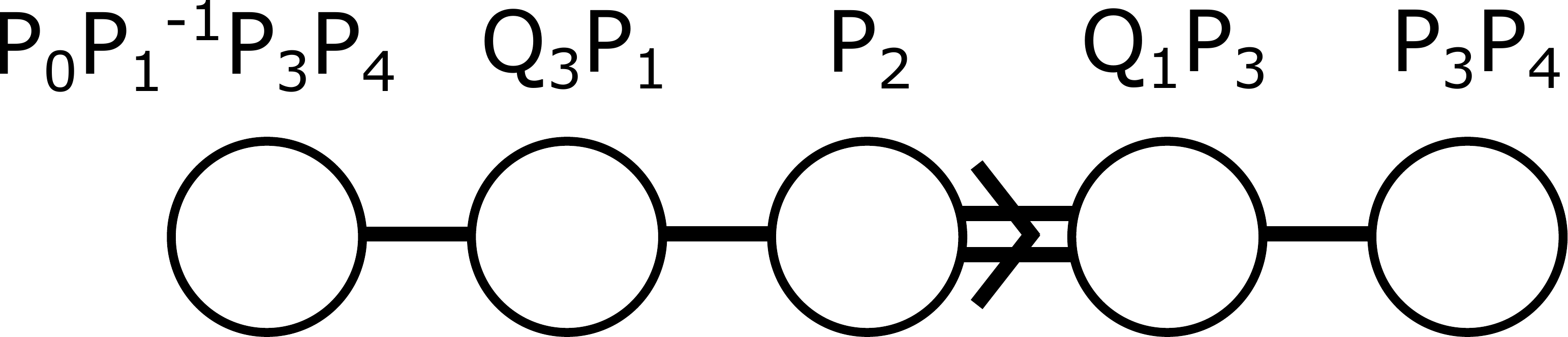}
\caption{The K\"ahler parameters of the fibers which form the affine $F_4$ Dynkin diagram.}
\label{fig:f4dynkin}
\end{figure}

Since the theories are realized on the web diagrams with the trivalent gluing it is possible to compute the partition functions using the topological vertex. For that we first parameterize the length of the lines in the web diagram as in Figure \ref{fig:f4para}.
The K\"ahler parameters of the fiber classes which form the affine $F_4$ Dynkin diagram are also depicted in Figure \ref{fig:f4dynkin}.

\begin{figure}[t]
\centering
\subfigure[]{\label{fig:halfsu6}
\includegraphics[width=5cm]{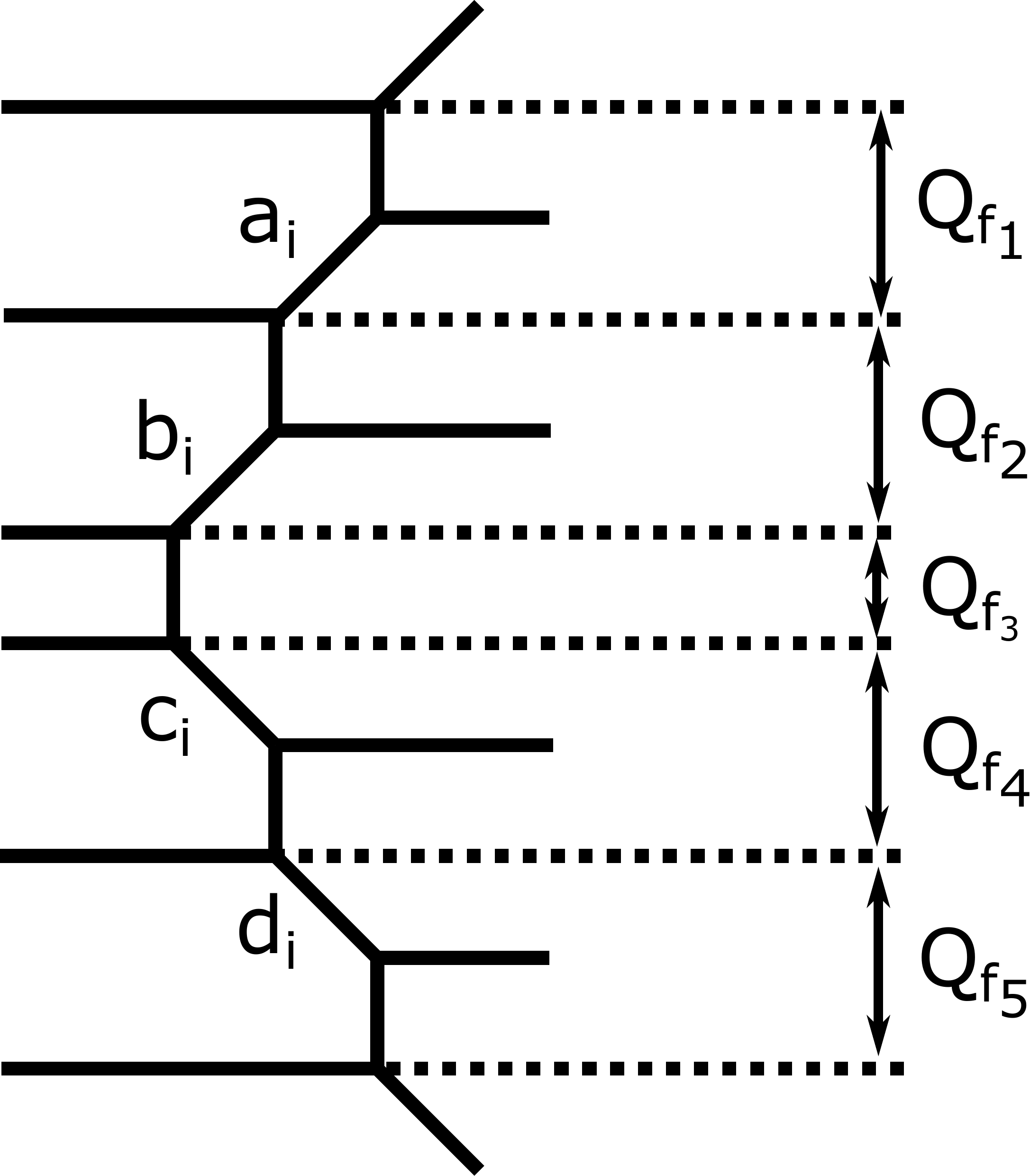}}
\hspace{1cm}
\subfigure[]{\label{fig:su6}
\includegraphics[width=7cm]{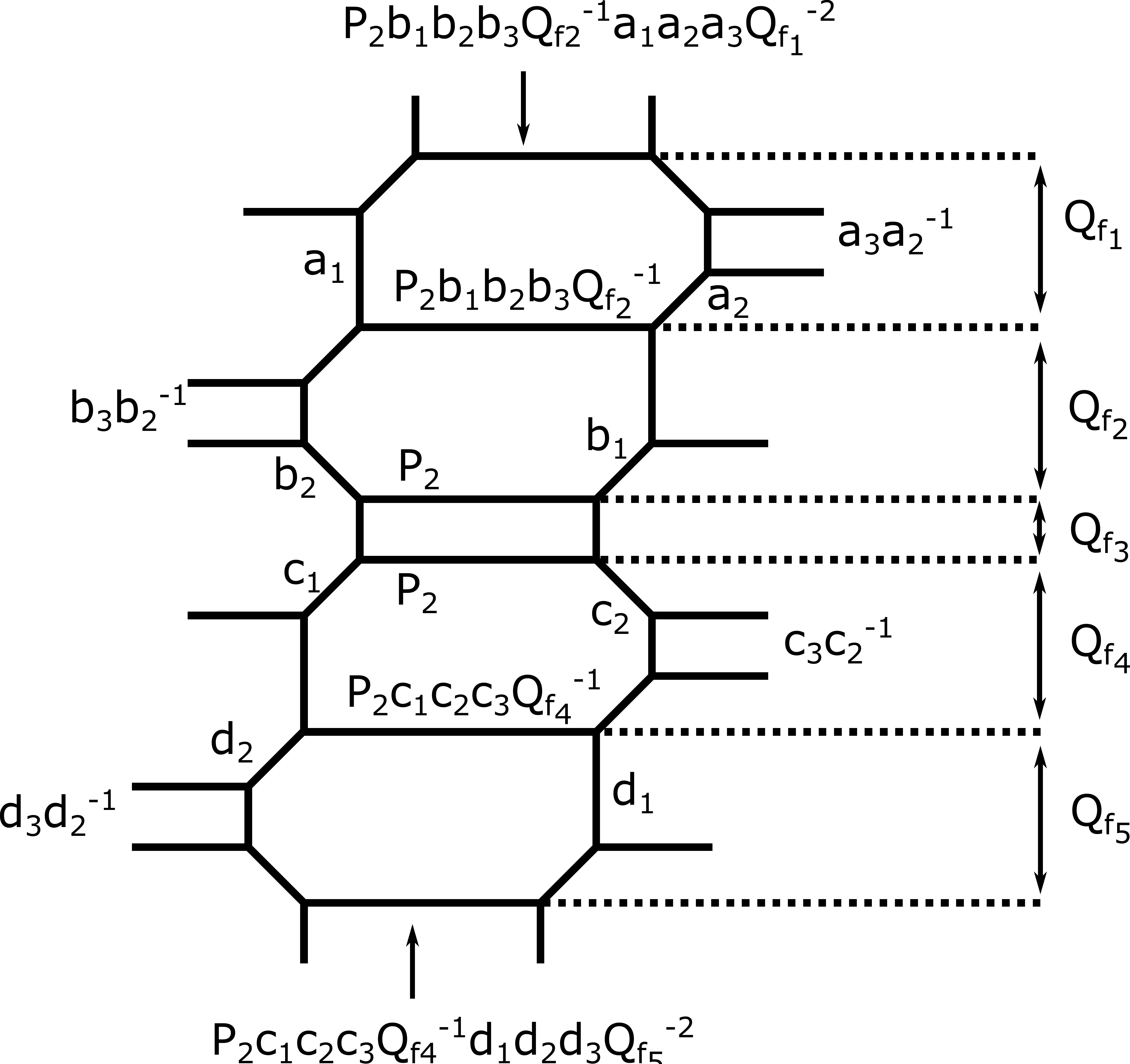}}
\caption{(a). Local geometry near the $\SU(6)$ part of each quiver tail of the affine $E_6$ Dynkin quiver theory. (b). Local geometry which describes the $\SU(6)_0$ gauge theory with $12$ flavors.}
\label{fig:su6fig}
\end{figure}

As we have done in the case of the $G_2$ theory in section \ref{sec:g2}, the parametrization of the gluing lines written in Figure \ref{fig:f4para1} can be understood from the Higgsing condition imposed on the web diagram for the original affine $E_6$ Dynkin quiver theory. From the viewpoint of the central $\SU(6)$ gauge node, each quiver tail adds $4$ flavors to the $\SU(6)$ gauge node. The local diagram around the $\SU(6)$ part coming from each quiver tail is given in Figure \ref{fig:halfsu6}. We also assign K\"ahler parameters for some lines in the figure where $i$ is either $1,2$ or $3$ representing the three quiver tails. 
Hence the central $\SU(6)$ node is locally described by the $\SU(6)$ gauge theory with $4 \times 3 = 12$ flavors with the zero Chern-Simons level. A 5-brane web diagram of the $\SU(6)$ gauge theory is depicted in Figure \ref{fig:su6}, where the K\"ahler parameter for the two middle lines is set to be $P_2$. 
Then we consider the tuning condition which realizes the $\left[(3, 3, 2), (3, 2, 2)\right]$ Higgsing leading to $\ff_4^{(1)}$ on $(-2)$-curve from the web diagram of the affine $E_6$ Dynkin quiver theory. 
The $\SU(2)$ Higgsing for one of the quiver tails imposes the condition,
\begin{equation}
a_i = Q_{f_1} \qquad \text{or} \qquad d_i = Q_{f_5}.
\end{equation}
On the other hand, the $\SU(3)$ Higgsing for one of the quiver tails imposes the condition,
\begin{equation}
a_i = Q_{f_1}, \; b_i = Q_{f_2}, \quad \text{or}\qquad c_i = Q_{f_4}, \; d_i = Q_{f_5}.
\end{equation}
Therefore, the $\left[(3, 3, 2), (3, 2, 2)\right]$  Higgsing is achieved by tuning the K\"ahler parameters in Figure \ref{fig:su6} as
\begin{equation}
a_1 = a_2 = a_3 = Q_{f_1}, \quad b_1 = b_2 = Q_{f_2}, \quad c_1 = Q_{f_4}, \quad d_1 = d_2 = d_3 = Q_{f_5}.
\end{equation}
Then the lengths of the horizontal lines in Figure \ref{fig:su6} become
\begin{align}
P_2b_3Q_{f_1}Q_{f_2} &= P_2P_3^2P_4Q_1Q_2,\\
P_2b_2Q_{f_2} &= P_2P_3Q_1Q_2,\\
P_2c_2c_3 &= P_2Q_1Q_3,\\
P_2c_2c_3Q_{f_5} &= P_2P_3P_4Q_1Q_3.
\end{align}
from the top to the bottom except for the middle two lines. Here we used the relation
\begin{equation}\label{su6root}
Q_{f_1} = P_3P_4, \quad Q_{f_2} = Q_1P_3, \quad Q_{f_3} = Q_B, \quad Q_{f_4} = Q_1P_3,\quad Q_{f_5} = P_3P_4,
\end{equation}
which follows from the parameterization in Figure \ref{fig:f4para}.

Given the parametrization in Figure \ref{fig:f4para}, we can apply the topological vertex to the web diagram in Figure \ref{fig:f4on2} using the trivalent gauging method. The only thing which we still need to determine is the framing factors for the gluing lines. Comparing the web diagram in Figure \ref{fig:e6e6e6} with the one in Figure \ref{fig:f4on2}, the $(p, q)$-charges of the lines attached to the gluing lines do not change after the Higgsing. Therefore the contribution of the framing factors will not change by the Higgsing. The framing factors for the gluing lines before the Higgsing can be read off from the web diagram of the $\SU(6)$ gauge theory with $12$ flavors in Figure \ref{fig:su6}. With this information taken into account, 
the application of the topological vertex to the web diagram in Figure \ref{fig:f4on2} gives rise to 
\begin{equation}\label{part.6df4}
\begin{split}
&Z_{\ff_4^{(1},(-2)}^{\text{6d}} \cr
&= \sum_{\{\mu_i\}}(-P_2P_3^2P_4Q_1Q_2)^{|\mu_1|}(-P_2P_3Q_1Q_2)^{|\mu_2|}(-P_2)^{|\mu_3| + |\mu_4|}(-P_2Q_1Q_3)^{|\mu_5|}(-P_2P_3P_4Q_1Q_3)^{|\mu_6|}\cr
&\hspace{1cm}Z_{\{\mu_i\}}^{\text{SU(6)$_L$}}\left(Q_B, \{Q_a\}, \{P_b\}\right) Z_{\{\mu_i\}}^{\text{SU(6)$_R$}}\left(Q_B, \{Q_a\}, \{P_b\}\right) f_{\mu_1}(q)^{-1}f_{\mu_3}(q)f_{\mu_4}(q)^{-1}f_{\mu_6}(q)\cr
&\hspace{1cm}Z^{\ff_4}_{1, \{\mu_i\}}\left(Q_B, \{Q_a\}, \{P_b\}\right)Z^{\ff_4}_{2, \{\mu_i\}}\left(Q_B, \{Q_a\}, \{P_b\}\right)Z^{\ff_4}_{3, \{\mu_i\}}\left(Q_B, \{Q_a\}, \{P_b\}\right),
\end{split}
\end{equation}
where we defined
\begin{align}
Z_{\{\mu_i\}}^{\text{SU(6)$_L$}}\left(Q_B, \{Q_a\}, \{P_b\}\right) &=q^{\frac{1}{2}\sum_{i=1}^6||\mu_i||^2}\left(\prod_{i=1}^6\tilde{Z}_{\mu_i}(q)\right)\prod_{1\leq i < j \leq 6}\mathcal{I}^-_{\mu_i, \mu_j}\left(Q_{f_i}Q_{f_{i+1}}\cdots Q_{f_{j-1}}\right),\label{SU6L}\\
Z_{\{\mu_i\}}^{\text{SU(6)$_R$}}\left(Q_B, \{Q_a\}, \{P_b\}\right) &=q^{\frac{1}{2}\sum_{i=1}^6||\mu_i^t||^2}\left(\prod_{i=1}^6\tilde{Z}_{\mu_i^t}(q)\right)\prod_{1\leq i < j \leq 6}\mathcal{I}^-_{\mu_i, \mu_j}\left(Q_{f_i}Q_{f_{i+1}}\cdots Q_{f_{j-1}}\right), \label{SU6R}
\end{align}
using \eqref{su6root}, and
\begin{align}
Z^{\ff_4}_{1,\{\mu_i\}}\left(Q_B, \{Q_a\}, \{P_b\}\right)&=
q^{\frac{1}{2}\sum_{i=1}^6||\mu_i^t||^2}\left(\prod_{i=1}^6\tilde{Z}_{\mu_i^t}(q)\right)/Z_{\{\mu_i\}}^{\text{SU(6)$_R$}}\left(Q_B, \{Q_a\}, \{P_b\}\right)\cr
&\quad\mathcal{I}^-_{\mu_1, \mu_6}\left(Q_BQ_1^2P_3^4P_4^2\right)\mathcal{I}^-_{\mu_2,\mu_5}\left(Q_BQ_1^2P_3^2\right)\mathcal{I}^-_{\mu_3,\mu_4}\left(Q_B\right),\\
Z^{\ff_4}_{2, \{\mu_i\}}\left(Q_B, \{Q_a\}, \{P_b\}\right) &=
q^{\frac{1}{2}\sum_{i=1}^6||\mu_i^t||^2}\left(\prod_{i=1}^6\tilde{Z}_{\mu_i^t}(q)\right)/Z_{\{\mu_i\}}^{\text{SU(6)$_R$}}\left(Q_B, \{Q_a\}, \{P_b\}\right)\cr
&\quad\mathcal{I}^{-}_{\mu_3,\mu_4}\left(Q_B\right)\mathcal{I}^{-}_{\mu_3,\mu_5}\left(Q_BQ_1P_3\right)\mathcal{I}^{-}_{\mu_4,\mu_5}\left(Q_1P_3\right)\mathcal{I}^{-}_{\mu_2,\mu_6}\left(Q_BQ_1^2P_3^3P_4\right)\cr
&\sum_{\nu_1, \nu_2}\left(-P_3\right)^{|\nu_1| + |\nu_2|}q^{\frac{1}{2}\left(||\nu_1||^2 + ||\nu_1^t||^2 + ||\nu_2||^2 + ||\nu_2^t||^2\right)}\left(\prod_{i=1}^2\tilde{Z}_{\nu_i}(q)\tilde{Z}_{\nu_i^t}(q)\right)\cr
&\quad\mathcal{I}^{-}_{\mu_1,\nu_2}\left(Q_BQ_1^2P_3^2P_4^2\right)\mathcal{I}^{-}_{\mu_2,\nu_1}\left(Q_BQ_1^2P_3\right)\mathcal{I}^{+}_{\mu_2,\nu_2}\left(Q_BQ_1^2P_3P_4\right)\cr
&\quad\mathcal{I}^+_{\nu_1,\nu_2}\left(P_4\right)\mathcal{I}^-_{\nu_1, \mu_6}\left(P_3^2P_4\right)\mathcal{I}^+_{\nu_2, \mu_6}\left(P_3^2\right)\cr
&\quad\mathcal{I}^{+}_{\mu_3,\nu_1}\left(Q_BQ_1\right)\mathcal{I}^{+}_{\mu_4,\nu_1}\left(Q_1\right)\mathcal{I}^{+}_{\nu_1,\mu_5}\left(P_3\right),\\
Z^{\ff_4}_{3, \{\mu_i\}}\left(Q_B, \{Q_a\}, \{P_b\}\right) &=q^{\frac{1}{2}\sum_{i=1}^6||\mu_i^t||^2}\left(\prod_{i=1}^6\tilde{Z}_{\mu_i^t}(q)\right)/Z_{\{\mu_i\}}^{\text{SU(6)$_R$}}\left(Q_B, \{Q_a\}, \{P_b\}\right)\cr
&\quad\mathcal{I}^-_{\mu_1, \mu_6}\left(Q_BQ_1^2P_3^4P_4^2\right)\mathcal{I}^-_{\mu_2, \mu_3}\left(Q_1P_3\right)\mathcal{I}^-_{\mu_2,\mu_4}\left(Q_BQ_1P_3\right)\cr
&\quad\mathcal{I}^-_{\mu_2,\mu_5}\left(Q_BQ_1^2P_3^2\right)\mathcal{I}^-_{\mu_3, \mu_4}\left(Q_B\right)\mathcal{I}^-_{\mu_3, \mu_5}\left(Q_BQ_1P_3\right)\mathcal{I}^-_{\mu_4, \mu_5}\left(Q_1P_3\right)\cr
&\sum_{\lambda_{1,2,3,4}} q^{\frac{1}{2}\sum_{i=1}^4\left(||\lambda_i||^2 + ||\lambda_i^t||\right)}\left(\prod_{i=1}^4\tilde{Z}_{\lambda_i}(q)\tilde{Z}_{\lambda_i^t}(q)\right)\cr
&\quad\left(-Q_2^{-1}Q_3P_1\right)^{|\lambda_1|}\left(-P_1\right)^{|\lambda_2|}\left(-Q_2^{-1}Q_3P_0\right)^{|\lambda_3|}\left(-P_0\right)^{|\lambda_4|}\cr
&\quad\mathcal{I}^+_{\mu_1, \lambda_3}\left(Q_1Q_2^{-2}Q_3P_1P_3\right)\mathcal{I}^-_{\mu_1, \lambda_1}\left(Q_1Q_2^{-1}P_3^2P_4\right)\mathcal{I}^-_{\mu_1, \lambda_2}\left(Q_BQ_1Q_3P_3^2P_4\right)\cr
&\quad\mathcal{I}^+_{\mu_1, \lambda_4}\left(Q_BQ_1Q_3P_1^{-1}P_3^3P_4^2\right)\mathcal{I}^+_{\lambda_3, \lambda_1}\left(Q_2Q_3^{-1}P_1^{-1}P_3P_4\right)\cr
&\quad\mathcal{I}^+_{\lambda_3, \lambda_2}\left(Q_BQ_2^2P_1^{-1}P_3P_4\right)\mathcal{I}^-_{\lambda_3, \lambda_4}\left(Q_BQ_2^2P_1^{-2}P_3^2P_4^2\right)^2\cr
&\quad\mathcal{I}^+_{\lambda_3, \mu_6}\left(Q_BQ_1Q_2^2Q_3^{-1}P_1^{-1}P_3^3P_4^2\right)\mathcal{I}^-_{\lambda_1, \lambda_2}\left(Q_BQ_2Q_3\right)^2\cr
&\quad\mathcal{I}^+_{\lambda_1, \lambda_4}\left(Q_BQ_2Q_3P_1^{-1}P_3P_4\right)\mathcal{I}^-_{\lambda_1, \mu_6}\left(Q_BQ_1Q_2P_3^2P_4\right)\cr
&\quad\mathcal{I}^+_{\lambda_2, \lambda_4}\left(P_1^{-1}P_3P_4\right)\mathcal{I}^-_{\lambda_2, \mu_6}\left(Q_1Q_3^{-1}P_3^2P_4\right)\mathcal{I}^+_{\lambda_4, \mu_6}\left(Q_1Q_3^{-1}P_1P_3\right)\cr
&\quad\mathcal{I}^+_{\mu_2, \lambda_1}\left(Q_1Q_2^{-1}P_3\right)\mathcal{I}^+_{\mu_2, \lambda_2}\left(Q_BQ_1Q_3P_3\right)\mathcal{I}^+_{\lambda_1, \mu_3}\left(Q_2\right)\cr
&\quad\mathcal{I}^+_{\lambda_1, \mu_4}\left(Q_BQ_2\right)\mathcal{I}^+_{\lambda_1, \mu_5}\left(Q_BQ_1Q_2P_3\right)\mathcal{I}^+_{\mu_3, \lambda_2}\left(Q_BQ_3\right)\cr
&\quad\mathcal{I}^+_{\mu_4, \lambda_2}\left(Q_3\right)\mathcal{I}^+_{\lambda_2, \mu_5}\left(Q_1Q_3^{-1}P_3\right).
\end{align}

Again the partition function \eqref{part.6df4} may contain an extra factor. To see the extra factor contribution we assign mass parameters for the length between parallel external legs. The web diagram in Figure \ref{fig:f4para2} contains two parallel external legs and the web diagram in Figure \ref{fig:f4para3} has two bunches of parallel external legs and hence we parameterize the lengths between the parallel external legs by three mass parametesr $M'_1, M'_2, M'_3$ as
\be
P_3^3 = M'_1, \qquad Q_1Q_2^{-3}Q_3^2P_0P_1P_3 =M'_2, \qquad Q_1Q_3^{-1}P_0P_1P_3 = M'_3.
\ee
There is one more mass parameter associated to the gluing. From the web diagram in Figure \ref{fig:su6}, the external lines attached to the top gluing line are also parallel to each other and we assign another mass parameter $M'_0$ to
\be
P_2P_3^2P_4Q_1Q_2 = M'_0. 
\ee
Note that the external lines attached to the bottom gluing line are also parallel to each other and indeed the length between them is parameterized by mass parameters due to the relation $Q_2^{3}Q_3^{-3}P_3^3 = M'^{-1}_1M'_2M'_3$.
 On the other hand the Coulomb branch moduli dependence can be determined by \eqref{QCB}. The class of the elliptic fiber is also given by the general relation \eqref{elliptic}, which becomes 
\be\label{Qtauf4}
Q_{\tau} = (P_0P_1^{-1}P_3P_4)(Q_3P_1)^2P_2^3(Q_1P_3)^4(P_3P_4)^2 = M'^3_0M'_2,
\ee
for the current case.  
 
Then the extra factor contained in \eqref{part.6df4} is characterized by a factor which depends on only the mass parameters and the partition function of \eqref{part.6df4} can be written as
\be\label{hatpart.6df4}
Z_{\ff_{4}^{(1)},(-2)}^{\text{6d}} = \hat{Z}_{\ff_{4}^{(1)},(-2)}^{\text{6d}}\left(\{A'_a\}, \{M'_i\}\right)Z_{\text{extra}}^{\ff_{4}^{(1)},(-2)}\left(M'_0, M'_1, M'_2, M'_3,\right).
\ee 
We claim that the partition function $\hat{Z}_{\ff_{4}^{(1)},(-2)}^{\text{6d}}\left(\{A'_a\}, \{M'_i\}\right)$ in \eqref{hatpart.6df4} gives rise to the partition function of the 6d $F_4$ gauge theory with three flavors and a tensor multiplet on $T^2 \times \mathbb{R}^4$ up to an extra factor. In this case also the extra factor in \eqref{hatpart.6df4} is non-trivial as the diagram in Figure \ref{fig:f4on2} has parallel external lines.

\paragraph{5d $F_4$ gauge theory with $3$ flavors.}
We now take the 5d limit which amounts to changing the diagram in Figure \ref{fig:f4on2} into the one in Figure \ref{fig:f4w3f} and make a comparison with the 5d partition function of the 5d $F_4$ gauge theory with three flavors. In terms of the K\"ahler parameters in Figure \ref{fig:f4para}, the 5d limit is realized by taking $P_0 \to 0$ with the other K\"ahler parameters fixed. In this limit the K\"ahler parameter for the elliptic class \eqref{Qtauf4} becomes zero, which implies $R_{\text{6d}} \to 0$ from the relation \eqref{QR6d}. 
Hence the application of the limit $P_0 \to 0$ to the partition function \eqref{part.5df4} yields the partition function of the 5d $F_4$ gauge theory with three flavors on $S^1\times \mathbb{R}^4$ up to an extra factor. 

We can parameterize the partition function by the gauge theory parameters. 
The Coulomb branch moduli $A_i = e^{-a_i}\; (i=1, 2, 3, 4)$ of the $F_4$ gauge theory may be assigned as with \eqref{QfAl}, and they are given by 
\begin{equation}\label{f4Qroot}
Q_3P_1 = A_1^2A_2^{-1}, \quad P_2 = A_1^{-1}A_2^2A_3^{-2}, \quad Q_1P_3 = A_2^{-1}A_3^2A_4^{-1}, \quad P_3P_4 = A_3^{-1}A_4^{2}.
\end{equation}
On the other hand, strings associated with the K\"ahler parameters $Q_1, Q_2, Q_3$ yield weights of the fundamental representation. It is possible to read off the Dynkin labels of the weights from the intersection numbers between the curves and the compact surfaces in the dual geometry and they are
\begin{equation}
Q_1\;:\;[0, -1, 2, -1], \qquad Q_2\;:\;[1, -1, 1, -1], \qquad Q_3\;:\;[1, -1, 1, -1].
\end{equation}
Hence we parameterize 
\begin{equation}\label{f4Qweight}
Q_1 = A_2^{-1}A_3A_4^{-1}M_1^{-1}, \quad Q_2 = A_1A_2^{-1}A_3A_4^{-1}M_2, \quad Q_3 = A_1A_2^{-1}A_3A_4^{-1}M_3,
\end{equation}
where $M_i = e^{-m_i}\; (i=1, 2, 3)$. Then the length between parallel external lines in Figure \ref{fig:f4para2} becomes
\be
P_3^3 = M_1^3,
\ee
which only depends on the mass parameter as expected. For determining the instanton fugacity $M_0$ we make use of the effective prepotential \eqref{prepotential} and compare the derivative with repsect to $a_2$ with the area of the middle face in Figure \ref{fig:f4w3f}. The explicit evaluation of the derivative of the prepotential with respect to $a_2$ becomes 
\be
\frac{\partial\mathcal{F}}{\partial a_2} = (-a_1 + 2a_2 - 2a_3)(2a_2 - 2a_3 + m_0 - 2m_2 - 2m_3).\label{areaf4}
\ee
The first factor in \eqref{areaf4} is the length of the fiber class of the middle face and the second factor corresponds to the length for $Q_B$. Hence we have
\begin{equation}
Q_B = A_2^2A_3^{-2}M_0M_2^{-2}M_3^{-2},
\end{equation}
where $M_0 = e^{-m_0}$. It turns out that the length between horizontal external parallel lines in Figure \ref{fig:f4w3f} is $m_0$. 

Then the application of the limit $P_0 \to 0$ to \eqref{part.6df4} gives
\begin{align}\label{part.5df4}
Z^{\text{5d}}_{\ff_4+3\text{F}} &= Z^{\text{6d}}_{\ff_4,(-2)}\Big|_{P_0 = 0}\cr
&=\hat{Z}^{\text{5d}}_{\ff_4+3\text{F}}\left(\{A_b\}, \{M_i\}\right)Z_{\text{extra}}^{\ff_4+3\text{F}}\left(M_0, M_1, M_2, M_3\right).
\end{align}
We argue that $\hat{Z}^{\text{5d}}_{\text{$\ff_4$+$3$F}}\left(\{A_b\}, \{M_i\}\right)$ is the partition function of the 5d $F_4$ gauge theory with $3$ hypermultiplets in the fundamental representation on $S^1 \times \mathbb{R}^4$ up to an extra factor. 

We can give support for the claim by 
comparing the perturbative part of the partition function \eqref{part.5df4} with the known result. From the universal formula for the perturbative part of the Nekrasov partition function in appendix \ref{sec:Nek}, it is possible to express the perturbative part of the partition function of the 5d $F_4$ gauge theory with three flavors by the following five factors,
\begin{align}\label{5df4pert}
Z^{\text{5d pert}}_{\mathfrak{f}_4+3\text{F}} = Z^{\mathfrak{f}_4}_{\text{cartan}}Z^{\mathfrak{f}_4}_{\text{roots}}Z^{\mathfrak{f}_4}_{\text{flavor 1}}Z^{\mathfrak{f}_4}_{\text{flavor 2}}Z^{\mathfrak{f}_4}_{\text{flavor 3}},
\end{align}
where each factor is given by
\begin{align}
Z^{\mathfrak{f}_4}_{\text{cartan}} = \text{PE}\left[\frac{4q}{(1-q)^2}\right],
\end{align}
\begin{equation}\label{f4root}
\begin{split}
Z^{\mathfrak{f}_4}_{\text{roots}}=&\text{PE}\left[\frac{2q}{(1-q)^2}\left(\frac{A_1^2}{A_2}+\frac{A_4^2 A_1}{A_2}+\frac{A_3 A_1}{A_2}+\frac{A_4A_1}{A_3}+\frac{A_1}{A_4}+\frac{A_2 A_1}{A_3^2}+\frac{A_3^2 A_1}{A_2A_4^2}+A_1+\frac{A_3^2}{A_2}\right.\right.\cr
&\hspace{1cm}+\frac{A_4^2}{A_3}+\frac{A_2 A_4^2}{A_3^2}+\frac{A_3A_4}{A_2}+A_4+\frac{A_2}{A_3}+\frac{A_3^2}{A_2A_4}+\frac{A_3}{A_4}+\frac{A_2}{A_4^2}+\frac{A_4^2}{A_1}+\frac{A_2}{A_1}+\frac{A_3}{A_1}\cr
&\hspace{1cm}\left.\left.+\frac{A_2A_4}{A_3 A_1}+\frac{A_2}{A_4 A_1}+\frac{A_2^2}{A_3^2 A_1}+\frac{A_3^2}{A_4^2 A_1}\right)\right],
\end{split}
\end{equation}
\begin{equation}\label{f4flavor1}
\begin{split}
Z^{\mathfrak{f}_4}_{\text{flavor 1}} =& \text{PE}\left[-\frac{q}{(1-q)^2}\left(\frac{A_3^2 M_1}{A_2 A_4}+\frac{A_3^2}{A_2 A_4 M_1}+\frac{A_4 A_3 M_1}{A_2}+\frac{A_3M_1}{A_1}+\frac{A_1 A_3 M_1}{A_2}+\frac{A_3 M_1}{A_4}\right.\right.\cr
&\hspace{1cm}+\frac{A_4 A_3}{A_2 M_1}+\frac{A_3}{A_1M_1}+\frac{A_1 A_3}{A_2 M_1}+\frac{A_3}{A_4 M_1}+A_4 M_1+\frac{A_1 M_1}{A_4}+\frac{A_2 M_1}{A_1A_4}+\frac{A_4}{M_1}\cr
&\hspace{1cm}+\frac{A_1}{A_4 M_1}+\frac{A_2}{A_1 A_4 M_1}+\frac{A_4^2 M_1}{A_3}+\frac{A_2M_1}{A_3}+\frac{A_1 A_4 M_1}{A_3}+\frac{A_2 A_4 M_1}{A_1 A_3}+\frac{A_4^2}{A_3M_1}\cr
&\hspace{1cm} \left.\left.+\frac{A_2}{A_3M_1}+\frac{A_1 A_4}{A_3 M_1}+\frac{A_2 A_4}{A_1 A_3 M_1} + 2M_1\right)\right],
\end{split}
\end{equation}
\begin{equation}\label{f4flavor2}
\begin{split}
Z^{\mathfrak{f}_4}_{\text{flavor 2}} =& \text{PE}\left[-\frac{q}{(1-q)^2}\left(\frac{A_3^2 M_2}{A_2 A_4}+\frac{A_4 A_3 M_2}{A_2}+\frac{A_3 M_2}{A_1}+\frac{A_1 A_3M_2}{A_2}+\frac{A_1 A_3 M_2}{A_2 A_4}+\frac{A_3 M_2}{A_4}\right.\right.\cr
&\hspace{1cm}+\frac{A_3 M_2}{A_4^2}+\frac{A_3}{A_1M_2}+\frac{A_1 A_3}{A_2 M_2}+\frac{A_3}{A_4 M_2}+\frac{A_4 M_2}{A_1}+\frac{A_1 A_4 M_2}{A_2}+A_4M_2\cr
&\hspace{1cm}+\frac{A_1 M_2}{A_4}+\frac{A_2 M_2}{A_1 A_4}+\frac{A_4}{M_2}+\frac{A_4^2 M_2}{A_3}+\frac{A_2M_2}{A_3}+\frac{A_1 A_4 M_2}{A_3}\cr
&\hspace{1cm}\left.\left.+\frac{A_2 A_4 M_2}{A_1 A_3}+\frac{A_2 M_2}{A_4A_3}+\frac{A_2}{A_3 M_2}+\frac{A_1 A_4}{A_3 M_2}+\frac{A_2 A_4 M_2}{A_3^2}+ 2M_2\right)\right],
\end{split}
\end{equation}
\begin{equation}\label{f4flavor3}
\begin{split}
Z^{\mathfrak{f}_4}_{\text{flavor 3}} =& \text{PE}\left[-\frac{q}{(1-q)^2}\left(\frac{A_3^2 M_3}{A_2 A_4}+\frac{A_4 A_3 M_3}{A_2}+\frac{A_3 M_3}{A_1}+\frac{A_1 A_3M_3}{A_2}+\frac{A_1 A_3 M_3}{A_2 A_4}+\frac{A_3 M_3}{A_4}\right.\right.\cr
&\hspace{1cm}+\frac{A_3 M_3}{A_4^2}+\frac{A_3}{A_1M_3}+\frac{A_1 A_3}{A_2 M_3}+\frac{A_3}{A_4 M_3}+\frac{A_4 M_3}{A_1}+\frac{A_1 A_4 M_3}{A_2}+A_4M_3\cr
&\hspace{1cm}+\frac{A_1 M_3}{A_4}+\frac{A_2 M_3}{A_1 A_4}+\frac{A_4}{M_3}+\frac{A_4^2 M_3}{A_3}+\frac{A_2M_3}{A_3}+\frac{A_1 A_4 M_3}{A_3}+\frac{A_2 A_4 M_3}{A_1 A_3}\cr
&\hspace{1cm}\left.\left.+\frac{A_2 M_3}{A_4
   A_3}+\frac{A_2}{A_3 M_3}+\frac{A_1 A_4}{A_3 M_3}+\frac{A_2 A_4 M_3}{A_3^2} + 2M_3\right)\right].
\end{split}
\end{equation}
The perturbative partition function is different depending on phases. The phase was determined by the choice \eqref{f4Qroot} and \eqref{f4Qweight}. 

\begin{figure}[t]
\centering
\includegraphics[width=5cm]{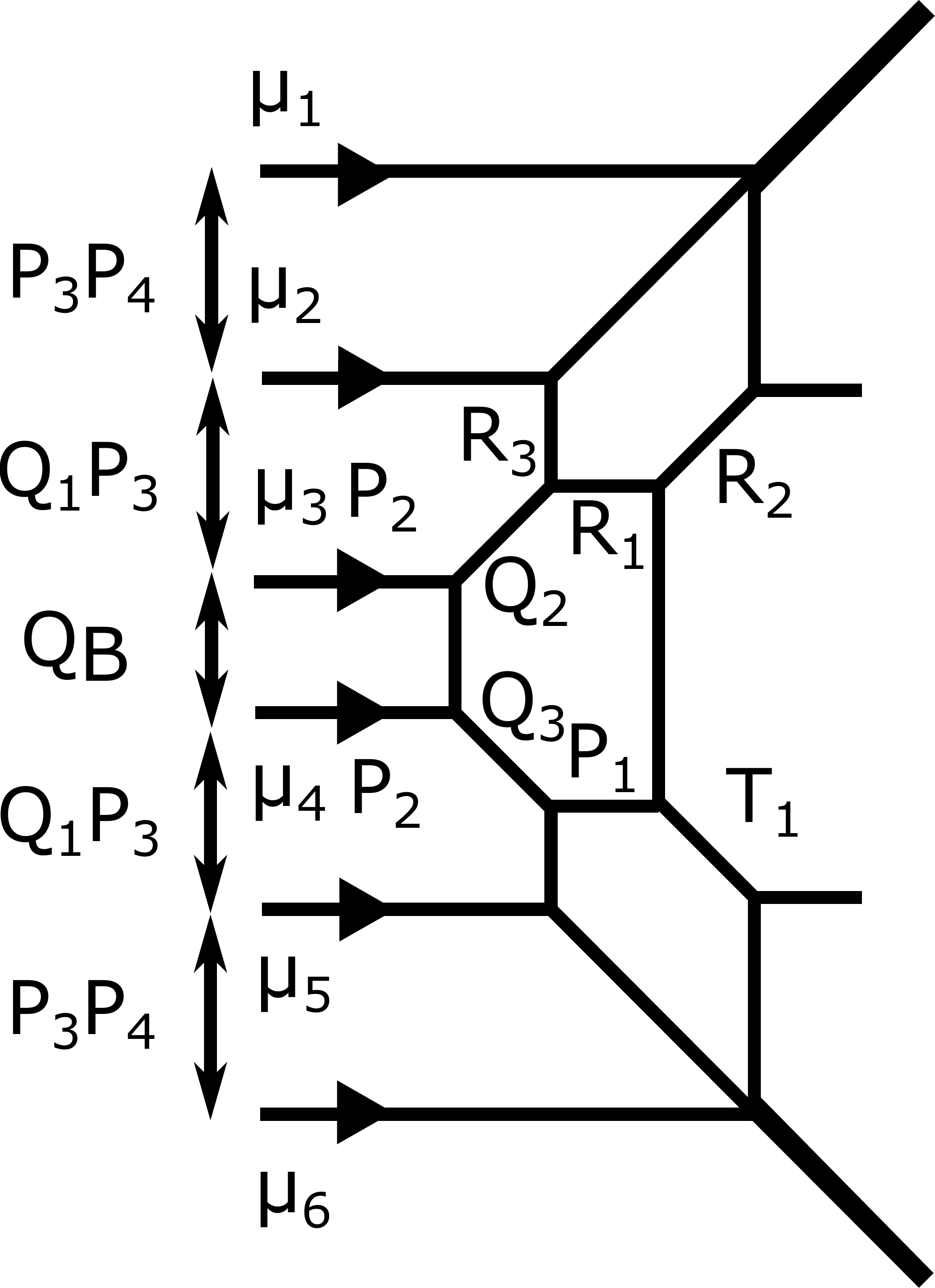}
\caption{A parameterization of the web which is obtaiend after applying the 5d limit to the diagram in  Figure \ref{fig:f4para3}.}
\label{fig:f4para4}
\end{figure}

We will compare the perturbative partition function \eqref{5df4pert} with the perturbative part of \eqref{part.5df4}. 
Since the partition function may contain an extra factor we focus on terms which are independent of the Coulomb branch moduli. The perturbative part is obtained by applying the limit $M_0 \to 0$ or $Q_B \to 0$ to \eqref{part.5df4}, which breaks the diagram in Figure \ref{fig:f4w3f} into the upper half part and the lower half part. 
For explicitly evaluating \eqref{part.5df4}, it is useful to introduce parameters $R_1, R_2, R_3, T_1, T_2$ as in Figure \ref{fig:f4para4}.
The newly introduced parameters are related by other parameters as
\begin{equation}
\begin{split}
&R_1 = Q_2^{-1}Q_3P_1, \quad R_2 = Q_1Q_2^{-1}P_3, \quad R_3 = Q_2Q_3^{-1}P_1^{-1}P_3P_4, 
\quad T_1 = P_1^{-1}P_3P_4.
\end{split}
\end{equation}
We computed the Plethystic logarithm of upper half part until the order $P_2^3Q_2^6R_1^4R_2^4R_3^4$ and the lower half part until the order $P_2^2P_3^2P_4^4Q_1^4Q_3^2T_1^2$\footnote{Here we choose $T_1$ as an independent parameter instead of $P_1$ since the summation of $T_1$ is already taken in \eqref{part.6df4}.} and the result 
agrees with \eqref{5df4pert} except for a Coulomb branch independent part. The Coulomb branch independent factor in \eqref{part.5df4} reads 
\begin{align}
\text{PE}\left[\frac{q}{(1-q)^2}\left(3M_2^2 + M_2^3 + 4\frac{M_3}{M_1} + \frac{M_3^2}{M_1^2} - 3M_1 + 3M_2 - 3M_3\right)\right].
\end{align}
until the orders we computed. 


\subsection{6d/5d $E_6$ gauge theory with matter}
\label{sec:e6}
We then consider the partition function of the 6d $E_6$ gauge theory with four flavors and a tensor multiplet on $T^2 \times \mathbb{R}^4$ and its 5d limit.

\begin{figure}[t]
\centering
\subfigure[]{\label{fig:e6on2}
\includegraphics[width=6cm]{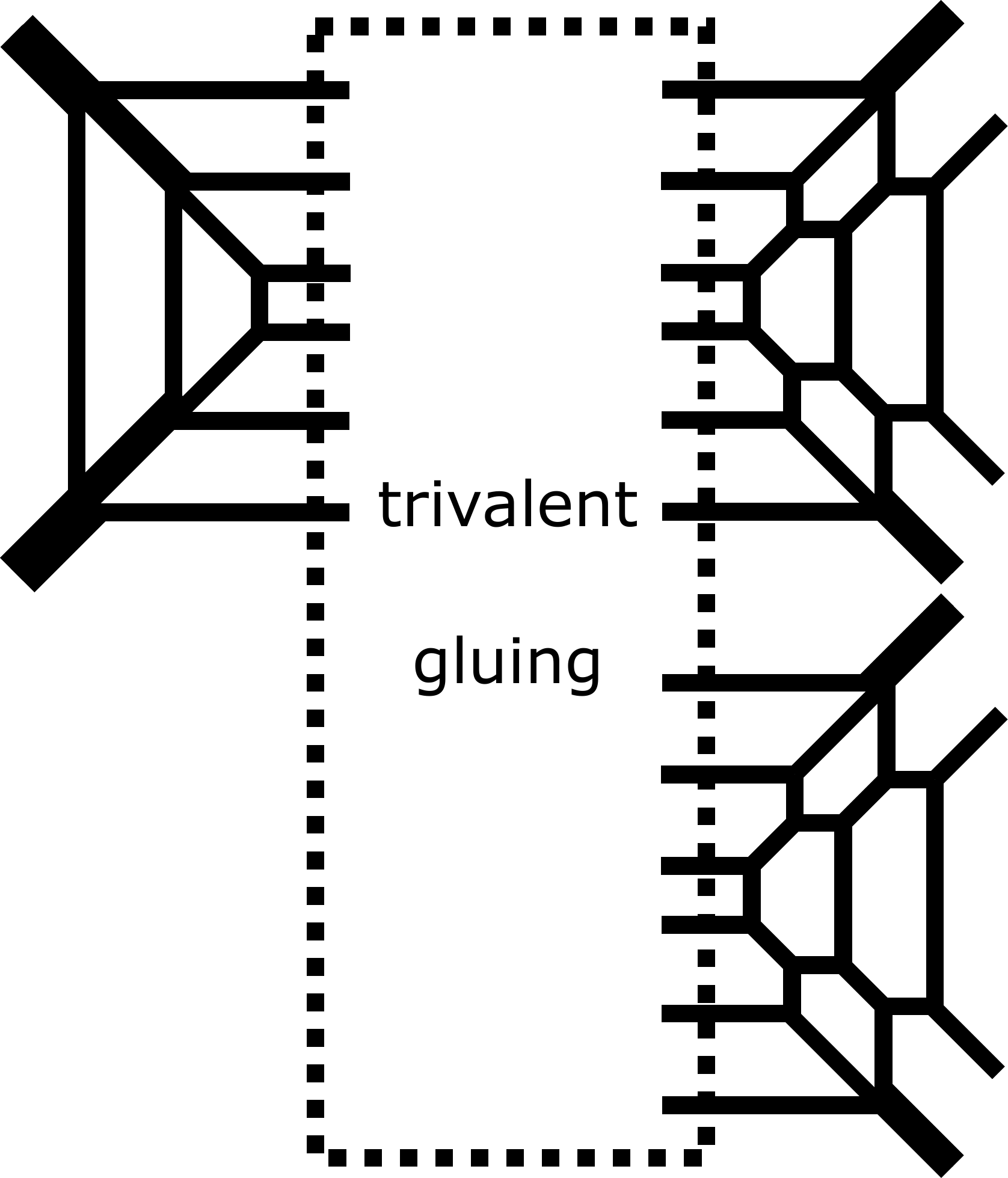}}
\hspace{1cm}
\subfigure[]{\label{fig:e6w4f}
\includegraphics[width=6cm]{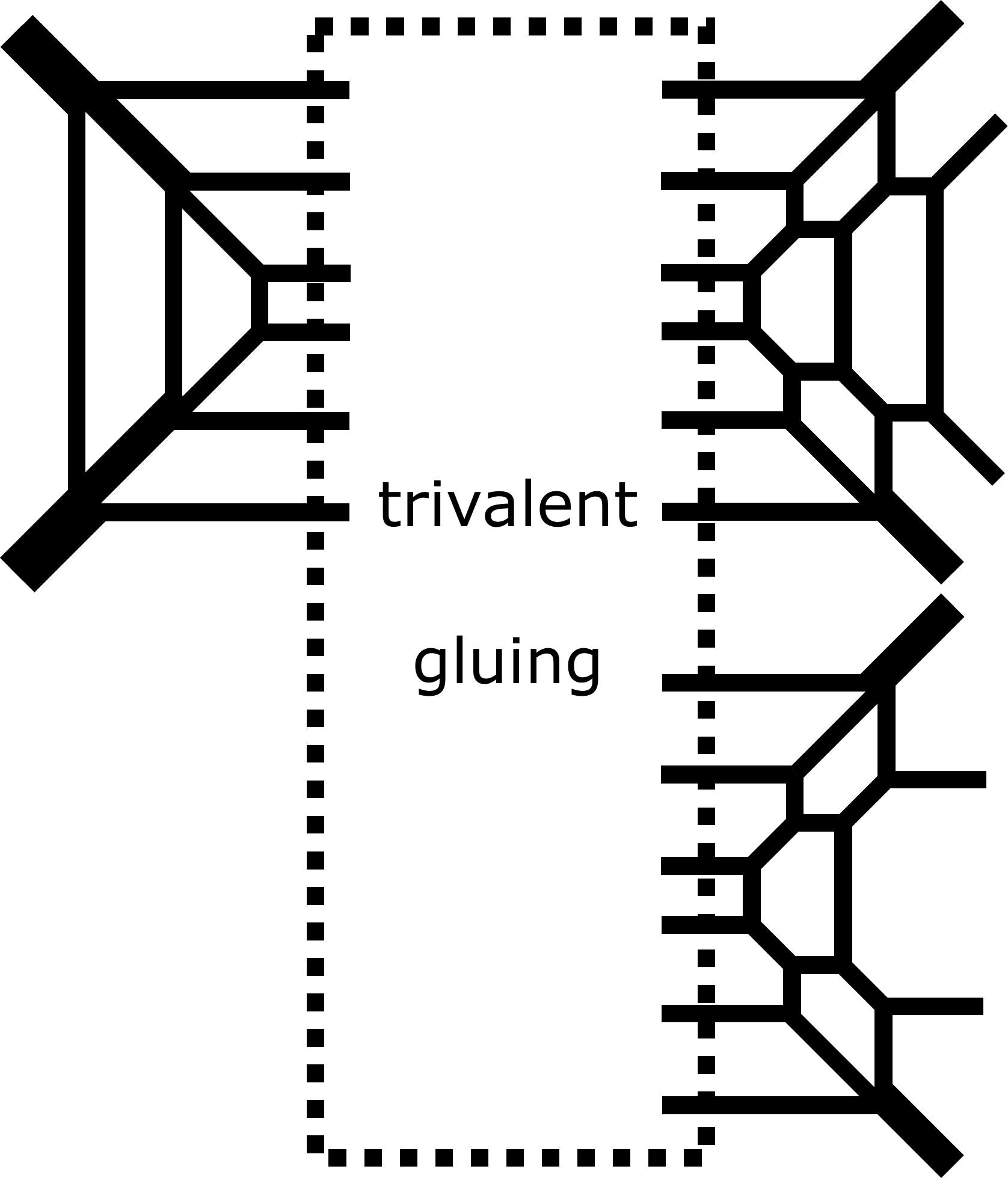}}
\caption{(a). The web diagram for the 6d theory given by $\fe_6^{(1)}$ on $(-2)$-curve. (b). The web diagram for 5d $E_6$ gaue theory with $4$ flavors. }
\label{fig:e6}
\end{figure}

\paragraph{$\fe_6^{(1)}$ on $(-2)$-curve.}
The 6d $E_6$ gauge theory with four flavors and a tensor multiplet compactified on a circle arises as a low energy theory on the $\left[(3,2,2), (3,2,2)\right]$ Higgs branch of the 
the 6d theory $(E_6, E_6)_2$ on $S^1$ as in Table \ref{tb:e6_2}. A 5d gauge theory description of the original $(E_6, E_6)_2$ theory on $S^1$ before the Higgsing is again the affine $E_6$ Dynkin quiver theory given in 
\eqref{e6quiver} and the quiver theory is realized on the web diagram in Figure \ref{fig:e6e6e6}. Then applying the $\left[(3,2,2), (3,2,2)\right]$ Higgsing to the diagram in Figure \ref{fig:e6e6e6} yields the diagram in Figure \ref{fig:e6on2}.
The fiber classes of the seven faces form the affine $E_6$ Dynkin diagram in the sense that the intersection numbers computed by \eqref{fScartan} gives the Cartan matrix of $\fe_6^{(1)}$. When we decouple the fiber associated to the affine node, the remaining fibers form the $E_6$ Dynkin diagram and the resulting diagram is depicted in Figure \ref{fig:e6w4f}. The decoupling corresponds to the 5d limit and the theory realized on the web diagram in Figure \ref{fig:e6w4f} is the 5d $E_6$ gauge theory with four flavors. 

\begin{figure}[t]
\centering
\subfigure[]{\label{fig:e6para1}
\includegraphics[width=4cm]{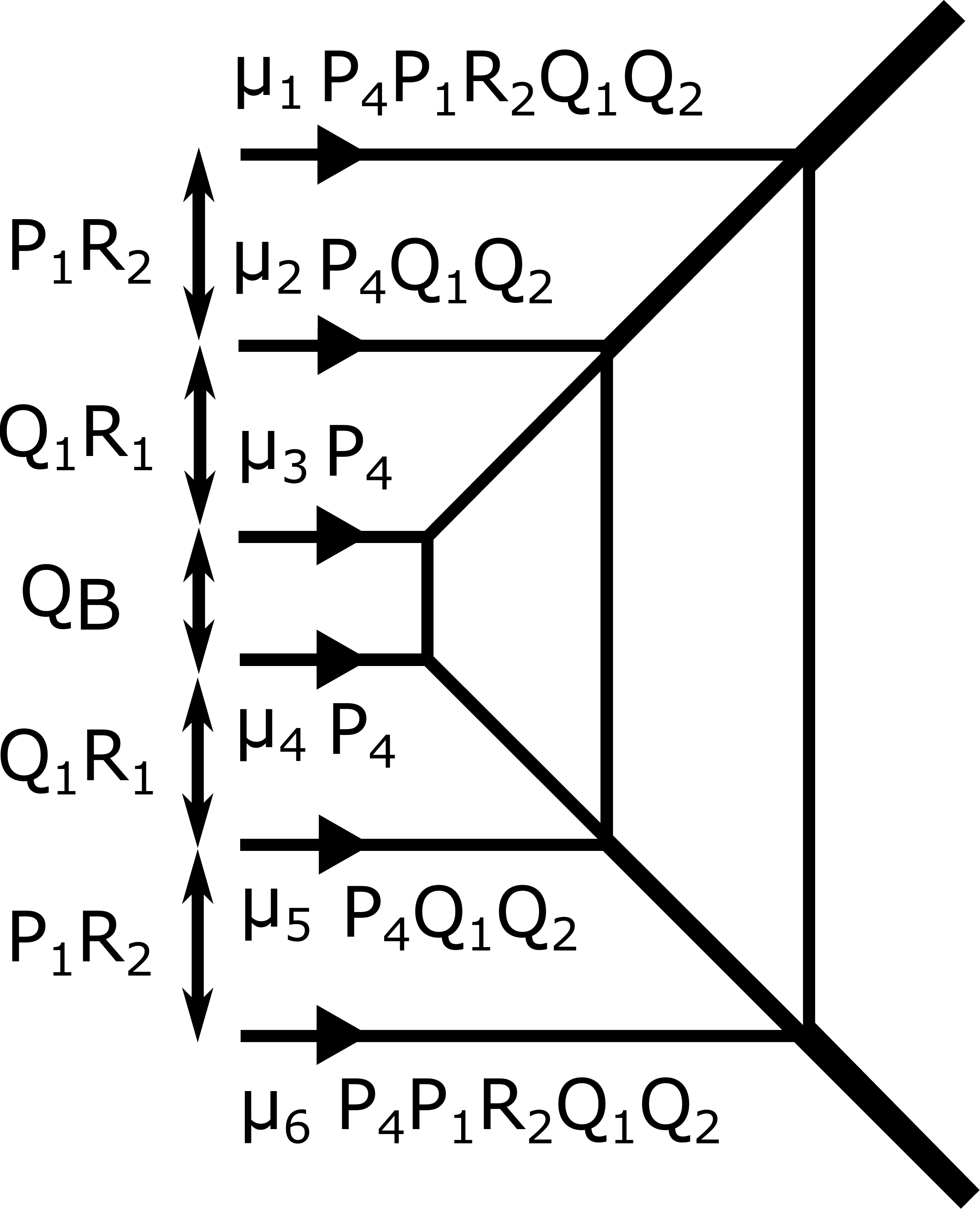}}
\hspace{1cm}
\subfigure[]{\label{fig:e6para2}
\includegraphics[width=4cm]{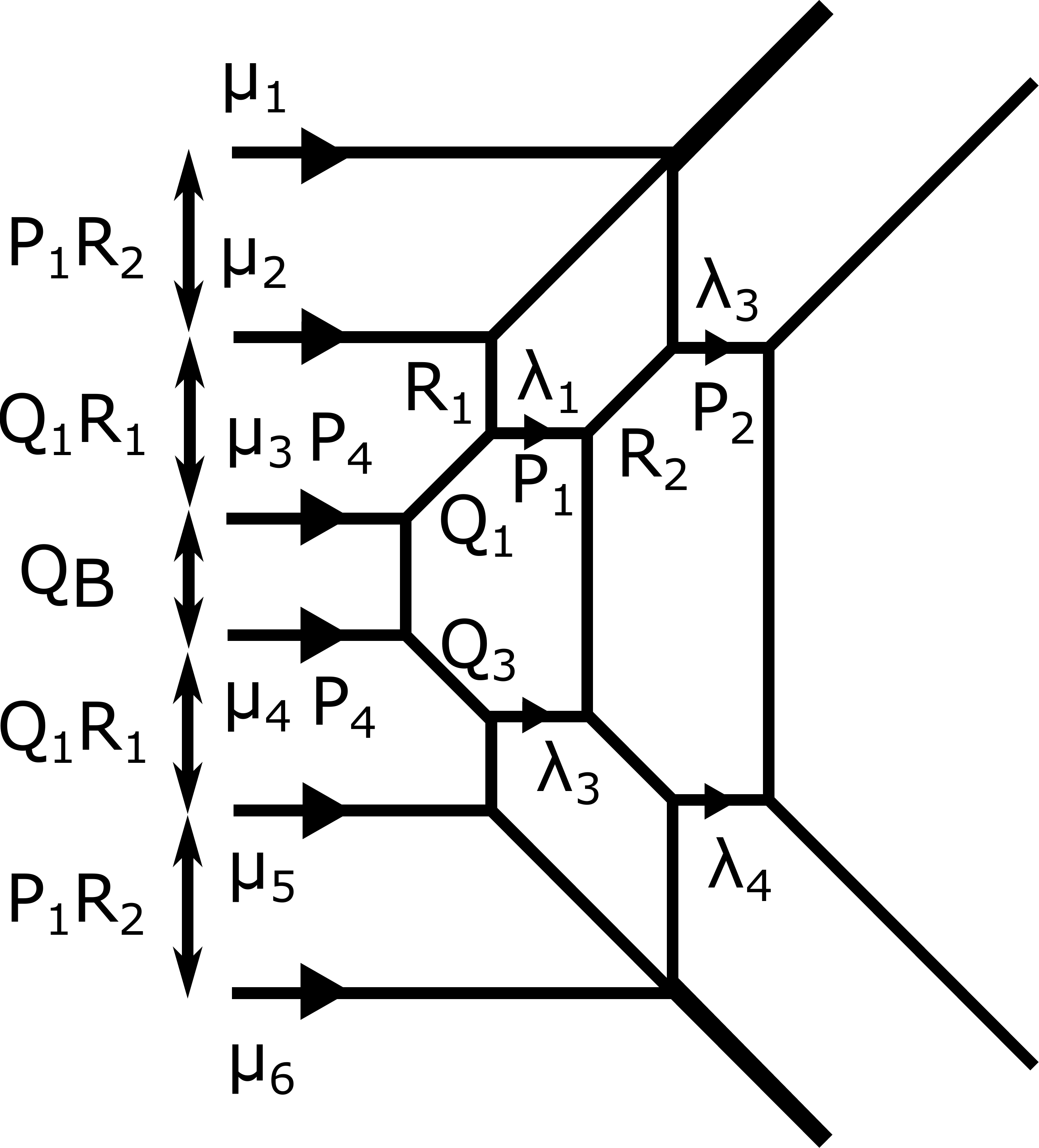}}
\hspace{1cm}
\subfigure[]{\label{fig:e6para3}
\includegraphics[width=4cm]{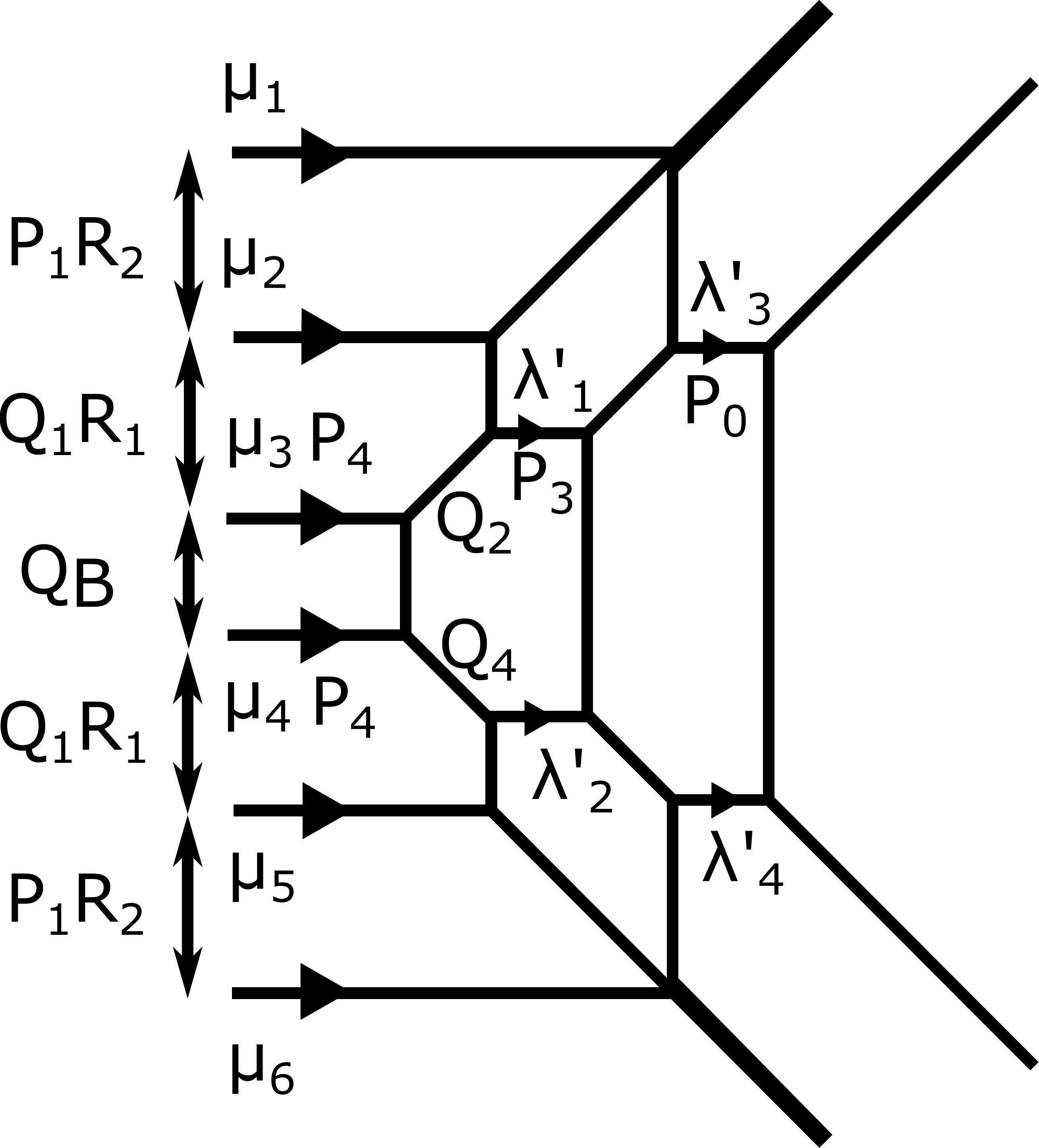}}
\caption{Parameterization of the web diagram for the geometry $\fe_6^{(1)}$ on $(-2)$-curve in Figure \ref{fig:e6on2}.}
\label{fig:e6para}
\end{figure}

\begin{figure}[t]
\centering
\includegraphics[width=6cm]{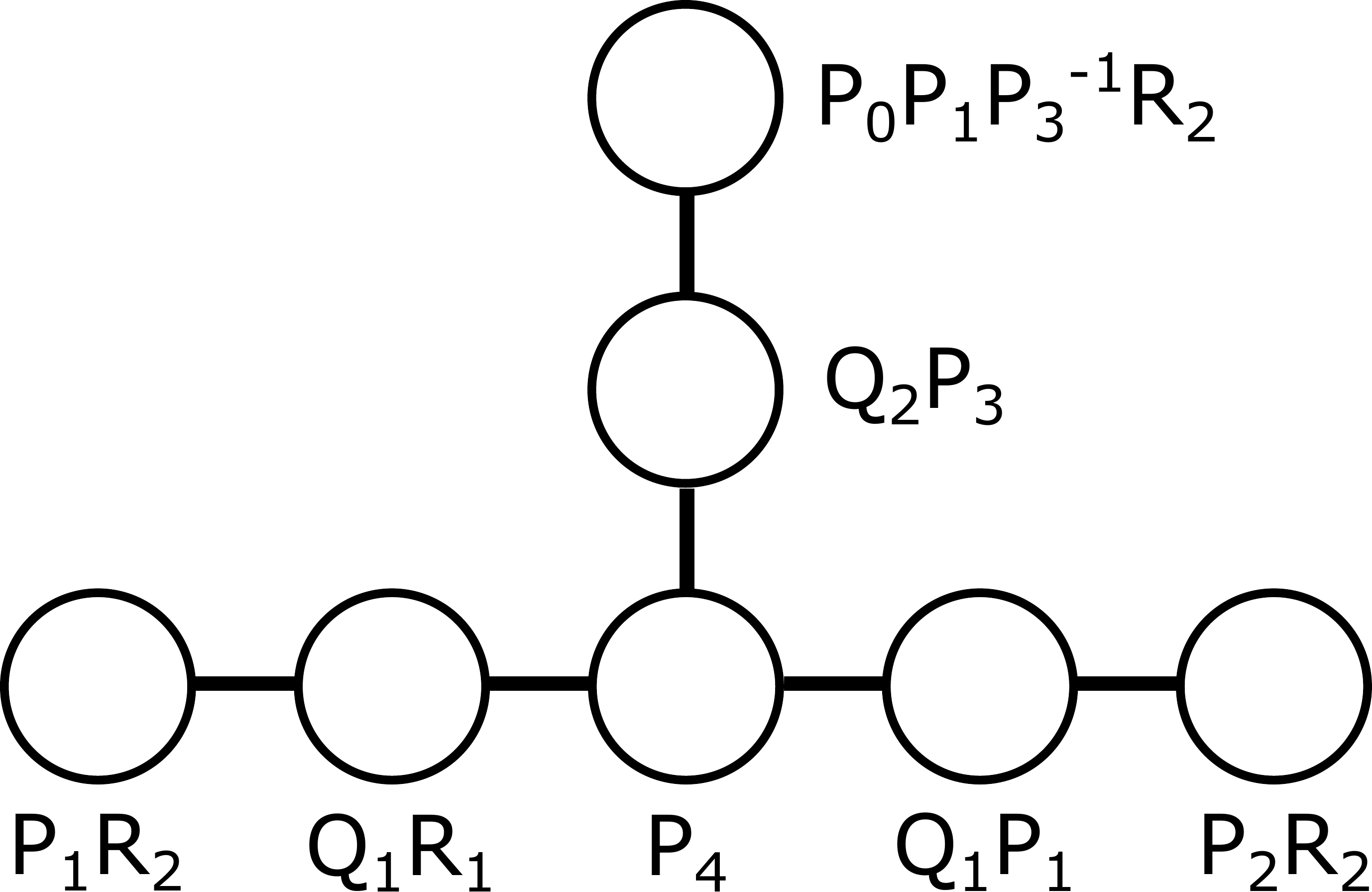}
\caption{The K\"ahler parameters of the fibers which form the affine $E_6$ Dynkin diagram.}
\label{fig:e6kahler}
\end{figure}

We consider 
applying the topological vertex as well as the trivalent gluing prescription to the diagram in Figure \ref{fig:e6on2}. 
For that we first parametrize the lengths of lines in the diagram in Figure \ref{fig:e6on2} as in Figure \ref{fig:e6para}. 
In the parameterization in Figure \ref{fig:e6para}, the K\"ahler parameters of the fiber classes which form the affine $E_6$ Dynkin diagram are given in Figure \ref{fig:e6kahler}.
The parameterization of the gluing lines can be determined by the Higgsing argument done in section \ref{sec:f4} since the web diagram in Figure \ref{fig:e6on2} is also obtained from the Higgsing of the web diagram of the affine $E_6$ Dynkin quiver theory. From the diagram in Figure \ref{fig:su6} we consider the $\left[(3,2,2), (3,2,2)\right]$ Higgsing and it is is achieved by the tuning
\begin{align}\label{e6tuning}
a_1=a_2=a_3 = Q_{f_1}, \quad b_1 = Q_{f_2}, \quad c_1 = Q_{f_4}, \quad d_1 = d_2 = d_3 = Q_{f_5}.
\end{align}
Then the application of the tuning condition \eqref{e6tuning} to the length of the color D5-branes in Figure \ref{fig:su6} gives rise to the parameterization of the gluing lines written in Figure \ref{fig:e6para1}. 

It is now possible to apply the topological vertex to the diagram in Figure \ref{fig:e6on2}. The framing factors of the gluing lines do not change by the Higgsing in this case also and they can be read off from the framing factors of the six color branes of the diagram in Figure \ref{fig:su6}. Then the 
topological vertex computation gives rise to 
\begin{equation}\label{part.6de6}
\begin{split}
&Z^{\text{6d}}_{\fe_6, (-2)}\cr
&= \sum_{\{\mu_i\}}(-P_1P_4R_2Q_1Q_2)^{|\mu_1|}(-P_4Q_1Q_2)^{|\mu_2|}(-P_4)^{|\mu_3| + |\mu_4|}(-P_4Q_1Q_2)^{|\mu_5|}(-P_1P_4R_2Q_1Q_2)^{|\mu_6|}\cr
&\hspace{0.5cm}Z_{\{\mu_i\}}^{\text{SU(6)$_L$}}\left(Q_B, \{Q_a\}, \{P_b\}, \{R_c\}\right) Z_{\{\mu_i\}}^{\text{SU(6)$_R$}}\left(Q_B, \{Q_a\}, \{P_b\}, \{R_c\}\right)\cr
&\hspace{0.5cm}f_{\mu_1}(q)^{-1}f_{\mu_3}(q)f_{\mu_4}(q)^{-1}f_{\mu_6}(q)\cr
&\hspace{0.5cm}Z^{\fe_6}_{1, \{\mu_i\}}\left(Q_B, \{Q_a\}, \{P_b\}, \{R_c\}\right)Z^{\fe_6}_{2, \{\mu_i\}}\left(Q_B, \{Q_a\}, \{P_b\}, \{R_c\}\right)Z^{\fe_6}_{3, \{\mu_i\}}\left(Q_B, \{Q_a\}, \{P_b\}, \{R_c\}\right),
\end{split}
\end{equation}
where $Z_{\{\mu_i\}}^{\text{SU(6)$_L$}}\left(Q_B, \{Q_a\}, \{P_b\}, \{R_c\}\right)$ and $Z_{\{\mu_i\}}^{\text{SU(6)$_R$}}\left(Q_B, \{Q_a\}, \{P_b\}, \{R_c\}\right)$ are \eqref{SU6L} and \eqref{SU6R} respectively with
\begin{align}
Q_{f_1} = P_1R_2, \quad Q_{f_2} = Q_1R_1, \quad Q_{f_3} = Q_B, \quad Q_{f_4} = Q_1R_1, \quad Q_{f_5} = P_1R_2.
\end{align}
The other factors of \eqref{part.6de6} are
\begin{align}
Z^{\fe_6}_{1,\{\mu_i\}}\left(Q_B, \{Q_a\}, \{P_b\}, \{R_c\}\right)&=q^{\frac{1}{2}\sum_{i=1}^6||\mu_i^t||^2}\left(\prod_{i=1}^6\tilde{Z}_{\mu_i^t}(q)\right)/Z_{\{\mu_i\}}^{\text{SU(6)$_R$}}\left(Q_B, \{Q_a\}, \{P_b\}, \{R_c\}\right)\cr
&\quad\mathcal{I}^-_{\mu_1, \mu_6}\left(Q_BQ_1^2P_1^2R_1^2R_2^2\right)\mathcal{I}^-_{\mu_2,\mu_5}\left(Q_BQ_1^2R_1^2\right)\mathcal{I}^-_{\mu_3,\mu_4}\left(Q_B\right),\\
Z^{\fe_6}_{2, \{\mu_i\}}\left(Q_B, \{Q_a\}, \{P_b\}\right) &=q^{\frac{1}{2}\sum_{i=1}^6||\mu_i^t||^2}\left(\prod_{i=1}^6\tilde{Z}_{\mu_i^t}(q)\right)/Z_{\{\mu_i\}}^{\text{SU(6)$_R$}}\left(Q_B, \{Q_a\}, \{P_b\}, \{R_c\}\right)\cr
&\quad\mathcal{I}^-_{\mu_1, \mu_6}\left(Q_BQ_1^2P_1^2R_1^2R_2^2\right)\mathcal{I}^-_{\mu_2, \mu_3}\left(Q_1R_1\right)\mathcal{I}^-_{\mu_2,\mu_4}\left(Q_BQ_1R_1\right)\cr
&\quad\mathcal{I}^-_{\mu_2,\mu_5}\left(Q_BQ_1^2R_1^2\right)\mathcal{I}^-_{\mu_3, \mu_4}\left(Q_B\right)\mathcal{I}^-_{\mu_3, \mu_5}\left(Q_BQ_1R_1\right)\mathcal{I}^-_{\mu_4, \mu_5}\left(Q_1R_1\right)\cr
&\sum_{\lambda_{1,2,3,4}} q^{\frac{1}{2}\sum_{i=1}^4\left(||\lambda_i||^2 + ||\lambda_i^t||^2\right)}\left(\prod_{i=1}^4\tilde{Z}_{\lambda_i}(q)\tilde{Z}_{\lambda_i^t}(q)\right)\cr
&\quad\left(-P_1\right)^{|\lambda_1|}\left(-Q_1Q_3^{-1}P_1\right)^{|\lambda_2|}\left(-P_2\right)^{|\lambda_3|}\left(-Q_1Q_3^{-1}P_2\right)^{|\lambda_4|}\cr
&\quad\mathcal{I}^+_{\mu_1, \lambda_3}\left(P_1R_1\right)\mathcal{I}^-_{\mu_1, \lambda_1}\left(P_1R_1R_2\right)\mathcal{I}^-_{\mu_1, \lambda_2}\left(Q_BQ_1Q_3P_1R_1R_2\right)\cr
&\quad\mathcal{I}^+_{\mu_1, \lambda_4}\left(Q_BQ_3^2P_1R_1R_2^2\right)\mathcal{I}^+_{\lambda_3, \lambda_1}\left(R_2\right)\mathcal{I}^+_{\lambda_3, \lambda_2}\left(Q_BQ_1Q_3R_2\right)\cr
&\quad\mathcal{I}^-_{\lambda_3, \lambda_4}\left(Q_BQ_3^2R_2^2\right)^2\mathcal{I}^+_{\lambda_3, \mu_6}\left(Q_BQ_1^2P_1R_1R_2^2\right)\mathcal{I}^-_{\lambda_1, \lambda_2}\left(Q_BQ_1Q_3\right)^2\cr
&\quad\mathcal{I}^+_{\lambda_1, \mu_4}\left(Q_BQ_3^2R_2\right)\mathcal{I}^-_{\lambda_1, \mu_6}\left(Q_BQ_1^2P_1R_1R_2\right)\mathcal{I}^+_{\lambda_2, \lambda_4}\left(Q_1^{-1}Q_3R_2\right)\cr
&\quad\mathcal{I}^-_{\lambda_2, \mu_6}\left(Q_1Q_3^{-1}P_1R_1R_2\right)\mathcal{I}^+_{\lambda_4, \mu_6}\left(Q_1^2Q_3^{-2}P_1R_1\right)\mathcal{I}^+_{\mu_2, \lambda_1}\left(R_1\right)\cr
&\quad\mathcal{I}^+_{\mu_2, \lambda_2}\left(Q_BQ_1Q_3R_1\right)\mathcal{I}^+_{\lambda_1, \mu_3}\left(Q_1\right)\mathcal{I}^+_{\lambda_1, \mu_4}\left(Q_BQ_1\right)\mathcal{I}^+_{\mu_4, \lambda_2}\left(Q_3\right)\cr
&\quad\mathcal{I}^+_{\lambda_1, \mu_5}\left(Q_BQ_1^2Q_3R_1\right)\mathcal{I}^+_{\mu_3, \lambda_2}\left(Q_BQ_3\right)\mathcal{I}^+_{\lambda_2, \mu_5}\left(Q_1Q_3^{-1}R_1\right),\cr
Z^{\fe_6}_{3, \{\mu_i\}}\left(Q_B, \{Q_a\}, \{P_b\}\right) &=q^{\frac{1}{2}\sum_{i=1}^6||\mu_i^t||^2}\left(\prod_{i=1}^6\tilde{Z}_{\mu_i^t}(q)\right)/Z_{\{\mu_i\}}^{\text{SU(6)$_R$}}\left(Q_B, \{Q_a\}, \{P_b\}, \{R_c\}\right)\cr
&\quad\mathcal{I}^-_{\mu_1, \mu_6}\left(Q_BQ_1^2P_1^2R_1^2R_2^2\right)\mathcal{I}^-_{\mu_2, \mu_3}\left(Q_1R_1\right)\mathcal{I}^-_{\mu_2,\mu_4}\left(Q_BQ_1R_1\right)\cr
&\quad\mathcal{I}^-_{\mu_2,\mu_5}\left(Q_BQ_1^2R_1^2\right)\mathcal{I}^-_{\mu_3, \mu_4}\left(Q_B\right)\mathcal{I}^-_{\mu_3, \mu_5}\left(Q_BQ_1R_1\right)\mathcal{I}^-_{\mu_4, \mu_5}\left(Q_1R_1\right)\cr
&\sum_{\lambda'_{1,2,3,4}} q^{\frac{1}{2}\sum_{i=1}^4\left(||\lambda'_i||^2 + ||\lambda'^t_i||^2\right)}\left(\prod_{i=1}^4\tilde{Z}_{\lambda'_i}(q)\tilde{Z}_{\lambda'^t_i}(q)\right)\cr
&\quad\left(-P_3\right)^{|\lambda'_1|}\left(-Q_2Q_4^{-1}P_3\right)^{|\lambda'_2|}\left(-P_0\right)^{|\lambda'_3|}\left(-Q_2Q_4^{-1}P_0\right)^{|\lambda'_4|}\cr
&\quad\mathcal{I}^+_{\mu_1, \lambda'_3}\left(P_3R_3\right)\mathcal{I}^-_{\mu_1, \lambda'_1}\left(P_3R_3R_4\right)\mathcal{I}^-_{\mu_1, \lambda'_2}\left(Q_BQ_2Q_4P_3R_3R_4\right)\cr
&\quad\mathcal{I}^+_{\mu_1, \lambda'_4}\left(Q_BQ_4^2P_3R_3R_4^2\right)\mathcal{I}^+_{\lambda'_3, \lambda'_1}\left(R_4\right)\mathcal{I}^+_{\lambda'_3, \lambda'_2}\left(Q_BQ_2Q_4R_4\right)\cr
&\quad\mathcal{I}^-_{\lambda'_3, \lambda'_4}\left(Q_BQ_4^2R_4^2\right)^2\mathcal{I}^+_{\lambda'_3, \mu_6}\left(Q_BQ_2^2P_3R_3R_4^2\right)\mathcal{I}^-_{\lambda'_1, \lambda'_2}\left(Q_BQ_2Q_4\right)^2\cr
&\quad\mathcal{I}^+_{\lambda'_1, \mu_4}\left(Q_BQ_4^2R_4\right)\mathcal{I}^-_{\lambda'_1, \mu_6}\left(Q_BQ_2^2P_3R_3R_4\right)\mathcal{I}^+_{\lambda'_2, \lambda'_4}\left(Q_2^{-1}Q_4R_4\right)\cr
&\quad\mathcal{I}^-_{\lambda'_2, \mu_6}\left(Q_2Q_4^{-1}P_3R_3R_4\right)\mathcal{I}^+_{\lambda'_4, \mu_6}\left(Q_2^2Q_4^{-2}P_3R_3\right)\mathcal{I}^+_{\mu_2, \lambda'_1}\left(R_3\right)\cr
&\quad\mathcal{I}^+_{\mu_2, \lambda'_2}\left(Q_BQ_2Q_4R_3\right)\mathcal{I}^+_{\lambda'_1, \mu_3}\left(Q_2\right)\mathcal{I}^+_{\lambda'_1, \mu_4}\left(Q_BQ_2\right)\mathcal{I}^+_{\mu_4, \lambda'_2}\left(Q_4\right)\cr
&\quad\mathcal{I}^+_{\lambda'_1, \mu_5}\left(Q_BQ_2^2Q_4R_3\right)\mathcal{I}^+_{\mu_3, \lambda'_2}\left(Q_BQ_4\right)\mathcal{I}^+_{\lambda'_2, \mu_5}\left(Q_2Q_4^{-1}R_3\right),
\end{align}
where we used
\begin{align}
R_3 = Q_1Q_2^{-1}R_1,\qquad R_4 = P_1P_3^{-1}R_2.
\end{align}

The partition function in \eqref{part.6de6} may contain an extra factor which is independent of Coulomb branch moduli. The parameterization of the Coulomb branch moduli is given in \eqref{QCB}. Mass parameters can be assigned to the lengths between parallel external lines in Figure \ref{fig:e6para}. From the Figure \ref{fig:e6para}, we can explicitly see parallel external lines and we parameterize
\be
P_1P_2R_1 = M'_1, \quad Q_1^3Q_3^{-3}P_1P_2R_1 = M'_2, \quad Q_1Q_2^{-1}P_0P_3R_1 = M'_3, \quad Q_1Q_2^2Q_4^{-3}P_0P_3R_1 = M'_4.
\ee
Also, the web diagram in Figure \ref{fig:su6} implies that the external lines attached to the top gluing line are parallel to each other and we assign
\be
Q_1Q_2P_1P_4R_2 = M'_0. 
\ee
The class of the elliptic fiber from \eqref{elliptic} becomes
\be\label{elliptice6}
Q_{\tau} = (P_1R_2)(Q_1R_1)^2(P_4^3)(Q_1P_1)^2(P_2R_2)(Q_2P_3)^2(P_0P_1P_3^{-1}R_2) = M'^3_0M'_1M'_3.
\ee

Then the partition function of \eqref{part.6de6} can be written as
\be\label{hatpart.6de6}
Z_{\fe_{6}^{(1)},(-2)}^{\text{6d}} = \hat{Z}_{\fe_{6}^{(1)},(-2)}^{\text{6d}}\left(\{A'_a\}, \{M'_i\}\right)Z_{\text{extra}}^{\fe_{6}^{(1)},(-2)}\left(M'_0, M'_1, M'_2, M'_3, M'_4\right).
\ee 
where $Z_{\text{extra}}^{\fe_{6}^{(1)},(-2)}\left(M'_0, M'_1, M'_2, M'_3, M'_4\right)$ is an extra factor which is independent from the Coulomb branch moduli. 
We argue that the partition function $\hat{Z}_{\fe_{6}^{(1)},(-2)}^{\text{6d}}\left(\{A'_a\}, \{M'_i\}\right)$ in \eqref{hatpart.6de6} yields the partition function of the 6d $E_6$ gauge theory with four flavors and a tensor multiplet on $T^2 \times \mathbb{R}^4$ up to an extra factor. 

\paragraph{5d $E_6$ gauge theory with $4$ flavors.}
The 5d limit 
which yields the diagram in Figure \ref{fig:e6w4f} is again realized by taking the limit $P_0 \to 0$ with the other K\"ahler parameters fixed. The K\"ahler parameter for the elliptic class \eqref{elliptice6} also becomes $Q_\tau \to 0$ as desired. Therefore applying the limit to \eqref{hatpart.6de6} yields the partition function of the 5d $E_6$ gauge theory with four flavors on $S^1 \times \mathbb{R}^4$ up to an extra factor. 

We 
rewrite the 5d partition function 
by the gauge theory parameters. 
After taking the limit the fiber classes of the six faces form the $E_6$ Dynkin diagram. The K\"ahler parameters for the fiber classes depend only on the Coulomb branch moduli $A_i = e^{-a_i}\; (i=1, 2, \cdots, 6)$ through the relation \eqref{QfAl} and they are given by 
\begin{equation} \label{e6simple}
\begin{split}
&P_1R_2 = A_1^2A_2^{-1}, \quad Q_1R_1 = A_1^{-1}A_2^2A_3^{-1}, \quad P_4 = A_2^{-1}A_3^2A_4^{-1}A_6^{-1}, \quad Q_1P_1 = A_3^{-1}A_4^2A_5^{-1},\cr
&P_2R_2 =A_4^{-1}A_5^2 , \quad Q_2P_3 =A_3^{-1}A_6^2.
\end{split}
\end{equation}
$Q_1, Q_2, Q_3, Q_4$ are related to the mass of matter and the Dynkin label of the corresponding weights can be obtained by the negative of the intersection numbers as
\begin{equation}
\begin{split}
&Q_1\;:\;[-1, 1, -1, 1, 0, 0], \quad Q_2\;:\;[-1, 1, -1, 0, 0, 1], \quad Q_3\;:\;[-1, 1, -1, 1, 0, 0],\cr
&Q_4\;:\;[-1, 1, -1, 0, 0, 1].
\end{split}
\end{equation}
Hence we introduce four mass parameters $M_i\; (i=1, 2, 3, 4)$ by 
\begin{equation}\label{e6weight}
\begin{split}
&Q_1 = A_1^{-1}A_2A_3^{-1}A_4M_1^{-1}, \quad Q_2 = A_1^{-1}A_2A_3^{-1}A_6M_2, \quad Q_3 = A_1^{-1}A_2A_3^{-1}A_4M_3^{-1}, \cr
&Q_4 = A_1^{-1}A_2A_3^{-1}A_6M_4.
\end{split}
\end{equation}
For indentifying the instanton fugacity we use the effective prepotential \eqref{prepotential}. The middle face in the $\SU(6)$ gauging is locally an $\mathbb{F}_0$ in the dual geometry and the area of the face corresponds to the derivative of the effective prepotential with respect to $a_2$ which is given by
\be
\frac{\partial\mathcal{F}}{\partial a_3} = (-a_2 + 2a_3 - a_4 - a_6)(-2a_2 + 2a_3 + m_0 - 2m_2 - 2m_4).
\ee
Hence we parameterize
\be\label{m0e6}
Q_B = A_2^{-2}A_3^2M_0M_2^{-2}M_4^{-2}.
\ee
for the instanton fugacity $M_0 = e^{-m_0}$. The parameterization implies that the length between the horizontal external lines in Figure \ref{fig:f4para3} is $m_0$. 

Applying the limit $P_0 \to 0$ to \eqref{part.6de6}, the partition function with the gauge theory parameterization becomes
\begin{align}\label{part.5de6}
Z_{\fe_6+4\text{F}}^{\text{5d}} &= Z^{\text{6d}}_{\fe_6,(-2)}\Big|_{P_0 = 0}\cr
&=\hat{Z}^{\text{5d}}_{\fe_6 + 4\text{F}}\left(\{A_b\}, \{M_i\}\right)Z_{\text{extra}}^{\fe_6 + 4\text{F}}\left(M_0, M_1, M_2, M_3, M_4\right). 
\end{align}
We claim that $\hat{Z}^{\text{5d}}_{\fe_6 + 4\text{F}}\left(\{A_b\}, \{M_i\}\right)$ gives the partition function of the 5d $E_6$ gauge theory with four flavors on $S^1 \times \mathbb{R}^4$ up to an extra factor.

Under this parameterization we compare the perturbative part of \eqref{part.5de6} with the universal formula of the perturbative part of the Nekrasov partition function of the $E_6$ gauge theory with $4$ flavors. The perturbative part depends on the phase of the gauge theory and it is determined by the choice \eqref{e6simple} and \eqref{e6weight}. Using the result in section \ref{sec:Nek}, the explicit form is given by 
\begin{align}\label{5de6pert}
Z^{\text{5d pert}}_{\mathfrak{e}_6+4\text{F}} = Z^{\mathfrak{e}_6}_{\text{cartan}}Z^{\mathfrak{e}_6}_{\text{roots}}Z^{\mathfrak{e}_6}_{\text{flavor 1}}Z^{\mathfrak{e}_6}_{\text{flavor 2}}Z^{\mathfrak{e}_6}_{\text{flavor 3}}Z^{\mathfrak{e}_6}_{\text{flavor 4}},
\end{align}
where each factor is 
\begin{align}
Z^{\mathfrak{e}_6}_{\text{cartan}} = \text{PE}\left[\frac{6q}{(1-q)^2}\right],
\end{align}
\begin{equation}
\begin{split}
Z^{\mathfrak{e}_6}_{\text{roots}} =& \text{PE}\left[\frac{2q}{(1-q)^2}\left(\frac{A_1^2}{A_2}+\frac{A_4 A_1}{A_2}+\frac{A_3 A_5 A_1}{A_2 A_4}+\frac{A_5 A_6 A_1}{A_3}+\frac{A_6A_1}{A_4}+\frac{A_4 A_6 A_1}{A_3 A_5}+\frac{A_2 A_1}{A_3}\right.\right.\cr
&\hspace{1cm}+\frac{A_3 A_1}{A_2 A_5}+\frac{A_5A_1}{A_6}+\frac{A_3 A_1}{A_4 A_6}+\frac{A_4 A_1}{A_5A_6}+\frac{A_5^2}{A_4}+\frac{A_6^2}{A_3}+\frac{A_2 A_4}{A_3}+\frac{A_4 A_5}{A_3}+\frac{A_2A_5}{A_4}\cr
&\hspace{1cm}+\frac{A_5 A_6}{A_2}+\frac{A_3 A_6}{A_2 A_4}+\frac{A_4 A_6}{A_2A_5}+A_6+\frac{A_4^2}{A_3 A_5}+\frac{A_2}{A_5}+\frac{A_3}{A_6}+\frac{A_3 A_5}{A_2A_6}+\frac{A_3^2}{A_2 A_4 A_6}\cr
&\hspace{1cm}+\frac{A_3 A_4}{A_2 A_5 A_6}+\frac{A_4}{A_1}+\frac{A_3 A_5}{A_4A_1}+\frac{A_2 A_5 A_6}{A_3 A_1}+\frac{A_2 A_6}{A_4 A_1}+\frac{A_2 A_4 A_6}{A_3 A_5A_1}+\frac{A_2^2}{A_3 A_1}\cr
&\hspace{1cm}\left.\left.++\frac{A_3}{A_5 A_1} + \frac{A_2 A_5}{A_6 A_1}+\frac{A_2 A_3}{A_4 A_6A_1}+\frac{A_2 A_4}{A_5 A_6 A_1}\right)\right],
\end{split}
\end{equation}
\begin{equation}
\begin{split}
Z^{\mathfrak{e}_6}_{\text{flavor 1}} =& \text{PE}\left[-\frac{q}{(1-q)^2}\left(\frac{A_5 A_2 M_1}{A_3}+\frac{A_2 M_1}{A_1}+\frac{A_2 M_1}{A_4}+\frac{A_4 A_2 M_1}{A_3A_5}+\frac{A_4 A_2}{A_1 A_3 M_1}+\frac{A_6 A_2}{A_3 M_1}\right.\right.\cr
&\hspace{1cm}+\frac{A_2}{A_6 M_1}+A_1 M_1+\frac{A_5M_1}{A_1}+\frac{A_4 A_6 M_1}{A_3}+\frac{A_5 A_6 M_1}{A_4}+\frac{A_6 M_1}{A_5}+\frac{A_4M_1}{A_6}\cr
&\hspace{1cm}+\frac{A_3 A_5 M_1}{A_4 A_6}+\frac{A_3 M_1}{A_5 A_6}+\frac{A_1 A_4}{A_3 M_1}+\frac{A_1A_5}{A_4 M_1}+\frac{A_5}{M_1}+\frac{A_6}{A_1 M_1}+\frac{A_3}{A_4 M_1}+\frac{A_4}{A_5M_1}\cr
&\hspace{1cm}\left.\left.+\frac{A_3}{A_1 A_6 M_1}+\frac{A_3 M_1}{A_2}+\frac{A_1 A_5 M_1}{A_2}+\frac{A_1 A_4 M_1}{A_5A_2}+\frac{A_1 A_6}{A_2 M_1}+\frac{A_1 A_3}{A_6 A_2 M_1}\right)\right],
\end{split}
\end{equation}
\begin{equation}
\begin{split}
Z^{\mathfrak{e}_6}_{\text{flavor 2}} =& \text{PE}\left[-\frac{q}{(1-q)^2}\left(\frac{A_5 A_2 M_2}{A_3}+\frac{A_6 A_2 M_2}{A_1 A_3}+\frac{A_2 M_2}{A_1}+\frac{A_2M_2}{A_4}+\frac{A_4 A_2 M_2}{A_3 A_5}+\frac{A_6 A_2}{A_3 M_2}\right.\right.\cr
&\hspace{1cm}+\frac{A_2}{A_6 M_2}+A_1M_2+\frac{A_5 M_2}{A_1}+\frac{A_4 A_6 M_2}{A_3}+\frac{A_5 A_6 M_2}{A_4}+\frac{A_1 A_6M_2}{A_3}+\frac{A_6 M_2}{A_5}\cr
&\hspace{1cm}+\frac{A_3 M_2}{A_1 A_4}+\frac{A_4 M_2}{A_1 A_5}+\frac{A_1M_2}{A_6}+\frac{A_4 M_2}{A_6}+\frac{A_3 A_5 M_2}{A_4 A_6}+\frac{A_3 M_2}{A_5A_6}+\frac{A_5}{M_2}+\frac{A_3}{A_4 M_2}\cr
&\hspace{1cm}\left.\left.+\frac{A_4}{A_5 M_2}+\frac{A_3 M_2}{A_2}+\frac{A_1 A_5M_2}{A_2}+\frac{A_1 A_3 M_2}{A_4 A_2}+\frac{A_1 A_4 M_2}{A_5 A_2}+\frac{A_1 A_6}{A_2 M_2}\right)\right],
\end{split}
\end{equation}
and $Z^{\mathfrak{e}_6}_{\text{flavor 3}}$, $Z^{\mathfrak{e}_6}_{\text{flavor 4}}$ are the same as $Z^{\mathfrak{e}_6}_{\text{flavor 1}}$ and $Z^{\mathfrak{e}_6}_{\text{flavor 2}}$ with $M_1, M_2$ replaced with $M_3, M_4$ respectively. 

The perturbative part of the partition funciton \eqref{part.5de6} 
may be obtained by taking the limit $M_0 \to 0$ or $Q_B \to 0$. In this case the diagram in Figure \ref{fig:e6w4f} splits into the upper part and the lower part. The separated two parts are identical to each other and hence the partition function from the upper part will yield the square root of the root contribution $Z^{\mathfrak{e}_6}_{\text{roots}}$ and the contribution of the two flavors $Z^{\mathfrak{e}_6}_{\text{flavor 1}}, Z^{\mathfrak{e}_6}_{\text{flavor 2}}$. On the other hand the partition function from the lower part will give rise to the remaining square root of the root contribution $Z^{\mathfrak{e}_6}_{\text{roots}}$ and the contribution of the other two flavors $Z^{\mathfrak{e}_6}_{\text{flavor 3}}, Z^{\mathfrak{e}_6}_{\text{flavor 4}}$. Namely we should obtain 
\begin{equation}\label{e6upperhalf}
Z_{\text{upper half}}^{\fe_6} = \sqrt{Z^{\mathfrak{e}_6}_{\text{roots}}}Z^{\mathfrak{e}_6}_{\text{flavor 1}}Z^{\mathfrak{e}_6}_{\text{flavor 2}}Z_{\text{extra 1}}^{\fe_6+4\text{F}},
\end{equation}
where $Z_{\text{extra 1}}^{\fe_6+4\text{F}}$ is a contribution from a possible extra factor from the upper half of the diagram. We evaluated the partition function of the upper part by using \eqref{part.5de6} and found the agreement with \eqref{e6upperhalf} 
until the order $P_1^2P_2^2P_3^2P_4^2R_1^2R_2^2Q_1^2Q_2^2$ and the extra factor is given by 
\begin{equation}
Z_{\text{extra 1}}^{\fe_6+4\text{F}} = \text{PE}\left[\frac{q}{(1-q)^2}\left(4\frac{M_2}{M_1} + M_1^3 + 2M_1^2M_2 + \frac{M_2^2}{M_1^2}\right)\right].
\end{equation}
until the orders we computed. 

\subsection{6d/5d $E_7$ gauge theory with matter}
\label{sec:e7}
As for final examples for gauge theory with an exceptional gauge group we consider 6d/5d $E_7$ gauge theories with matter. The fundamental representation of $E_7$ is pseudo-real and $E_7$ gauge theories can have half-hypermultiplets. The cases with half-hypermultiplets may be intersecting in the sense that the application of the blow up formula of \cite{Kim:2019uqw, Gu:2020fem, Kim:2020hhh} may need a subtle treatment due to the absence of unity blowup equations. Hence 
we here compute the partition functions of two $E_7$ theories. One is the 6d $E_7$ gauge theory with one flavor and a half-hypermultiplet as well as a tensor multiplet. The other is the case with the maximal number of flavors, namely the 6d $E_7$ gauge theory with three flavors. We will write $n$ hypermultiplets and a half-hypermultiplet in the fundamental representation as $\frac{n+1}{2}$ flavors for simplicity. 

\begin{figure}[t]
\centering
\subfigure[]{\label{fig:e7on5}
\includegraphics[width=6cm]{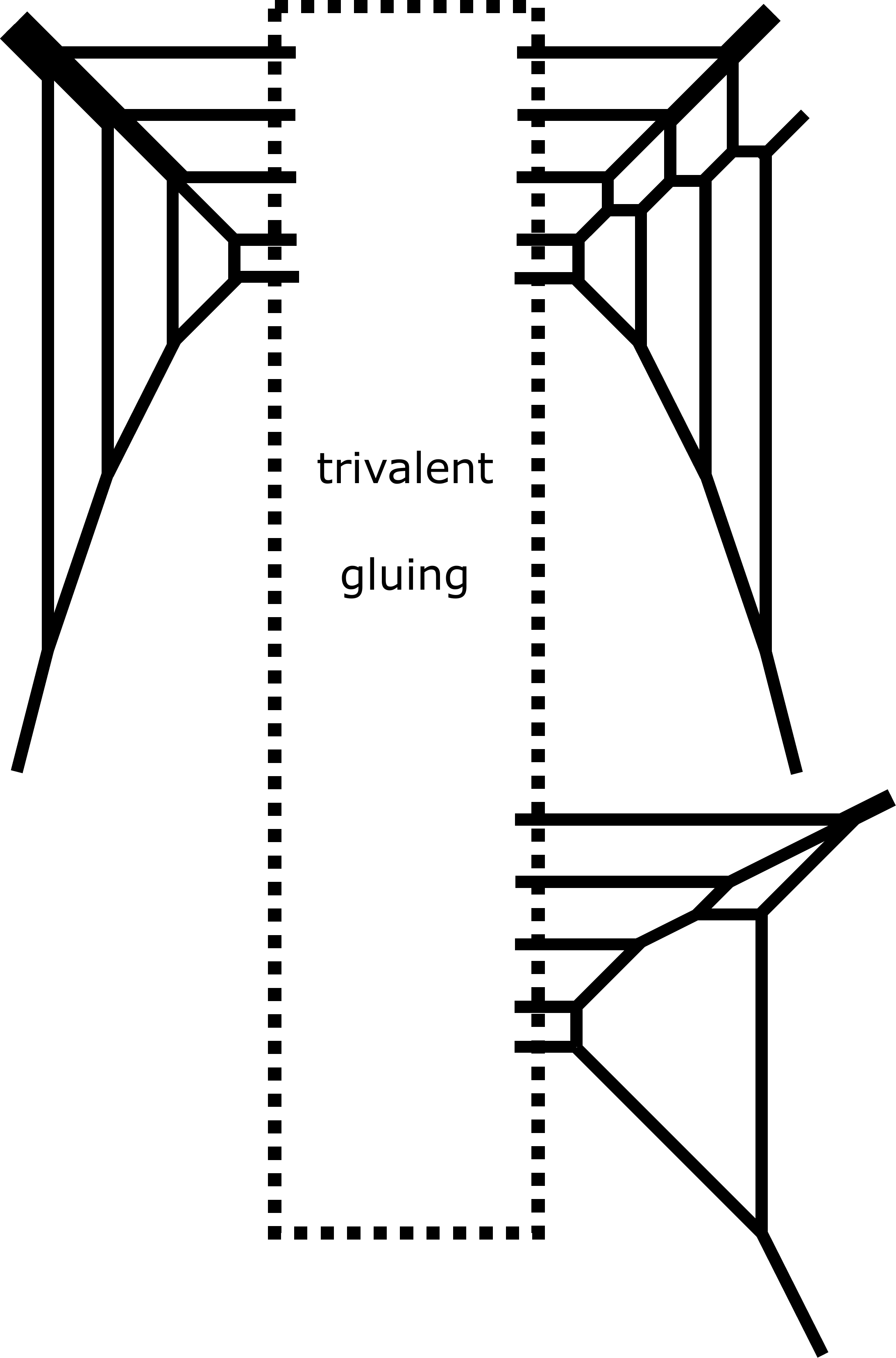}}
\hspace{1cm}
\subfigure[]{\label{fig:e7w32f}
\includegraphics[width=6cm]{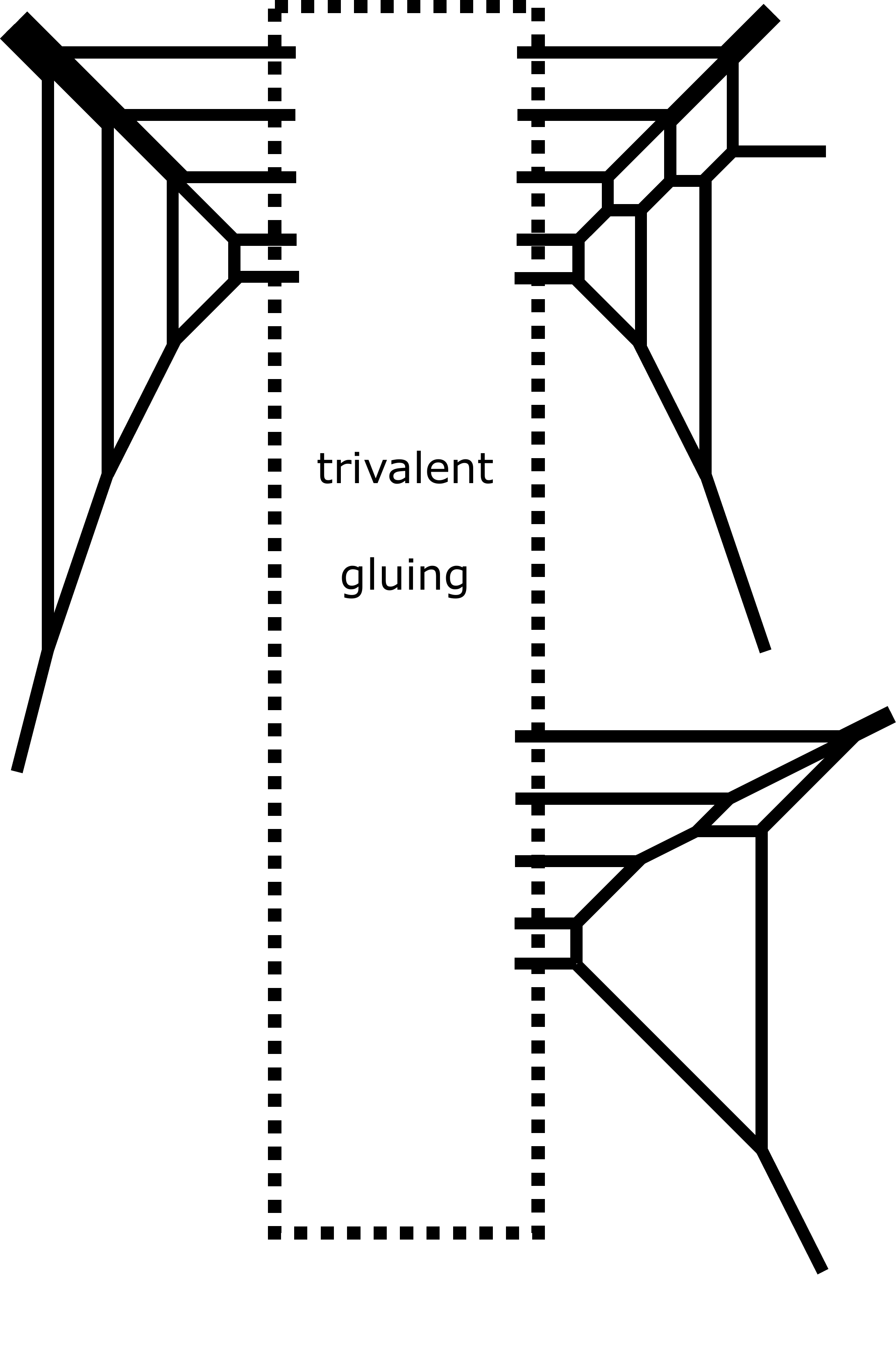}}
\caption{(a). The web diagram for the theory given by $\fe_7^{(1)}$ on $(-5)$-curve. (b). The web diagram for 5d $E_7$ gaue theory with one fundamental hypermultiplet and a fundamental half-hypermultiplet.  }
\label{fig:e7half}
\end{figure}

\paragraph{$\fe_7^{(1)}$ on $(-5)$-cruve.}
We start from the partition function of the 6d $E_7$ gauge theory with $\frac{3}{2}$ flavors on $T^2 \times \mathbb{R}^4$. The 6d $E_7$ gauge theory with $\frac{3}{2}$ flavors on $S^1$ is realized from the geometry of $\fe_7^{(1)}$ on $(-5)$-curve. 
From Table \ref{tb:e7_1} the $E_7$ theory is obtained from the $(4, 3, 2)$ Higgsing of the theory $(E_7, \underline{E_7})_1$ on $S^1$. 
The web diagram of the original theory $(E_7, \underline{E_7})_1$ on $S^1$ is depicted in Figure \ref{fig:ge7e7}. Then applying the Higgsing $(4, 3, 2)$ to the diagram in Figure \ref{fig:ge7e7} yields the web diagram in Figure \ref{fig:e7on5} and hence the diagram in Figure \ref{fig:e7on5} realizes the geometry $\fe_7^{(1)}$ on $(-5)$-curve. 
The fiber classes of the eight faces form the affine $E_7$ Dynkin diagram. Then we can also consider the 5d limit by decoupling the fiber class corresponding to the affine node of the affine $E_7$ Dynkin diagram, which leads to the $E_7$ Dynkin diagram formed by the remaining fiber classes. 
In terms of the web diagram in Figure \ref{fig:e7on5} the limit corroesponds to decoupling 
the rightmost face in the upper-right diagram in Figure \ref{fig:e7on5}, which gives rise to the diagram in Figure \ref{fig:e7w32f}. 
Since it corresponds to the 5d limit, the diagram in Figure \ref{fig:e7w32f} realizes the 5d $E_7$ gauge theory with a hypermultiplet in the fundamental representation and a half-hypermultiplet in the fundamental representation. 

\begin{figure}[t]
\centering
\subfigure[]{\label{fig:e7halfpara1}
\includegraphics[width=4cm]{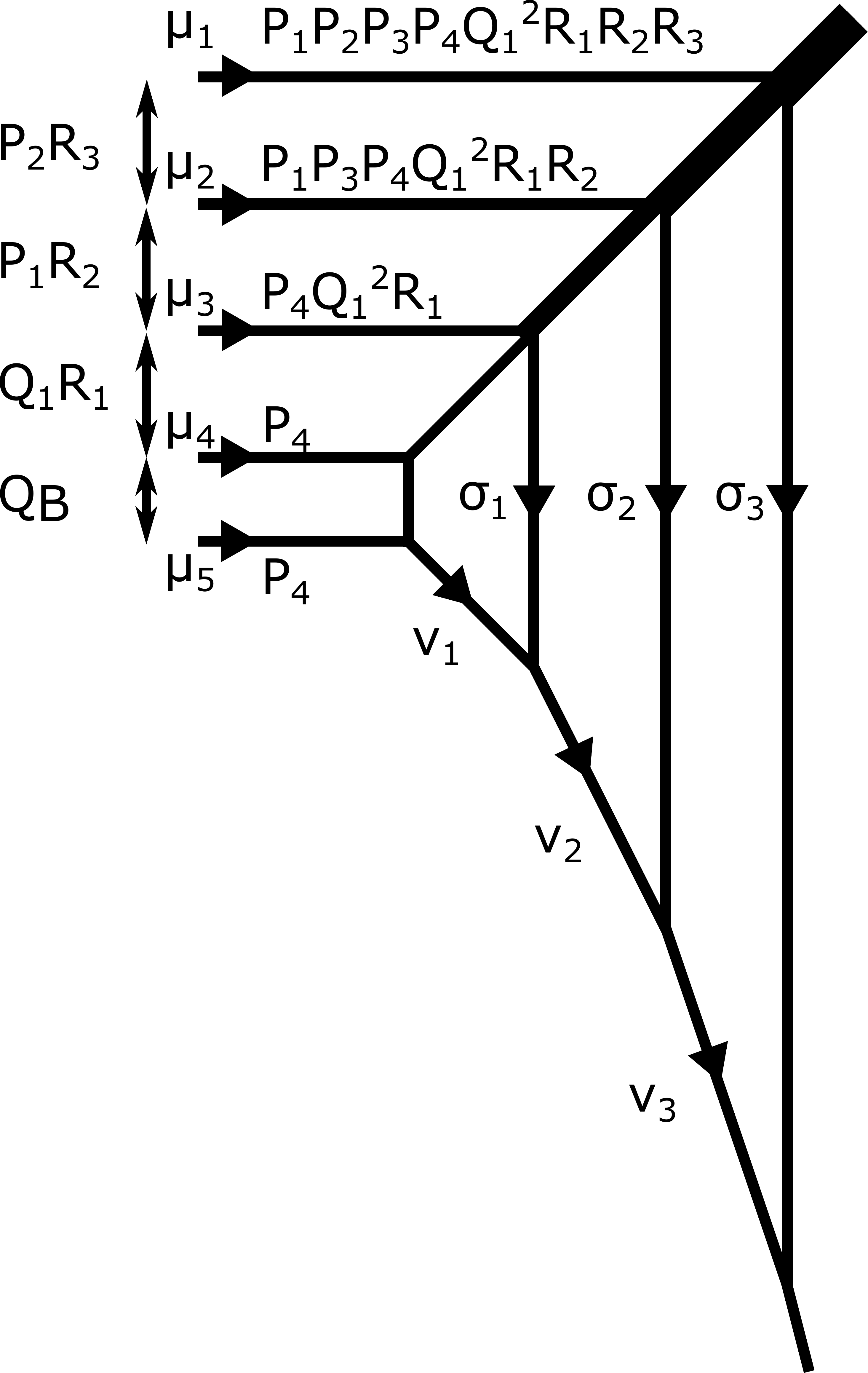}}
\hspace{1cm}
\subfigure[]{\label{fig:e7halfpara2}
\includegraphics[width=4cm]{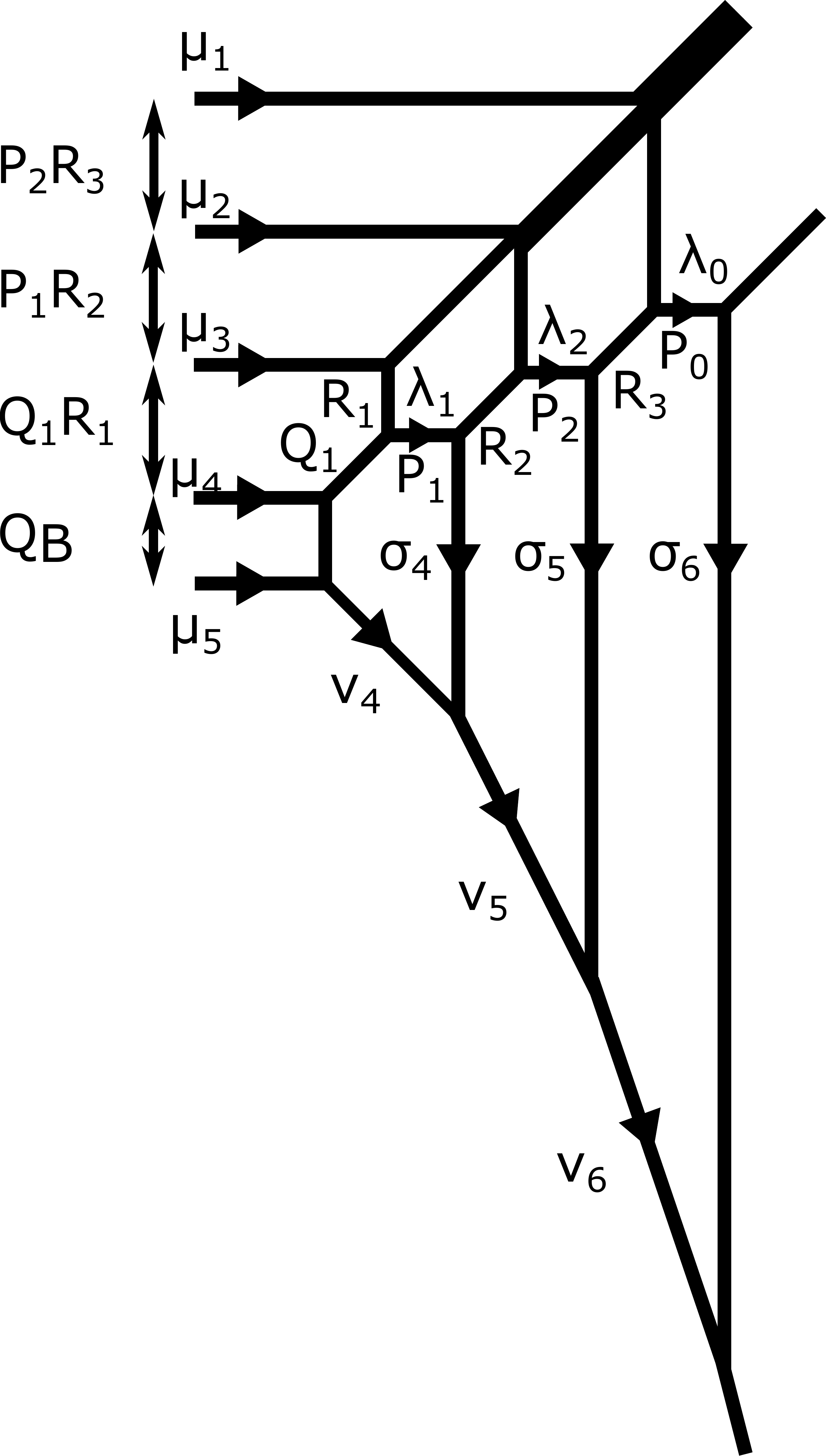}}
\hspace{1cm}
\subfigure[]{\label{fig:e7halfpara3}
\includegraphics[width=4cm]{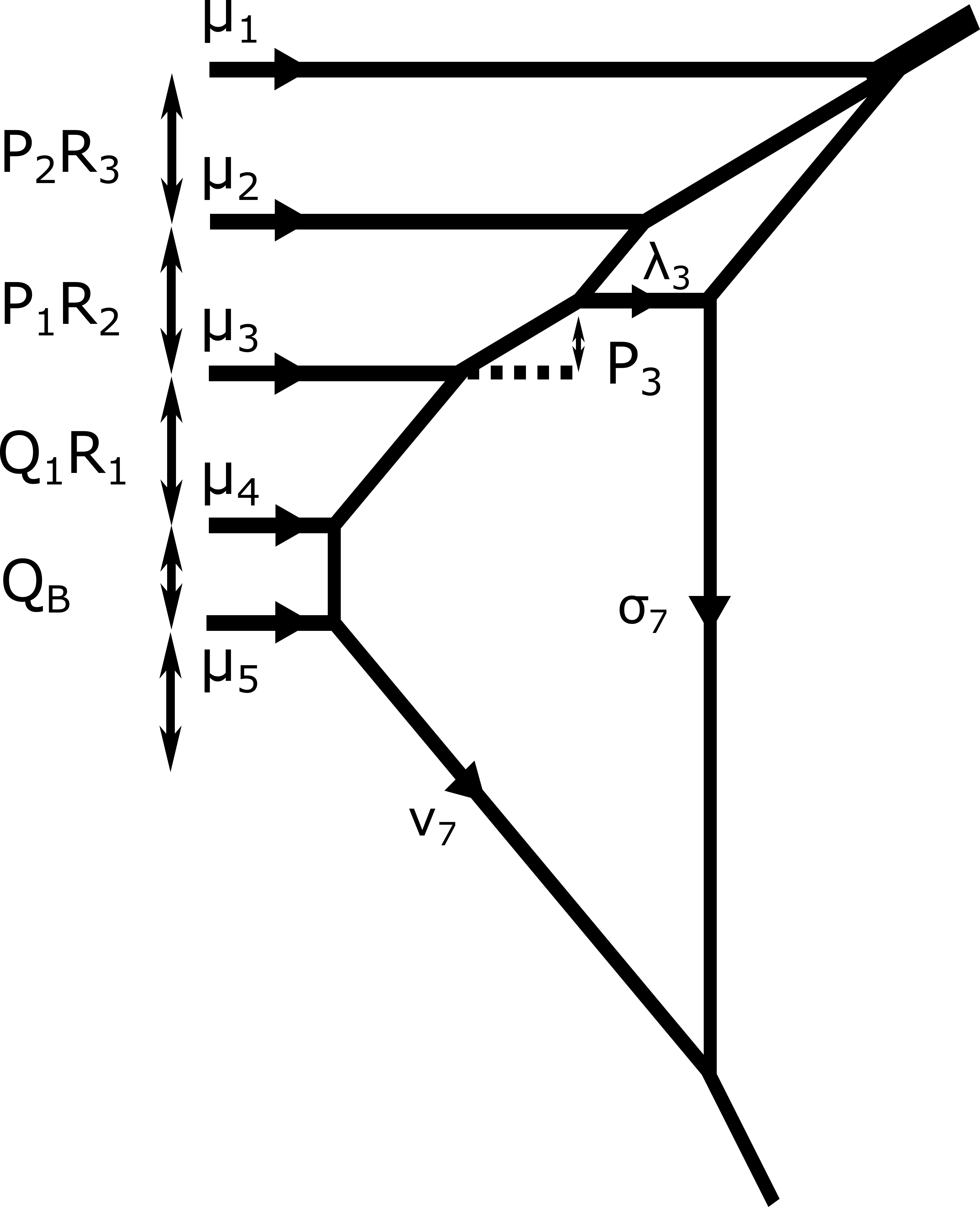}}
\caption{Parameterization of the web diagram for the geometry $\fe_7^{(1)}$ on $(-5)$-curve in Figure \ref{fig:e7on5}.}
\label{fig:e7halfpara}
\end{figure}

\begin{figure}[t]
\centering
\includegraphics[width=6cm]{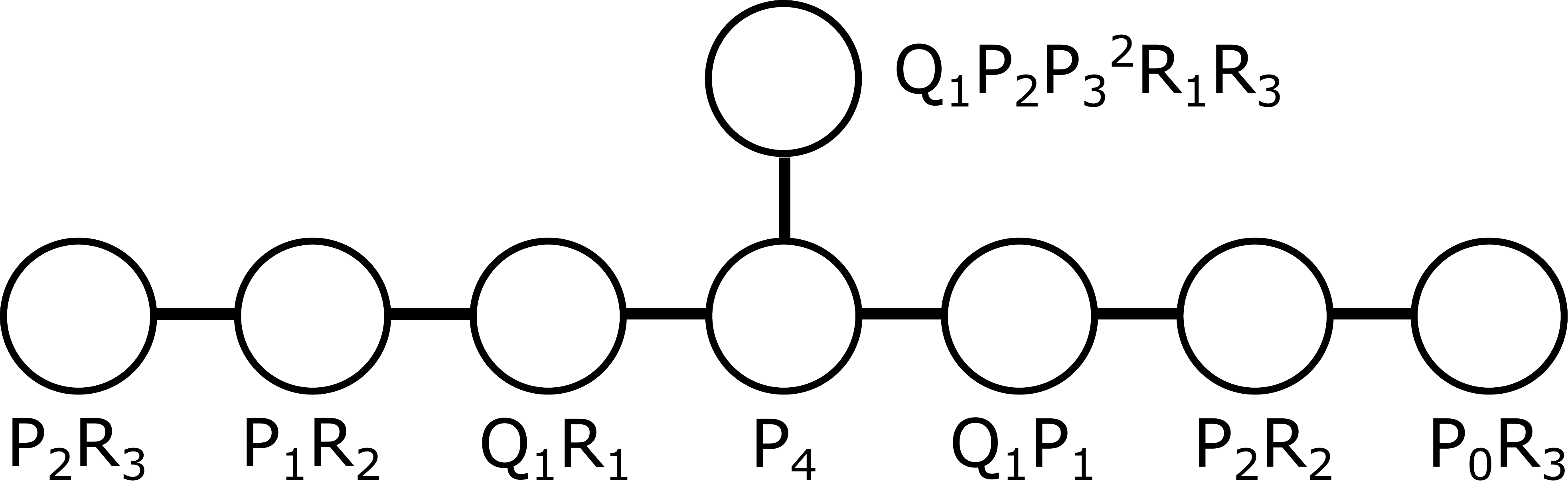}
\caption{The K\"ahler parameters of the fibers which form the affine $E_7$ Dynkin diagram.}
\label{fig:e7kahler}
\end{figure}

Since the web diagram in Figure \ref{fig:e7on5} is made by the trivalent gluing, it is possible to apply the topological vertex to the diagram in Figure \ref{fig:e7on5} and compute the partition function of 
the 6d $E_7$ gauge theory with a hypermultiplet in the fundamental representation, a half-hypermultiplet in the fundamental representation and a tensor multiplet on $T^2 \times \mathbb{R}^4$. For that we first parameterize the lengths of the 5-branes as in Figure \ref{fig:e7halfpara}.
The fiber classes of the web diagram in Figure \ref{fig:e7on5} have formed the affine $E_7$ Dynkin diagram and the parameterization for each node which follows from the one in Figure \ref{fig:e7halfpara} is depicted in Figure \ref{fig:e7kahler}.

\begin{figure}[t]
\centering
\subfigure[]{\label{fig:halfsu51}
\includegraphics[width=4cm]{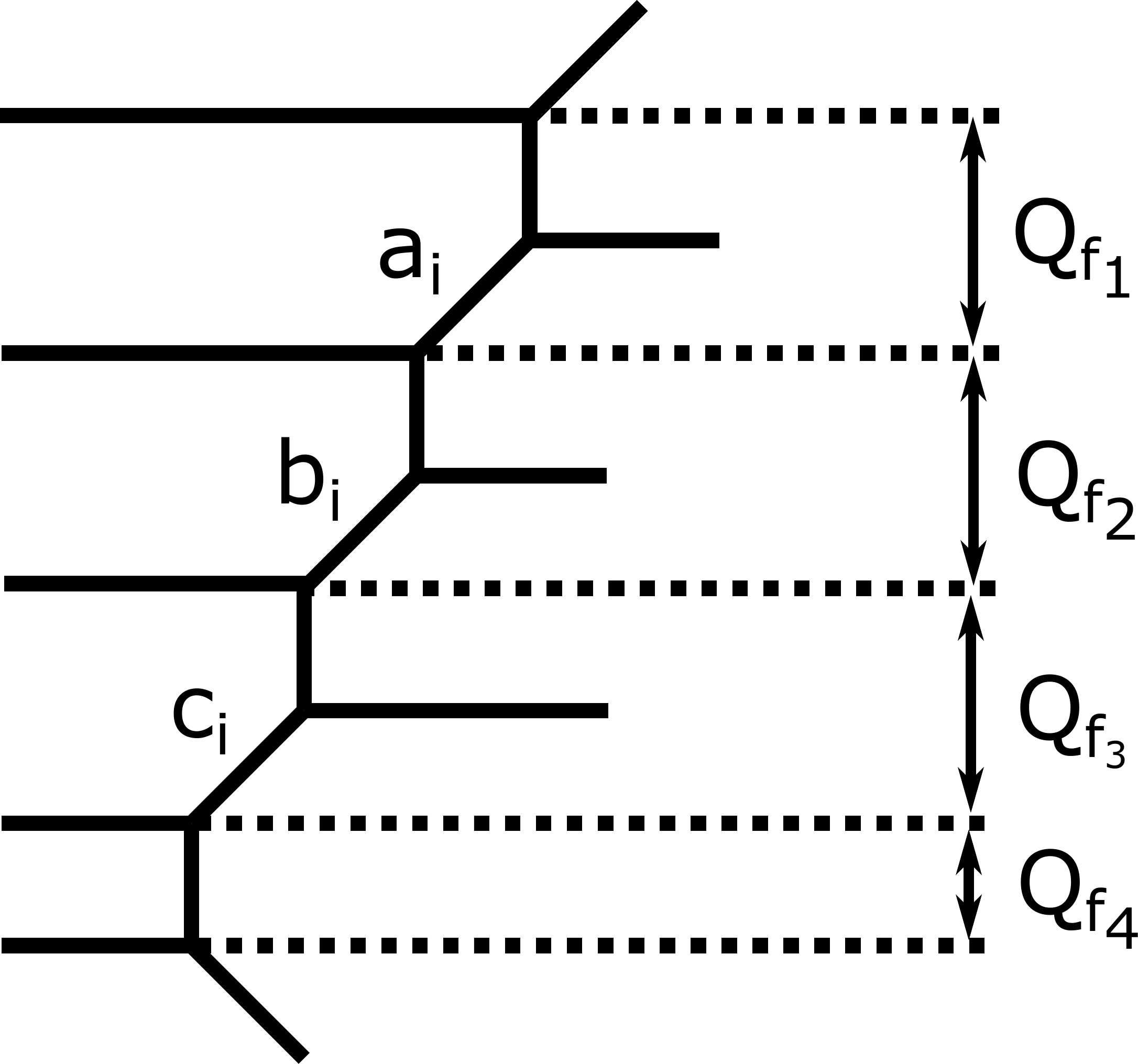}}
\hspace{1cm}
\subfigure[]{\label{fig:halfsu52}
\includegraphics[width=5cm]{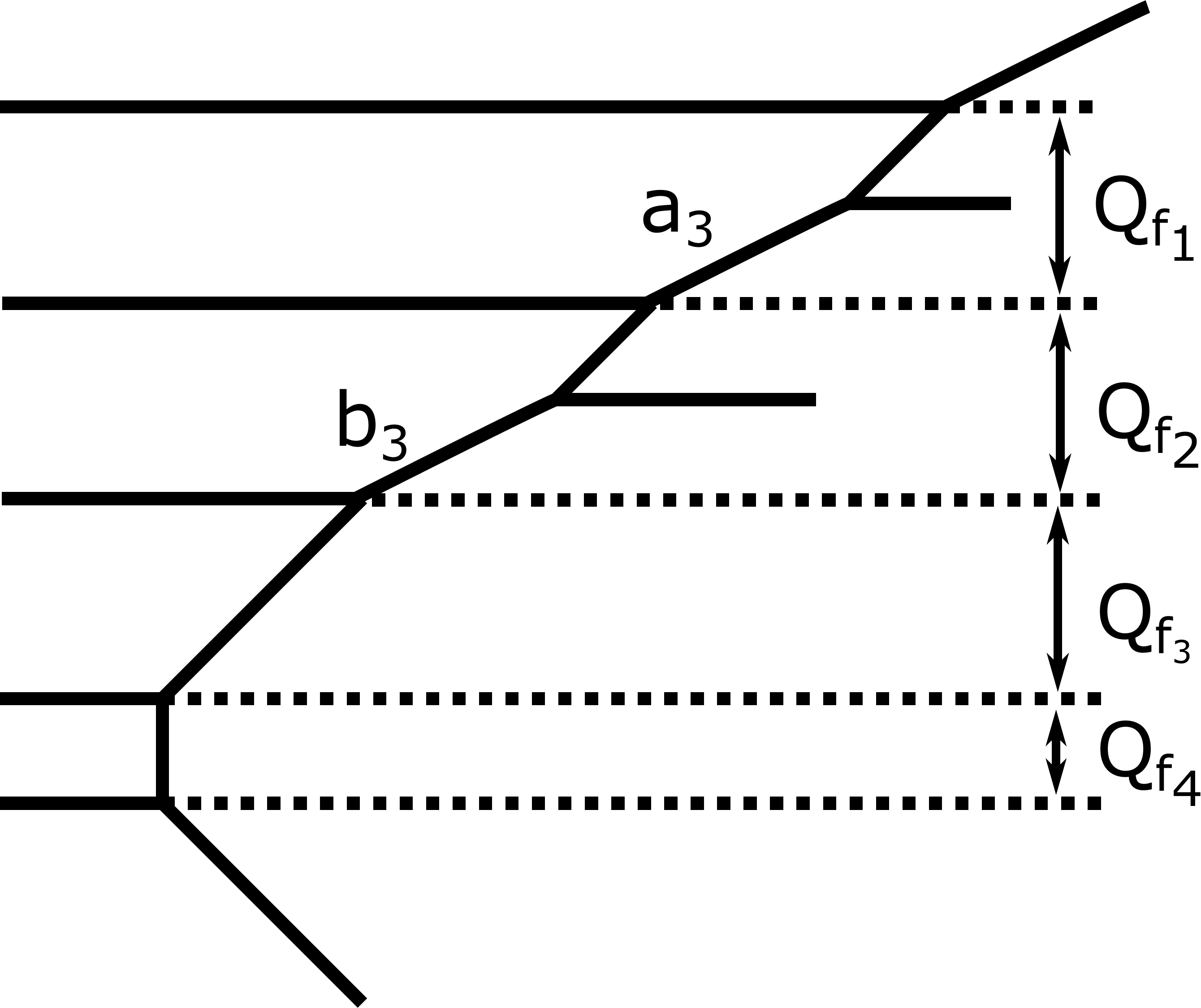}}
\caption{(a). Local geometry near the $\SU(5)$ part of the two rank $6$ theories in \eqref{halfe7quiver}. $i$ labels the two rank $6$ theories and $i=1, 2$. (b). Local geometry near the $\SU(5)$ part of the rank $2$ theory in \eqref{halfe7quiver}.}
\label{fig:halfsu5}
\end{figure}

\begin{figure}[t]
\centering
\includegraphics[width=7cm]{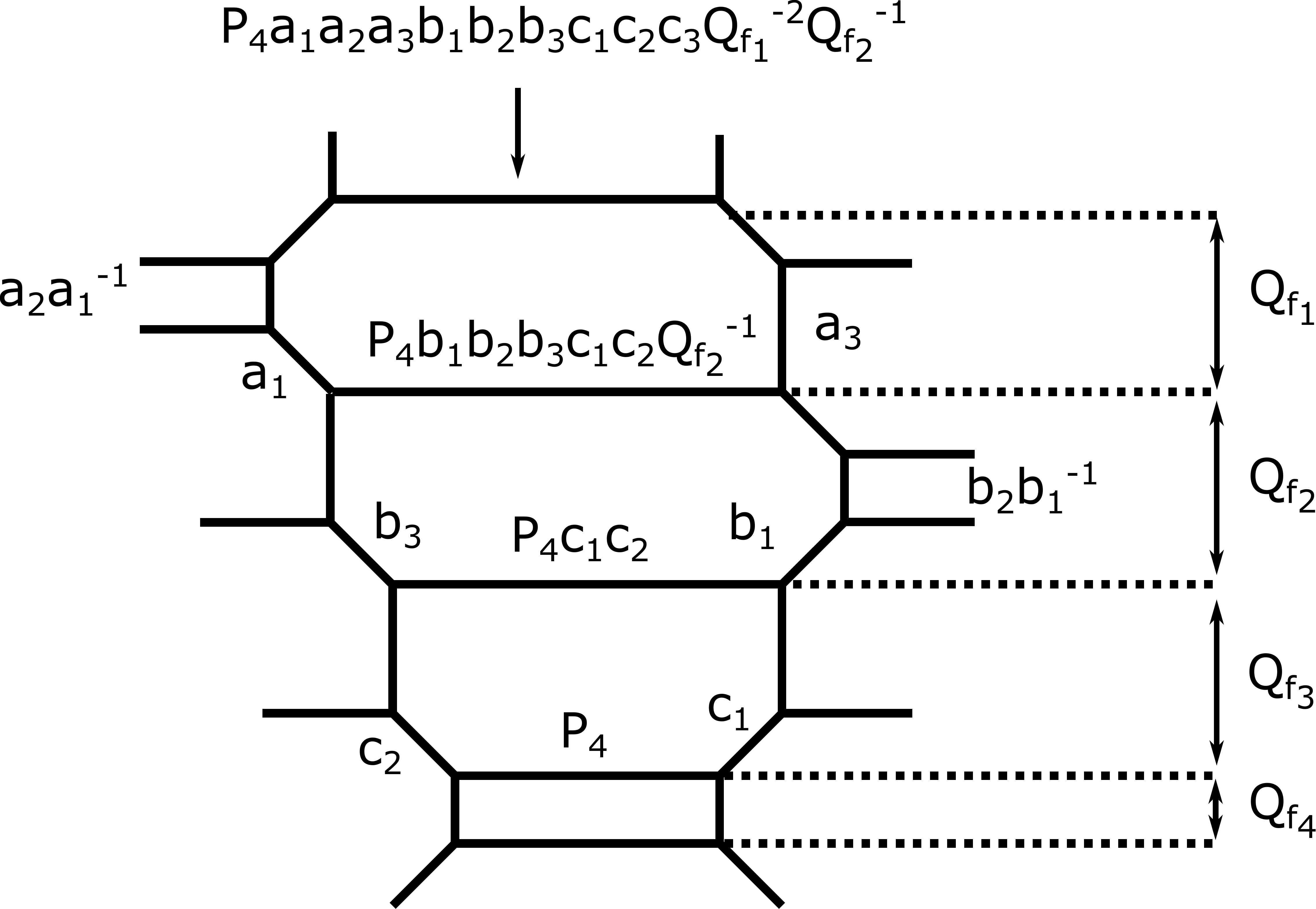}
\caption{Local geometry for the middle $\SU(5)$ gauge node of \eqref{halfe7quiver}.}
\label{fig:su5}
\end{figure}

We can determine the parametrization for the gluing lines in Figure \ref{fig:e7halfpara} by focusing on the local structure of the trivalent gauging. From the web diagram in Figure \ref{fig:ge7e7}, a 5d description of the theory \eqref{ge7e7} is given by
\be
(\text{rank }6)-{\overset{\overset{\text{\normalsize$(\text{rank }2)$}}{\textstyle\vert}}{\SU(5)}}- (\text{rank }6),\label{halfe7quiver}
\ee
where the rank $6$ theories correspond to the theories from the upper diagrams in Figure \ref{fig:ge7e7} and the rank $2$ theory is given by the lower diagram in Figure \ref{fig:ge7e7}. The $\SU(5)$ part in Figure \ref{fig:ge7e7} consists of four faces corresponding to four Coulomb branch moduli. 
The lowest face of the $\SU(5)$ 
corresponds to the $\mathbb{F}_0$ surface in the geometry of $\fe_7^{(1)}$ on $(-8)$-curve. Hence the local structure of the $\SU(5)$ gauging is given by combining two copies of the diagram in Figure \ref{fig:halfsu51} and the diagram in Figure \ref{fig:halfsu52} with the constraint that the lowest face becomes $\mathbb{F}_0$.  
The $Q_{f_i}, (i=1, \cdots, 4)$ in Figure \ref{fig:halfsu5} are the Coulomb branch moduli of the $\SU(5)$. The combined diagram for the $\SU(5)$ is described by the diagram in Figure \ref{fig:su5}. 
Then the $(4, 3, 2)$ Higgsing can be done by tuning the parameters as 
\begin{equation}\label{432half}
\begin{split}
&a_1 = a_2 = a_3 = Q_{f_1}, \quad b_1 = b_2 = Q_{f_2}, \quad c_1 = Q_{f_3}.
\end{split}
\end{equation}
After the tuning the Coulomb branch moduli of the $\SU(5)$ become the ones in Figure \ref{fig:e7halfpara1}, namely
\begin{equation}\label{su5CB}
Q_{f_1} = P_2R_3, \quad Q_{f_2} =  P_1R_2, \quad Q_{f_3}  = Q_1R_1, \quad Q_{f_4} = Q_B.  
\end{equation}
Using the tuning condition \eqref{432half} as well as the parametrization of the Coulomb branch moduli in \eqref{su5CB}, we obtain the K\"ahler parameters for the gluing lines in Figure \ref{fig:su5} 
written in Figure \ref{fig:e7halfpara1}.

We can then apply the topological vertex to the diagram in Figure \ref{fig:e7on5} using the parameterization in Figure \ref{fig:e7halfpara}. Since the charge of 5-branes attached to the gluing lines does not change by the $(4, 3, 2)$ Higgsing, we can use the framing factors which can be read from the diagram before the Higgsing in Figure \ref{fig:su5}. Then the partition function computed by applying the topological vertex to 
the diagram in Figure \ref{fig:e7on5} is 
\begin{equation}\label{part.6de7half}
\begin{split}
&Z^{\text{6d}}_{\fe_7, (-5)} \cr
&=\sum_{\{\mu_i\}}\left(-Q_1^2P_1P_2P_3P_4R_1R_2R_3\right)^{|\mu_1|}\left(-Q_1^2P_1P_3P_4R_1R_2\right)^{|\mu_2|}\left(-Q_1^2P_4R_1\right)^{|\mu_3|}\left(-P_4\right)^{|\mu_4| + |\mu_5|}\cr
&\hspace{0.5cm}Z_{\{\mu_i\}}^{\text{SU(5)$_L$}}\left(Q_B, \{Q_a\}, \{P_b\}, \{R_c\}\right) Z_{\{\mu_i\}}^{\text{SU(5)$_R$}}\left(Q_B, \{Q_a\}, \{P_b\}, \{R_c\}\right)f_{\mu_1}(q)^{-1}f_{\mu_3}(q)f_{\mu_4}(q)f_{\mu_5}(q)^{-1}\cr
&\hspace{0.5cm}\tilde{Z}^{\fe_7}_{1, \{\mu_i\}}\left(Q_B, \{Q_a\}, \{P_b\}, \{R_c\}\right)\tilde{Z}^{\fe_7}_{2, \{\mu_i\}}\left(Q_B, \{Q_a\}, \{P_b\}, \{R_c\}\right)\tilde{Z}^{\fe_7}_{3, \{\mu_i\}}\left(Q_B, \{Q_a\}, \{P_b\}, \{R_c\}\right),
\end{split}
\end{equation}
where we defined
\begin{align}
Z_{\{\mu_i\}}^{\text{SU(5)$_L$}}\left(Q_B, \{Q_a\}, \{P_b\}\right) &=q^{\frac{1}{2}\sum_{i=1}^5||\mu_i||^2}\left(\prod_{i=1}^5\tilde{Z}_{\mu_i}(q)\right)\cr
&\quad\prod_{1\leq i < j \leq 5}\mathcal{I}^-_{\mu_i, \mu_j}\left(Q_{f_i}Q_{f_{i+1}}\cdots Q_{f_{j-1}}\right),\label{SU5L}\\
Z_{\{\mu_i\}}^{\text{SU(5)$_R$}}\left(Q_B, \{Q_a\}, \{P_b\}\right) &=q^{\frac{1}{2}\sum_{i=1}^5||\mu_i^t||^2}\left(\prod_{i=1}^5\tilde{Z}_{\mu_i^t}(q)\right)\cr
&\quad\prod_{1\leq i < j \leq 5}\mathcal{I}^-_{\mu_i, \mu_j}\left(Q_{f_i}Q_{f_{i+1}}\cdots Q_{f_{j-1}}\right), \label{SU5R}
\end{align}
with $Q_{f_i}, \;(i=1, \cdots, 4)$ given in \eqref{su5CB}. Furthermore the factors $\tilde{Z}^{\fe_7}_{j, \{\mu_i\}}\left(Q_B, \{Q_a\}, \{P_b\}, \{R_c\}\right),$  $(j=1, 2, 3)$ are the partition functions from the diagrams in Figure \ref{fig:e7halfpara1}, \ref{fig:e7halfpara2} and \ref{fig:e7halfpara3} with the trivalent gluing prescription used. Each factor is given by
\begin{align}
&\tilde{Z}^{\fe_7}_{1,\{\mu_i\}}\left(Q_B, \{Q_a\}, \{P_b\}, \{R_c\}\right)\cr
&= q^{\frac{1}{2}\sum_{i=1}^5||\mu_i^t||^2}\left(\prod_{i=1}^5\tilde{Z}_{\mu_i^t}(q)\right)\mathcal{I}^-_{\mu_4,\mu_5}\left(Q_B\right)/Z_{\{\mu_i\}}^{\text{SU(5)$_R$}}\left(Q_B, \{Q_a\}, \{P_b\}, \{R_c\}\right)\cr
&\quad \sum_{\nu_{1,2,3}, \sigma_{1,2,3}, \eta, \eta'}q^{-\frac{1}{2}\sum_{i=1}^3(||\nu_i^t||^2 - 2||\nu_i||^2)}\left(\prod_{i=1}^3\tilde{Z}_{\nu_i^t}(q)\right)q^{\sum_{k=1}^3\frac{k}{2}(||\sigma_k^t||^2 - ||\sigma_k||^2)}(P_1R_2)^{|\nu_2|}(P_2R_3)^{|\nu_3|}\cr
&\quad s_{\nu_1}\left(Q_1R_1q^{-\rho - \mu_5}, Q_BQ_1R_1q^{-\rho - \mu_4}\right)s_{\nu_2/\eta'}(q^{-\rho-\nu_1})s_{\nu_3/\eta}(q^{-\rho - \nu_2})\cr
&\quad s_{\sigma_3}(-Q_BQ_1^2R_1^2P_1^3R_2^3P_2^4R_3^4q^{-\rho-\mu_1})s_{\sigma_3}(q^{-\rho - \nu_3^t})s_{\sigma_2}(Q_BQ_1^2R_1^2P_1^3R_2^3q^{-\rho-\mu_2})s_{\sigma_2/\eta}(q^{-\rho - \nu_2^t})\cr
&\quad s_{\sigma_1}(-Q_BQ_1^2R_1^2q^{-\rho - \mu_3})s_{\sigma_1/\eta'}(q^{-\rho - \nu_1^t}),\nn\\\\
&\tilde{Z}^{\fe_7}_{2,\{\mu_i\}}\left(Q_B, \{Q_a\}, \{P_b\}, \{R_c\}\right)\cr
&= q^{\frac{1}{2}\sum_{i=1}^5||\mu_i^t||^2}\left(\prod_{i=1}^5\tilde{Z}_{\mu_i^t}(q)\right)/Z_{\{\mu_i\}}^{\text{SU(5)$_R$}}\left(Q_B, \{Q_a\}, \{P_b\}, \{R_c\}\right)\cr
&\quad\mathcal{I}^-_{\mu_3,\mu_4}\left(Q_1R_1\right)\mathcal{I}^-_{\mu_3,\mu_5}\left(Q_BQ_1R_1\right)\mathcal{I}^-_{\mu_4,\mu_5}\left(Q_B\right)\cr
&\quad\sum_{\substack{\lambda_{0, 1, 2}, \nu_{4, 5, 6}, \sigma_{4, 5, 6}\\ \eta_{1,2,3,4,5,6,7,8,9}}} q^{\frac{1}{2}\sum_{i=0}^2\left(||\lambda_i||^2 + ||\lambda_i^t||^2\right)}\left(\prod_{i=0}^2\tilde{Z}_{\lambda_i}(q)\tilde{Z}_{\lambda_i^t}(q)\right)\left(\prod_{i=0}^2\left(-P_i\right)^{|\lambda_i|}\right)\cr
&\quad\mathcal{I}^+_{\mu_1,\lambda_0}\left(P_1P_2R_1\right)\mathcal{I}^-_{\mu_1,\lambda_2}\left(P_1P_2R_1R_3\right)\mathcal{I}^+_{\lambda_0, \lambda_2}\left(R_3\right)\cr
&\quad\mathcal{I}^+_{\mu_2, \lambda_2}\left(P_1R_1\right)\mathcal{I}^-_{\mu_2, \lambda_1}\left(P_1R_1R_2\right)\mathcal{I}^+_{\lambda_2, \lambda_1}\left(R_2\right)\mathcal{I}^+_{\mu_3, \lambda_1}(R_1)\mathcal{I}^+_{\lambda_1, \mu_4}\left(Q_1\right)\mathcal{I}^+_{\lambda_1, \mu_5}\left(Q_BQ_1\right)\cr
&\quad q^{-\frac{1}{2}\sum_{i=4}^6(||\nu_i^t||^2 - 2||\nu_i||^2)}\left(\prod_{i=4}^6\tilde{Z}_{\nu_i^t}(q)\right)q^{\sum_{k=1}^3\frac{k}{2}(||\sigma_{k+3}^t||^2 - ||\sigma_{k+3}||^2)}(Q_1P_1)^{|\nu_4|}(R_2P_2)^{|\nu_5|}(R_3P_0)^{|\nu_6|}\cr
&\quad s_{\nu_4/\eta5}(q^{-\rho - \mu_5})s_{\nu_5/\eta_4}(q^{-\rho - \nu_4})s_{\nu_6/\eta_2}(q^{-\rho - \nu_5}) s_{\sigma_6}(-Q_BQ_1^2P_1R_2^3P_2^2R_3^4P_0^3q^{-\rho -\nu_6^t})s_{\sigma_6}(q^{-\rho - \lambda_0})\cr
&\quad s_{\sigma_5/\eta_2}(q^{-\rho - \nu_5^t})s_{\sigma_5/\eta_1}(q^{-\rho - \lambda_2})(Q_BQ_1^2P_1R_2^3P_2^2)^{|\sigma_5|} s_{\sigma_4/\eta_4}(q^{-\rho - \nu_4^t})s_{\sigma_4/\eta_3}(q^{-\rho - \lambda_1})(-Q_BQ_1^2P_1)^{|\sigma_4|}\cr
&\quad s_{\eta_6}(P_1P_2R_1R_3q^{-\rho - \mu_1})s_{\eta_1^t/\tau^t}(-R_3q^{-\rho - \lambda_0})s_{\eta_7}(P_1R_1R_2q^{-\rho - \mu_2})s_{\eta_3^t/\eta_7^t}(-R_2q^{-\rho - \lambda_2})\cr
&\quad s_{\eta_5/\eta_8}(Q_Bq^{-\rho - \mu_4}) s_{\eta_8^t/\eta_9}(-Q_BQ_1q^{-\rho - \lambda_1})s_{\eta_9^t}(Q_BQ_1R_1q^{-\rho - \mu_3}),\\
&\tilde{Z}^{\fe_7}_{3,\{\mu_i\}}\left(Q_B, \{Q_a\}, \{P_b\}, \{R_c\}\right)\cr
&=q^{\frac{1}{2}\sum_{i=1}^5||\mu_i^t||^2}\left(\prod_{i=1}^5\tilde{Z}_{\mu_i^t}(q)\right)/Z_{\{\mu_i\}}^{\text{SU(5)$_R$}}\left(Q_B, \{Q_a\}, \{P_b\}, \{R_c\}\right)\cr
&\quad\mathcal{I}^-_{\mu_2,\mu_3}\left(P_1R_2\right)\mathcal{I}^-_{\mu_2,\mu_4}\left(Q_1P_1R_1R_2\right)\mathcal{I}^-_{\mu_2,\mu_5}\left(Q_BQ_1P_1R_1R_2\right)\mathcal{I}^-_{\mu_3,\mu_4}\left(Q_1R_1\right)\mathcal{I}^-_{\mu_3,\mu_5}\left(Q_BQ_1R_1\right)\cr
&\quad\mathcal{I}^-_{\mu_4, \mu_5}\left(Q_B\right)\sum_{\substack{\lambda_{3}, \nu_7, \sigma_7 \\ \eta_{1,2,3,4}}} q^{||\lambda_3||^2}\tilde{Z}_{\lambda_3}(q)\left(P_2R_3\right)^{|\lambda_3|} q^{-\frac{||\nu_7^t||^2}{2} + ||\nu_7||^2}\tilde{Z}_{\nu_7^t}(q)(Q_1R_1P_3^2P_2R_3)^{|\nu_7|}\cr
&\quad q^{\frac{1}{2}(||\sigma_7^t||^2 - ||\sigma_7||^2)}s_{\nu_7/\eta_1}(q^{-\rho - \mu_5})s_{\sigma_7}(Q_BQ_1^2R_1^2P_3^3P_2R_3q^{-\rho - \nu_7^t})s_{\sigma_7}(q^{-\rho - \lambda_4}, P_1R_2P_2R_3P_3^{-1}q^{-\rho - \mu_1})\cr
&\quad s_{\eta_1/\eta_2}(Q_Bq^{-\rho -\mu_4})s_{\eta_2/\eta_3}(Q_BQ_1R_1q^{-\rho - \mu_3})s_{\eta_3^t/\eta_4}(-Q_BQ_1R_1P_3q^{-\rho - \lambda_4})s_{\eta_4^t}(Q_BQ_1R_1P_1R_2q^{-\rho - \mu_2})\cr
&\quad\mathcal{I}^-_{\mu_1, \lambda_3}\left(P_1P_2P_3^{-1}R_2R_3\right)\mathcal{I}_{\mu_2, \lambda_3}^+\left(P_1P_3^{-1}R_2\right)\mathcal{I}_{\lambda_3, \mu_3}^+\left(P_3\right)\mathcal{I}_{\lambda_3, \mu_4}^+\left(Q_1P_3R_1\right)\mathcal{I}_{\lambda_3, \mu_5}^+\left(Q_BQ_1P_3R_1\right).\cr
\end{align}

The partition function \eqref{part.6de7half} may contain an extra factor. The extra factor may be determined by parameterizing the lengths in terms of Coulomb branch moduli and mass parameters. The Coulomb branch moduli dependence on each K\"ahler parameter is given by \eqref{QCB}. The theory contains two mass parameters and we can associate them to the length between parallel external legs. From the length between parallel external legs in Figure \ref{fig:e7halfpara2}, we assign
\be
P_0P_1P_2R_1 = M'_1.
\ee
The diagram in Figure \ref{fig:su5} suggests that the external lines attached to the top gluing line are parallel to each other and we parameterize the length by
\be
Q_1^2P_1P_2P_3P_4R_1R_2R_3 = M'_0
\ee
With this parameterization the K\"ahler parameter for the elliptic class given by \eqref{elliptic} becomes
\be\label{Qtaue7half}
Q_{\tau} = (P_2R_3)(P_1R_2)^2(Q_1R_1)^3(P_4)^4(Q_1P_1)^3(P_2R_2)^2(P_0R_3)(Q_1P_2P_3^2R_1R_3)^2 = M'^4_0M'_1,
\ee
and it only depends on the mass parameters. 

Then the partition function of \eqref{part.6de7half} can be written as
\be\label{hatpart.6de7half}
Z_{\fe_{7}^{(1)},(-5)}^{\text{6d}} = \hat{Z}_{\fe_{7}^{(1)},(-5)}^{\text{6d}}\left(\{A'_a\}, \{M'_i\}\right)Z_{\text{extra}}^{\fe_{7}^{(1)},(-5)}\left(M'_0, M'_1\right).
\ee 
Here $Z_{\text{extra}}^{\fe_{7}^{(1)},(-5)}\left(M'_0, M'_1\right)$ is an extra factor in \eqref{part.6de7half} which is independent from the Coulomb branch moduli. 
We claim that the partition function $\hat{Z}_{\fe_{7}^{(1)},(-5)}^{\text{6d}}\left(\{A'_a\}, \{M'_i\}\right)$ in \eqref{hatpart.6de7half} gives the partition function of the 6d $E_7$ gauge theory with $\frac{3}{2}$ flavors and a tensor multiplet on $T^2 \times \mathbb{R}^4$ up to an extra factor. 

\paragraph{5d $E_7$ gauge theory with $\frac{3}{2}$ flavors.}
We then apply the 5d limit to the partition function \eqref{part.6de7half}. The 5d limit is given by decouping the fiber class associated to the affine node in Figure \ref{fig:e7kahler} and the resulting web diagram is depicted in Figure \ref{fig:e7w32f}. In terms of the K\"ahler parameter assignment in Figure \ref{fig:e7halfpara} the limit is given by $P_0 \to 0$ with the other K\"ahler parameters fixed. The K\"ahler parameter of the elliptic class \eqref{Qtaue7half} becomes zero in the limit. Hence the application of the limit $P_0 \to 0$ to \eqref{part.6de7half} gives rise to the partition function of the 5d $E_7$ gauge theory with $\frac{3}{2}$ flavors on $S^1 \times \mathbb{R}^4$ up to an extra factor. 

To identify the extra factor and also make a comparison later we rewrite the partition function by the gauge theory parameters in the 5d $E_7$ gauge theory. The Coulomb branch moduli dependence is determined by \eqref{QCB}. In particular 
since the fiber classes forming the $E_7$ Dynkin diagram in Figure \ref{fig:e7kahler} correspond to simple roots of the $E_7$ Lie algebra, 
K\"ahler parameters for the fiber classes are parameterized by the Coulomb branch moduli $A_i = e^{-a_i}, \; (i=1, \cdots, 7)$ as
\begin{equation}\label{e7CB}
\begin{split}
&P_2R_2 = A_1^2A_2^{-1}, \quad Q_1P_1 = A_1^{-1}A_2^2A_3^{-1}, \quad P_4 = A_2^{-1}A_3^2A_4^{-1}A_7^{-1}, \quad Q_1R_1 = A_3^{-1}A_4^2A_5^{-1}, \cr
&P_1R_2 = A_4^{-1}A_5^2A_6^{-1}, \quad P_2R_3 = A_5^{-1}A_6^2, \quad Q_1P_2P_3^2R_1R_3 = A_3^{-1}A_7.
\end{split}
\end{equation}
On the other hand a string with length characterized by the K\"ahler parameter $Q_1$ gives rise to a hypermultiplet corresponding to a weight of the fundamental representation of $E_7$. The Dynkin label may be read off from the intersection number between the curve and the complex surfaces. The Dynkin label of the weights for $Q_1$ is
\begin{equation}
Q_1\;:\;[0, 1, -1, 1, -1, 0, 0].
\end{equation}
Hence we parameterize them by
\begin{equation}\label{e7halfmass}
Q_1 = A_2A_3^{-1}A_4A_5^{-1}M_1.
\end{equation}
with the mass parameter $M_1 = e^{-m_1}$. The remaining mass parameter is the instanton fugacity and it can be determined from the analysis of the effective prepotential given in \eqref{prepotential}. Note that the lowest face in the diagram of the trivalent $\SU(5)$ gauging is $\mathbb{F}_0$. The volume of the surface or the area of the corresponding face is given by taking a derivative of the effective prepotential with respect to $a_3$ and it is given by
\be
\frac{\partial F}{\partial a_3} = (-a_2 + 2a_3 - a_4 - a_7)(2a_3 - a_4 + m_0 - 3m_1).
\ee
Then the instanton fugacity $M_0 = e^{-m_0}$ is given by 
\begin{equation}\label{e7halfinst}
Q_B = A_3^2A_4^{-1}M_0M_1^{^-3}. 
\end{equation}

The application of the limit $P_0 \to 0$ to \eqref{part.6de6} 
with the gauge theory parameterization becomes
\begin{align}\label{part.5de7w32f}
Z_{\fe_7+\frac{3}{2}\text{F}}^{\text{5d}} &= Z^{\text{6d}}_{\fe_7,(-5)}\Big|_{P_0 = 0}\cr
&=\hat{Z}^{\text{5d}}_{\fe_7 + \frac{3}{2}\text{F}}\left(\{A_b\}, \{M_i\}\right)Z_{\text{extra}}^{\fe_7+ \frac{3}{2}\text{F}}\left(M_0, M_1\right). 
\end{align}
We argue that $\hat{Z}^{\text{5d}}_{\fe_7 + \frac{3}{2}\text{F}}\left(\{A_b\}, \{M_i\}\right)$ gives the partition function of the 5d $E_7$ gauge theory with $\frac{3}{2}$ flavors on $S^1 \times \mathbb{R}^4$ up to an extra factor. 

\begin{figure}[t]
\centering
\subfigure[]{\label{fig:e7on2}
\includegraphics[width=6cm]{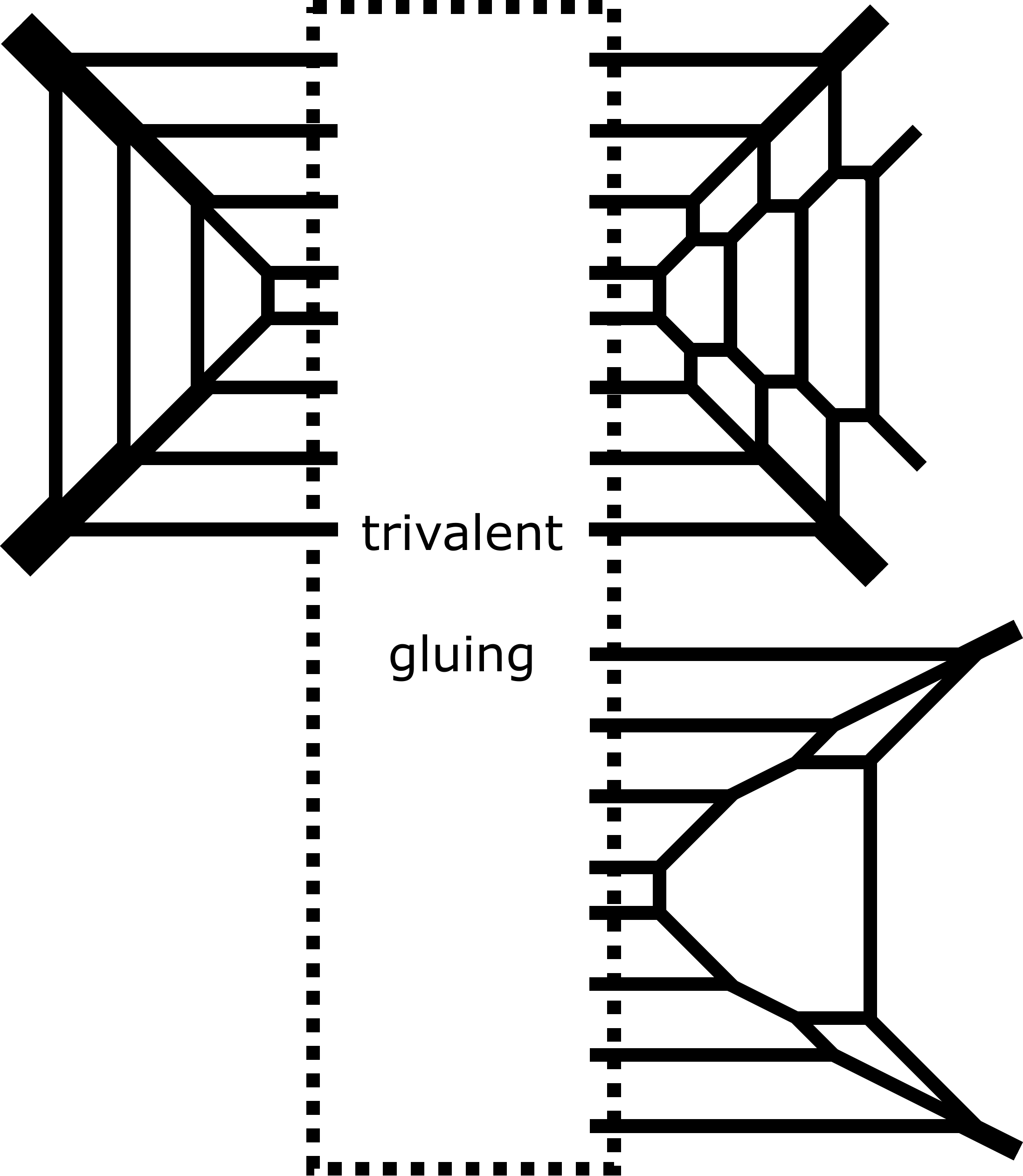}}
\hspace{1cm}
\subfigure[]{\label{fig:e7w3f}
\includegraphics[width=6cm]{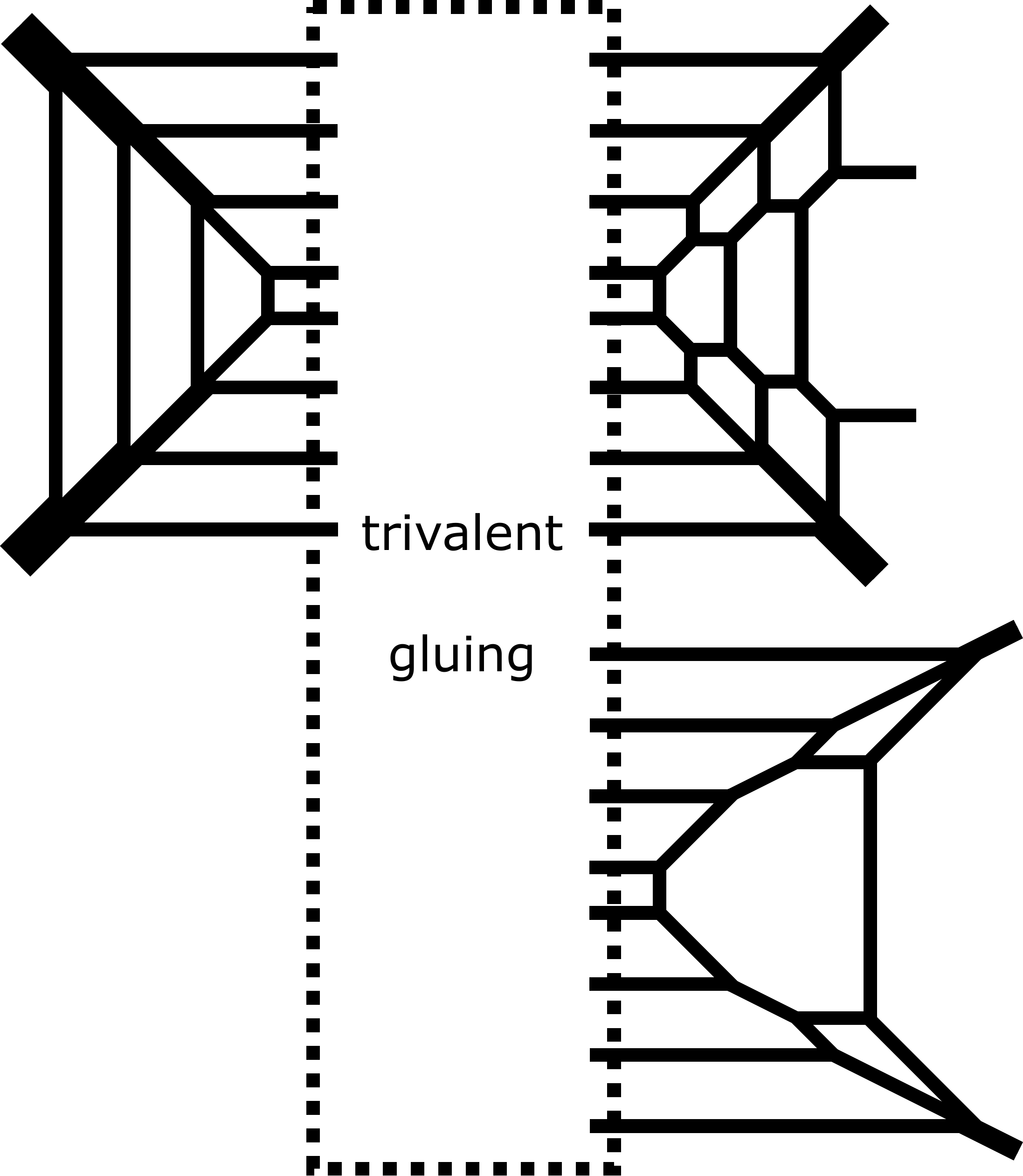}}
\caption{(a). The web diagram for the theory given by $\fe_7^{(1)}$ on $(-2)$-curve. (b). The web diagram for 5d $E_7$ gaue theory with $3$ flavors but one mass parameter is turned off.  }
\label{fig:e7}
\end{figure}

To check the validity of the partition function \eqref{part.5de7w32f}, we compare it with the perturbative part of the Nekrasov partition function of the 5d $E_7$ gauge theory with $\frac{3}{2}$ flavors using the universal formula in section \ref{sec:Nek}. 
For the phase which gives the diagram in Figure \ref{fig:e7w3f}, the perturbative partition function is given by 
\begin{align}\label{5de7halfpert}
Z^{\text{5d pert}}_{\mathfrak{e}_7+\frac{3}{2}\text{F}} = Z^{\mathfrak{e}_7}_{\text{cartan}}Z^{\mathfrak{e}_7}_{\text{roots}}Z^{\mathfrak{e}_7}_{\text{flavor}}Z^{\mathfrak{e}_7}_{\text{$\frac{1}{2}$ flavor}},
\end{align}
where each factor is 
\begin{align}\label{e7cartan}
Z^{\mathfrak{e}_7}_{\text{cartan}} = \text{PE}\left[\frac{7q}{(1-q)^2}\right],
\end{align}
\begin{align}\label{e7root}
Z^{\mathfrak{e}_7}_{\text{roots}} &= \text{PE}\left[\frac{2q}{(1-q)^2}\left(\frac{A_1^2}{A_2}+\frac{A_5 A_1}{A_2}+\frac{A_3 A_6 A_1}{A_2 A_4}+\frac{A_4 A_6 A_1}{A_2A_5}+\frac{A_6 A_7 A_1}{A_3}+\frac{A_7 A_1}{A_4}+\frac{A_4 A_7 A_1}{A_3 A_5}\right.\right.\cr
&\quad\qquad+\frac{A_5 A_7
A_1}{A_3 A_6}+\frac{A_2 A_1}{A_3}+\frac{A_3 A_1}{A_2 A_5}+\frac{A_4 A_1}{A_2 A_6}+\frac{A_3 A_5
   A_1}{A_2 A_4 A_6}+\frac{A_6 A_1}{A_7}+\frac{A_3 A_1}{A_4 A_7}+\frac{A_4 A_1}{A_5 A_7}\cr
&\quad\qquad+\frac{A_5
   A_1}{A_6 A_7}+A_1+\frac{A_6^2}{A_5}+\frac{A_7^2}{A_3}+\frac{A_3}{A_2}+\frac{A_2
   A_5}{A_3}+\frac{A_4 A_6}{A_3}+\frac{A_5 A_6}{A_4}+\frac{A_2 A_6}{A_4}+\frac{A_2 A_4 A_6}{A_3
   A_5}\cr
&\quad\qquad+\frac{A_4 A_7}{A_3}+\frac{A_5 A_7}{A_4}+\frac{A_6 A_7}{A_2}+\frac{A_6 A_7}{A_5}+\frac{A_3
   A_7}{A_2 A_4}+\frac{A_4 A_7}{A_2 A_5}+\frac{A_5 A_7}{A_2 A_6}+\frac{A_7}{A_6}+\frac{A_4^2}{A_3
   A_5}\cr
&\quad\qquad+\frac{A_2}{A_5}+\frac{A_5^2}{A_4 A_6}+\frac{A_2 A_4}{A_3 A_6}+\frac{A_4 A_5}{A_3
   A_6}+\frac{A_2 A_5}{A_4 A_6}+\frac{A_4}{A_7}+\frac{A_3 A_5}{A_4 A_7}+\frac{A_3 A_6}{A_2
   A_7}+\frac{A_3 A_6}{A_5 A_7}\cr
&\quad\qquad+\frac{A_3^2}{A_2 A_4 A_7}+\frac{A_3 A_4}{A_2 A_5 A_7}+\frac{A_3}{A_6
   A_7}+\frac{A_3 A_5}{A_2 A_6 A_7}+\frac{A_2}{A_1}+\frac{A_5}{A_1}+\frac{A_3 A_6}{A_4
   A_1}+\frac{A_4 A_6}{A_5 A_1}\cr
&\quad\qquad+\frac{A_2 A_6 A_7}{A_3 A_1}+\frac{A_2 A_7}{A_4 A_1}+\frac{A_2 A_4
   A_7}{A_3 A_5 A_1}+\frac{A_2 A_5 A_7}{A_3 A_6 A_1}+\frac{A_2^2}{A_3 A_1}+\frac{A_3}{A_5
   A_1}+\frac{A_4}{A_6 A_1}\cr
&\quad\qquad\left.\left.+\frac{A_3 A_5}{A_4 A_6 A_1}+\frac{A_2 A_6}{A_7 A_1}+\frac{A_2 A_3}{A_4
   A_7 A_1}+\frac{A_2 A_4}{A_5 A_7 A_1}+\frac{A_2 A_5}{A_6 A_7 A_1}\right)\right],
\end{align}
\begin{align}\label{e7flavor}
Z^{\mathfrak{e}_7}_{\text{flavor}} &= \text{PE}\left[-\frac{q}{(1-q)^2}\left(\frac{A_4 M_1 A_2}{A_1 A_3}+\frac{A_5 M_1 A_2}{A_1 A_4}+\frac{A_6 M_1 A_2}{A_3}+\frac{A_6 M_1
   A_2}{A_1 A_5}+\frac{A_7 M_1 A_2}{A_3}\right.\right.\cr
&\quad\qquad+\frac{A_4 M_1 A_2}{A_3 A_5}+\frac{A_5 M_1 A_2}{A_3
   A_6}+\frac{M_1 A_2}{A_1 A_6}+\frac{M_1 A_2}{A_7}+\frac{A_4 A_2}{A_1 A_3 M_1}+\frac{A_5 A_2}{A_1
   A_4 M_1}+\frac{A_7 A_2}{A_3 M_1}\cr
&\quad\qquad+\frac{A_2}{A_7 M_1}+\frac{A_1 A_4 M_1}{A_3}+\frac{A_1 A_5
   M_1}{A_4}+\frac{A_6 M_1}{A_1}+\frac{A_1 A_6 M_1}{A_5}+A_6 M_1+\frac{A_5 A_7 M_1}{A_3}\cr
&\quad\qquad+\frac{A_6
   A_7 M_1}{A_4}+\frac{A_4 A_6 A_7 M_1}{A_3 A_5}+\frac{A_7 M_1}{A_1}+\frac{A_7 M_1}{A_5}+\frac{A_4
   A_7 M_1}{A_3 A_6}+\frac{A_5 A_7 M_1}{A_4 A_6}+\frac{A_3 M_1}{A_4}\cr
&\quad\qquad+\frac{A_4 M_1}{A_5}+\frac{A_1
   M_1}{A_6}+\frac{A_5 M_1}{A_6}+\frac{A_3 M_1}{A_1 A_7}+\frac{A_5 M_1}{A_7}+\frac{A_3 A_6 M_1}{A_4
   A_7}+\frac{A_4 A_6 M_1}{A_5 A_7}+\frac{A_3 M_1}{A_5 A_7}\cr
&\quad\qquad+\frac{A_4 M_1}{A_6 A_7}+\frac{A_3 A_5
   M_1}{A_4 A_6 A_7}+\frac{A_1 A_4}{A_3 M_1}+\frac{A_1 A_5}{A_4 M_1}+\frac{A_1 A_6}{A_5
   M_1}+\frac{A_6}{M_1}+\frac{A_7}{A_1 M_1}+\frac{A_3}{A_4 M_1}\cr
&\quad\qquad+\frac{A_4}{A_5 M_1}+\frac{A_5}{A_6
   M_1}+\frac{A_3}{A_1 A_7 M_1}+\frac{A_4 M_1}{A_2}+\frac{A_3 A_5 M_1}{A_4 A_2}+\frac{A_1 A_6
   M_1}{A_2}+\frac{A_3 A_6 M_1}{A_5 A_2}\cr
&\quad\qquad\left.\left.+\frac{A_1 A_7 M_1}{A_2}+\frac{A_3 M_1}{A_6 A_2}+\frac{A_1
   A_5 M_1}{A_6 A_2}+\frac{A_1 A_3 M_1}{A_7 A_2}+\frac{A_4}{M_1 A_2}+\frac{A_1 A_7}{M_1
   A_2}+\frac{A_1 A_3}{A_7 M_1 A_2}\right)\right]
\end{align}
\begin{align}\label{e7halfflavor}
Z^{\mathfrak{e}_7}_{\text{$\frac{1}{2}$ flavor}} &=\text{PE}\left[-\frac{q}{(1-q)^2}\left(\frac{A_4 A_1}{A_3}+\frac{A_5 A_1}{A_4}+\frac{A_6 A_1}{A_5}+\frac{A_7
   A_1}{A_2}+\frac{A_1}{A_6}+\frac{A_3 A_1}{A_2 A_7}+\frac{A_4}{A_2}+\frac{A_3 A_5}{A_2
   A_4}\right.\right.\cr
&\quad\qquad+\frac{A_3 A_6}{A_2 A_5}+A_6+\frac{A_5 A_7}{A_3}+\frac{A_6 A_7}{A_4}+\frac{A_4 A_6 A_7}{A_3
   A_5}+\frac{A_2 A_7}{A_3}+\frac{A_4 A_7}{A_3 A_6}+\frac{A_5 A_7}{A_4
   A_6}\cr
&\quad\qquad+\frac{A_3}{A_4}+\frac{A_4}{A_5}+\frac{A_3}{A_2
   A_6}+\frac{A_5}{A_6}+\frac{A_2}{A_7}+\frac{A_5}{A_7}+\frac{A_2 A_4}{A_3 A_1}+\frac{A_2 A_5}{A_4
   A_1}+\frac{A_2 A_6}{A_5 A_1}+\frac{A_7}{A_1}\cr
&\quad\qquad\left.\left.+\frac{A_2}{A_6 A_1}+\frac{A_3}{A_7 A_1}\right)\right],
\end{align}

To compare \eqref{5de7halfpert} with \eqref{part.5de7w32f}, we apply the $M_0 \to 0$ or $Q_B \to 0$ limit to \eqref{part.5de7w32f} in order to extract the perturbative part of the partition function. Then the diagram in Figure \ref{fig:e7w32f} splits into the upper half part and the lower half part. The lower half part is exactly the half of the diagram of the 5d pure $E_7$ gauge theory obtained in \cite{Hayashi:2017jze}, and hence it yields the square root of the root contribution. Then the upper half part should give the remaining contribution in \eqref{5de7halfpert} except for the Cartan contribution which is not captured from the topological vertex. Hence we should obtain
\begin{align}
Z_{\text{upper half}}^{\fe_7 + \frac{3}{2}F} &= \sqrt{Z^{\mathfrak{e}_7}_{\text{roots}}}Z^{\mathfrak{e}_7}_{\text{flavor}}Z^{\mathfrak{e}_7}_{\text{$\frac{1}{2}$ flavor}}Z_{\text{extra}}^{\fe_7 + \frac{3}{2}F},\label{e7halfupper}\\
Z_{\text{lower half}}^{\fe_7 + \frac{3}{2}F} &=\sqrt{Z^{\mathfrak{e}_7}_{\text{roots}}},\label{e7halflower}
\end{align}
where $Z_{\text{extra}}^{\fe_7 + \frac{3}{2}F}$ is a contribution from a possible extra factor. We evaluated the upper half part and the lower half part from \eqref{part.5de7w32f} in the limit $Q_B \to 0$ until the order $Q_1^4P_1^3P_2^3P_3^3P_4^2R_1^3R_2^3R_3^3$ and found precise agreement with \eqref{e7halfupper} and \eqref{e7halflower}. The upper half part has an extra factor and it is given by 
\be\label{e7extra}
Z_{\text{extra}}^{\fe_7} = \text{PE}\left[\frac{q}{(1-q)^2}\left(6M_1 + 4M_1^2\right)\right].
\ee
until the order we computed.

\paragraph{$\fe_7^{(1)}$ on $(-2)$-curve.}

The other $E_7$ theory we consider is the 6d $E_7$ gauge theory with $3$ flavors on $S^1$ and we 
compute its partition function 
on $T^2 \times \mathbb{R}^4$. The theory is realized from the geometry given by $\fe_7^{(1)}$ on $(-2)$-curve. 
The geometry of $\fe_7^{(1)}$ on $(-2)$-curve is obtained from the $\left[(4,3,2), (4,3,2)\right]$ Higgsing of 
the theory $(E_7, E_7)_2$ on $S^1$ as in Table \ref{tb:e7_2}. 
A 5d gauge theory description of the original theory before the Higgsing is given by the affine $E_7$ Dynkin quiver theory 
\eqref{e7quiver} and its realization using a web diagram is given in Figure \ref{fig:e7e7e7}. The application of the $\left[(4,3,2), (4,3,2)\right]$ Higgsing to the diagram in Figure \ref{fig:e7e7e7} should yield the web diagram corresponding to the geometry $\fe_7^{(1)}$ on $(-2)$-curve. The resulting web diagram after the Higgsing is depicted in Figure \ref{fig:e7on2}.
Furthermore decoupling 
the fiber class corresponding to the affine node for the affine $E_7$ Dynkin diagram is the 5d limit and the diagram after the limit is depicted in Figure \ref{fig:e7w3f}. Since the limit is the 5d limit of the 6d $E_7$ gauge theory with three flavors and a tensor mutliplet on $S^1$, the web diagram in Figure \ref{fig:e7w3f} yields the 5d $E_7$ gauge theory with three flavors. 


One subtle point about the theory realized on the web in Figure \ref{fig:e7on2} is that the diagram will not allow us to turn on a mass parameter for one of the $3$ flavors. 
Note that the diagram in Figure \ref{fig:e7on2} only has three mass parameters one of which is related to the radius of the $S^1$ compactification. One can also see this from the comparison with the diagram for $\fe_7^{(1)}$ on $(-5)$-curve given in Figure \ref{fig:e7on5}. The upper half part of the diagram in Figure \ref{fig:e7on5} yields the perturbative part of the Nekrasov partition function for $\frac{3}{2}$ flavors as in \eqref{e7halfupper}. The diagram in Figure \ref{fig:e7on2} is given by combinig two copies of the upper half part of the diagram in Figure \ref{fig:e7on5} and hence the resulting theory is expected to have two massive hypermultiplets in the fundamental representation and two half-hypermultiplets in the fundamental representation. 

\begin{figure}[t]
\centering
\subfigure[]{\label{fig:e7para1}
\includegraphics[width=4cm]{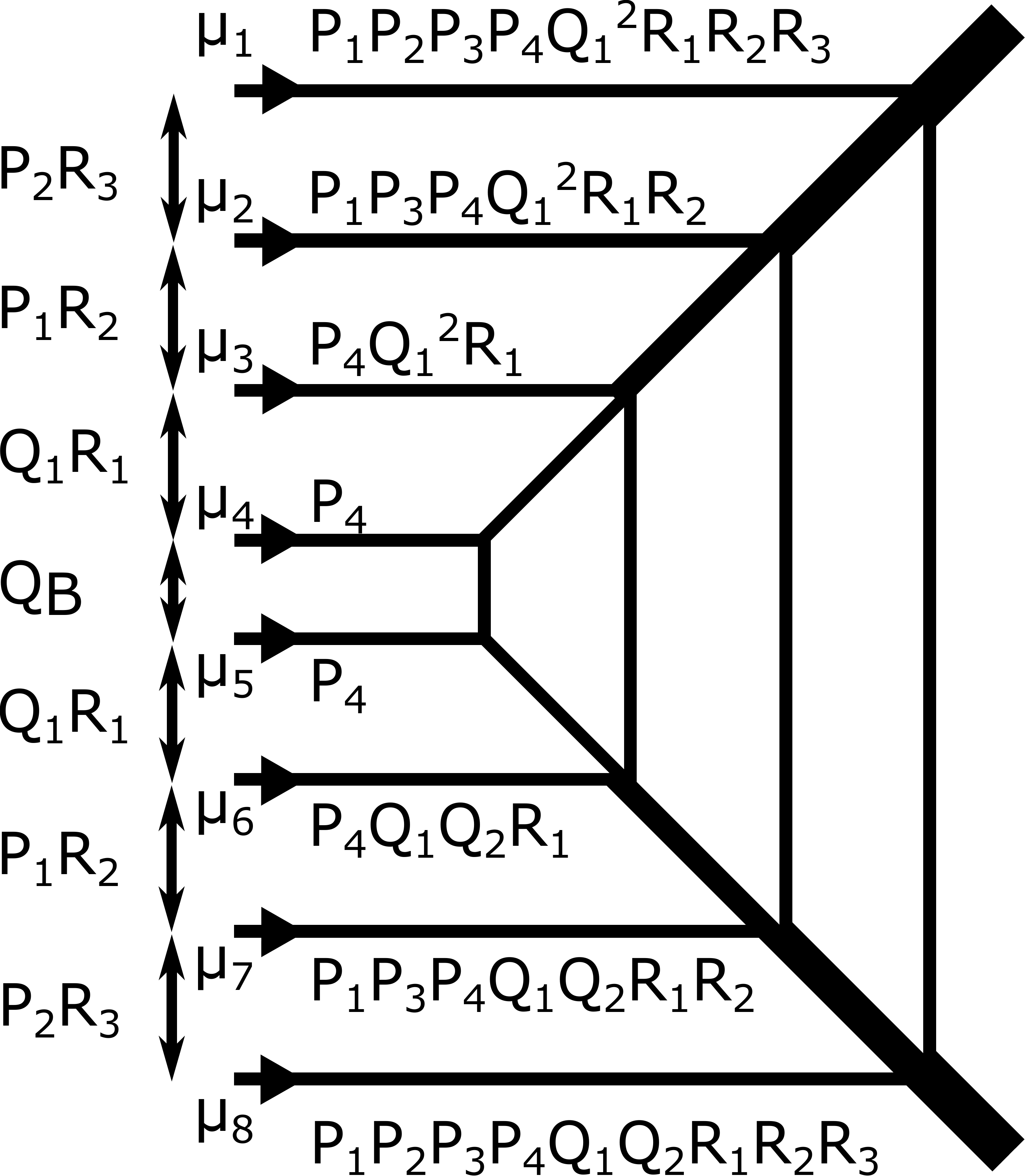}}
\hspace{1cm}
\subfigure[]{\label{fig:e7para2}
\includegraphics[width=4cm]{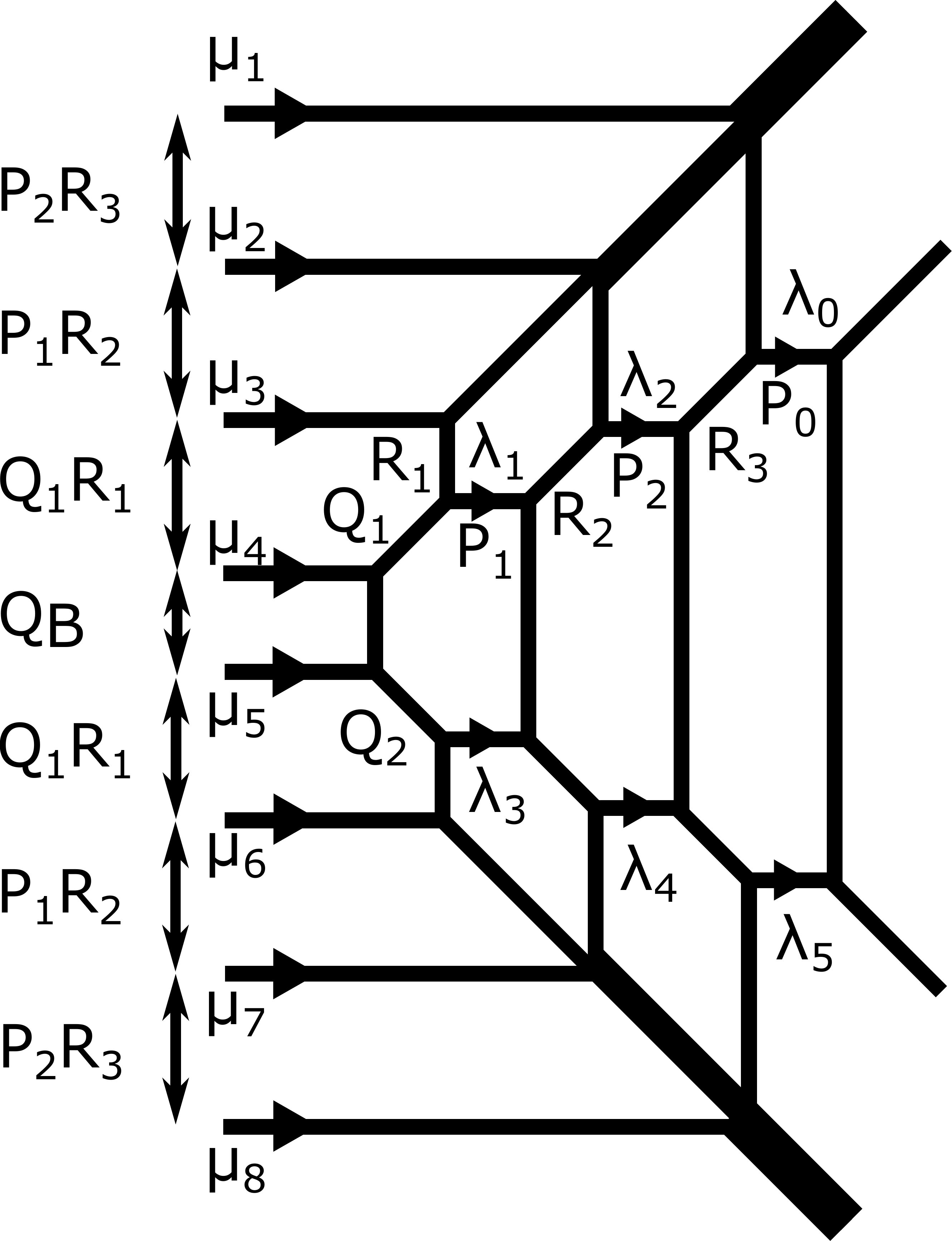}}
\hspace{1cm}
\subfigure[]{\label{fig:e7para3}
\includegraphics[width=4cm]{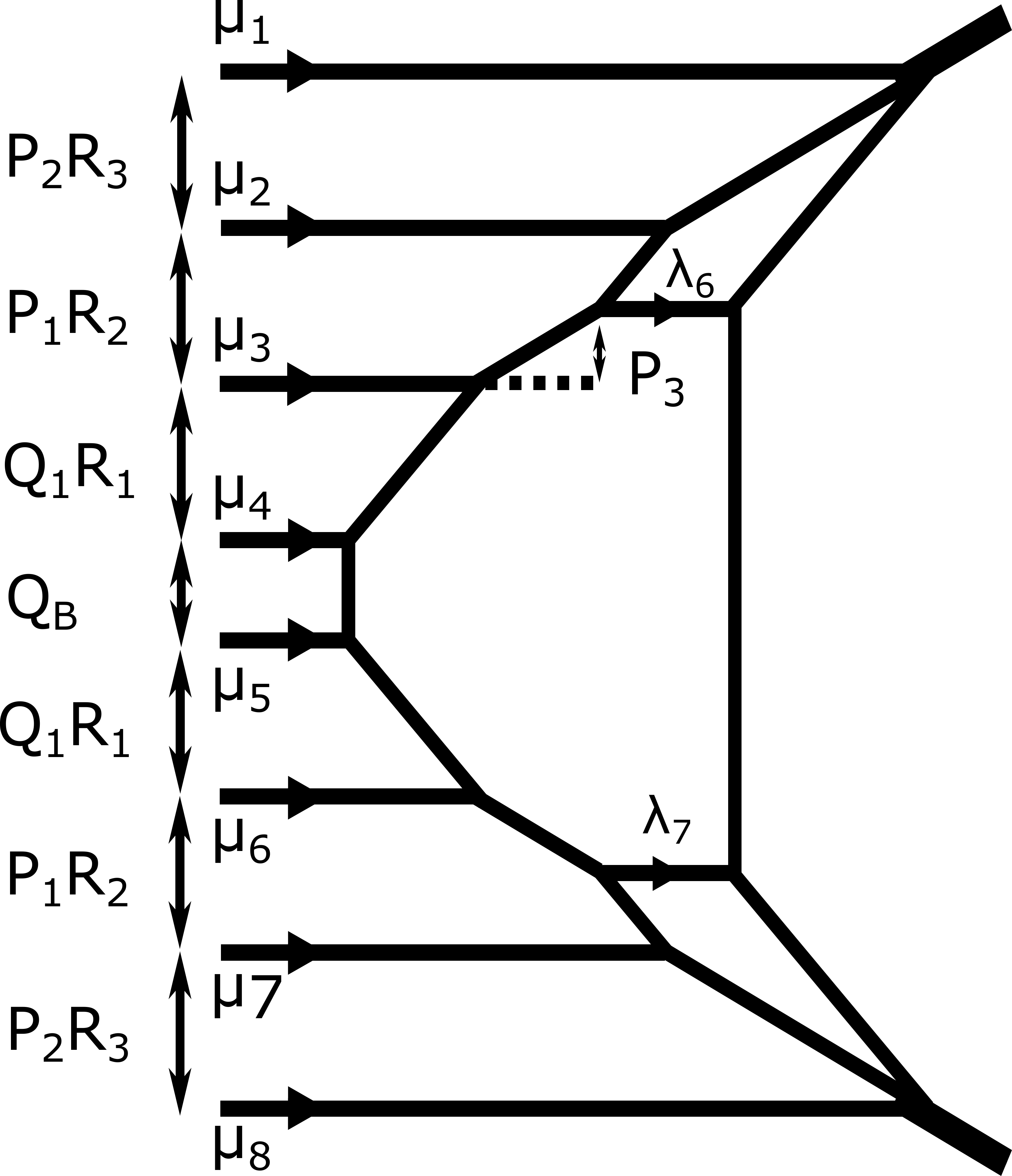}}
\caption{Parameterization of the web diagram for the geometry $\fe_7^{(1)}$ on $(-2)$-curve in Figure \ref{fig:e7on2}.}
\label{fig:e7para}
\end{figure}

\begin{figure}[t]
\centering
\subfigure[]{\label{fig:halfsu81}
\includegraphics[width=4cm]{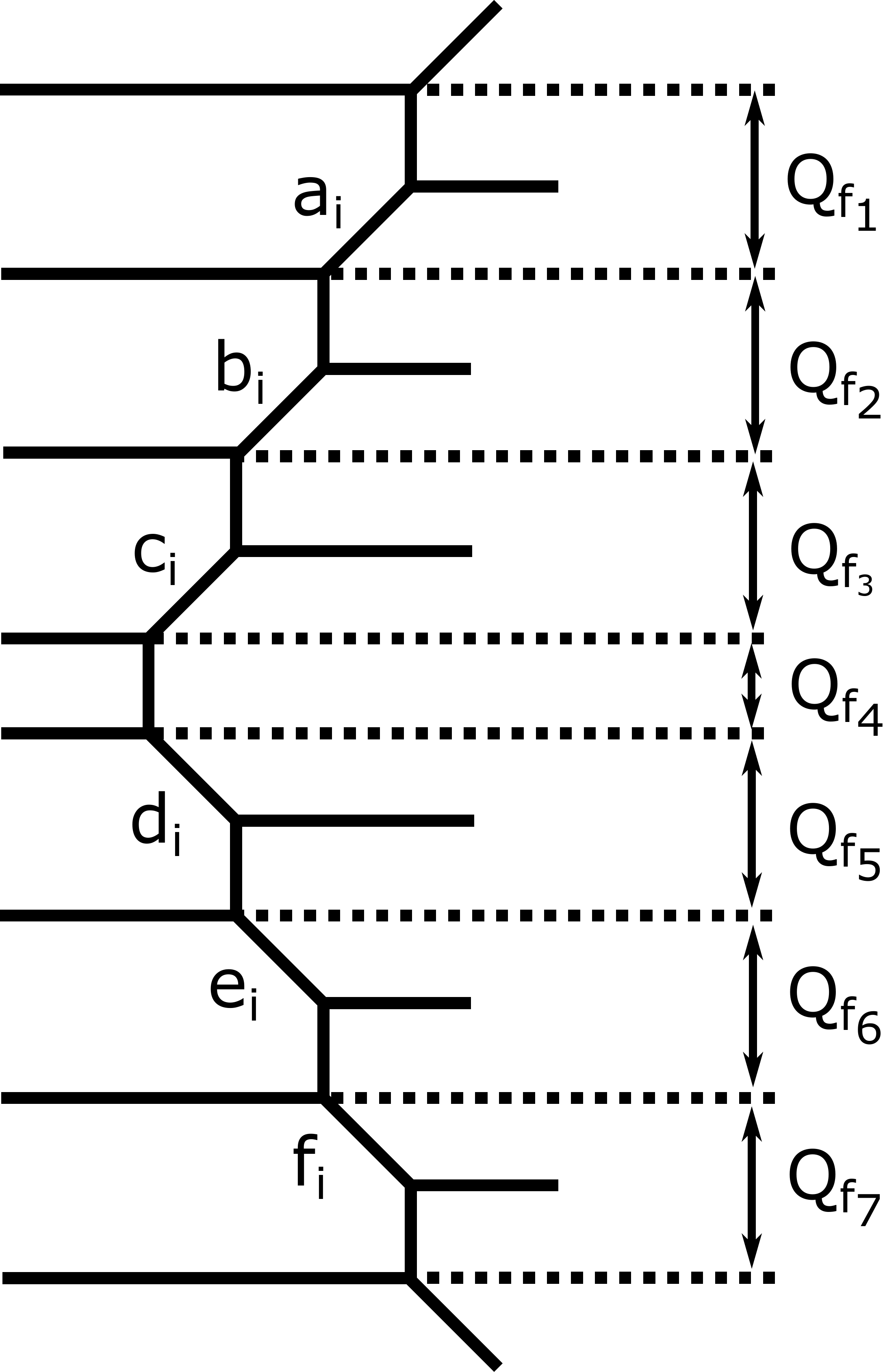}}
\hspace{1cm}
\subfigure[]{\label{fig:halfsu82}
\includegraphics[width=5cm]{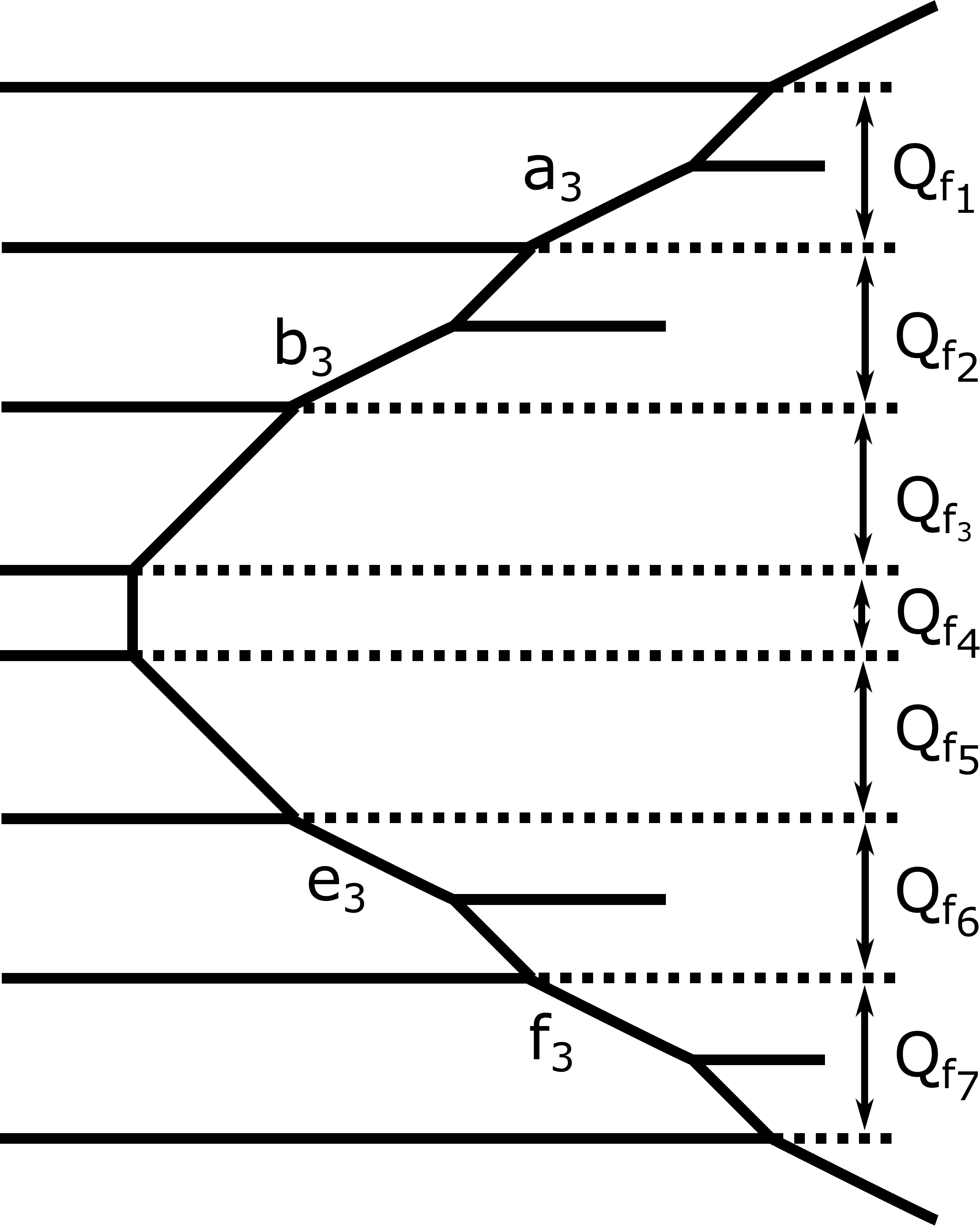}}
\caption{(a). Local geometry near the $\SU(8)$ part of the two long quiver tails of the affine $E_7$ Dynkin quiver theoriy \eqref{e7quiver}. $i$ labels the two long quiver tails and $i=1, 2$. (b). Local geometry near the $\SU(8)$ part of the short quiver tail of the affine $E_7$ Dynkin quiver theory \eqref{e7quiver}.}
\label{fig:halfsu8}
\end{figure}

\begin{figure}[t]
\centering
\includegraphics[width=7cm]{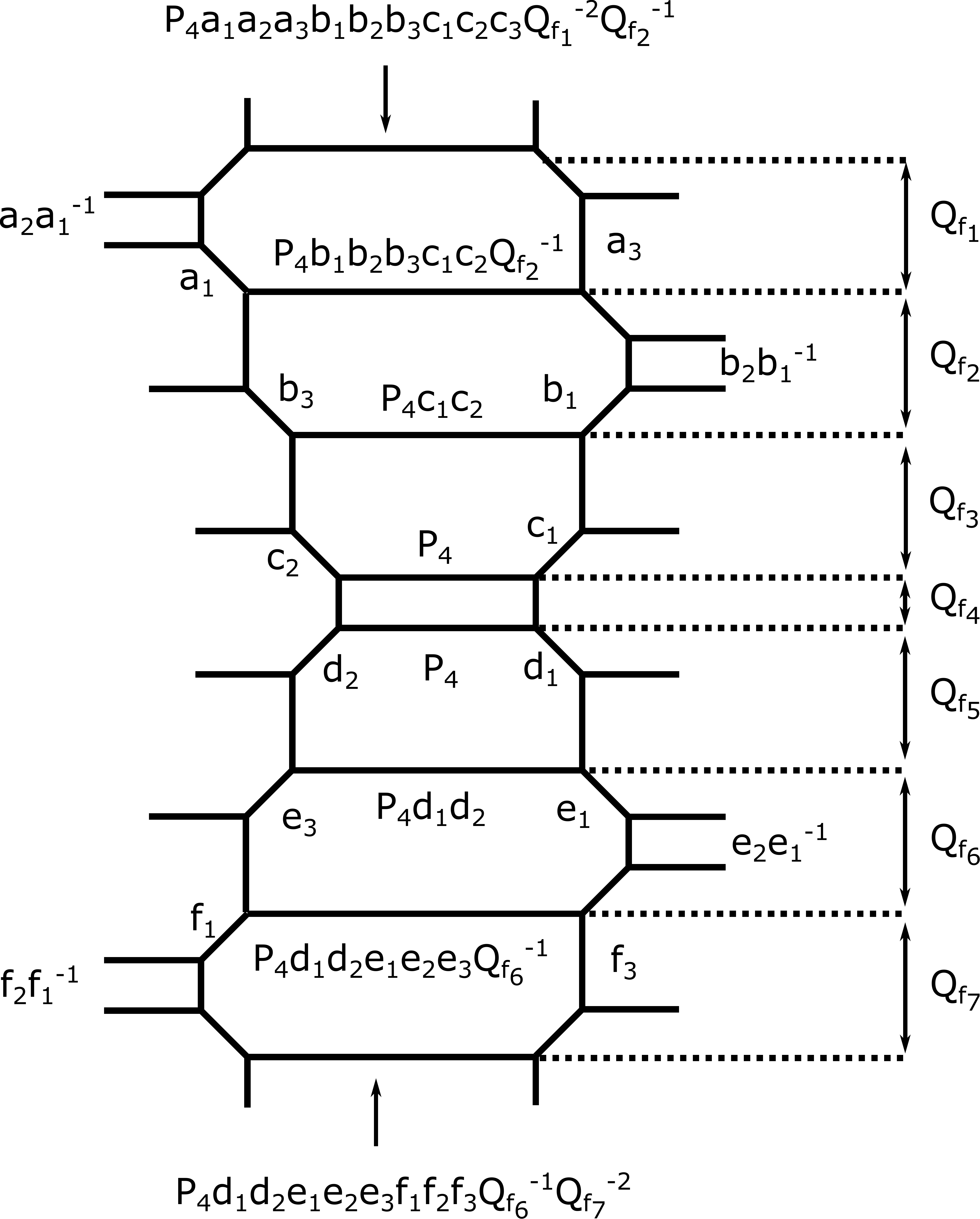}
\caption{Local geometry for the middle $\SU(8)$ gauge node in the affine $\fe_7$ Dynkin quiver.}
\label{fig:su8}
\end{figure}

We then consider applying the topological vertex as well as the trivalent gluing prescription to the diagram in Figure \ref{fig:e7on2} to compute the partition function. 
For that we choose the parameterization as in Figure \ref{fig:e7para}. 
The parameterization for each node of the affine $E_7$ Dynkin diagram which follows from Figure \ref{fig:e7para} is again the same one depicted in Figure \ref{fig:e7kahler}. The parametrization for the gluing lines can be obtained by looking at the middle $\SU(8)$ gauging part in the affine $E_7$ Dynkin quiver theory \eqref{e7quiver}. Before the Higgsing the local structure around the trivalent gluing is given by the $\SU(8)$ gauge theory with $6 + 6 + 4 = 16$ flavors with the Chern-Simons level zero. It consists of two copies of the diagram in Figure \ref{fig:halfsu81} and the diagram in Figure \ref{fig:halfsu82}.  
The $Q_{f_i}, (i=1, \cdots, 7)$ in Figure \ref{fig:halfsu8} are the Coulomb branch moduli of the middle $\SU(8)$. Then the local geometry around the $\SU(8)$ gauge theory with $16$ flavors with the parameterziation in Figure \ref{fig:halfsu8} is described by the diagram in Figure \ref{fig:su8}. 
Then the $\left[(4,3,2), (4,3,2)\right]$ Higgsing is achieved by tuning the parameters as 
\begin{equation}\label{432}
\begin{split}
&a_1 = a_2 = a_3 = Q_{f_1}, \quad b_1 = b_2 = Q_{f_2}, \quad c_1 = Q_{f_3}, \quad d_1 = Q_{f_5}, \quad e_1 = e_2 = Q_{f_6},\cr
&f_1 = f_2 = f_3 = Q_{f_7}. 
\end{split}
\end{equation}
and then the Coulomb branch of the $\SU(8)$ becomes the one in Figure \ref{fig:e7para1}, namely
\begin{equation}\label{su8CB}
Q_{f_1} = Q_{f_7} = P_2R_3, \quad Q_{f_2} = Q_{f_6} = P_1R_2, \quad Q_{f_3} = Q_{f_5} = Q_1R_1, \quad Q_{f_4} = Q_B.  
\end{equation}
Inserting the tuning \eqref{432} as well as \eqref{su8CB} into the K\"ahler parameters for the gluing lines in Figure \ref{fig:su8} gives rise to the parameterizaiton in Figure \ref{fig:e7para1}.

We can now apply the topological vertex formalism as well as the trivalent gluing prescription to the diagram in Figure \ref{fig:e7on2}. Since the tuning does not change the charges of the 5-branes attached to the gluing lines, the framing factors for the gluing lines do not change after the Higgsing. Therefore we use the same framing factors as those for the gluing lines in Figure \ref{fig:su8}. Then the partition function of the theory computed from the topological vertex is given by
\begin{equation}\label{part.6de7}
\begin{split}
&Z^{\text{6d}}_{\fe_7, (-2)} \cr
&=\sum_{\{\mu_i\}}\left(-Q_1^2P_1P_2P_3P_4R_1R_2R_3\right)^{|\mu_1|}\left(-Q_1^2P_1P_3P_4R_1R_2\right)^{|\mu_2|}\left(-Q_1^2P_4R_1\right)^{|\mu_3|}\left(-P_4\right)^{|\mu_4| + |\mu_5|}\cr
&\hspace{1.2cm}\left(-Q_1Q_2P_4R_1\right)^{|\mu_6|}\left(-Q_1Q_2P_1P_3P_4R_1R_2\right)^{|\mu_7|}\left(-Q_1Q_2P_1P_2P_3P_4R_1R_2R_3\right)^{|\mu_8|}\cr
&\hspace{0.5cm}Z_{\{\mu_i\}}^{\text{SU(8)$_L$}}\left(Q_B, \{Q_a\}, \{P_b\}, \{R_c\}\right) Z_{\{\mu_i\}}^{\text{SU(8)$_R$}}\left(Q_B, \{Q_a\}, \{P_b\}, \{R_c\}\right)\cr
&\hspace{0.5cm}f_{\mu_1}(q)^{-1}f_{\mu_3}(q)f_{\mu_4}(q)f_{\mu_5}(q)^{-1}f_{\mu_6}(q)^{-1}f_{\mu_8}(q)\cr
&\hspace{0.5cm}Z^{\fe_7}_{1, \{\mu_i\}}\left(Q_B, \{Q_a\}, \{P_b\}, \{R_c\}\right)Z^{\fe_7}_{2, \{\mu_i\}}\left(Q_B, \{Q_a\}, \{P_b\}, \{R_c\}\right)Z^{\fe_7}_{3, \{\mu_i\}}\left(Q_B, \{Q_a\}, \{P_b\}, \{R_c\}\right),
\end{split}
\end{equation}
where we defined
\begin{align}
Z_{\{\mu_i\}}^{\text{SU(8)$_L$}}\left(Q_B, \{Q_a\}, \{P_b\}\right) &=q^{\frac{1}{2}\sum_{i=1}^8||\mu_i||^2}\left(\prod_{i=1}^8\tilde{Z}_{\mu_i}(q)\right)\cr
&\quad\prod_{1\leq i < j \leq 8}\mathcal{I}^-_{\mu_i, \mu_j}\left(Q_{f_i}Q_{f_{i+1}}\cdots Q_{f_{j-1}}\right),\label{SU8L}\\
Z_{\{\mu_i\}}^{\text{SU(8)$_R$}}\left(Q_B, \{Q_a\}, \{P_b\}\right) &=q^{\frac{1}{2}\sum_{i=1}^8||\mu_i^t||^2}\left(\prod_{i=1}^8\tilde{Z}_{\mu_i^t}(q)\right)\cr
&\quad\prod_{1\leq i < j \leq 8}\mathcal{I}^-_{\mu_i, \mu_j}\left(Q_{f_i}Q_{f_{i+1}}\cdots Q_{f_{j-1}}\right), \label{SU8R}
\end{align}
with $Q_{f_i}, \;(i=1, \cdots, 7)$ given in \eqref{su8CB}. Furthermore the factors $Z^{\fe_7}_{j, \{\mu_i\}}\left(Q_B, \{Q_a\}, \{P_b\}, \{R_c\}\right),$  $(j=1, 2, 3)$ are the partition functions from the diagrams in Figure \ref{fig:e7para1}, \ref{fig:e7para2} and \ref{fig:e7para3} with the trivalent gluing prescription implemented, and they become
\begin{align}
&Z^{\fe_7}_{1,\{\mu_i\}}\left(Q_B, \{Q_a\}, \{P_b\}, \{R_c\}\right)\cr
&= q^{\frac{1}{2}\sum_{i=1}^8||\mu_i^t||^2}\left(\prod_{i=1}^8\tilde{Z}_{\mu_i^t}(q)\right)/Z_{\{\mu_i\}}^{\text{SU(8)$_R$}}\left(Q_B, \{Q_a\}, \{P_b\}, \{R_c\}\right)\cr
&\quad\mathcal{I}^-_{\mu_1, \mu_8}\left(Q_BQ_1^2P_1^2P_2^2R_1^2R_2^2R_3^2\right)\mathcal{I}^-_{\mu_2,\mu_7}\left(Q_BQ_1^2P_1^2R_1^2R_2^2\right)\mathcal{I}^-_{\mu_3,\mu_6}\left(Q_BQ_1^2R_1^2\right)\mathcal{I}^-_{\mu_4,\mu_5}\left(Q_B\right),\nn\\\\
&Z^{\fe_7}_{2,\{\mu_i\}}\left(Q_B, \{Q_a\}, \{P_b\}, \{R_c\}\right)\cr
&= q^{\frac{1}{2}\sum_{i=1}^8||\mu_i^t||^2}\left(\prod_{i=1}^8\tilde{Z}_{\mu_i^t}(q)\right)/Z_{\{\mu_i\}}^{\text{SU(8)$_R$}}\left(Q_B, \{Q_a\}, \{P_b\}, \{R_c\}\right)\cr
&\quad\mathcal{I}^-_{\mu_1, \mu_8}\left(Q_BQ_1^2P_1^2P_2^2R_1^2R_2^2R_3^2\right)\mathcal{I}^-_{\mu_2,\mu_7}\left(Q_BQ_1^2P_1^2R_1^2R_2^2\right)\mathcal{I}^-_{\mu_3,\mu_4}\left(Q_1R_1\right)\mathcal{I}^-_{\mu_3,\mu_5}\left(Q_BQ_1R_1\right)\cr
&\quad\mathcal{I}^-_{\mu_3,\mu_6}\left(Q_BQ_1^2R_1^2\right)\mathcal{I}^-_{\mu_4,\mu_5}\left(Q_B\right)\mathcal{I}^-_{\mu_4,\mu_6}\left(Q_BQ_1R_1\right)\mathcal{I}^-_{\mu_5,\mu_6}\left(Q_1R_1\right)\cr
&\quad\sum_{\lambda_{0, 1, 2, 3, 4, 5}} q^{\frac{1}{2}\sum_{i=0}^5\left(||\lambda_i||^2 + ||\lambda_i^t||^2\right)}\left(\prod_{i=0}^5\tilde{Z}_{\lambda_i}(q)\tilde{Z}_{\lambda_i^t}(q)\right)\cr
&\quad\left(\prod_{i=0}^2\left(-P_i\right)^{|\lambda_i|}\right)\left(-Q_1Q_2^{-1}P_1\right)^{|\lambda_3|}\left(-P_2R_2R'^{-1}_2\right)^{|\lambda_4|}\left(-P_0R_3R'^{-1}_3\right)^{|\lambda_5|}\cr
&\quad\mathcal{I}^+_{\mu_1,\lambda_0}\left(P_1P_2R_1\right)\mathcal{I}^-_{\mu_1,\lambda_2}\left(P_1P_2R_1R_3\right)\mathcal{I}^-_{\mu_1,\lambda_4}\left(Q_BQ_1Q_2P_1P_2R_1R_2R'_2R_3\right)\cr
&\quad\mathcal{I}^+_{\mu_1,\lambda_5}\left(Q_BQ_1Q_2P_1P_2R_1R_2R'_2R_3R'_3\right)\mathcal{I}^+_{\lambda_0, \lambda_2}\left(R_3\right)\mathcal{I}^+_{\lambda_0, \lambda_4}\left(Q_BQ_1Q_2R_2R'_2R_3\right)\cr
&\quad\mathcal{I}^-_{\lambda_0, \lambda_5}\left(Q_BQ_1Q_2R_2R'_2R_3R'_3\right)^2\mathcal{I}^+_{\lambda_0, \mu_8}\left(Q_BQ_1^2P_1P_2R_1R_2^2R_3^2\right)\mathcal{I}^-_{\lambda_2, \lambda_4}\left(Q_BQ_1Q_2R_2R'_2\right)^2\cr
&\quad\mathcal{I}^+_{\lambda_2, \lambda_5}\left(Q_BQ_1Q_2R_2R'_2R'_3\right)\mathcal{I}^-_{\lambda_2, \mu_8}\left(Q_BQ_1^2P_1P_2R_1R_2^2R_3\right)\mathcal{I}^+_{\lambda_4, \lambda_5}\left(R'_3\right)\mathcal{I}^-_{\lambda_4, \mu_8}\left(Q_1^3Q_2^{-3}P_1P_2R_1R'_3\right)\cr
&\quad\mathcal{I}^+_{\lambda_5, \mu_8}\left(Q_1^3Q_2^{-3}P_1P_2R_1\right)\mathcal{I}^+_{\mu_2, \lambda_2}\left(P_1R_1\right)\mathcal{I}^-_{\mu_2, \lambda_1}\left(P_1R_1R_2\right)\mathcal{I}^-_{\mu_2, \lambda_3}\left(Q_BQ_1Q_2P_1R_1R_2\right)\cr
&\quad\mathcal{I}^+_{\mu_2, \lambda_4}\left(Q_BQ_1Q_2P_1R_1R_2R'_2\right)\mathcal{I}^+_{\lambda_2, \lambda_1}\left(R_2\right)\mathcal{I}^+_{\lambda_2, \lambda_3}\left(Q_BQ_1Q_2R_2\right)\mathcal{I}^-_{\lambda_2, \mu_7}\left(Q_BQ_1^2P_1R_1R_2^2\right)\cr
&\quad\mathcal{I}^-_{\lambda_1, \lambda_3}\left(Q_BQ_1Q_2\right)^2\mathcal{I}^+_{\lambda_1, \lambda_4}\left(Q_BQ_1Q_2R'_2\right)\mathcal{I}^-_{\lambda_1, \mu_7}\left(Q_BQ_1^2P_1R_1R_2\right)\mathcal{I}^+_{\lambda_3, \lambda_4}\left(R'_2\right)\cr
&\quad\mathcal{I}^-_{\lambda_3, \mu_7}\left(Q_1^2Q_2^{-2}P_1R_1R'_2\right)\mathcal{I}^-_{\lambda_4, \mu_7}\left(Q_1^2Q_2^{-2}P_1R_1\right)\mathcal{I}^+_{\mu_3, \lambda_1}\left(R_1\right)\mathcal{I}^+_{\mu_3, \lambda_3}\left(Q_BQ_1Q_2R_1\right)\mathcal{I}^+_{\lambda_1, \mu_4}\left(Q_1\right)\cr
&\quad\mathcal{I}^+_{\lambda_1, \mu_5}\left(Q_BQ_1\right)\mathcal{I}^+_{\lambda_1, \mu_6}\left(Q_BQ_1^2R_1\right)\mathcal{I}^+_{\mu_4, \lambda_3}\left(Q_BQ_2\right)\mathcal{I}^+_{\mu_5, \lambda_3}\left(Q_2\right)\mathcal{I}^+_{\lambda_3, \mu_6}\left(Q_1Q_2^{-1}R_1\right)\\
&Z^{\fe_7}_{3,\{\mu_i\}}\left(Q_B, \{Q_a\}, \{P_b\}, \{R_c\}\right)\cr
&=q^{\frac{1}{2}\sum_{i=1}^8||\mu_i^t||^2}\left(\prod_{i=1}^8\tilde{Z}_{\mu_i^t}(q)\right)/Z_{\{\mu_i\}}^{\text{SU(8)$_R$}}\left(Q_B, \{Q_a\}, \{P_b\}, \{R_c\}\right)\cr
&\quad\mathcal{I}^-_{\mu_1, \mu_8}\left(Q_BQ_1^2P_1^2P_2^2R_1^2R_2^2R_3^2\right)\mathcal{I}^-_{\mu_2,\mu_3}\left(P_1R_2\right)\mathcal{I}^-_{\mu_2,\mu_4}\left(Q_1P_1R_1R_2\right)\mathcal{I}^-_{\mu_2,\mu_5}\left(Q_BQ_1P_1R_1R_2\right)\cr
&\quad\mathcal{I}^-_{\mu_2,\mu_6}\left(Q_BQ_1^2P_1R_1^2R_2\right)\mathcal{I}^-_{\mu_2,\mu_7}\left(Q_BQ_1^2P_1^2R_1^2R_2^2\right)\mathcal{I}^-_{\mu_3,\mu_4}\left(Q_1R_1\right)\mathcal{I}^-_{\mu_3,\mu_5}\left(Q_BQ_1R_1\right)\cr
&\quad\mathcal{I}^-_{\mu_3,\mu_6}\left(Q_BQ_1^2R_1^2\right)\mathcal{I}^-_{\mu_3,\mu_7}\left(Q_BQ_1^2R_1^2P_1R_2\right)\mathcal{I}^-_{\mu_4,\mu_5}\left(Q_B\right)\mathcal{I}^-_{\mu_4,\mu_6}\left(Q_BQ_1R_1\right)\cr
&\quad\mathcal{I}^-_{\mu_4,\mu_7}\left(Q_BQ_1P_1R_1R_2\right)\mathcal{I}^-_{\mu_5,\mu_6}\left(Q_1R_1\right)\mathcal{I}^-_{\mu_5,\mu_7}\left(Q_1P_1R_1R_2\right)\mathcal{I}^-_{\mu_6,\mu_7}\left(P_1R_2\right)\cr
&\quad\sum_{\lambda_{6,7}} q^{||\lambda_6||^2 + ||\lambda_7^t||^2}\left(\prod_{i=6}^7\tilde{Z}_{\lambda_i}(q)\tilde{Z}_{\lambda_i^t}(q)\right)\left(\prod_{i=6}^7\left(P_2R_3\right)^{|\lambda_i|}\right)\cr
&\quad\mathcal{I}^-_{\mu_1, \lambda_6}\left(P_1P_2P_3^{-1}R_2R_3\right)\mathcal{I}^-_{\mu_1, \lambda_7}\left(Q_BQ_1^2P_1P_2P_3R_1^2R_2R_3\right)\mathcal{I}^-_{\lambda_6, \lambda_7}\left(Q_BQ_1^2P_3^2R_1^2\right)^2\cr
&\quad\mathcal{I}^-_{\lambda_6, \mu_8}\left(Q_BQ_1^2P_1P_2P_3R_1^2R_2R_3\right)\mathcal{I}^-_{\lambda_7, \mu_8}\left(P_1P_2P_3^{-1}R_2R_3\right)\mathcal{I}_{\mu_2, \lambda_6}^+\left(P_1P_3^{-1}R_2\right)\cr
&\quad\mathcal{I}_{\mu_2, \lambda_7}^+\left(Q_BQ_1^2P_1P_3R_1^2R_2\right)\mathcal{I}_{\lambda_6, \mu_3}^+\left(P_3\right)\mathcal{I}_{\lambda_6, \mu_4}^+\left(Q_1P_3R_1\right)\mathcal{I}_{\lambda_6, \mu_5}^+\left(Q_BQ_1P_3R_1\right)\cr
&\quad\mathcal{I}_{\lambda_6, \mu_6}^+\left(Q_BQ_1^2P_3R_1^2\right)\mathcal{I}_{\lambda_6, \mu_7}^+\left(Q_BQ_1^2P_2P_3R_1^2R_2\right)\mathcal{I}_{\mu_3, \lambda_7}^+\left(Q_BQ_1^2P_3R_1^2\right)\mathcal{I}_{\mu_4, \lambda_7}^+\left(Q_BQ_1P_3R_1\right)\cr
&\quad\mathcal{I}_{\mu_5, \lambda_7}^+\left(Q_1P_3R_1\right)\mathcal{I}_{\mu_6, \lambda_7}^+\left(P_3\right)\mathcal{I}_{\lambda_7, \mu_7}^+\left(P_1P_3^{-1}R_2\right)
\end{align}
where we defined
\begin{equation}
R'_2 = Q_1^{-1}Q_2R_2, \quad R'_3 = Q_1^{-1}Q_2R_3
\end{equation}
for simplicity of the expresions.

The partition function \eqref{part.6de7} may contain an extra factor and we can identify it by parameterizing the K\"ahler parameters by mass parameters. The Coulomb branch moduli dependence comes from \eqref{QCB}. On the other hand we can assign mass parameters for the lengths between parallel external lines. The diagram in Figure \ref{fig:e7para2} contains two bunches of parallel external lines and we parameterize the lengths by
\be
P_0P_1P_2R_1 = M'_1, \qquad Q_1^4Q_2^{-4}P_0P_1P_2R_1= M'_2.
\ee
Furthermore the diagram in Figure \ref{fig:su8} implies that the external lines attached to the top gluing line are parallel to each other and we assign 
\be
Q_1^2P_1P_2P_3P_4R_1R_2R_3 = M'_0. 
\ee
In this case the parameterization for the K\"ahler parameter for the elliptic class \eqref{elliptic} becomes 
\be\label{Qtaue7}
Q_{\tau} = (P_2R_3)(P_1R_2)^2(Q_1R_1)^3(P_4)^4(Q_1P_1)^3(P_2R_2)^2(P_0R_3)(Q_1P_2P_3^2R_1R_3)^2 = M'^4_0M'_1,
\ee
which only depends on the mass parameters as expected. 

Using the parameterization the partition function of \eqref{part.6de7} can be written as
\be\label{hatpart.6de7}
Z_{\fe_{7}^{(1)},(-2)}^{\text{6d}} = \hat{Z}_{\fe_{7}^{(1)},(-2)}^{\text{6d}}\left(\{A'_a\}, \{M'_i\}\right)Z_{\text{extra}}^{\fe_{7}^{(1)},(-2)}\left(M'_0, M'_1, M'_2\right),
\ee 
where $Z_{\text{extra}}^{\fe_{7}^{(1)},(-2)}\left(M'_0, M'_1, M'_2\right)$ is an extra factor in \eqref{part.6de7} which is independent from the Coulomb branch moduli. 
We argue that the partition function $\hat{Z}_{\fe_{7}^{(1)},(-2)}^{\text{6d}}\left(\{A'_a\}, \{M'_i\}\right)$ in \eqref{hatpart.6de7} yields the partition function of the 6d $E_7$ gauge theory with three flavors and a tensor multiplet on $T^2 \times \mathbb{R}^4$ up to an extra factor. 

\paragraph{5d $E_7$ gauge theory with $3$ flavors.} We can 
apply the 5d limit to the partition function \eqref{part.6de7} to obtain a 5d partition function. The diagram after the limit is depicted in Figure \ref{fig:e7w3f} and it realizes the 5d $E_7$ gauge theory with $3$ flavors but one flavor is massless. In terms of the K\"ahler parameters, this limit can be achieved by taking $P_0 \to 0$ with the other K\"ahler parameters fixed. Hence applying the limit $P_0 \to 0$ to the partition function \eqref{part.6de7} gives rise to the partition function of the 5d $E_7$ gauge theory with three flavors up to an extra factor. 

Let us rewrite the partition function 
by the gauge theory parameters of the $E_7$ gauge theory. 
Since the parameterization for the fiber classes forming the $E_7$ Dynkin diagram is the same, the parameterization of the Coulomb branch moduli $A_i = e^{-a_i}\; (i=1, \cdots, 7)$ are again given by \eqref{e7CB}.   
As with the case of the 5d $E_7$ gauge theory with $\frac{3}{2}$ flavors, a string with length characterized by the K\"ahler parameter $Q_1$ or $Q_2$ gives rise to a hypermultiplet corresponding to a weight of the fundamental representation of $E_7$. 
From the intesrection number between the curves and the complex surfaces in the dual geometry, the Dynkin labels of the weights for $Q_1$ and $Q_2$ are
\begin{equation}
Q_1\;:\;[0, 1, -1, 1, -1, 0, 0], \qquad Q_2\;:\;[0, 1, -1, 1, -1, 0, 0].
\end{equation}
Hence we parameterize them by
\begin{equation}\label{e7mass}
Q_1 = A_2A_3^{-1}A_4A_5^{-1}M_1, \qquad Q_2 = A_2A_3^{-1}A_4A_5^{-1}M_2.
\end{equation}
Finally the instanton fugacity can be determined by evaluating the effective prepotential in this case. The middle face in the diagram for the $\SU(8)$ gauging is $\mathbb{F}_0$ and the volume of the surface or the area of the face computed from the effective prepotential becomes
\be
\frac{\partial \mathcal{F}}{\partial a_3}  = (-a_2 + 2a_3 - a_4 - a_7)(2a_3 - 2a_4 + m_0 - 3m_1 - 3m_2). 
\ee
Hence the parameterization for the instanton fugacity is given by
\begin{equation}\label{e7inst}
Q_B = A_3^2A_4^{-2}M_0M_1^{-3}M_2^{-3}.
\end{equation}
Then it turns out that the K\"ahler paramter for the length between the horizontal parallel external lines in Figure \ref{fig:e7para2} is $M_0$. 

Applying the 5d limit $P_0 \to 0$ to \eqref{part.6de7} 
with the gauge theory parameterization gives
\begin{align}\label{part.5de7}
Z_{\fe_7+ 3\text{F}}^{\text{5d}} &= Z^{\text{6d}}_{\fe_7,(-2)}\Big|_{P_0 = 0}\cr
&=\hat{Z}^{\text{5d}}_{\fe_7 + 3\text{F}}\left(\{A_b\}, \{M_i\}\right)Z_{\text{extra}}^{\fe_7+ 3\text{F}}\left(M_0, M_1, M_2\right). 
\end{align}
We claim that $\hat{Z}^{\text{5d}}_{\fe_7 + 3\text{F}}\left(\{A_b\}, \{M_i\}\right)$  in \eqref{part.5de7} gives the partition function of the 5d $E_7$ gauge theory with $3$ flavors on $S^1 \times \mathbb{R}^4$ up to an extra factor. 

Using the parameterization \eqref{e7CB}, \eqref{e7mass} and \eqref{e7inst} for the partition function \eqref{part.5de7}, we compare the partition function \eqref{part.5de7} with the perturbative part of the Nekrasov partition function of the 5d $E_7$ gauge theory with $3$ flavors. For the phase which gives the diagram in Figure \ref{fig:e7w3f}, the perturbative partition function is given by 
\begin{align}\label{5de7pert}
Z^{\text{5d pert}}_{\mathfrak{e}_7+3\text{F}} = Z^{\mathfrak{e}_7}_{\text{cartan}}Z^{\mathfrak{e}_7}_{\text{roots}}Z^{\mathfrak{e}_7}_{\text{flavor 1}}Z^{\mathfrak{e}_7}_{\text{flavor 2}}Z^{\mathfrak{e}_7}_{\text{flavor 3}},
\end{align}
where $Z^{\mathfrak{e}_7}_{\text{cartan}}$ is \eqref{e7cartan}, $Z^{\mathfrak{e}_7}_{\text{roots}}$ is \eqref{e7root}, 
and $Z^{\mathfrak{e}_7}_{\text{flavor $i$}}$ is given by \eqref{e7flavor} with $M_1$ being $M_i\; (i=1, 2)$ . Also $Z^{\mathfrak{e}_7}_{\text{flavor 3}} = \left(Z^{\mathfrak{e}_7}_{\text{$\frac{1}{2} $flavor}}\right)^2$ where $Z^{\mathfrak{e}_7}_{\text{$\frac{1}{2} $flavor}}$ is \eqref{e7halfflavor} and it is the contribution from a massless hypermultiplet or two half-hypermultiplets. 

It is possible to compare \eqref{5de7pert} with \eqref{part.5de7} by taking the limit $M_0 \to 0$ or $Q_B \to 0$. Note that the diagram in Figure \ref{fig:e7w3f} is symmetric under the reflection with respect to the horizontal axis. After taking the limit $Q_B \to 0$, the diagram splits into the upper half part and the lower half part and they shapes are identical to each other. Hence it is enough to compute the upper half part which gives a half of the contribution of \eqref{5de7pert} except for the Cartan part. 
Firstly due to the symmetry under the reflection with respect to the horizontal axis, the upper half part contains the square root of the root contributions. Since $Q_1$ is contained in the upper half part, $Z^{\mathfrak{e}_7+3\text{F}}_{\text{flavor 1}}$ can be reprodued from the partition function of the upper half diagram. Also $P_3$ in the upper half diagram is related to a contribution from a half-hypermultiplet, the square root of $Z^{\mathfrak{e}_7+3\text{F}}_{\text{flavor 3}}$ can be also reproduced from the partition function of the upper half diagram. Namely the partition function from the upper half diagram should give
\begin{equation}\label{e7compare1}
Z_{\text{upper half}}^{\fe_7 + 3\text{F}} = \sqrt{Z^{\mathfrak{e}_7}_{\text{roots}} }Z^{\mathfrak{e}_7}_{\text{flavor 1}} \sqrt{Z^{\mathfrak{e}_7}_{\text{flavor 3}}}Z_{\text{extra}}^{\fe_7 +3\text{F}},
\end{equation}
where $Z_{\text{extra}}^{\fe_7 + 3\text{F}}$ is a contribution from an extra factor. Since the upper half of the diagram is exactly the same as the upper half diagram of the web diagram in Figure \ref{fig:e7on5}, the partition function for the upper half part computed from \eqref{part.5de7} in the $Q_B \to 0$ limit should also give rise to \eqref{e7halfupper}, which we checked until the order $Q_1^4P_1^3P_2^3P_3^3P_4^2R_1^3R_2^3R_3^3$. We can see that \eqref{e7compare1} is exactly the same as \eqref{e7halfupper} and the extra factor in \eqref{e7compare1} is again given by \eqref{e7extra} until the order we computed. 

\subsection{5d marginal pure $\SU(n)$ gauge theories with $n=3, 4$}
\label{sec:puremarginal}
\begin{figure}[t]
\centering
\subfigure[]{\label{fig:pureSU3k9}
\includegraphics[width=6cm]{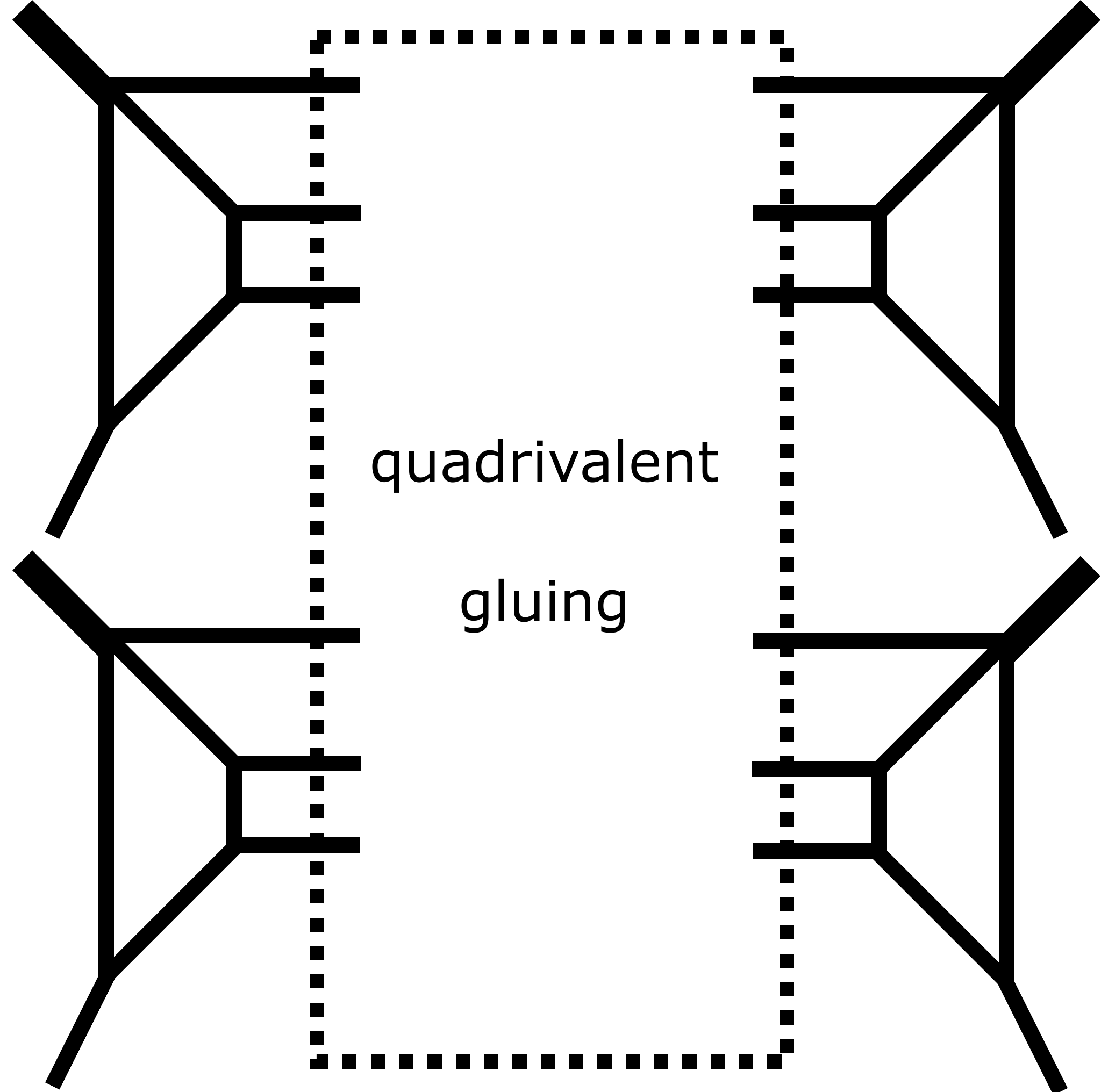}}
\hspace{1cm}
\subfigure[]{\label{fig:pureSU4k8}
\includegraphics[width=5cm]{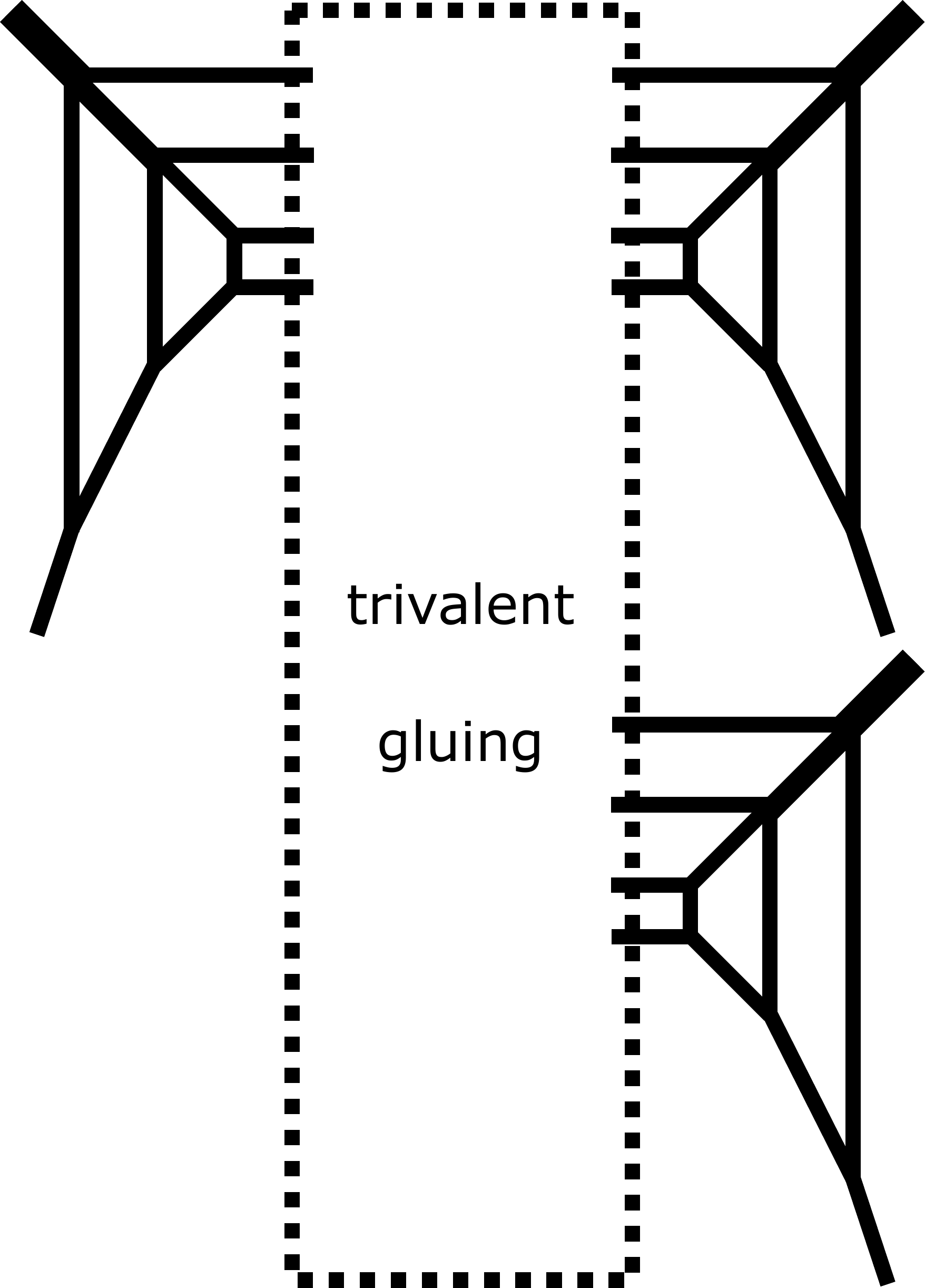}}
\caption{(a). The web diagram for the 5d $SU(3)$ gauge theory with the CS level $\kappa =9$. (b). The web diagram for the 5d $SU(4)$ gauge theory with the CS level $\kappa = 8$.}
\label{fig:pureSU}
\end{figure}

The Higgsings discussed in section \ref{sec:CMT} yield another interesting class of theories which are pure gauge theories obtained by twisted compactifications of 6d theories. Such theories include the 5d pure $\SU(3)$ gauge theory with the CS level $9$ and the 5d pure $\SU(4)$ gauge theory with the CS level $8$ \cite{Jefferson:2017ahm}. The pure $\SU(3)$ gauge theory with $\kappa = 9$ arises as a low energy theory on the Higgs branch labeled by $(2,2,2,2)$ in Table \ref{tb:d4_1} of the 
6d theory $(D_4, \underline{D_4})_1$ on $S^1$. On the other hand the pure $\SU(4)$ gauge theory with $\kappa = 8$ is realized on the Higgs branch labeled by $(3, 3, 3)$ in Table \ref{tb:e6_1} of 6d theory $(E_6, \underline{E_6})_1$ on $S^1$. 
In this section we compute the partition functions of the 5d pure $\SU(3)$ gauge theory with the CS level $\kappa = 9$ and the 5d pure $SU(4)$ gauge theory with the CS level $\kappa = 8$ on $S^1 \times \mathbb{R}^4$. The web diagrams of the theories can be obtained by applying the Higgsings to the diagram in Figure \ref{fig:gd4d4} for the marginal pure $\SU(3)$ gauge theory or the diagram in Figure \ref{fig:ge6e6} for the marginal pure $\SU(4)$ gauge theory. Both of the web diagrams are depicted in Figure \ref{fig:pureSU}. 

\begin{figure}[t]
\centering
\includegraphics[width=4.5cm]{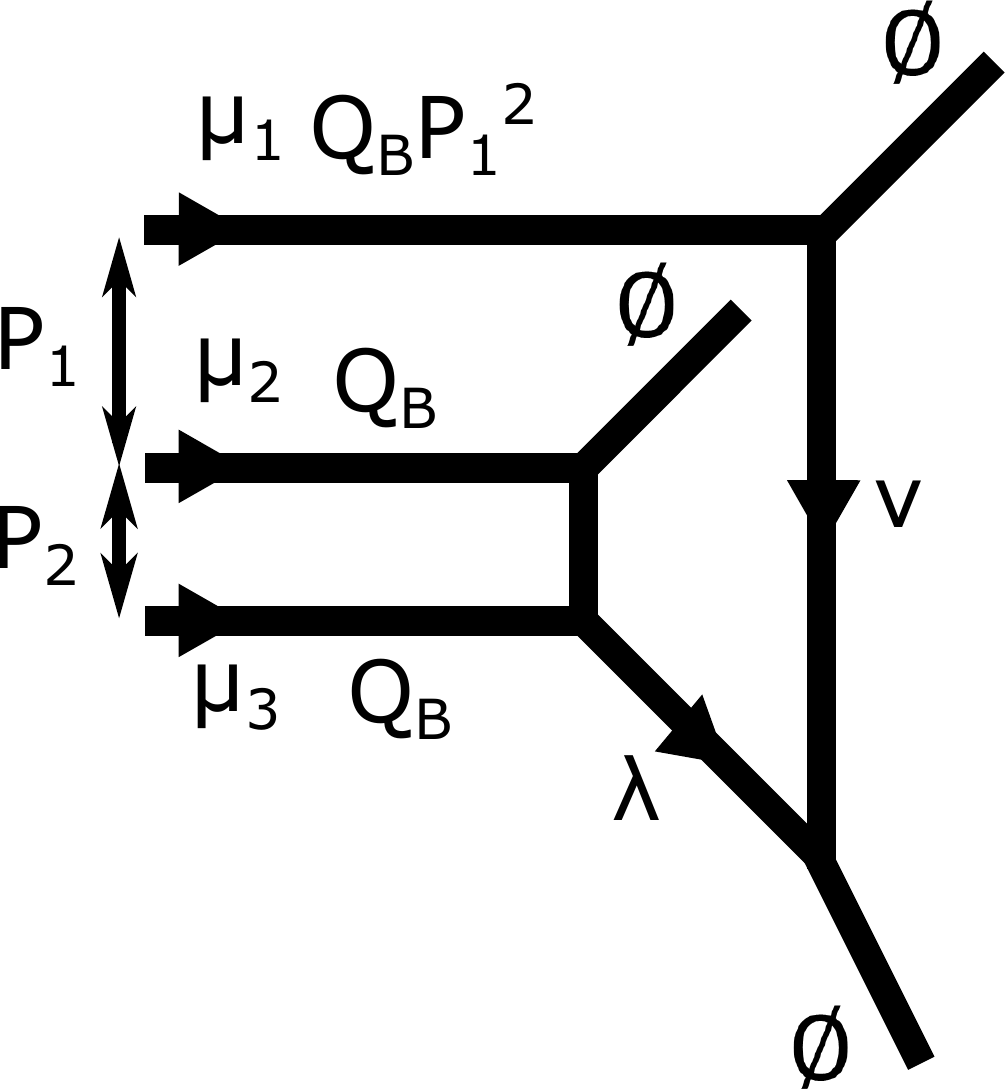}
\caption{Assignment of Young diagrams and K\"ahler parameters for one of the four diagrams in Figure \ref{fig:pureSU3k9}. The parameterization of the other diagrams is also the same one as this. }
\label{fig:su3para}
\end{figure}

\begin{figure}[t]
\centering
\subfigure[]{\label{fig:halfsu3}
\includegraphics[width=4cm]{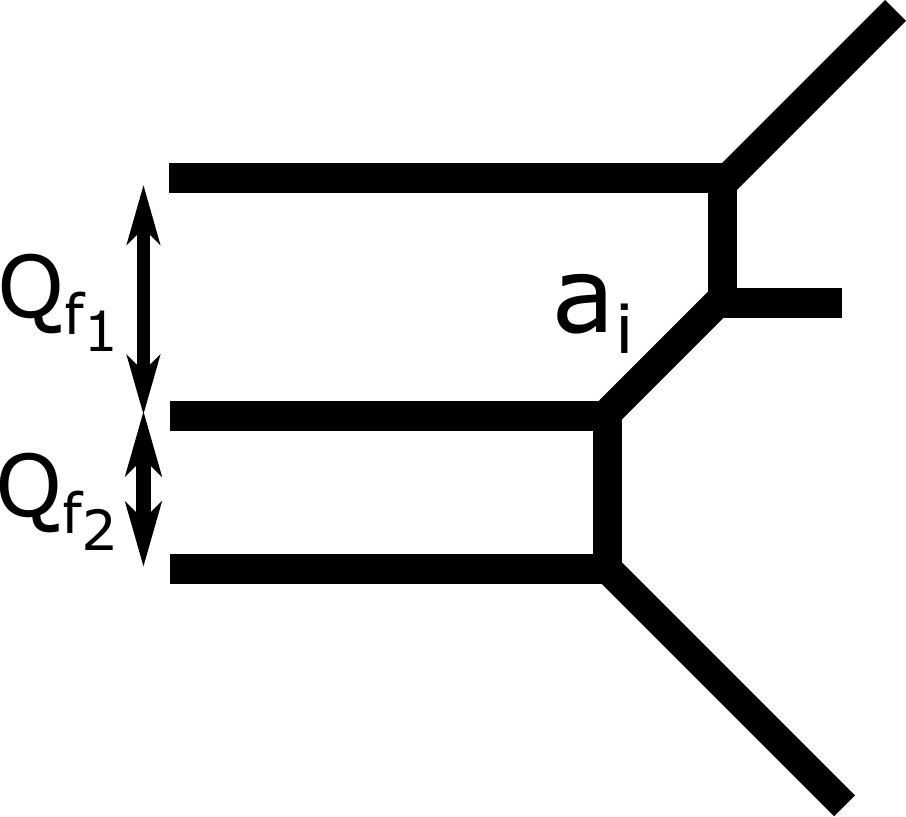}}
\hspace{1cm}
\subfigure[]{\label{fig:su3}
\includegraphics[width=7cm]{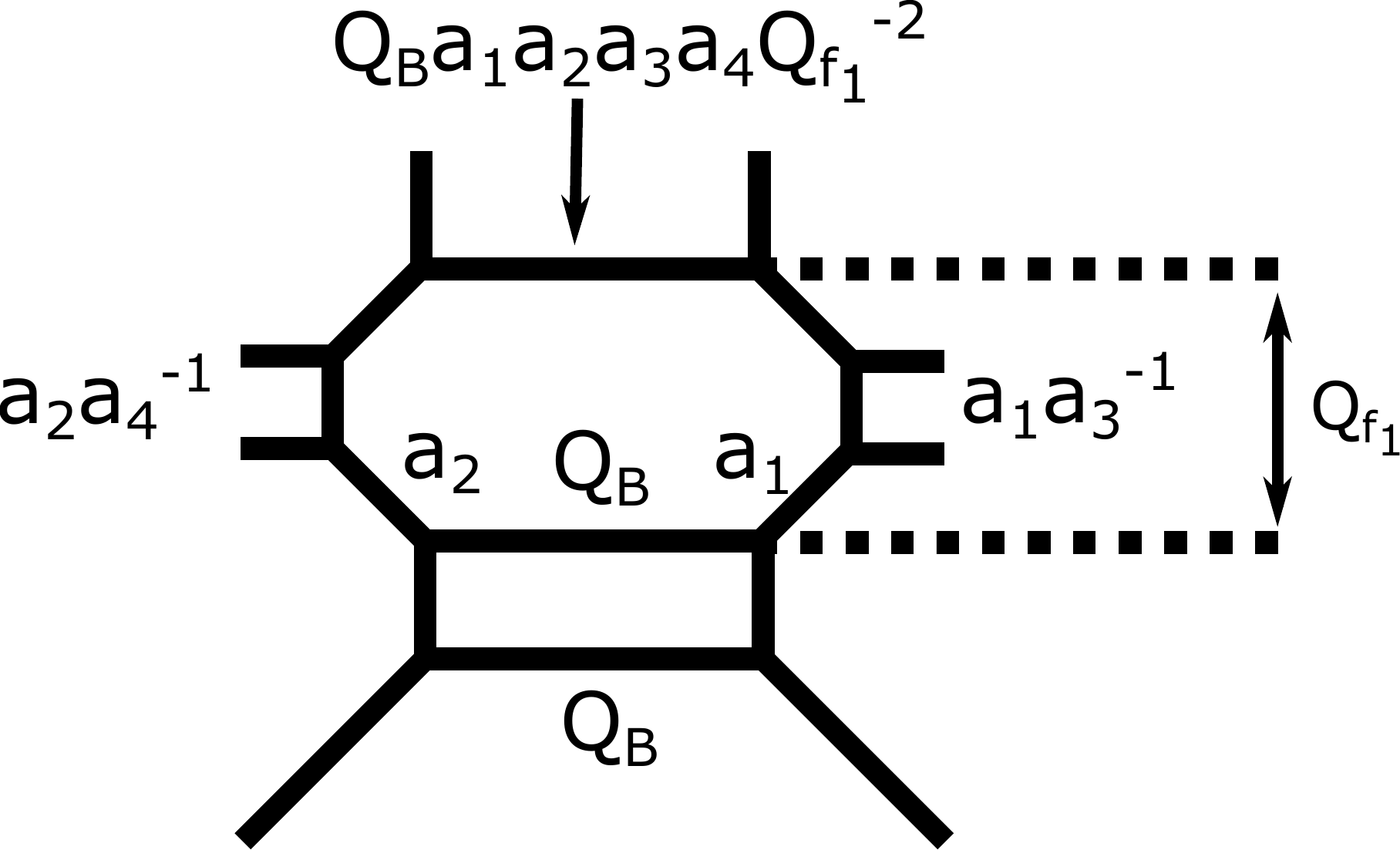}}
\caption{(a). The local diagram for one of the four parts which form the quadrivalent $\SU(3)$ gauging. $i$ runs from $1$ to $4$ representing the four parts. (b). The local diagram which describes the $\SU(3)_{-1}$ gauge theory with $4$ flavors.}
\label{fig:su3fig}
\end{figure}

\paragraph{5d pure $\SU(3)$ gauge theory with $\kappa = 9$.} We begin with the computation of the partition function of the pure $SU(3)$ gauge theory with the CS level $\kappa = 9$. For applying the topological vertex to the diagram in Figure \ref{fig:pureSU3k9}, we parameterize the lengths of the lines in the diagram as in Figure \ref{fig:su3para}. 
The parameterization of the gluing lines is determined by applying the tuning condition of the Higgsing to the parameterization of the  gluing lines before the Higgsing. Before the Higgsing the gluing part is locally described by the $\SU(3)$ gauge theory with four flavors. The local structure for one of the four pieces of the quadrivalent gauging is depicted in Figure \ref{fig:halfsu3}.
The local diagram for the $\SU(3)$ part is given in Figure \ref{fig:su3}. The shape of the diagram is determined by requiring that the lower face in Figure \ref{fig:su3} corresponds to $\mathbb{F}_0$ in the dual geometry. The Higgsing is achieved by imposing the condition 
\be
a_1 = a_2 = a_3 = a_4 = Q_{f_1},
\ee
and  the parameters $Q_{f_1}, Q_{f_2}$ are related to the parameters in Figure \ref{fig:su3para} by
\be
Q_{f_1} = P_1, \qquad Q_{f_2} = P_2.\label{kahlersu3k9}
\ee
Then the length of the top gluing line becomes 
\be
Q_Ba_1a_2a_3a_4Q_{f_1}^{-2} = Q_BP_1^2.
\ee

We then apply the topological vertex to the diagram in Figure \ref{fig:pureSU3k9} with the parameterization in Figure \ref{fig:su3para}. By comparing the diagram in Figure \ref{fig:halfsu3} with each of the diagrams in Figure \ref{fig:pureSU3k9} the 5-brane charge attached to the gluing horizontal lines does not change after the Higgsing and hence the framing factors for the gluing lines after the Higgsing are the same as those read off from Figure \ref{fig:su3}. The partition function computed by the topological vertex yields 
\begin{equation}\label{puresu3k9}
\begin{split}
Z_{SU(3)_9} =&\sum_{\mu_1, \mu_2, \mu_3} \left(-Q_BP_1^2\right)^{\left|\mu_1\right|} \left(-Q_B\right)^{\left|\mu_2\right| + \left|\mu_3\right|}Z_{\{\mu_i\}}^{SU(3)_L}(P_1, P_2)Z_{\{\mu_i\}}^{SU(3)_R}(P_1, P_2)\cr
&f_{\mu_1}^{-1}(q)f_{\mu_2}(q)f_{\mu_3}^{-1}(q)Z_{ \{\mu_i\}}^{SU(3)}(P_1, P_2)^4,
\end{split} 
\end{equation}
where we defined
\begin{align}
Z_{\{\mu_i\}}^{SU(3)_L}(P_1, P_2) &= q^{\frac{1}{2}(\sum_{i=1}^3||\mu_i||^2)}\left(\prod_{i=1}^3\tilde{Z}_{\mu_i}(q)\right)\prod_{1\leq i < j \leq 3}\mathcal{I}^-_{\mu_i, \mu_j}(Q_{f_i}\cdots Q_{f_{j-1}}),\\
Z_{\{\mu_i\}}^{SU(3)_R}(P_1, P_2) &= q^{\frac{1}{2}(\sum_{i=1}^3||\mu_i^t||^2)}\left(\prod_{i=1}^3\tilde{Z}_{\mu_i^t}(q)\right)\prod_{1\leq i < j \leq 3}\mathcal{I}^-_{\mu_i, \mu_j}(Q_{f_i}\cdots Q_{f_{j-1}}),
\end{align}
with $Q_{f_1}$ and $Q_{f_2}$ given by \eqref{kahlersu3k9}. The factor $Z_{\{\mu_i\}}^{SU(3)}(P_1, P_2)$ is the partition function from the diagram in Figure \ref{fig:su3para} divided by $Z_{\{\mu_i\}}^{SU(3)_R}(P_1, P_2)$ for the quadrivalent gluing prescription 
and the exponent of the factor is due to the fact that the four diagrams give identical contributions. The explict form of $Z_{\{\mu_i\}}^{SU(3)}(P_1, P_2)$ is 
\begin{equation}\label{su3factor}
\begin{split}
Z_{\{\mu_i\}}^{SU(3)}(P_1, P_2) &= q^{\frac{1}{2}\sum_{i=1}^3||\mu_i^t||^2}\left(\prod_{i=1}^3\tilde{Z}_{\mu_i^t}(q)\right)\mathcal{I}_{\mu_2, \mu_3}^-(P_2)/Z_{\{\mu_i\}}^{SU(3)_R}(P_1, P_2)\cr
&\sum_{\nu, \lambda}q^{\frac{1}{2}||\nu^t||^2 - \frac{1}{2}||\nu||^2 - \frac{1}{2}||\lambda^t||^2 + ||\lambda||^2}\tilde{Z}_{\lambda^t}(q)s_{\nu}\left(-P_1^2P_2q^{-\rho - \mu_1}\right)s_{\nu}\left(q^{-\rho - \lambda^t}\right)\cr
&\qquad s_{\lambda}\left(P_1q^{-\rho - \mu_3}, P_1P_2q^{-\rho -\mu_2}\right).
\end{split}
\end{equation}

In terms of the gauge theory parameterization, $P_1, P_2$ are related to the Coulomb branch moduli $A_1, A_2, A_3,$ $(A_1A_2A_3=1)$ of the $SU(3)$ gauge theory,
\be
P_1 = A_1A_2^{-1}, \qquad P_2 = A_2A_3^{-1}. 
\ee
On the other hand $Q_B$ is proportional to the instanton fugacity $M_0$. The instanton fugacity is the Coulomb branch independent part of $Q_B$. The intersection numbers between the curve with K\"ahler parameter $Q_B$ and the top surface and the bottom surface in Figure \ref{fig:pureSU3k9} are $4$ and $-2$ respectively. Hence the instanton fugacity is given by 
\be
Q_B =M_0P_1^{-2}.
\ee
In fact $M_0$ is related to the length of the top gluing line. This is consistent with the fact that the external lines attached to the top gluing line are parallel to each other, which can be seen from the diagram in Figure \ref{fig:su3}. From \eqref{fsu32} the K\"ahler parameter $Q_{f_{\fsu(3)^{(2)}}}$ for the fiber $f_{\fsu(3)^{(2)}}$ is 
\be
Q_{f_{\fsu(3)^{(2)}}} = Q_BP_1^2 = M_0.
\ee
Hence $M_0$ is related to the mass of a fractional KK mode.

Since the partition function \eqref{puresu3k9} may contain an extra factor, we have 
\be\label{hatpuresu3k9}
Z_{SU(3)_9} = \hat{Z}_{SU(3)_9}(A_1, A_2, A_3, M_0)Z_{\text{extra}}^{SU(3)_9}(M_0),
\ee
and we argue that $\hat{Z}_{SU(3)_9}(A_1, A_2, A_3, M_0)$ is the partition function of the 5d pure $\SU(3)$ gauge theory with the CS level $\kappa = 9$ on $S^1 \times \mathbb{R}_4$ up to an extra factor. 

The perturbative part of the partition function \eqref{puresu3k9} is obtained by taking the limit $Q_B \to 0$. Then the quardrivalent gluing is cut off since nonzero contribution comes from $|\mu_i| = \emptyset$ for $i=1, \cdots ,4$. 
The perturbative contribution can come from the part $Z_{\{\mu_i\}}^{SU(3)_L}(P_1, P_2)Z_{\{\mu_i\}}^{SU(3)_R}(P_1, P_2)$ in \eqref{puresu3k9} and then the contribution from \eqref{su3factor} should be one. Indeed we checked that \eqref{su3factor} is trivial for $|\mu_i| = \emptyset, (i=1, \cdots ,4)$ until the order $P_1^5P_2^5$. 

We can also compute the one-instanton part and the Plethystic exponential of the one-instanton part until the order $P_1^3P_2^3$ becomes
\begin{equation} \label{instsu3cs9}
\begin{split}
\frac{\hat{Z}_{SU(3)_9}(A_1, A_2, M_0)}{\hat{Z}_{SU(3)_9}(A_1, A_2, 0)} &=\text{PE}\left[\frac{2Q_Bq}{(1-q)^2}\left(1 + P_1 +P_1^3 + 2P_2 + 3P_1P_2 + 3P_1^2P_2 + 3P_1^3P_2 + 3P_2^2 \right.\right.\cr 
&\qquad\left.\left.+ 5P_1P_2^2+6P_1^2P_2^2 + 6P_1^3P_2^2 + 4P_2^3 + 7P_1P_2^3 + 9P_1^2P_2^3 + 10P_1^3P_2^3\right)\right].
\end{split}
\end{equation}
The partition function \eqref{instsu3cs9} perfectly agrees with the results in \cite{Hayashi:2020hhb, Kim:2020hhh, Kim:2021cua} which were computed in  different methods. 
We also have the extra factor and it is given by 
\be
Z_{\text{extra}}^{SU(3)_9}(M_0) = \text{PE}\left[\frac{q}{(1-q)^2}8Q_BP_1^2\right],
\ee
until the orders we computed. Since $Q_BP_1^2$ is the K\"ahler parameter for the top gluing line, the extra factor is associated to the parallel external lines which can be explicitly seen from the diagram in Figure \ref{fig:su3}.

\begin{figure}[t]
\centering
\includegraphics[width=4cm]{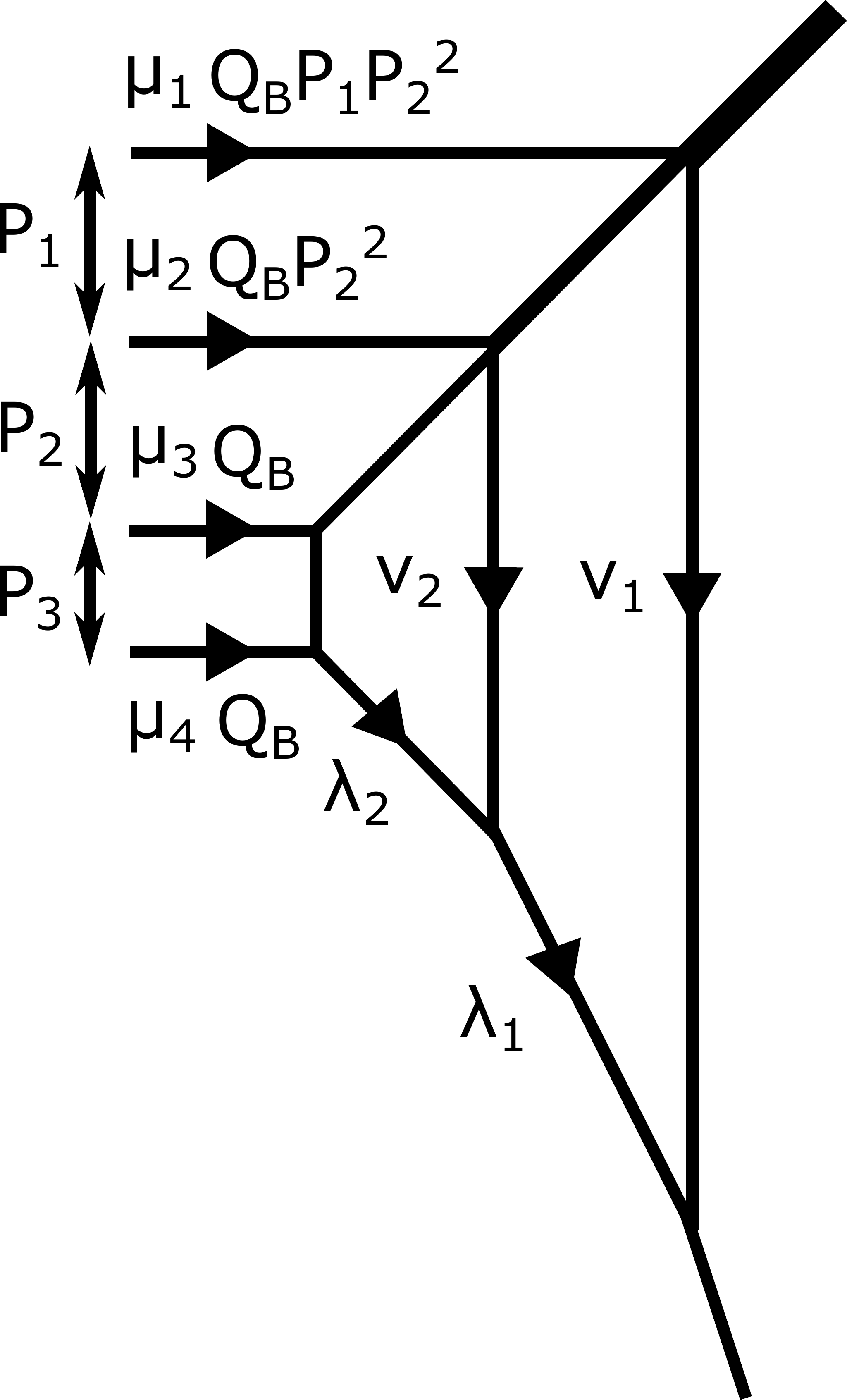}
\caption{Assignment of Young diagrams and K\"ahler parameters for one of the three diagrams in Figure \ref{fig:pureSU4k8}. The parameterization of the other diagrams is also the same one as this. }
\label{fig:su4para}
\end{figure}

\begin{figure}[t]
\centering
\subfigure[]{\label{fig:purehalfsu4}
\includegraphics[width=5cm]{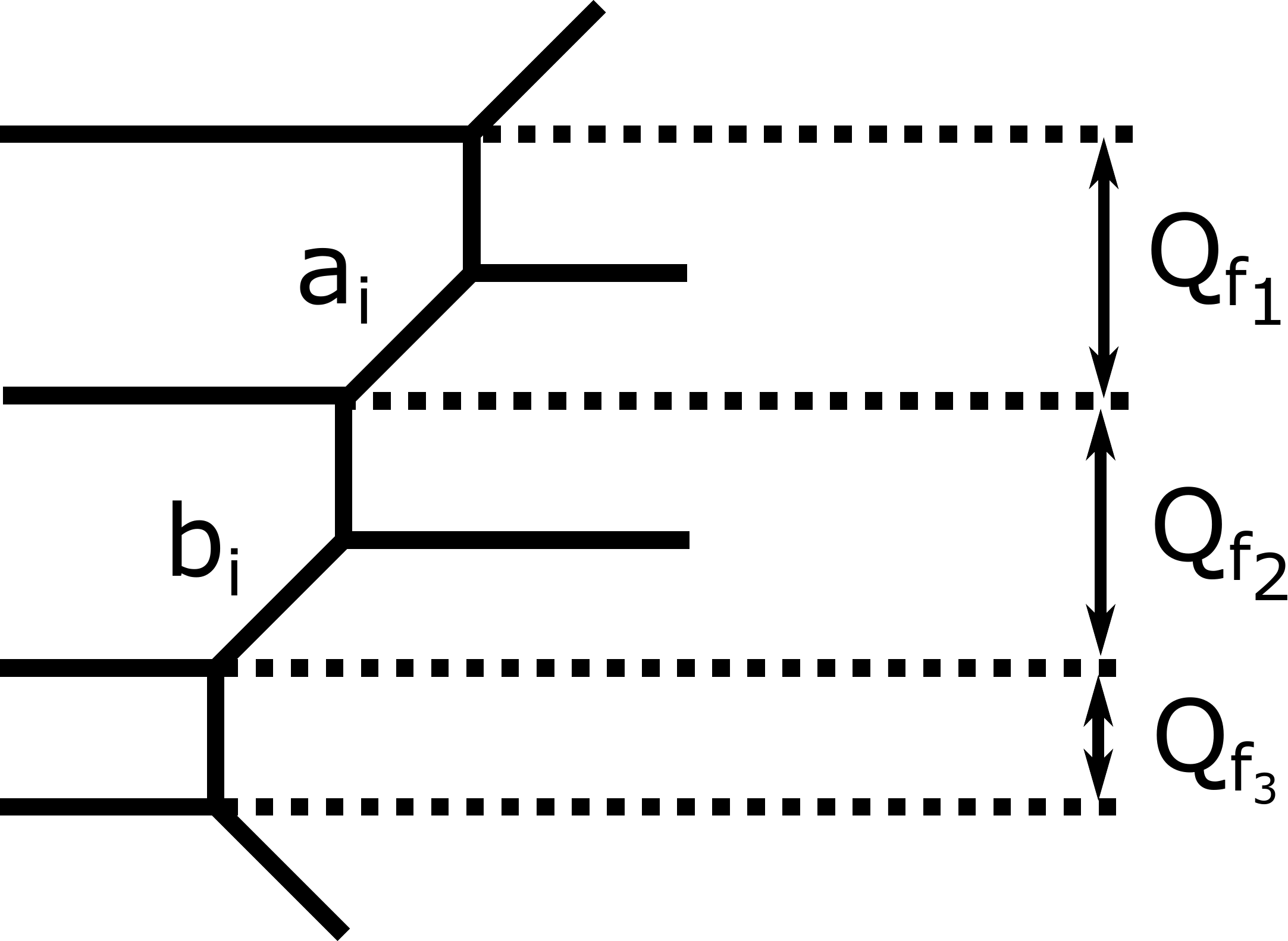}}
\hspace{0.5cm}
\subfigure[]{\label{fig:puresu4}
\includegraphics[width=8cm]{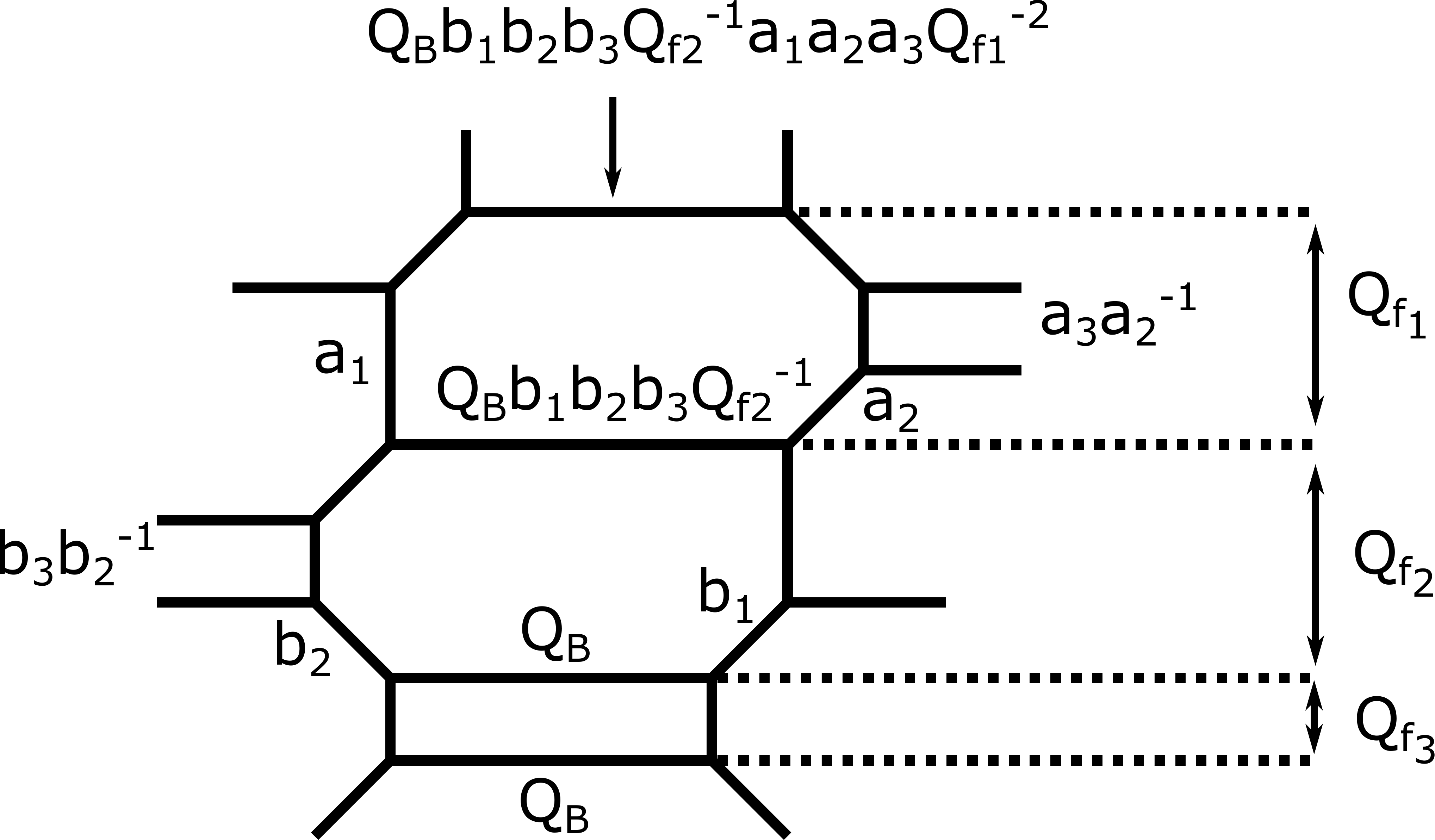}}
\caption{(a). The local diagram for one of the tree parts which form the trivalent $\SU(4)$ gauging. $i$ runs from $1$ to $3$ representing the three parts. (b). The local diagram which describes the $\SU(4)_{-1}$ gauge theory with $6$ flavors.}
\label{fig:puresu4fig}
\end{figure}

\paragraph{$\SU(4)$ with $\kappa = 8$.} Next we compute the partition function of the 5d pure $\SU(4)$ gauge theory with the CS level $\kappa = 8$ on $S^1 \times \mathbb{R}^4$. We parameterize the lengths of the diagram in Figure \ref{fig:pureSU4k8} as in Figure \ref{fig:su4para}.  
The length of the gluing line is again determined by following the Higgsing procedure. Before the Higgsing the gluing part is locally described by the $\SU(4)$ gauge theory with six flavors. The local diagram for one of the three diagrams for the trivalent $\SU(4)$ gauging is depicted in Figure \ref{fig:purehalfsu4}.  
Then gluing three copies of the diagram in Figure \ref{fig:purehalfsu4} yields the diagram in Figure \ref{fig:puresu4}. The CS level of the $SU(4)$ gauge theory is fixed by requiring that the bottom surface in Figure \ref{fig:puresu4} corresponds to $\mathbb{F}_0$ in the dual geometry. Then the $(3, 3, 3)$ Higgsing is realized by imposing the condition 
\be
a_1 = a_2 = a_3 = Q_{f_1}, \qquad b_1 = b_2 = b_3 = Q_{f_2},
\ee 
and the parameters $Q_{f_1}, Q_{f_2}, Q_{f_3}$ are related to the parameters in Figure \ref{fig:su4para} by
\be
Q_{f_1} = P_1, \qquad Q_{f_2} = P_2, \qquad Q_{f_3} = P_3. \label{kahlersu4k8}
\ee
Then the length of the gluing lines after the tuning becomes 
\begin{align}
Q_Bb_1b_2b_3Q_{f_2}^{-1}a_1a_2a_3Q_{f_1}^{-2} &= Q_BP_1P_2^2,\\
Q_Bb_1b_2b_3Q_{f_2}^{-1} &= Q_BP_2^2.
\end{align}

We then apply the topological vertex to the diagram in Figure \ref{fig:pureSU4k8} with the parameterization in Figure \ref{fig:su4para}. Again the framing factors for the gluing lines do not change by the Higgsing and we read off the framing factors for the gluing lines from the diagram in Figure \ref{fig:puresu4fig}. The partition function computed by the topological vertex gives 
\begin{equation}\label{puresu4k8}
\begin{split}
Z_{SU(4)_8} =&\sum_{\{\mu_i\}} \left(-Q_BP_1P_2^2\right)^{\left|\mu_1\right|} \left(-Q_BP_2^2\right)^{\left|\mu_2\right|}\left(-Q_B\right)^{|\mu_3| + |\mu_4|}Z_{\{\mu_i\}}^{SU(4)_L}\left(\{P_a\}\right)Z_{\{\mu_i\}}^{SU(4)_R}\left(\{P_a\}\right)\cr
&f_{\mu_1}^{-1}(q)f_{\mu_3}(q)f_{\mu_4}^{-1}(q)Z_{ \{\mu_i\}}^{SU(4)}(P_1, P_2, P_3)^3,
\end{split} 
\end{equation}
where we defined
\begin{align}
Z_{\{\mu_i\}}^{SU(4)_L}(P_1, P_2, P_3) &= q^{\frac{1}{2}(\sum_{i=1}^4||\mu_i||^2)}\left(\prod_{i=1}^4\tilde{Z}_{\mu_i}(q)\right)\prod_{1\leq i < j \leq 4}\mathcal{I}^-_{\mu_i, \mu_j}(Q_{f_i}\cdots Q_{f_{j-1}}),\\
Z_{\{\mu_i\}}^{SU(4)_R}(P_1, P_2, P_3) &= q^{\frac{1}{2}(\sum_{i=1}^4||\mu_i^t||^2)}\left(\prod_{i=1}^4\tilde{Z}_{\mu_i^t}(q)\right)\prod_{1\leq i < j \leq 4}\mathcal{I}^-_{\mu_i, \mu_j}(Q_{f_i}\cdots Q_{f_{j-1}}),
\end{align}
with $Q_{f_1}, Q_{f_2}, Q_{f_3}$ given by \eqref{kahlersu4k8}. The factor $Z_{ \{\mu_i\}}^{SU(4)}(P_1, P_2, P_3)$ is the contribution from the diagram in Figure \ref{fig:su4para} with the trivalent gluing prescription implemented and the explicit expression is 
\begin{equation}\label{su4factor}
\begin{split}
Z_{\{\mu_i\}}^{SU(4)}(P_1, P_2, P3) &= q^{\frac{1}{2}\sum_{i=1}^4||\mu_i^t||^2}\left(\prod_{i=1}^4\tilde{Z}_{\mu_i^t}(q)\right)\mathcal{I}_{\mu_3, \mu_4}^-(P_4)/Z_{\{\mu_i\}}^{SU(4)_R}(P_1, P_2, P_3)\cr
&\sum_{\nu_1, \nu_2, \lambda_1, \lambda_2}q^{||\nu_1^t||^2 - ||\nu_1||^2 + \frac{1}{2}||\nu_2^t||^2 - \frac{1}{2}||\nu_2||^2 - \sum_{i=1}^2\left(\frac{1}{2}||\lambda_i^t||^2 - ||\lambda_i||^2\right)}\prod_{i=1}^2\tilde{Z}_{\lambda_i^t}(q)\cr
&s_{\nu_1}\left(P_1^3P_2^2P_3q^{-\rho - \mu_1}\right)s_{\nu_1}\left(q^{-\rho - \lambda_1^t}\right)s_{\nu_2}\left(-P_2^2P_3q^{-\rho - \mu_2}\right)s_{\nu_2/\eta}\left(q^{-\rho - \lambda_2^t}\right)\cr
&s_{\lambda_1/\eta}\left(q^{-\rho - \lambda_2}\right)s_{\lambda_2}\left(P_2q^{-\rho - \mu_4}, P_2P_3q^{-\rho -\mu_3}\right)\left(P_1\right)^{|\lambda_1|}.
\end{split}
\end{equation}

Let us rewrite the K\"ahler parameters by the gauge theory parameters of the pure $\SU(4)$ gauge theory. 
$P_1, P_2, P_3$ are related to the Coulomb branch moduli $A_1, A_2, A_3, A_4,$ $(A_1A_2A_3A_4=1)$ by
\begin{align}
P_1 = A_1A_2^{-1}, \qquad P_2 = A_2A_3^{-1}, \qquad P_3 = A_3A_4^{-1}. 
\end{align}
$Q_B$ is proportional to the instanton fugacity $M_0$ and $M_0$ is the Coulomb branch independent part of $Q_B$. Since the intersection numbers between the curve with the K\"ahler parameter and the top, middle bottom surface in Figure \ref{fig:pureSU4k8} are $0, 3, -2$ respectively, we identify the instanton fugacity as
\be
Q_B = M_0P_1^{-1}P_2^{-2}.
\ee
Again $M_0$ is related to the length of the top gluing line, which is consistent with the fact that the external lines attached to the top gluing line are parallel to each other as can be seen from the diagram in Figure \ref{fig:puresu4}.  From \eqref{fso83} the K\"ahler parameter $Q_{f_{\fso(8)^{(3)}}}$ for the fiber $f_{\fso(8)^{(3)}}$ is 
\be
Q_{f_{\fso(8)^{(3)}}} = Q_BP_2^2P_1 = M_0.
\ee
Hence $M_0$ is also related to the mass of a fractional KK mode.

The partition function \eqref{puresu3k9} may contain an extra factor and hence it can be written as 
\be\label{hatpuresu4k8}
Z_{SU(4)_8} = \hat{Z}_{SU(4)_8}(A_1, A_2, A_3, A_4, M_0)Z_{\text{extra}}^{SU(4)_8}(M_0),
\ee
and we argue that $\hat{Z}_{SU(4)_8}(A_1, A_2, A_3, A_4, M_0)$ is the partition function of the 5d pure $SU(4)$ gauge theory with the CS level $\kappa = 8$ on $S^1 \times \mathbb{R}_4$ up to an extra factor. 

Then the perturbative part of the partition funciton \eqref{puresu4k8} is given in the limit $Q_B \to 0$. In this case, the perturbative part of the pure $\SU(4)$ gauge theory with $\kappa = 8$ can be obtained from the $Z_{\{\mu_i\}}^{SU(4)_L}\left(\{P_a\}\right)Z_{\{\mu_i\}}^{SU(4)_R}\left(\{P_a\}\right)$. Then the contribution from \eqref{su4factor} needs to be one. We checked that \eqref{su4factor} with $\mu_i = \emptyset, (i=1, 2, 3, 4)$ becomes indeed trivial until the order $P_1^3P_2^3P_3^3$. 

It is also possible to compute the one-instanton part and the Plethystic exponential of the one-instanton part until the order $P_1^2P_2^2P_3^2$ becomes
\begin{equation}\label{instsu4cs8}
\begin{split}
&\frac{\hat{Z}_{SU(4)_8}(A_1, A_2, A_3, A_4, M_0)}{\hat{Z}_{SU(4)_8}(A_1, A_2, A_3, A_4, 0)}\cr
&=\text{PE}\left[\frac{2Q_Bq}{(1-q)^2}\left(1+ P_2 + P_1P_2 + P_2^2 + P_1^2P_2^2 + 2P_3 + 3P_2P_3 + 3P_1P_2P_3 + 3P_2^2P_3+ 4P_1P_2^2P_3\right.\right.\cr
&\qquad\qquad\quad\left.\left.  + 3P_1^2P_2^2P_3 + 3P_3^2 + 5P_2P_3^2 + 5P_1P_2P_3^2 + 6P_2^2P_3^2 + 8P_1P_2^2P_3^2 + 6P_1^2P_2^2P_3^2\right)\right].
\end{split}
\end{equation}
The partition functon \eqref{instsu4cs8} perfectly agrees with the result obtained by a blow up method in the unrefined limit computed in \cite{Kim:2020hhh}. 
The partition function has an extra factor and it is given by 
\be
Z_{\text{extra}}^{SU(4)_8}(M_0)=\text{PE}\left[\frac{q}{(1-q)^2}9Q_BP_1P_2^2\right],
\ee
until the orders we computed. Since $Q_BP_1P_2^2$ is the K\"ahler parameter for the top gluing line, the extra factor is associated to the parallel external lines in Figure \ref{fig:puresu4}.

\bigskip

\section{Conclusion}\label{sec:concl}

In this paper we constructed web diagrams using the trivalent or the quadrivalent gluing for various 6d/5d theories. The theories were obtained from certain Higgsings of the 6d conformal matter theories of type $(G, G)_k$ with $G=D_4, E_6, E_7$ on $S^1$. The theories include gauge theories with exceptional gauge groups, in particular the 6d $G_2$ gauge theory with $4$ flavors, 6d $F_4$ gauge theories with  $3$ flavors, 6d $E_6$ gauge theories with $4$ flavors and also 6d $E_7$ gauge theory with $3$ flavors, 
each of which has contain a single tensor multiplet. By taking the 5d limit, we also obtained web diagrams for the 5d version of the theories. From Higgsings of $(G, \underline{G})_1$ on $S^1$ with $G=D_4, E_6, E_7$ we could construct web diagrams for the 5d $\SU(3)$ gauge theory with the CS level $9$, 5d $\SU(4)$ gauge theory with the CS level $8$ and the 6d $E_7$ gauge theory with $\frac{3}{2}$ flavors and a tensor multiplet. Since the web diagrams are realized by the trivalent or the quadrivalent gluing we could apply the topological vertex formalism to the web diagram and computed the Nekrasov partition functions of the aforementioned theories. We also took the 5d limit of the 6d gauge theories with exceptional gauge groups and performed the consistency checks by seeing if the topological vertex computations correctly reproduce the perturbative part of the partition functions. We indeed found perfect matchings with the known perturbative contributions until the orders we computed. For the partition functions of the marginal pure $\SU(3)$ and $\SU(4)$ gauge theories we checked that the instanton parts also agree with the results which were recently obtained in other methods in \cite{Hayashi:2020hhb, Kim:2020hhh, Kim:2021cua}.

In general 5d gauge theories can possess the limited number of flavors for having a UV completion. The upper bound of the number of flavors for having a 5d UV completion has been obtained in \cite{Jefferson:2017ahm, Bhardwaj:2019xeg} and the numbers are $5$ for $G_2$, $3$ for $F_4$, $4$ for $E_6$, and $3$ for $E_7$. In this paper we constructed web diagrams for 5d $G_2$ gauge theory with $4$ flavors, 5d $F_4$ gauge theory with $3$ flavors, 5d $E_6$ gauge theory with $4$ flavors and 5d $E_7$ gauge theory with $3$ flavors (with one mass parameter turned off). Furthermore, a 5-brane web for the 5d $G_2$ gauge theory with $5$ flavors has been constructed in \cite{Hayashi:2018lyv}. For an $E_8$ gauge theory we cannot have a fundamental flavor and a web diagram for the pure $E_8$ gauge theory has been obtained in \cite{Hayashi:2017jze}. Hence we now have a web diagram description for all the gauge theories with a single exceptional gauge group $G_2$, $F_4$, $E_6,$ $E_7$ or $E_8 $ and the maximal number of flavors for having a 5d UV completion. We can also compute the Nekrasov partition function of all of the theories using the topological vertex. Note that the web diagram we constructed for the 5d $E_7$ gauge theory with $3$ flavors can incorporate only two mass parameters for the three flavors. It would be interesting to construct a web diagram for the 5d $E_7$ gauge theory with three flavors with all the mass parameters turned on.

As for the computations of the Nekrasov partition functions we have used the unrefined topological vertex. The application of the refined topological vertex to web diagrams with trivalent gluing has been already presented in \cite{Hayashi:2017jze}. It would be interesting to extend the computations to the one for computing the refined Nekrasov partition functions by utilizing the method. 
Also our checks for the obtained partition functions of the gauge theories with an exceptional gauge group have focused on the perturbative part of the 5d partition functions. It would be also interesting to check the instanton part  of the 5d partition functions or the 6d partition functions themselves by comparing our result with the results computed in other methods \cite{Benvenuti:2010pq, Hanany:2012dm, Keller:2012da, Haghighat:2014vxa, Cremonesi:2014xha, DelZotto:2016pvm, Kim:2018gjo, DelZotto:2018tcj, Kim:2019uqw, Gu:2019dan, Gu:2020fem, Kim:2020hhh}. 

\bigskip

\acknowledgments
H.H. thanks the University of Tokyo, APCTP and Simons Center for Geometry and Physics for hospitality where a part of the work is done. The work of H.H. is supported in part by JSPS KAKENHI Grant Number JP18K13543. The research of H.K. is supported by the POSCO Science Fellowship of POSCO TJ Park Foundation and the National Research Foundation of Korea (NRF) Grant 2018R1D1A1B07042934. 
\bigskip

\appendix 

\section{Brane webs and Higgs branches of $(E_8, E_8)$ conformal matter on a circle}
\label{sec:e8}
In section \ref{sec:CMT}, we have considered theories 
obtained by applying certain Higgsings to the theories $(D_4, D_4)_2$, $(E_6, E_6)_2$ and $(E_7, E_7)$ on $S^1$. 
We focused on such examples since 
the Higgsings yield gauge theories with a single exceptional gauge group and matter whose partition functions we have computed in section \ref{sec:PF}. For completeness, we consider Higgsings of the theory $(E_8, \underline{E_8})_1$ on $S^1$ in this appendix.

\begin{figure}[t]
\centering
\subfigure[]{\label{fig:e8e8}
\includegraphics[width=6cm]{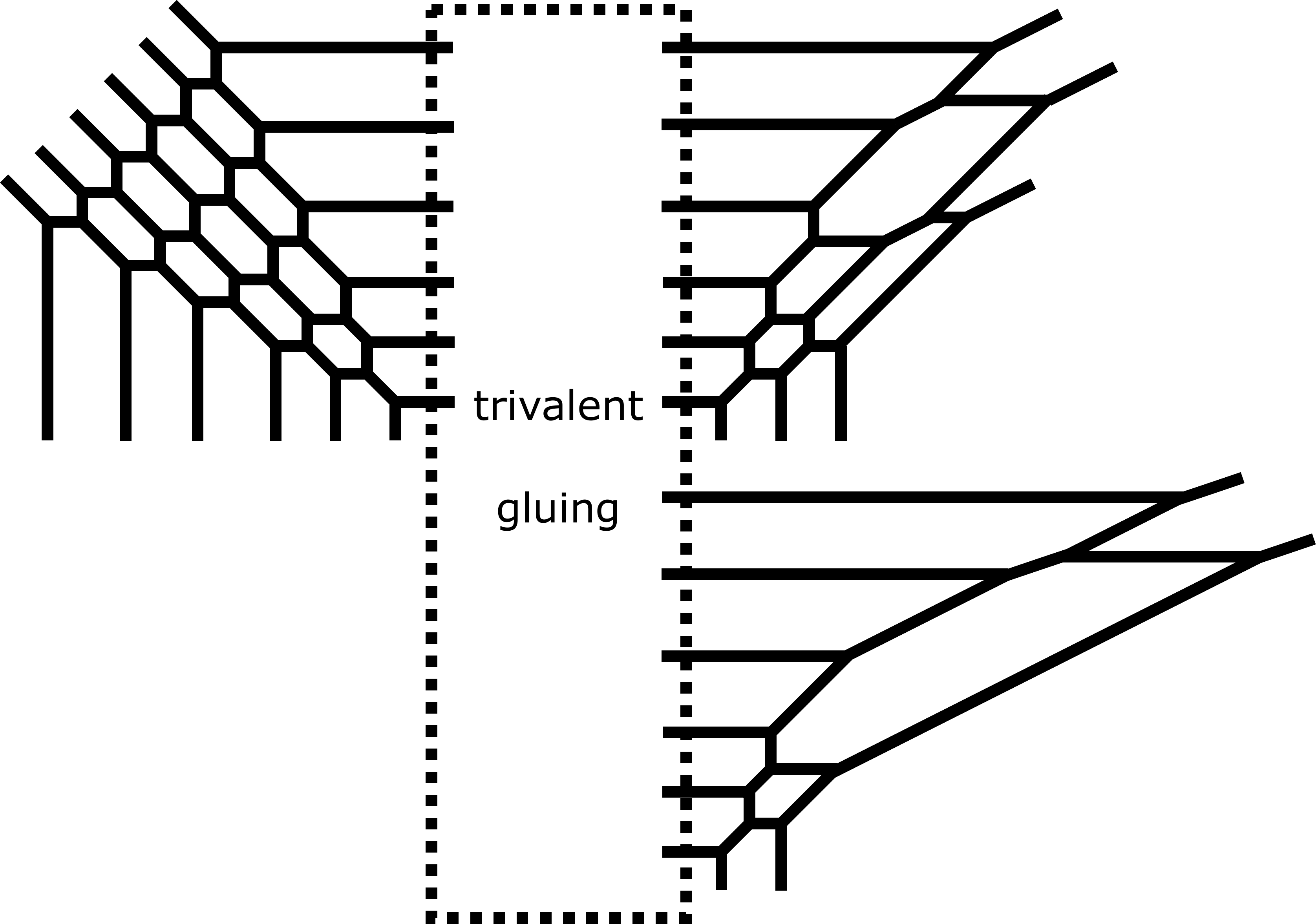}}
\hspace{1cm}
\subfigure[]{\label{fig:ge8e8}
\includegraphics[width=5cm]{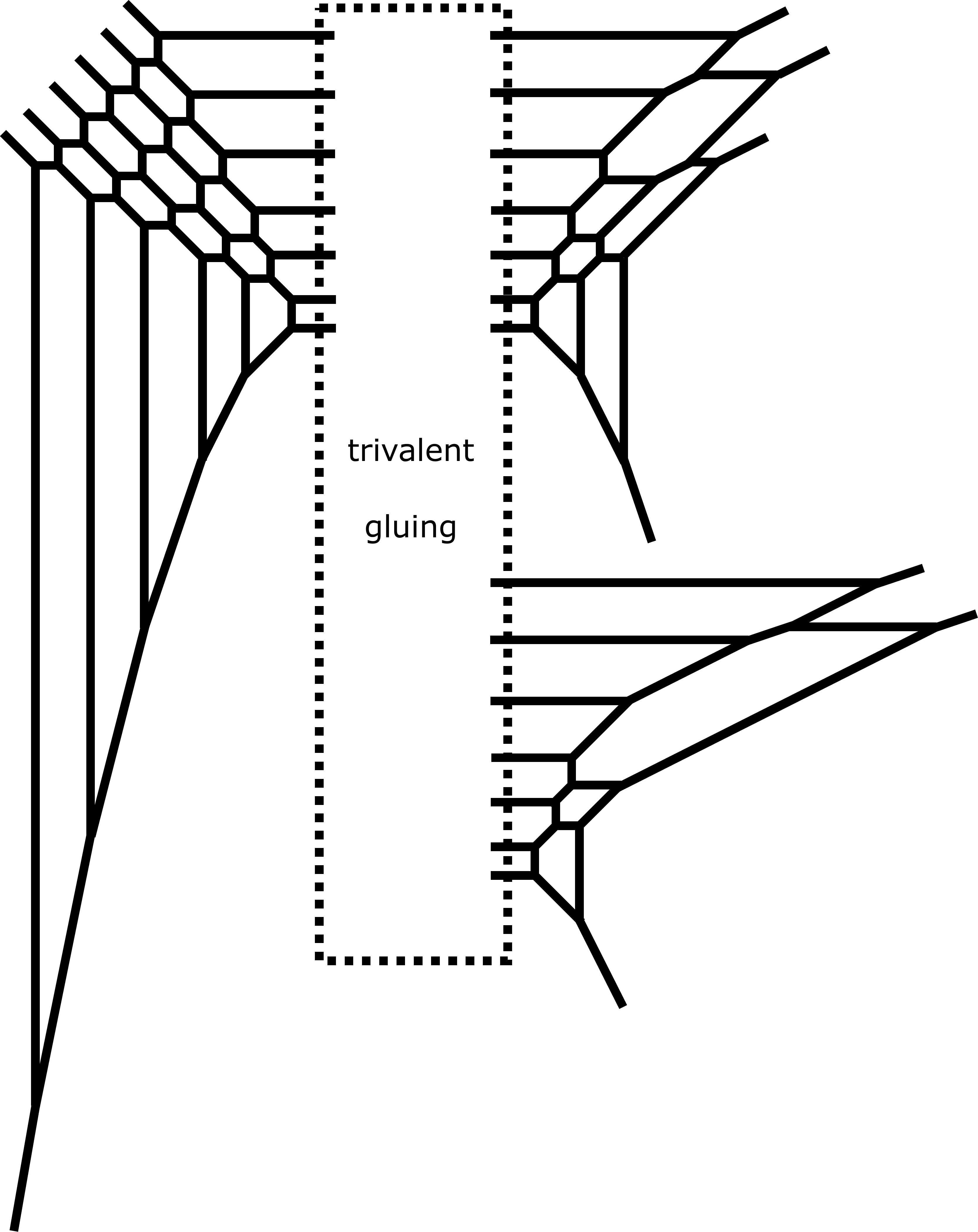}}
\caption{(a): The web diagram for the theory \eqref{e8e8} or \eqref{min.affineE8}. (b): The web diagram for the theory \eqref{ge8e8}.}
\label{fig:e8e8g}
\end{figure}
The minimal $(E_8, E_8)$ conformal matter theory on a circle is given by 
\be
\us\os1^{\fsp(0)^{(1)}}_{\left[\fe_8^{(1)}\right]}-\os2^{\fsu(1)^{(1)}}-\os2^{\fsu(2)^{(1)}}-\os3^{\fg_2^{(1)}}-\os1^{\fsp(0)^{(1)}}- \os5^{\ff_4^{(1)}}-\os1^{\fsp(0)^{(1)}}-\os3^{\fg_2^{(1)}}-\os2^{\fsu(2)^{(1)}}-\os2^{\fsu(1)^{(1)}}-\us\os1^{\fsp(0)^{(1)}}_{\left[\fe_8^{(1)}\right]}.\label{e8e8}
\ee
The theory has a 5d gauge theory description and it is given by the following affine $E_8$ Dynkin quiver theory
\be
\SU(1)  - \SU(2) - \SU(3)_0 - \SU(4)_0 - \SU(5)_0 -{\overset{\overset{\text{\normalsize$\SU(3)_0$}}{\textstyle\vert}}{\SU(6)_0}}- \SU(4)_0 - \SU(2)_0.\label{min.affineE8}
\ee
From the gauge theory description \eqref{min.affineE8}, it is possible to write down a web diagram of the theory using the trivalent gluing and the web diagram is depicted in Figure \ref{fig:e8e8}. When we gauge one of the $E_8$ flavor symmetry of the minimal $(E_8, E_8)$ conformal matter theory, the resulting theory which we denote by $(E_8, \underline{E_8})$ on a circle becomes 
\be
\us\os1^{\fsp(0)^{(1)}}_{\left[\fe_8^{(1)}\right]}-\os2^{\fsu(1)^{(1)}}-\os2^{\fsu(2)^{(1)}}-\os3^{\fg_2^{(1)}}-\os1^{\fsp(0)^{(1)}}-\os5^{\ff_4^{(1)}}-\os1^{\fsp(0)^{(1)}}-\os3^{\fg_2^{(1)}}-\os2^{\fsu(2)^{(1)}}-\os2^{\fsu(1)^{(1)}}-\os1^{\fsp(0)^{(1)}}-\os12^{\fe_8^{(1)}}.\label{ge8e8}
\ee
and its web diagram realization is given in Figure \ref{fig:ge8e8}. 

We study Higgsings of the theory \eqref{ge8e8} in this appendix. We focus on Higgsings which can be realized from the web diagram in Figure \ref{fig:ge8e8}. From the web diagram we can explicitly see an $\SU(6) \times \SU(3) \times \SU(2)$ flavor symmetry from the parallel external lines going in the upper direction. Hence we consider Higgsings which break a part of the $\SU(6) \times \SU(3) \times \SU(2)$ flavor symmetry. Such Higgsings can be labeled by $(a_1, a_2, a_3)$ where $a_1, a_2, a_3$ are associated to a breaking of $\SU(6), \SU(3), \SU(2)$ respectively. As before, $a_i = (n_1, n_2, \cdots)$ implies a Higgsing which breaks $\SU(n_1) \times \SU(n_2) \times \cdots$ in $\SU(6) \times \SU(3) \times \SU(2)$. It is possible to obtain a weighted Dynkin diagram of the $\fe_8$ algebra from the Higgsing label $(a_1, a_2, a_3)$. Then from the relation between a weighted Dynkin diagram and a B-C label, we can utilize the result of \cite{Heckman:2016ssk} to read off the resulting theory after the Higgsing as we have done in section \ref{sec:d4d4}, \ref{sec:e6e6} and \ref{sec:e7e7}. The end result is summarized in Table \ref{tb:e8_1}. 
\begin{center}
\begin{longtable}{c|c|c}
\caption{ \label{tb:e8_1} Higgsings labeled by $(a_1, a_2, a_3)$ for the theory \eqref{ge8e8}. }\\
$\text{B-C Label}$&$\text{twist}$&$\text{Higgsing}$
\\[3 pt]
\hline
\multicolumn{3}{c}{$\text{Theory}$}
\\
\hline\hline
\rule[-10pt]{0pt}{30pt}
$ 0$&$1$ & $(0, 0, 0)$ \\[3pt]
\hline
\multicolumn{3}{c}{$\us\os1^{\fsp(0)^{(1)}}_{\left[\fe_8^{(1)}\right]}-\os2^{\fsu(1)^{(1)}}-\os2^{\fsu(2)^{(1)}}-\os3^{\fg_2^{(1)}}- \os1^{\fsp(0)^{(1)}}-\os5^{\ff_4^{(1)}}-\os1^{\fsp(0)^{(1)}}-\os3^{\fg_2^{(1)}}-\os2^{\fsu(2)^{(1)}}-\os2^{\fsu(1)^{(1)}}- \os1^{\fsp(0)^{(1)}}-\os12^{\fe_8^{(1)}}$}
\\[10 pt]
\hline\hline
\rule[-10pt]{0pt}{30pt}
$ A_1$&$1$ &$(2, 0, 0), (0, 2, 0), (0,0, 2)$ \\[3pt]
\hline
\multicolumn{3}{c}{$\us\os1^{\fsp(0)^{(1)}}_{\left[\fe_7^{(1)}\right]}-\os2^{\fsu(2)^{(1)}}-\os3^{\fg_2^{(1)}}-\os1^{\fsp(0)^{(1)}}- \os5^{\ff_4^{(1)}}-\os1^{\fsp(0)^{(1)}}-\os3^{\fg_2^{(1)}}-\os2^{\fsu(2)^{(1)}}-\os2^{\fsu(1)^{(1)}}-\os1^{\fsp(0)^{(1)}}-\os12^{\fe_8^{(1)}}$}
\\[10 pt]
\hline\hline
\rule[-10pt]{0pt}{30pt}
$ 2A_1$&$1$ &$(2, 2, 0), (2, 0, 2), (0,2, 2), ((2,2), 0, 0)$ \\[3pt]
\hline
\multicolumn{3}{c}{$\us\os1^{\fsp(1)^{(1)}}_{\left[\fso(13)^{(1)}\right]}-\os3^{\fg_2^{(1)}}-\os1^{\fsp(0)^{(1)}}-\os5^{\ff_4^{(1)}}- \os1^{\fsp(0)^{(1)}}-\os3^{\fg_2^{(1)}}-\os2^{\fsu(2)^{(1)}}-\os2^{\fsu(1)^{(1)}}-\os1^{\fsp(0)^{(1)}}-\os12^{\fe_8^{(1)}}$}
\\[10 pt]
\hline\hline
\rule[-10pt]{0pt}{30pt}
$3A_1$&$1$ &$(2, 2, 2), ((2,2), 2, 0), ((2, 2),0, 2), ((2, 2, 2), 0, 0)$ \\[3pt]
\hline
\multicolumn{3}{c}{$\us\os1^{\fsp(0)^{(1)}}_{\left[\ff_4^{(1)}\right]}-\us\os3^{\fg_2^{(1)}}_{\left[\fsp(1)^{(1)}\right]}-\os1^{\fsp(0)^{(1)}}-\os5^{\ff_4^{(1)}}-\os1^{\fsp(0)^{(1)}}-\os3^{\fg_2^{(1)}}-\os2^{\fsu(2)^{(1)}}-\os2^{\fsu(1)^{(1)}}-\os1^{\fsp(0)^{(1)}}-\os12^{\fe_8^{(1)}}$}
\\[10 pt]
\hline\hline
\rule[-10pt]{0pt}{30pt}
$A_2$&$1$ &$(3, 0, 0), (0, 3, 0)$ \\[3pt]
\hline
\multicolumn{3}{c}{$\us\os1^{\fsp(0)^{(1)}}_{\left[\fe_6^{(1)}\right]}-\os3^{\fsu(3)^{(1)}}-\os1^{\fsp(0)^{(1)}}-\os5^{\ff_4^{(1)}}- \os1^{\fsp(0)^{(1)}}-\os3^{\fg_2^{(1)}}-\os2^{\fsu(2)^{(1)}}-\os2^{\fsu(1)^{(1)}}-\os1^{\fsp(0)^{(1)}}-\os12^{\fe_8^{(1)}}$}
\\[10 pt]
\hline\hline
\rule[-10pt]{0pt}{30pt}
$A_2$&$Z_2$ &$((2, 2, 2), 0, 2)$ \\[3pt]
\hline
\multicolumn{3}{c}{$\us\os1^{\fsp(0)^{(1)}}_{\left[\fe_6^{(2)}\right]}-\os3^{\fsu(3)^{(2)}}-\os1^{\fsp(0)^{(1)}}-\os5^{\ff_4^{(1)}}- \os1^{\fsp(0)^{(1)}}-\os3^{\fg_2^{(1)}}-\os2^{\fsu(2)^{(1)}}-\os2^{\fsu(1)^{(1)}}-\os1^{\fsp(0)^{(1)}}-\os12^{\fe_8^{(1)}}$}
\\[10 pt]
\hline\hline
\rule[-10pt]{0pt}{30pt}
$4A_1$&$1$ &$((2, 2), 2, 2), ((2, 2, 2), 2, 0)$ \\[3pt]
\hline
\multicolumn{3}{c}{$\us\os2^{\fg_2^{(1)}}_{\left[\fsp(4)^{(1)}\right]}-\os1^{\fsp(0)^{(1)}}-\os5^{\ff_4^{(1)}}-\os1^{\fsp(0)^{(1)}}- \os3^{\fg_2^{(1)}}-\os2^{\fsu(2)^{(1)}}-\os2^{\fsu(1)^{(1)}}-\os1^{\fsp(0)^{(1)}}-\os12^{\fe_8^{(1)}}$}
\\[10 pt]
\hline\hline
\rule[-10pt]{0pt}{30pt}
$A_2 + A_1 $&$1$ &$(3, 2, 0), (3, 0, 2), (2, 3, 0), (0, 3, 2), ((3, 2), 0, 0)$ \\[3pt]
\hline
\multicolumn{3}{c}{$\us\os2^{\fsu(3)^{(1)}}_{\left[\fsu(6)^{(1)}\right]}-\os1^{\fsp(0)^{(1)}}-\os5^{\ff_4^{(1)}}-\os1^{\fsp(0)^{(1)}}- \os3^{\fg_2^{(1)}}-\os2^{\fsu(2)^{(1)}}-\os2^{\fsu(1)^{(1)}}-\os1^{\fsp(0)^{(1)}}-\os12^{\fe_8^{(1)}}$}
\\[10 pt]
\hline\hline
\rule[-10pt]{0pt}{30pt}
$A_2 + A_1 $&$Z_2$ &$((2, 2, 2), 2, 2)$ \\[3pt]
\hline
\multicolumn{3}{c}{$\us\os2^{\fsu(3)^{(2)}}_{\left[\fsu(6)^{(2)}\right]}-\os1^{\fsp(0)^{(1)}}-\os5^{\ff_4^{(1)}}-\os1^{\fsp(0)^{(1)}}- \os3^{\fg_2^{(1)}}-\os2^{\fsu(2)^{(1)}}-\os2^{\fsu(1)^{(1)}}-\os1^{\fsp(0)^{(1)}}-\os12^{\fe_8^{(1)}}$}
\\[10 pt]
\hline\hline
\rule[-10pt]{0pt}{30pt}
$A_2 + 2A_1 $&$1$ &$(3, 2, 2), ((3,2), 2, 0), ((3, 2), 0, 2)$ \\[3pt]
\hline
\multicolumn{3}{c}{$\us\os2^{\fsu(2)^{(1)}}_{\left[\fso(7)^{(1)}\right]}-\us\os1^{\fsp(0)^{(1)}}_{\left[\fsu(2)^{(1)}\right]}- \os5^{\ff_4^{(1)}}-\os1^{\fsp(0)^{(1)}}-\os3^{\fg_2^{(1)}}-\os2^{\fsu(2)^{(1)}}-\os2^{\fsu(1)^{(1)}}-\os1^{\fsp(0)^{(1)}}-\os12^{\fe_8^{(1)}}$}
\\[10 pt]
\hline\hline
\rule[-10pt]{0pt}{30pt}
$A_3$&$1$ &$(4, 0, 0)$ \\[3pt]
\hline
\multicolumn{3}{c}{$\us\os1^{\fsp(1)^{(1)}}_{\left[\fso(11)^{(1)}\right]}-\os4^{\fso(9)^{(1)}}-\os1^{\fsp(0)^{(1)}}-\os3^{\fg_2^{(1)}}-\os2^{\fsu(2)^{(1)}}-\os2^{\fsu(1)^{(1)}}-\os1^{\fsp(0)^{(1)}}-\os12^{\fe_8^{(1)}}$}
\\[10 pt]
\hline\hline
\rule[-10pt]{0pt}{30pt}
$A_2 + 3A_1$&$1$ &$((3, 2), 2, 2)$ \\[3pt]
\hline
\multicolumn{3}{c}{$\us\os2^{\fsu(1)^{(1)}}_{\left[\fsu(2)^{(1)}\right]}-\us\os1^{\fsp(0)^{(1)}}_{\left[\fg_2^{(1)}\right]}-\os5^{\ff_4^{(1)}}-\os1^{\fsp(0)^{(1)}}-\os3^{\fg_2^{(1)}}-\os2^{\fsu(2)^{(1)}}-\os2^{\fsu(1)^{(1)}}-\os1^{\fsp(0)^{(1)}}-\os12^{\fe_8^{(1)}}$}
\\[10 pt]
\hline\hline
\rule[-10pt]{0pt}{30pt}
$2A_2 $&$1$ &$(3, 3, 0)$ \\[3pt]
\hline
\multicolumn{3}{c}{$\us\os1^{\fsp(0)^{(1)}}_{\left[\fg_2^{(1)}\right]}-\us\os5^{\ff_4^{(1)}}_{\us|_{\text{\normalsize$\us\os1^{\fsp(0)^{(1)}}_{\left[\fg_2^{(1)}\right]}$}}}-\os1^{\fsp(0)^{(1)}}-\os3^{\fg_2^{(1)}}-\os2^{\fsu(2)^{(1)}}-\os2^{\fsu(1)^{(1)}}-\os1^{\fsp(0)^{(1)}}-\os12^{\fe_8^{(1)}}$}
\\[10 pt]
\hline\hline
\rule[-10pt]{0pt}{30pt}
$2A_2 $&$Z_2$ &$((3, 3), 0, 0)$ \\[3pt]
\hline
\multicolumn{3}{c}{$\us\os1^{\fsp(0)^{(1)}}_{\left[\fg_2^{(1)}\right]}\os\leftarrow^{2}\os5^{\ff_4^{(1)}}-\os1^{\fsp(0)^{(1)}}- \os3^{\fg_2^{(1)}}-\os2^{\fsu(2)^{(1)}}-\os2^{\fsu(1)^{(1)}}-\os1^{\fsp(0)^{(1)}}-\os12^{\fe_8^{(1)}}$}
\\[10 pt]
\hline\hline
\rule[-10pt]{0pt}{30pt}
$2A_2+A_1 $&$1$ &$(3, 3, 2), ((3, 3), 2, 0), ((3, 3), 0, 2), ((3, 2), 3, 0)$ \\[3pt]
\hline
\multicolumn{3}{c}{$\us\os1^{\fsp(0)^{(1)}}_{\left[\fg_2^{(1)}\right]}-\us\os4^{\ff_4^{(1)}}_{\left[\fsp(1)^{(1)}\right]}-\os1^{\fsp(0)^{(1)}}-\os3^{\fg_2^{(1)}}-\os2^{\fsu(2)^{(1)}}-\os2^{\fsu(1)^{(1)}}-\os1^{\fsp(0)^{(1)}}-\os12^{\fe_8^{(1)}}$}
\\[10 pt]
\hline\hline
\rule[-10pt]{0pt}{30pt}
$A_3+A_1 $&$1$ &$(4, 2, 0), (4, 0, 2), ((4, 2), 0, 0)$ \\[3pt]
\hline
\multicolumn{3}{c}{$\us\os1^{\fsp(0)^{(1)}}_{\left[\fso(7)^{(1)}\right]}-\us\os4^{\fso(9)^{(1)}}_{\left[\fsu(2)^{(1)}\right]}- \os1^{\fsp(0)^{(1)}}-\os3^{\fg_2^{(1)}}-\os2^{\fsu(2)^{(1)}}-\os2^{\fsu(1)^{(1)}}-\os1^{\fsp(0)^{(1)}}-\os12^{\fe_8^{(1)}}$}
\\[10 pt]
\hline\hline
\rule[-10pt]{0pt}{30pt}
$2A_2+2A_1 $&$1$ &$((3, 3), 2, 2), ((3, 2), 3, 2)$ \\[3pt]
\hline
\multicolumn{3}{c}{$\us\os3^{\ff_4^{(1)}}_{\left[\fsp(2)^{(1)}\right]}-\os1^{\fsp(0)^{(1)}}-\os3^{\fg_2^{(1)}}-\os2^{\fsu(2)^{(1)}}- \os2^{\fsu(1)^{(1)}}-\os1^{\fsp(0)^{(1)}}-\os12^{\fe_8^{(1)}}$}
\\[10 pt]
\hline\hline
\rule[-10pt]{0pt}{30pt}
$D_4(a_1) $&$Z_2$ &$((4, 2), 0, 2)$ \\[3pt]
\hline
\multicolumn{3}{c}{$\us\os1^{\fsp(0)^{(1)}}_{\left[\fso(8)^{(2)}\right]}-\os4^{\fso(8)^{(2)}}-\os1^{\fsp(0)^{(1)}}-\os3^{\fg_2^{(1)}}-\os2^{\fsu(2)^{(1)}}-\os2^{\fsu(1)^{(1)}}-\os1^{\fsp(0)^{(1)}}-\os12^{\fe_8^{(1)}}$}
\\[10 pt]
\hline\hline
\rule[-10pt]{0pt}{30pt}
$D_4(a_1) $&$Z_3$ &$((3, 3), 3, 0)$ \\[3pt]
\hline
\multicolumn{3}{c}{$\us\os1^{\fsp(0)^{(1)}}_{\left[\fso(8)^{(3)}\right]}-\os4^{\fso(8)^{(3)}}-\os1^{\fsp(0)^{(1)}}-\os3^{\fg_2^{(1)}}-\os2^{\fsu(2)^{(1)}}-\os2^{\fsu(1)^{(1)}}-\os1^{\fsp(0)^{(1)}}-\os12^{\fe_8^{(1)}}$}
\\[10 pt]
\hline\hline
\rule[-10pt]{0pt}{30pt}
$A_3+2A_1 $&$1$ &$(4, 2, 2), ((4, 2), 2, 0)$ \\[3pt]
\hline
\multicolumn{3}{c}{$\us\os3^{\fso(9)^{(1)}}_{\left[\fsp(2)^{(1)}\oplus \fsu(2)^{(1)}\right]}-\os1^{\fsp(0)^{(1)}}-\os3^{\fg_2^{(1)}}-\os2^{\fsu(2)^{(1)}}-\os2^{\fsu(1)^{(1)}}-\os1^{\fsp(0)^{(1)}}-\os12^{\fe_8^{(1)}}$}
\\[10 pt]
\hline\hline
\rule[-10pt]{0pt}{30pt}
$D_4(a_1) + A_1 $&$Z_2$ &$((4, 2), 2, 2)$ \\[3pt]
\hline
\multicolumn{3}{c}{$\us\os3^{\fso(8)^{(2)}}_{\left[\fsu(2)^{(1)}\oplus \fsu(2)^{(1)}\right]}-\os1^{\fsp(0)^{(1)}}-\os3^{\fg_2^{(1)}}-\os2^{\fsu(2)^{(1)}}-\os2^{\fsu(1)^{(1)}}-\os1^{\fsp(0)^{(1)}}-\os12^{\fe_8^{(1)}}$}
\\[10 pt]
\hline\hline
\rule[-10pt]{0pt}{30pt}
$D_4(a_1) + A_1 $&$Z_3$ &$((3, 3), 3, 2)$ \\[3pt]
\hline
\multicolumn{3}{c}{$\us\os3^{\fso(8)^{(3)}}_{\left[\fsu(2)^{(1)}\right]}-\os1^{\fsp(0)^{(1)}}-\os3^{\fg_2^{(1)}}-\os2^{\fsu(2)^{(1)}}- \os2^{\fsu(1)^{(1)}}-\os1^{\fsp(0)^{(1)}}-\os12^{\fe_8^{(1)}}$}
\\[10 pt]
\hline\hline
\rule[-10pt]{0pt}{30pt}
$A_3 + A_2 $&$1$ &$(4, 3, 0)$ \\[3pt]
\hline
\multicolumn{3}{c}{$\us\os3^{\fso(7)^{(1)}}_{\left[\fsp(2)^{(1)}\right]}\hdashrule[0.5ex]{0.5cm}{1pt}{0.5mm}\us\os1^{\fsp(0)^{(1)}}_{\left[\fso(2)^{(1)}\right]}-\os3^{\fg_2^{(1)}}-\os2^{\fsu(2)^{(1)}}- \os2^{\fsu(1)^{(1)}}-\os1^{\fsp(0)^{(1)}}-\os12^{\fe_8^{(1)}}$}
\\[10 pt]
\hline\hline
\rule[-10pt]{0pt}{30pt}
$A_4 $&$1$ &$(5, 0, 0)$ \\[3pt]
\hline
\multicolumn{3}{c}{$\us\os2^{\fsu(4)^{(1)}}_{\left[\fsu(5)^{(1)}\right]}-\os2^{\fsu(3)^{(1)}}-\os2^{\fsu(2)^{(1)}}-\os2^{\fsu(1)^{(1)}}- \os1^{\fsp(0)^{(1)}}-\os12^{\fe_8^{(1)}}$}
\\[10 pt]
\hline\hline
\rule[-10pt]{0pt}{30pt}
$A_3 + A_2 + A_1 $&$1$ &$(4, 3, 2), ((4, 2), 3, 0)$ \\[3pt]
\hline
\multicolumn{3}{c}{$\us\os3^{\fg_2^{(1)}}_{\left[\fsp(1)^{(1)}\right]}-\us\os1^{\fsp(0)^{(1)}}_{\left[\fsu(2)^{(1)}\right]}-\os3^{\fg_2^{(1)}}-\os2^{\fsu(2)^{(1)}}-\os2^{\fsu(1)^{(1)}}-\os1^{\fsp(0)^{(1)}}-\os12^{\fe_8^{(1)}}$}
\\[10 pt]
\hline\hline
\rule[-10pt]{0pt}{30pt}
$D_4(a_1) + A_2 $&$Z_2$ &$((4, 2), 3, 2)$ \\[3pt]
\hline
\multicolumn{3}{c}{$\os3^{\fsu(3)^{(2)}}-\us\os1^{\fsp(0)^{(1)}}_{\left[\fsu(3)^{(2)}\right]}-\os3^{\fg_2^{(1)}}-\os2^{\fsu(2)^{(1)}}- \os2^{\fsu(1)^{(1)}}-\os1^{\fsp(0)^{(1)}}-\os12^{\fe_8^{(1)}}$}
\\[10 pt]
\hline\hline
\rule[-10pt]{0pt}{30pt}
$A_4 +A_1$&$1$ &$(5, 2, 0), (5, 0, 2)$ \\[3pt]
\hline
\multicolumn{3}{c}{$
\begin{array}{c}
\us\os2^{\fsu(3)^{(1)}}_{\left[\fsu(3)^{(1)}\right]}-\us\os2^{\fsu(3)^{(1)}}_{\left[N_f = 1\right]}-\os2^{\fsu(2)^{(1)}}-\os2^{\fsu(1)^{(1)}}-\os1^{\fsp(0)^{(1)}}-\os12^{\fe_8^{(1)}}\\[10pt]
\text{flavor:}\quad \fsu(3)^{(1)} \oplus \fu(1)^{(1)}
\end{array}
$}
\\[10 pt]
\hline\hline
\rule[-10pt]{0pt}{30pt}
$A_4 +2A_1$&$1$ &$(5, 2, 2)$ \\[3pt]
\hline
\multicolumn{3}{c}{$
\begin{array}{c}
\us\os2^{\fsu(2)^{(1)}}_{\left[N_f = 1\right]}-\us\os2^{\fsu(3)^{(1)}}_{\left[\fsu(2)^{(1)}\right]}-\os2^{\fsu(2)^{(1)}}-\os2^{\fsu(1)^{(1)}}-\os1^{\fsp(0)^{(1)}}-\os12^{\fe_8^{(1)}}\\[10pt]
\text{flavor:}\quad \fsu(2)^{(1)} \oplus \fu(1)^{(1)}
\end{array}
$}
\\[10 pt]
\hline\hline
\rule[-10pt]{0pt}{30pt}
$A_4  + A_2$&$1$ &$(5, 3, 0)$ \\[3pt]
\hline
\multicolumn{3}{c}{$
\begin{array}{c}
\us\os2^{\fsu(2)^{(1)}}_{\left[N_f = 2\right]}-\os2^{\fsu(2)^{(1)}}-\us\os2^{\fsu(2)^{(1)}}_{\left[N_f = 1\right]}- \os2^{\fsu(1)^{(1)}}-\os1^{\fsp(0)^{(1)}}-\os12^{\fe_8^{(1)}}\\[10pt]
\text{flavor:}\quad \fsu(2)^{(1)} \oplus \fsu(2)^{(1)}
\end{array}
$}
\\[10 pt]
\hline\hline
\rule[-10pt]{0pt}{30pt}
$A_4  + A_2 + A_1$&$1$ &$(5, 3, 2)$ \\[3pt]
\hline
\multicolumn{3}{c}{$
\begin{array}{c}
\os2^{\fsu(1)^{(1)}}-\us\os2^{\fsu(2)^{(1)}}_{\left[N_f = 1\right]}-\us\os2^{\fsu(2)^{(1)}}_{\left[N_f = 1\right]}- \os2^{\fsu(1)^{(1)}}-\os1^{\fsp(0)^{(1)}}-\os12^{\fe_8^{(1)}}\\[10pt]
\text{flavor:}\quad \fsu(2)^{(1)}
\end{array}
$}
\\[10 pt]
\hline\hline
\rule[-10pt]{0pt}{30pt}
$A_5$&$1$ &$(6, 0, 0)$ \\[3pt]
\hline
\multicolumn{3}{c}{$\us\os2^{\fsu(2)^{(1)}}_{\left[\fg_2^{(1)}\right]}-\os2^{\fsu(1)^{(1)}}-\os1^{\fsp(0)^{(1)}}-\os12^{\fe_8^{(1)}}-\os1^{\fsp(0)^{(1)}}-\us\os2^{\fsu(1)^{(1)}}_{\left[\fsu(2)^{(1)}\right]}$}
\\[10 pt]
\hline\hline
\rule[-10pt]{0pt}{30pt}
$E_6(a_3)$&$Z_2$ &$(6, 0, 2)$ \\[3pt]
\hline
\multicolumn{3}{c}{$\us\os2^{\fsu(2)^{(1)}}_{\left[\fg_2^{(1)}\right]}-\os2^{\fsu(1)^{(1)}}-\os1^{\fsp(0)^{(1)}}-\os12^{\fe_8^{(1)}}\os\rightarrow^2\os1^{\fsp(0)^{(1)}}$}
\\[10 pt]
\hline\hline
\rule[-10pt]{0pt}{30pt}
$A_5 + A_1$&$1$ &$(6, 2, 0)$ \\[3pt]
\hline
\multicolumn{3}{c}{$\us\os2^{\fsu(1)^{(1)}}_{\left[\fsu(2)^{(1)}\right]}-\os2^{\fsu(1)^{(1)}}-\os1^{\fsp(0)^{(1)}}-\os12^{\fe_8^{(1)}}-\os1^{\fsp(0)^{(1)}}-\us\os2^{\fsu(1)^{(1)}}_{\left[\fsu(2)^{(1)}\right]}$}
\\[10 pt]
\hline\hline
\rule[-10pt]{0pt}{30pt}
$E_6(a_3) + A_1$&$Z_2$ &$(6, 2, 2)$ \\[3pt]
\hline
\multicolumn{3}{c}{$\us\os2^{\fsu(1)^{(1)}}_{\left[\fsu(2)^{(1)}\right]}-\os2^{\fsu(1)^{(1)}}-\os1^{\fsp(0)^{(1)}}-\os12^{\fe_8^{(1)}}\os\rightarrow^2\os1^{\fsp(0)^{(1)}}$}
\\[10 pt]
\hline\hline
\rule[-10pt]{0pt}{30pt}
$E_7(a_5) $&$Z_3$ &$(6, 3, 0)$ \\[3pt]
\hline
\multicolumn{3}{c}{$\us\os2^{\fsu(1)^{(1)}}_{\left[\fsu(2)^{(1)}\right]}-\os1^{\fsp(0)^{(1)}}-\os12^{\fe_8^{(1)}}\os\rightarrow^3\os1^{\fsp(0)^{(1)}}$}
\\[10 pt]
\hline\hline
\rule[-10pt]{0pt}{30pt}
$E_8(a_7) $&$Z_6$ &$(6, 3, 2)$ \\[3pt]
\hline
\multicolumn{3}{c}{$\os1^{\fsp(0)^{(1)}}\os\leftarrow^2\os12^{\fe_8^{(1)}}\os\rightarrow^3\os1^{\fsp(0)^{(1)}}$}
\\[3 pt]
\hline\hline
\end{longtable}
\end{center}

The Higgsings of the theory $(E_8, \underline{E_8})$ on a cirlce in Table \ref{tb:e8_1} exhibits new cases. For example let us consider the Higgsing $(6, 0, 2)$. The Higgsing $(6, 0, 2)$ of the theory $(E_8, \underline{E_8})$ without a circle compactification gives rise to 
\be\label{602e8}
\us\os2^{\fsu(2)}_{\left[\fg_2\right]}-\os2^{\fsu(1)}-\os1^{\fsp(0)}-\us\os10^{\fe_8}_{\left[n_{\text{inst}}=2\right]}.
\ee
The theory \eqref{602e8} includes a node with the $\fe_8$ gauge algebra on a $(-10)$-curve and $n_{\text{inst}} = 2$ is the number of small instantons. In general the number of small instantons for a node with the $\fe_8$ algebra is given by
\be 
\us\os{n}^{\fe_8}_{\left[n_{\text{inst}}=12-n\right]}, 
\ee
for $1 \leq n \leq 12$. We can move into a tensor branch of the theory and then $n_{\text{inst}}$ E-string theories are attached to the $\fe_8$ node. For example, on a tensor branch of the theory \eqref{602e8} we have 
\be\label{602e8v2}
\us\os2^{\fsu(2)}_{\left[\fg_2\right]}-\os2^{\fsu(1)}-\os1^{\fsp(0)}-\us\os12^{\fe_8}_{\us|_{\text{\normalsize$\os1^{\fsp(0)}$}}}-\os1^{\fsp(0)}.
\ee
A circle compactification of the theory \eqref{602e8v2} gives 
\be\label{602e8v3}
\us\os2^{\fsu(2)^{(1)}}_{\left[\fg_2^{(1)}\right]}-\os2^{\fsu(1)^{(1)}}-\os1^{\fsp(0)^{(1)}}-\us\os12^{\fe_8^{(1)}}_{\us|_{\text{\normalsize$\os1^{\fsp(0)^{(1)}}$}}}-\os1^{\fsp(0)^{(1)}},
\ee
which has $1 + 8 + 6 = 15$ Coulomb branch moduli and $2 + 1 = 3$ mass prameters in 5d. On the other hand the web diagram after applying the Higgsing $(6, 0, 2)$ gives a theory with $14$ Coulomb branch moduli and $3$ mass parameters. Note that the theory \eqref{602e8v2} has a $Z_2$ symmetry which exchanges two E-string theories which are only attached to the $(-12)$-curve. Hence it is possible to consider a circle compactification of \eqref{602e8v2} with the $Z_2$ twist, which gives
\be\label{602e8v4}
\us\os2^{\fsu(2)^{(1)}}_{\left[\fg_2^{(1)}\right]}-\os2^{\fsu(1)^{(1)}}-\os1^{\fsp(0)^{(1)}}-\os12^{\fe_8^{(1)}}\os\rightarrow^2\os1^{\fsp(0)^{(1)}}.
\ee
The theory has $14$ Coulomb branch moduli and $3$ mass parameters and the numbers agree with those from the Higgsed web diagram. Therefore we argue that the $(6, 0, 2)$ Higgsing of the theory $(E_8, \underline{E_8})$ on $S^1$ gives rise to \eqref{602e8v4}. The theories after the Higgsings of $(6, 2, 2)$ and $(6, 3, 0)$ can be determined in a similar way. 

In the case of the $(6, 3, 2)$ Higgsing the 6d Higgsings gives rise to 
\be\label{632e8v1}
\us\os{7}^{\fe_8}_{\left[n_{\text{inst} = 5}\right]}. 
\ee
On a tensor branch of the theory \eqref{632e8v1} five E-strings theories are attached to a $(-12)$-curve with the $\fe_8$ gauge algebra. Then there are several possible twists which exchange some of the E-strings. Since the $(6, 3, 0)$ Higgsing gives the $Z_3$ twist and the $(6, 0, 2)$ Higgsing gives the $Z_2$ twist we expect that the $(6, 3, 2)$ yields the combintation of the twists which becomes $Z_6$. Hence we propose that the theory after the $(6, 3, 2)$ Higgsing gives 
\be\label{23twistede8}
\os1^{\fsp(0)^{(1)}}\os\leftarrow^2\os12^{\fe_8^{(1)}}\os\rightarrow^3\os1^{\fsp(0)^{(1)}}.
\ee

\begin{figure}[t]
\centering
\includegraphics[width=7cm]{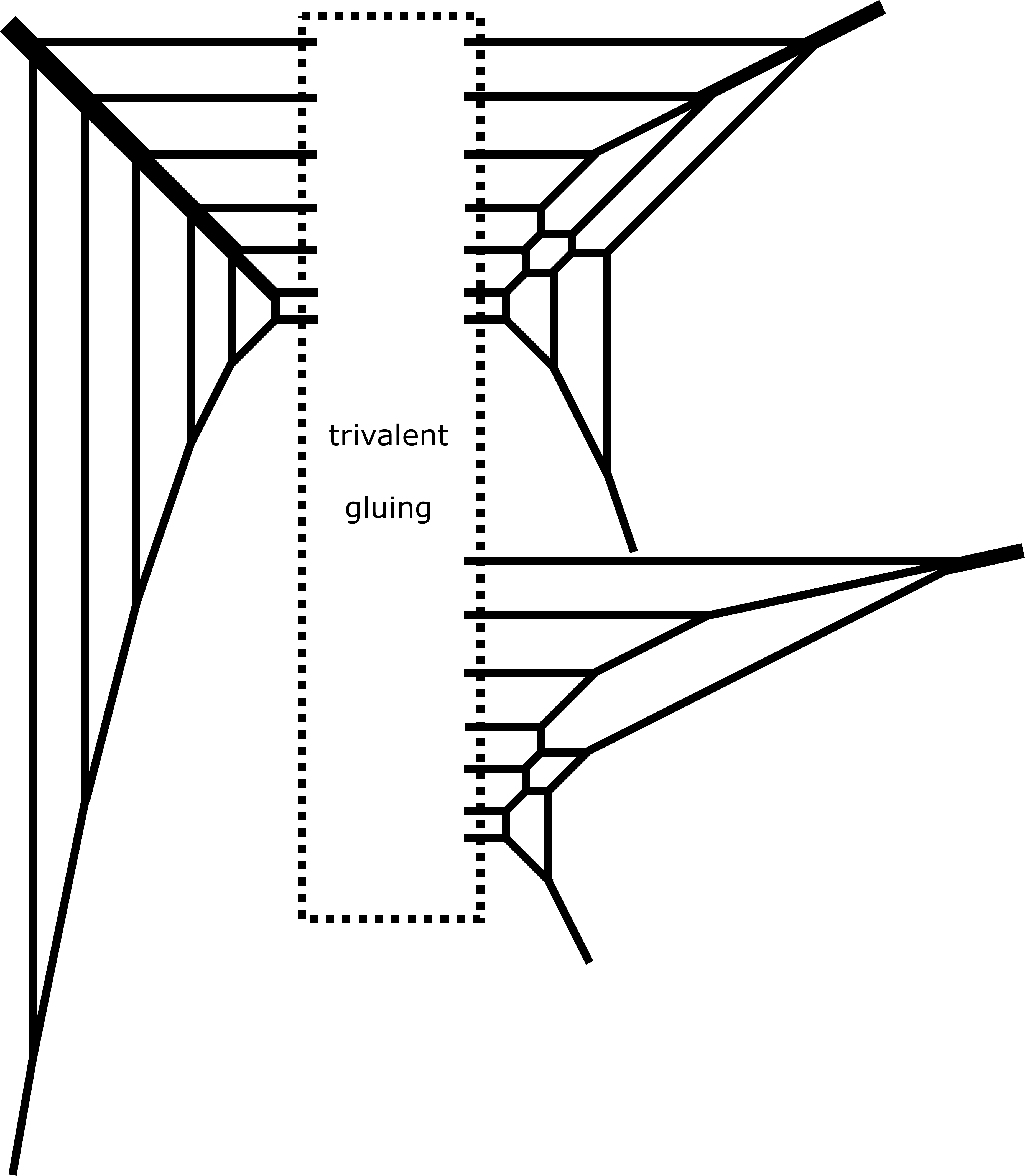}
\caption{The web diagram of the theory obtained after the Higgsing $(6, 3, 2)$.
}
\label{fig:e8632}
\end{figure}

\begin{figure}[t]
\centering
\includegraphics[width=8cm]{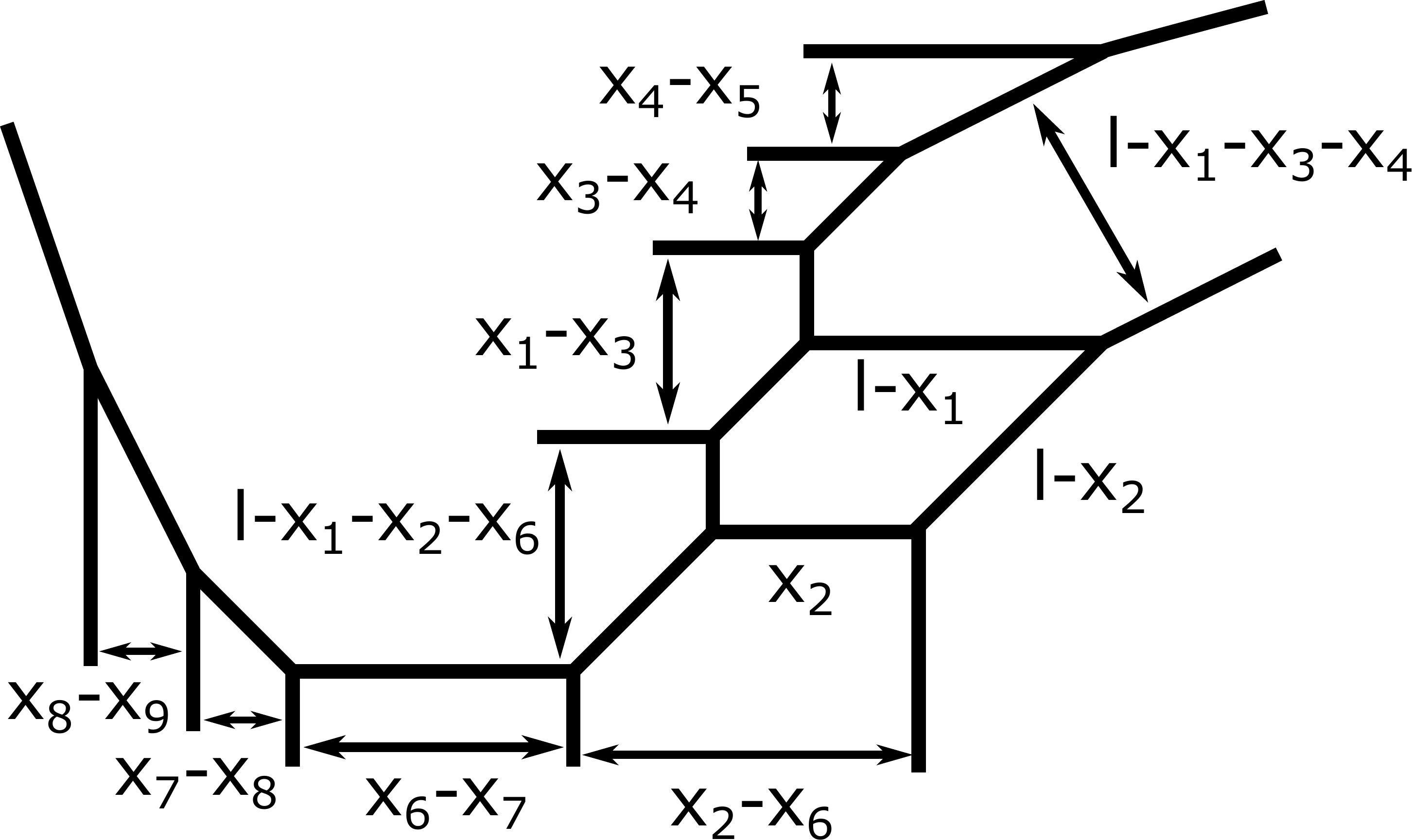}
\caption{The web diagram of dP$_9$ which corresponds to the E-string 
whose compact surface is contained in the lower-right diagram in Figure \ref{fig:e8632}. 
$l$ is the hyperplane class of $\mathbb{P}^2$ and $x_i\; (i=1, \cdots, 9)$ are the exceptional curve classes. A curve class next to a double arrow implies that the length of the double arrow is related to the volume of the curve class. }
\label{fig:e8lestring}
\end{figure}

We can give support for the proposal \eqref{23twistede8} by examining the web diagram of the Higgsed theory. The web diagram after applying the Higgsing $(6, 3, 2)$ to the theory $(E_8, \underline{E_8})$ on $S^1$ is drawn in Figure \ref{fig:e8632}. 
In order to see if the E-strings are connected to the $(-12)$-curve as in \eqref{23twistede8}, let us consider how the fiber \eqref{fiberI} on each base curve are related to each other. In the web diagram in Figure \ref{fig:e8632}, we have $10$ faces for the $10$ Coulomb branch moduli. Among the $10$ faces one in the upper-right diagram and another one in the lower-right diagram in Figure \ref{fig:e8632} are the two faces for the two E-string theories. We first focus on the E-string whose face is in the lower-right diagram in Figure \ref{fig:e8632}. By performing some flop transitions, it is possible to extract the web diagram for the E-string and the web diagram is depicted in Figure \ref{fig:e8lestring}.
The web diagram corresponds to a dP$_9$ surface and the torus fiber of the surface is given by
\be
f_{\fsp(0)^{(1)}_{lr}} = 3l - \sum_{i=1}^9x_i. 
\ee
The nine $(-2)$-curves written next to the double arrows in Figure \ref{fig:e8lestring} are related to fibers which form the affine $E_8$ Dynkin diagram and the fiber \eqref{fiberI} for the $\fe_{8}^{(1)}$ algebra is given by
\begin{equation}
\begin{split}
f_{\fe_8^{(1)}} &= (l-x_1 -x_3 - x_4) + 2(x_4 - x_5) + 3(x_3 - x_4) + 4(x_1 - x_3) \cr
&\hspace{2cm} + 5(l - x_1 - x_2 - x_6)+ 6(x_6 - x_7) + 4(x_7 - x_8) + 2(x_8 - x_9) + 3(x_2 - x_6)\cr
&=2\left(3l - \sum_{i=1}^9x_i\right).
\end{split}
\end{equation}
Hence we have the relation 
\be\label{sp0e81}
2f_{\fsp(0)^{(1)}_{lr}} \sim f_{\fe_8^{(1)}}. 
\ee

\begin{figure}[t]
\centering
\includegraphics[width=8cm]{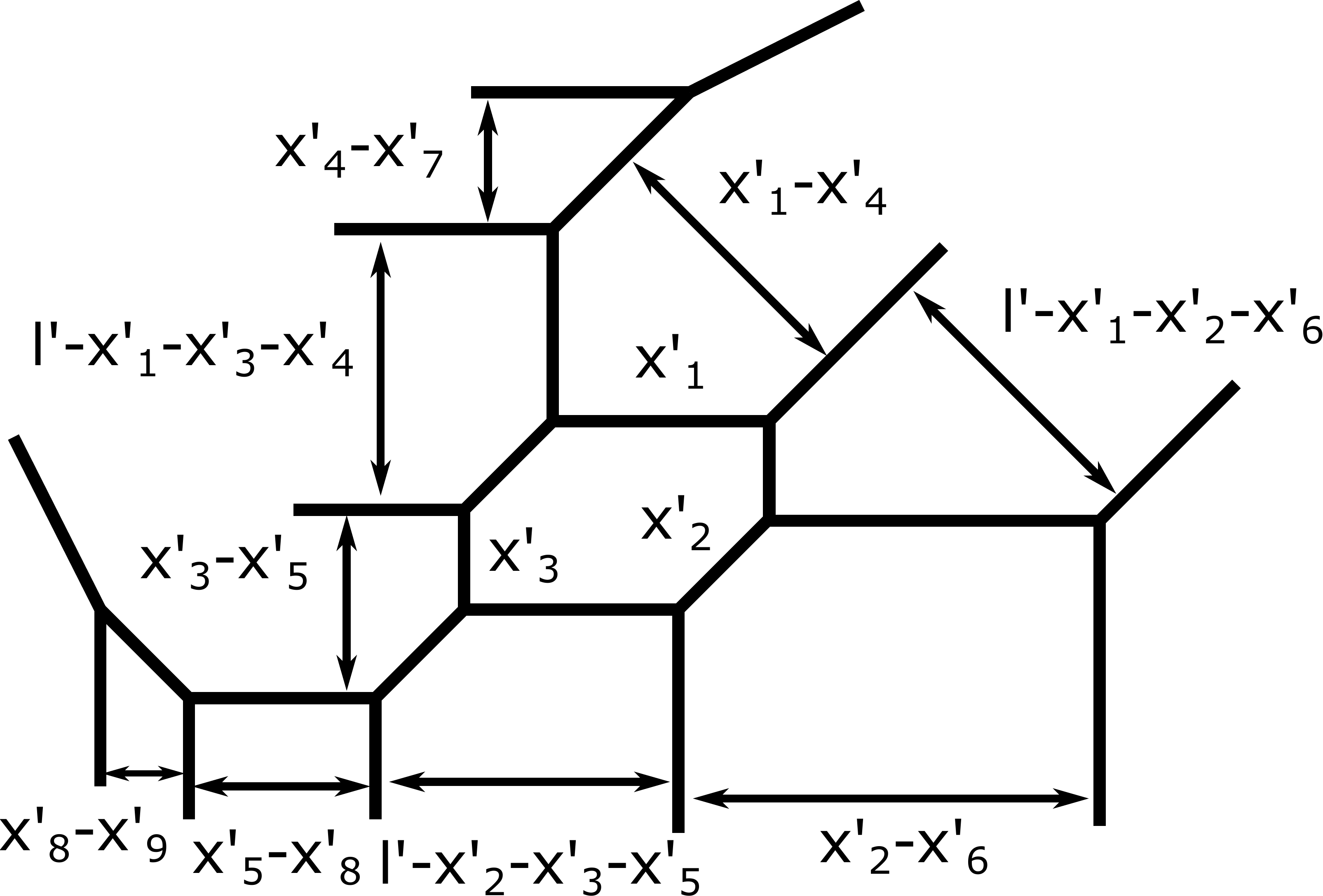}
\caption{The web diagram of dP$_9$ which corresponds to the E-string 
whose compact surface is contained in the upper-right diagram in Figure \ref{fig:e8632}. 
The notation for the curve classes is the same as the one in Figure \ref{fig:e8lestring}. The curve classes here are represented with a prime mark. }
\label{fig:e8restring}
\end{figure}

We then consider the E-string whose face is included in the upper-right diagram in Figure \ref{fig:e8632}. The web diagram for the E-string theory extracted from the diagram in Figure \ref{fig:e8632} is drawn in Figure \ref{fig:e8restring}. 
The torus fiber class of this dP$_9$ is again given by
\be
f_{\fsp(0)^{(1)}_{ur}} = 3l' - \sum_{i=1}^9x'_i. 
\ee
On the other hand, the nine $(-2)$-curves written next to the double arrows in Figure \ref{fig:e8restring} are fibers which form the affine $E_8$ Dynkin diagram of the $\fe_8^{(1)}$. The fiber \eqref{fiberI} of the $\fe_8^{(1)}$ gauge algebra then becomes
\begin{equation}
\begin{split}
f_{\fe_8^{(1)}} &= (l'-x'_1 - x'_2 - x'_6) + 2(x'_1 - x'_4) + 3(x'_4 - x'_7) + 4(l' - x'_1 - x'_3 - x'_4)\cr
&\hspace{1cm} + 5(x'_3 - x'_5) + 6(x'_5 - x'_8) + 4(l' - x'_2 - x'_3 - x'_5) + 2(x'_2 - x'_6) + 3(x'_8 - x'_9)\cr
&=3\left(3l' - \sum_{i=1}^9x'_i\right).
\end{split}
\end{equation}
Therefore we obtain
\be\label{sp0e82}
3f_{\fsp(0)^{(1)}_{ur}} \sim f_{\fe_8^{(1)}}.
\ee

From \eqref{sp0e81} and \eqref{sp0e82}, the fiber \eqref{fiberI} on each base curve is glued by
\be\label{e8fiberglue}
2f_{\fsp(0)^{(1)}_{lr}} \sim f_{\fe_8^{(1)}} \sim 3f_{\fsp(0)^{(1)}_{ur}} .
\ee
The relation is \eqref{e8fiberglue} is completely consistent with the gluing rule \eqref{gluingrule} for the theory \eqref{23twistede8}. The E-string with the fiber $f_{\fsp(0)^{(1)}_{lr}}$ is identified with the E-string on the left in \eqref{23twistede8} and the E-string with the fiber $f_{\fsp(0)^{(1)}_{ur}}$ is identified with the E-string on the right in \eqref{23twistede8}. This identifications are indeed expected. The $(6, 3, 0)$ Higgsing yields the theory 
\be\label{63twistede8}
\us\os2^{\fsu(1)^{(1)}}_{\left[\fsu(2)^{(1)}\right]}-\os1^{\fsp(0)^{(1)}}-\os12^{\fe_8^{(1)}}\os\rightarrow^3\os1^{\fsp(0)^{(1)}},
\ee
and the rightmost E-string in \eqref{63twistede8} comes from the upper-right diagram in Figure \ref{fig:ge8e8}. The $(6, 0, 2)$ Higgsing gives 
\be\label{62twistede8}
\us\os2^{\fsu(2)^{(1)}}_{\left[\fg_2^{(1)}\right]}-\os2^{\fsu(1)^{(1)}}-\os1^{\fsp(0)^{(1)}}-\os12^{\fe_8^{(1)}}\os\rightarrow^2\os1^{\fsp(0)^{(1)}},
\ee
and the rightmost E-string in \eqref{62twistede8} comes from the lower-right diagram in Figure \ref{fig:ge8e8}. These are the identifications which follows from \eqref{e8fiberglue}.

We can also consider Higgsings of the theory $(E_8, E_8)_2$ on a circle. When we focus on Higgsings of the flavor symmetry $\left[\SU(6) \times \SU(3) \times \SU(2)\right]^2$ which can be explicitly seen from the web diagram for $(E_8, E_8)_2$ on $S^1$, the Higgsings can be labeled by $[(a_1, a_2, a_3), (b_1, b_2, b_3)]$ where $(a_1, a_2, a_3)$ is associated to the Higgsings of one $\SU(6) \times \SU(3) \times \SU(2)$ and $(b_1, b_2, b_3)$ is associated to the Higgsings of the other $\SU(6) \times \SU(3) \times \SU(2)$. Note that the Higgsing $(a_1, a_2, a_3)$ of $(E_8, \underline{E_8})$, which is summarized in Table \ref{tb:e8_1}, does not Higgs the $\fe_8$ algebra. Hence the theory after applying the Higgsing $[(a_1, a_2, a_3), (b_1, b_2, b_3)]$ to the theory $(E_8, E_8)_2$ on a circle can be obtained by combining the result in Table \ref{tb:e8_1}. Namely if the Higgsing $[(a_1, a_2, a_3)]$ of $(E_8, \underline{E_8})$ on $S^1$ gives
\be
\cdots \os{n_3}^{\fg_3}-\os{n_2}^{\fg_2}-\os{(12-n_1)}^{\fe_8^{(1)}},
\ee
and the Higgsing $[(b_1, b_2, b_3)]$ of $(E_8, \underline{E_8})$ on $S^1$ gives
\be
\cdots \os{m_3}^{\fg_3}-\os{m_2}^{\fg_2}-\os{(12-m_1)}^{\fe_8^{(1)}},
\ee
then the theory after applying the Higgsing $[(a_1, a_2, a_3), (b_1, b_2, b_3)]$ to the theory $(E_8, E_8)_2$ on a circle becomes
\begin{align}
&\left[\cdots \os{n_3}^{\fg_3}-\os{n_2}^{\fg_2}-\os{\left(12-n_1\right)}^{\fe_8^{(1)}}\right] + \left[\os{\left(12-m_1\right)}^{\fe_8^{(1)}}-\os{m_2}^{\fg'_2}-\os{m_3}^{\fg'_3}\cdots \right]\cr
&\to \cdots \os{n_3}^{\fg_3}-\os{n_2}^{\fg_2}-\os{\left(12-n_1-m_1\right)}^{\fe_8^{(1)}}-\os{m_2}^{\fg'_2}-\os{m_3}^{\fg'_3}\cdots.\label{e8combine}
\end{align}
When $12 - n_1 - m_1 < 12$, then the node with $\fe_8^{(1)}$ is associated with $n_1 + m_1$ small instantons. On a tensor branch E-string theories are attached to the $\fe_8^{(1)}$ node and how they are attached also follows from the corresponding Higgsings in Table \ref{tb:e8_1}.

\bigskip
\section{Topological vertex formalism}
\label{sec:top}
In this appendix, we review some formulae of topological vertex which we have used for computing the partition functions of some 6d or 5d theories in section \ref{sec:PF}. 

\subsection{Topological vertex}
\label{sec:vertex}
Given a 5-brane web diagram which is dual to a toric Calabi-Yau threefold, it is possible to compute the Nekrasov partition function of a 5d theory realized on the web diagram using the topological vertex \cite{Iqbal:2002we, Aganagic:2003db, Iqbal:2003ix, Iqbal:2003zz, Eguchi:2003sj, Hollowood:2003cv}. For that, we decompose a 5-brane diagram into vertices with three lines. A Young diagram is assigned to each line and we assign a function called the topological vertex to each vertex of a web by
\begin{align}\label{topvertex}
\begin{tikzpicture}
\draw[->] (0,0)--(0,-1);
\draw[->] (0,0)--(-1,0.7);
\draw[->] (0,0)--(1,0);
\node at (-1, 0.7) [below] {$\mu$};
\node at (0,-1) [left] {$\lambda$};
\node at (1,0) [below] {$\nu$};
\node at (1.5,0) {$:$};
\node at (2, 0) [right] {$C_{\lambda\mu\nu} = q^{\frac{-||\mu^t||^2 + ||\mu||^2 + ||\nu||^2}{2}}\tilde{Z}_{\nu}(q)\sum_{\eta}s_{\lambda^t/\eta}(q^{-\rho -\nu})s_{\mu/\eta}(q^{-\nu^t-\rho}),$};
\end{tikzpicture}
\end{align}
where
\begin{align}
||\lambda||^2 &= \sum_{i}\lambda_i^2\qquad \text{for}\quad \lambda = (\lambda_1, \lambda_2, \cdots),\\
\tilde{Z}_{\nu}(q) &= \prod_{(i,j) \in \nu}\left(1 - q^{l_{\nu}(i, j) + a_{\nu}(i, j) + 1}\right)^{-1},
\end{align}
and we also defined
\be
l_{\nu}(i, j) = \nu_i - j, \qquad , a_{\nu}(i, j) = \nu_j^t - i.
\ee
$s_{\lambda/\eta}(x)$ is the skew Schur function. When the argument of a skew Schur function is $q^{-\rho - \nu}$, it is defined as 
\be
s_{\lambda/\eta}(q^{-\rho -\nu})=s_{\lambda/\eta}\left(q^{\frac{1}{2} - \nu_1}, q^{\frac{3}{2} - \nu_2}, \cdots, q^{\frac{L}{2} -\nu_L}, q^{\frac{L+1}{2}}, \cdots\right),
\ee
for $\nu = (\nu_1, \nu_2, \cdots, \nu_L)$. When the direction of an arrow of \eqref{topvertex} is the opposite then the corresponding Young diagram is transposed. For external lines we assign empty Young diagrams. When we glue two vertices along a line with and a Young diagram $\nu$,
\begin{align}\label{framing}
\begin{tikzpicture}
\draw[->] (0,0)--(0,-1);
\draw[->] (0,0)--(-1,0.7);
\draw[->] (0,0)--(1,0);
\node at (0,-1) [right] {$v_1$};
\node at (1,0) [below] {$\nu$};
\draw (1, 0)--(2,0);
\draw[->] (2,0)--(2,-1);
\draw[->] (2,0)--(3,0.7);
\node at (3, 0.7) [right] {$v_2$};
\draw[<->] (0,0.25)--(2,0.25);
\node at (1, 0.5) {$\ell$};
\end{tikzpicture}
\end{align}
we sum over the Young diagram $\nu$ with a weight,
\begin{align}
(-e^{-\ell})^{|\nu|}f_{\nu}(q)^{\det(v_1, v_2)},
\end{align}
where
\be
f_{\nu}(q) = (-1)^{|\nu|}q^{\frac{||\nu^t||^2 - ||\nu||^2}{2}}.
\ee
$v_1, v_2$ are $(p, q)$-charges for the 5-branes next to $v_1, v_2$ in \eqref{framing} and $\ell$ is the length of the glued line. The topological string partition function is obtained by summing over all the Young diagrams for the product of the topological vertices assigned to each vertex and the framing factors together with the K\"ahler parameters assigned to each internal line. 

When a 5-brane web diagram contains parallel external lines then we need to further subtract some factors from the topological string partition function computed for obtaining the Nekrasov partition function for the 5d theory realized on the 5-brane web. When a part of the 5-brane is given by \eqref{framing} where the vertical parallel lines are external lines, then the extra factor associated to the parallel external legs is given by
\begin{align}
\prod_{i, j=1}^{\infty}\left(1 - Qq^{i+j-1}\right)^{-1} = \text{PE}\left[\frac{qQ}{(1-q)^2}\right],
\end{align} 
where $\text{PE}(f(x_1, x_2, \cdots))$ stands for the Plethystic exponential,
\be
\text{PE}(x) = \text{Ext}\left(\sum_{n=1}^{\infty}\frac{f(x_1^n, x_2^n, \cdots)}{n}\right).\label{PE}
\ee
Let the topological string partition function be $Z_{\text{top}}$ obtained by applying the topological vertex to a diagram and let the product of all the extra factors be $Z_{\text{extra}}$, then the Nekrasov parititon function is given by
\be
\tilde{Z}_{\text{Nek}} = \frac{Z_{\text{top}}}{Z_{\text{extra}}}.\label{Nek}
\ee
Note that the Nekrasov partition function computed by the topological vertex does not contain the perturbative contribution of vector multiplets in the Cartan subalgebra. Such a factor has a universal form given by
\be
Z_{\text{Cartan}} 
 = \text{PE}\left[\frac{\text{rank}(G)q}{(1-q)^2}\right], \label{cartan}
\ee
where $\text{rank}(G)$ is the dimension of the Coulomb branch moduli space. The Nekrasov partition partition function is then given by the product of \eqref{Nek} and \eqref{cartan}. 

\begin{figure}[t]
\centering
\subfigure[]{\label{fig:nontoric1}
\includegraphics[width=5cm]{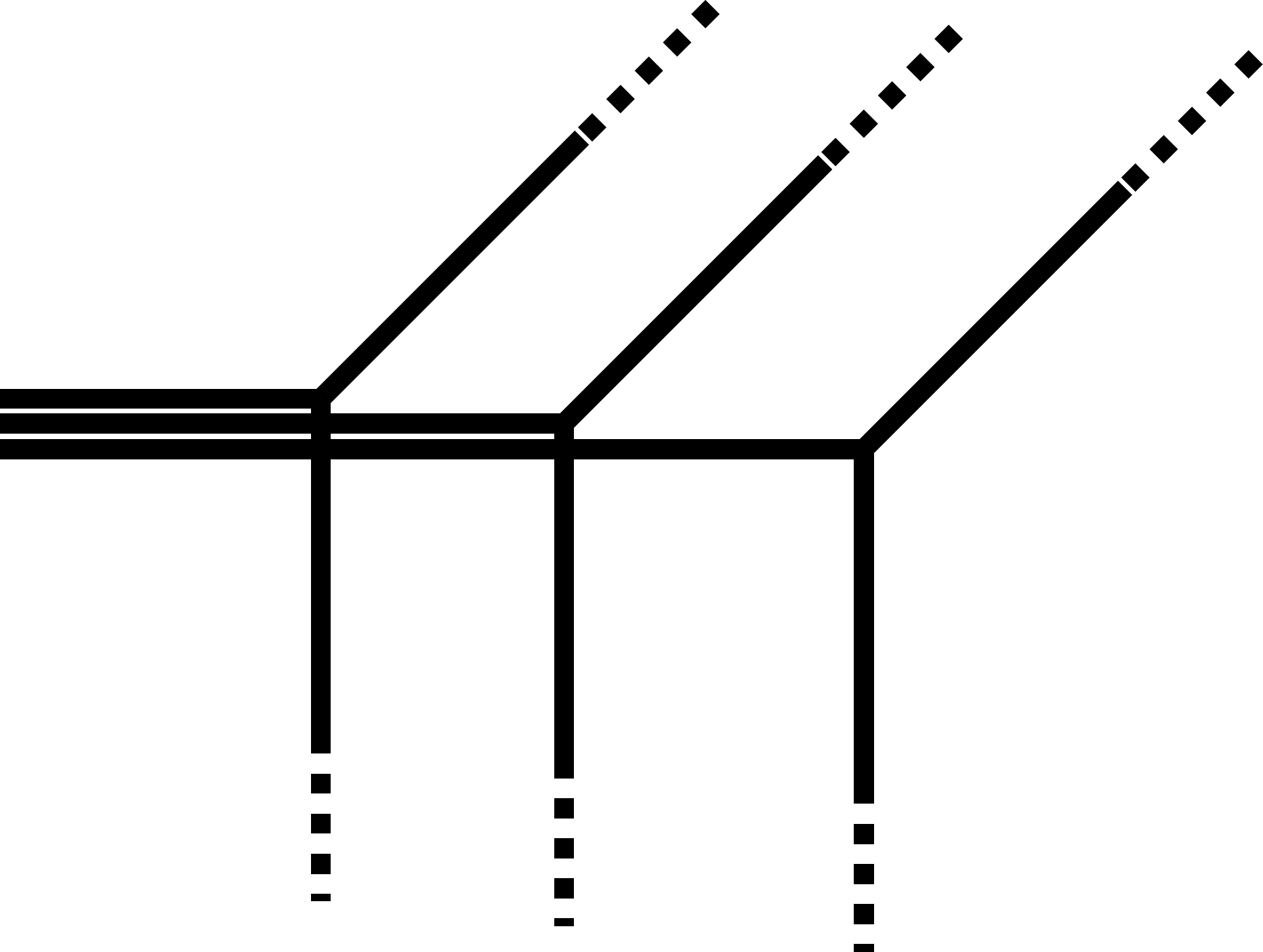}}
\hspace{1cm}
\subfigure[]{\label{fig:nontoric2}
\includegraphics[width=5cm]{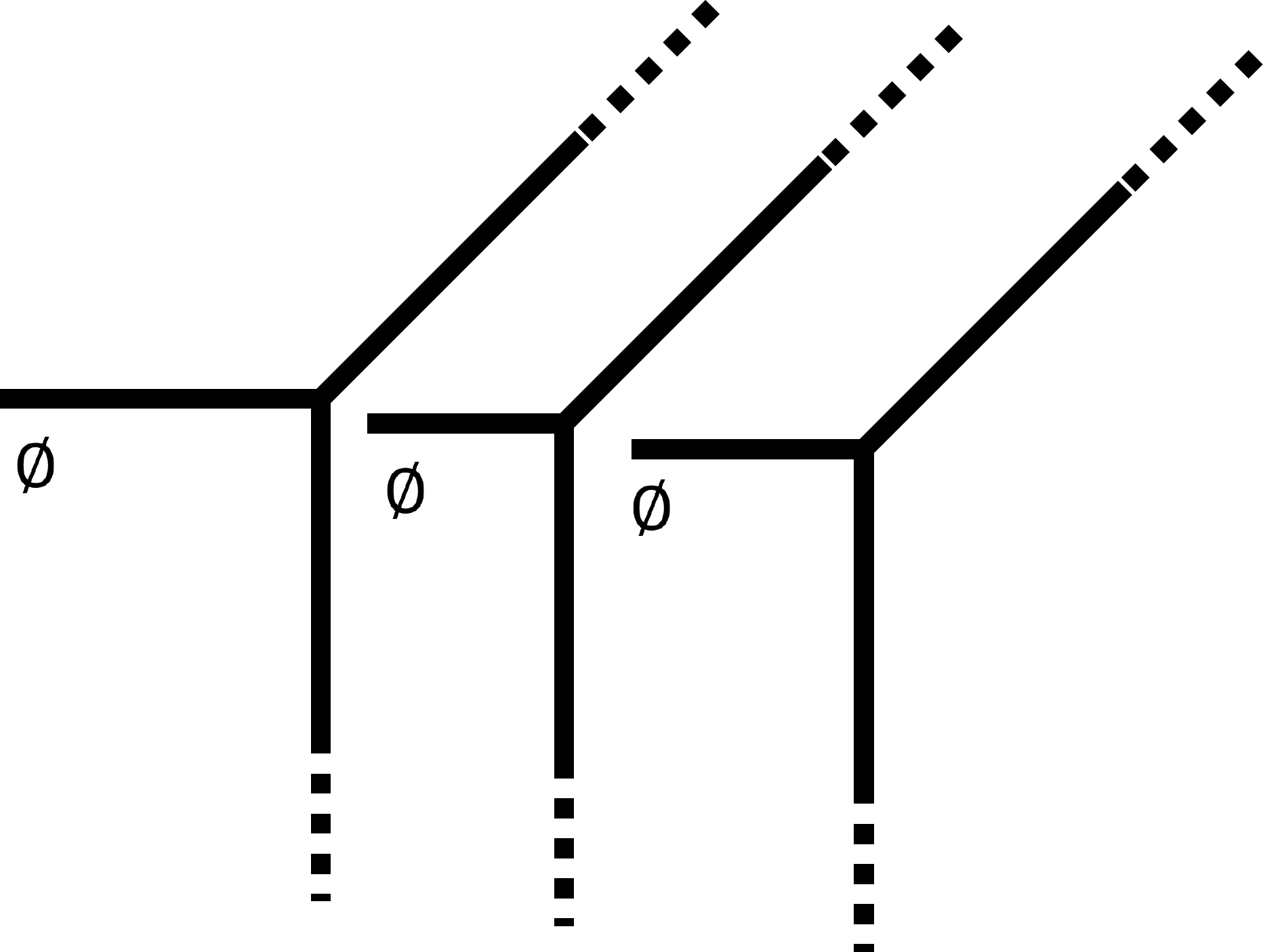}}
\caption{(a): A part of a non-toric web diagram. (b): The diagram which we used for applyling the topological vertex to the diagram in Figure \ref{fig:nontoric1}.}
\label{fig:nontoricapp}
\end{figure}
In fact it is also possible to apply the topological vertex to certain non-toric Calabi-Yau threefolds \cite{Hayashi:2013qwa, Hayashi:2014wfa, Kim:2015jba, Hayashi:2015xla}. Consider a diagram in Figure \ref{fig:nontoric1} where the three horizontal lines are at the same height and all of them are external lines. When we introduce 7-branes at each end of 5-branes, the three horizontal 5-branes are put on a single 7-brane. In this case some of the horizontal 5-branes jump over other 5-branes so that the configuration satisfies the s-rule. For applying the topological vertex to the diagram in Figure \ref{fig:nontoric1}, it turns out that we can use the diagram in Figure \ref{fig:nontoric2} where empty Young diagrams are assigned to the horizontal lines. Similarly when several $(p, q)$ 5-branes are put on a single 7-brane, we can use the same trick by assigning empty Young diagrams on each of the $(p, q)$ 5-brane.  

\subsection{Trivalent gluing prescription}
\label{sec:trivalent}

\begin{figure}[t]
\centering
\includegraphics[width=6cm]{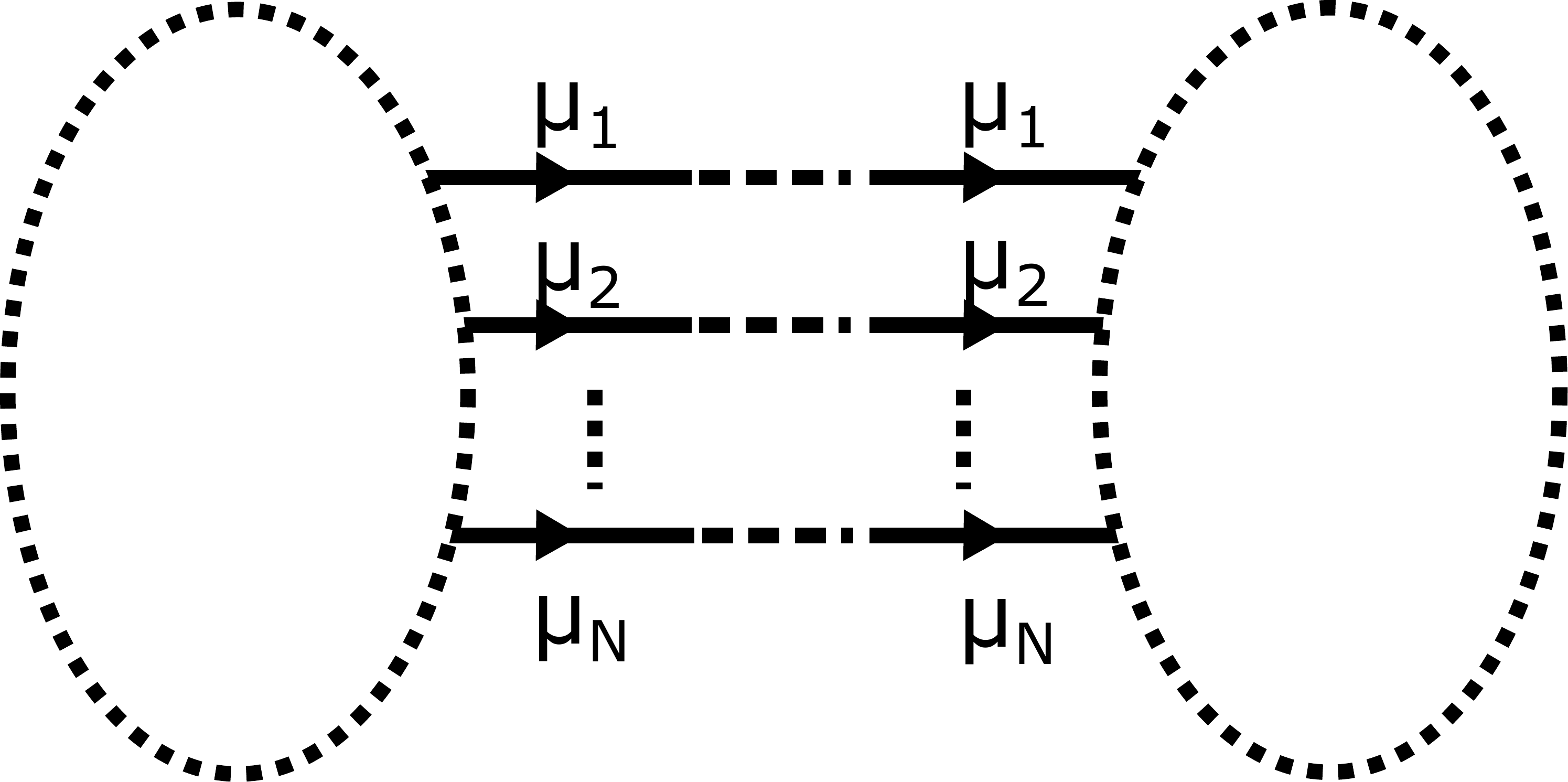}
\caption{The diagram of the usual $\SU(N)$ gauging.}
\label{fig:gluing1}
\end{figure}
\begin{figure}[t]
\centering
\subfigure[]{\label{fig:gluing2}
\includegraphics[width=6cm]{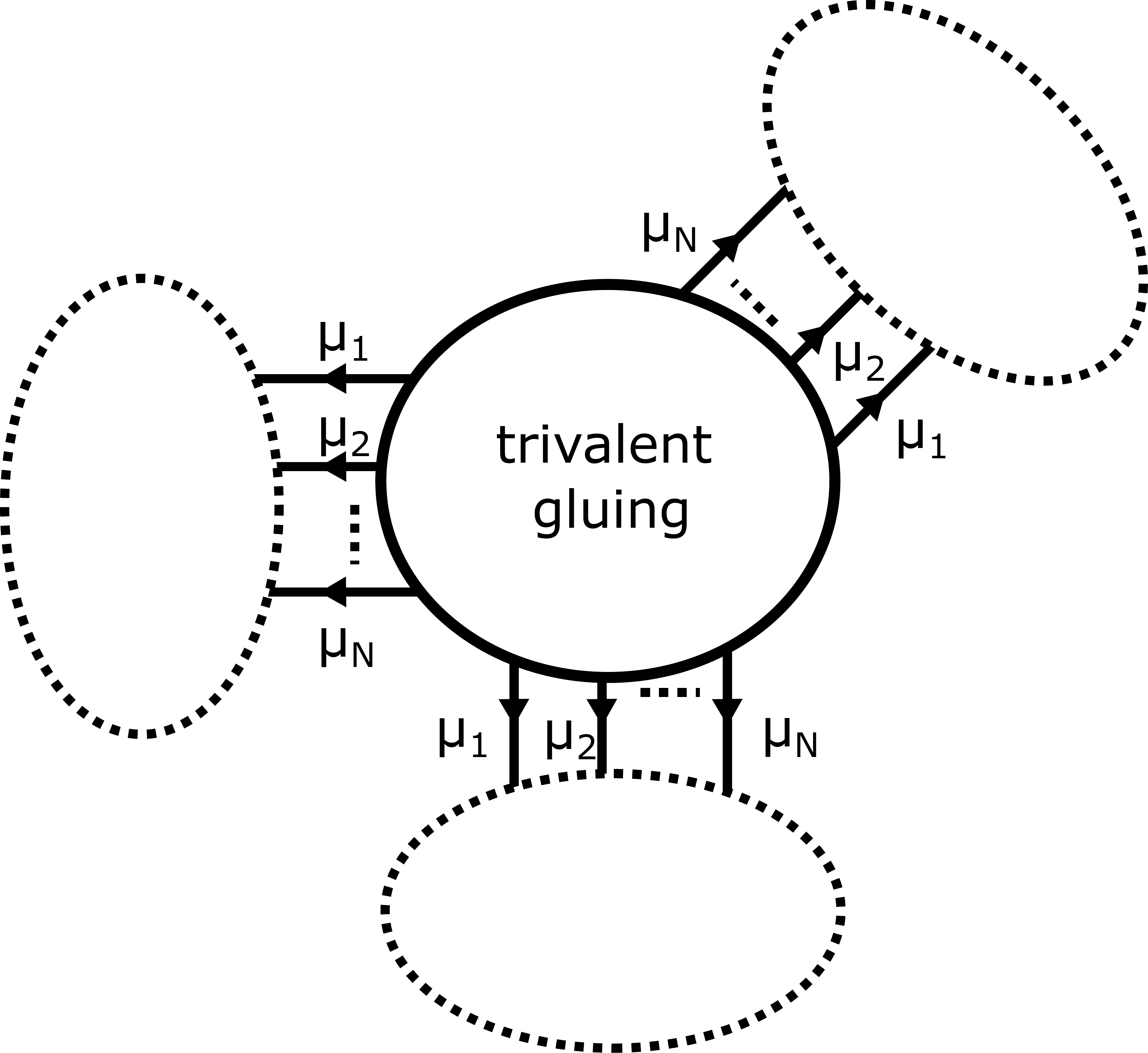}}
\hspace{1cm}
\subfigure[]{\label{fig:gluing3}
\includegraphics[width=6cm]{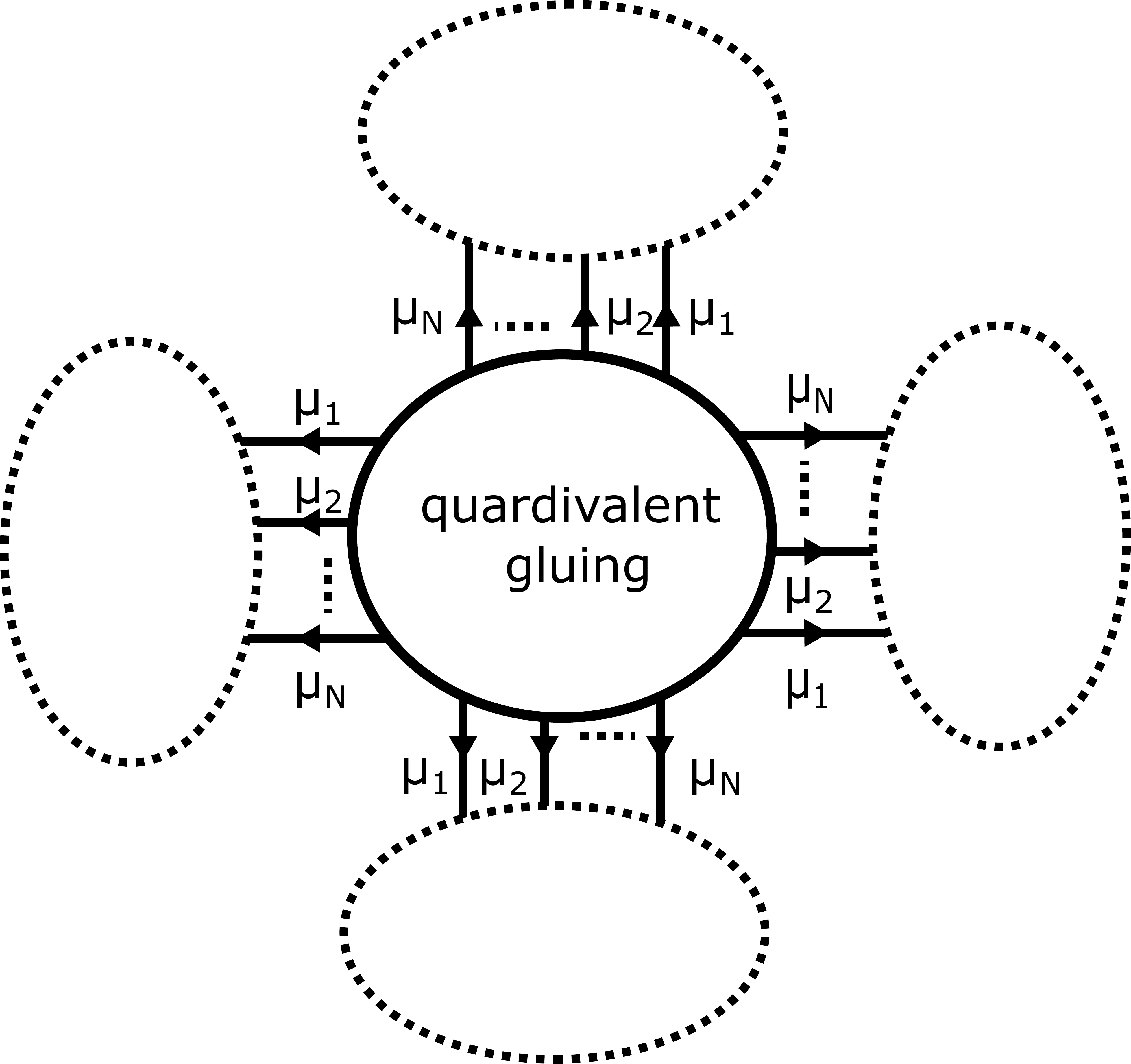}}
\caption{(a): A schematic drawing of the $\SU(N)$ trivalent gauging.  (b): A schematic drawing of the $\SU(N)$ quadrivalent gauging.}
\label{fig:gluing23}
\end{figure}
The topological vertex formalism has been further extended in \cite{Hayashi:2017jze} to digrams which consists of a trivalent or quadrivalent gluing of web diagrams. For connecting two 5-brane webs as in Figure \ref{fig:gluing1}, we can simply sum over the Young diagrams assigned to the gluing lines. In terms of the gauge theory description the gluing $N$ parallel lines means $\SU(N)$ gauging. It is in fact possible to extend the $\SU(N)$ gauging for the gluing of three diagrams or four diagrams. We will call the gluing of three diagrams as trivalent gluing/gauging and also we call the gluing of four diagrams as quadrivalent gluing/gauging. Such gluing may be schematically drawn as Figure \ref{fig:gluing2} for the trivalent gauging and Figure \ref{fig:gluing3} for the quadrivalent gauging. In this case simply summing over all the Young diagrams overcount the Gopakumar-Vafa invariants of the corresponding topological string partition funciton.

\begin{figure}[t]
\centering
\subfigure[]{\label{fig:each}
\includegraphics[width=6cm]{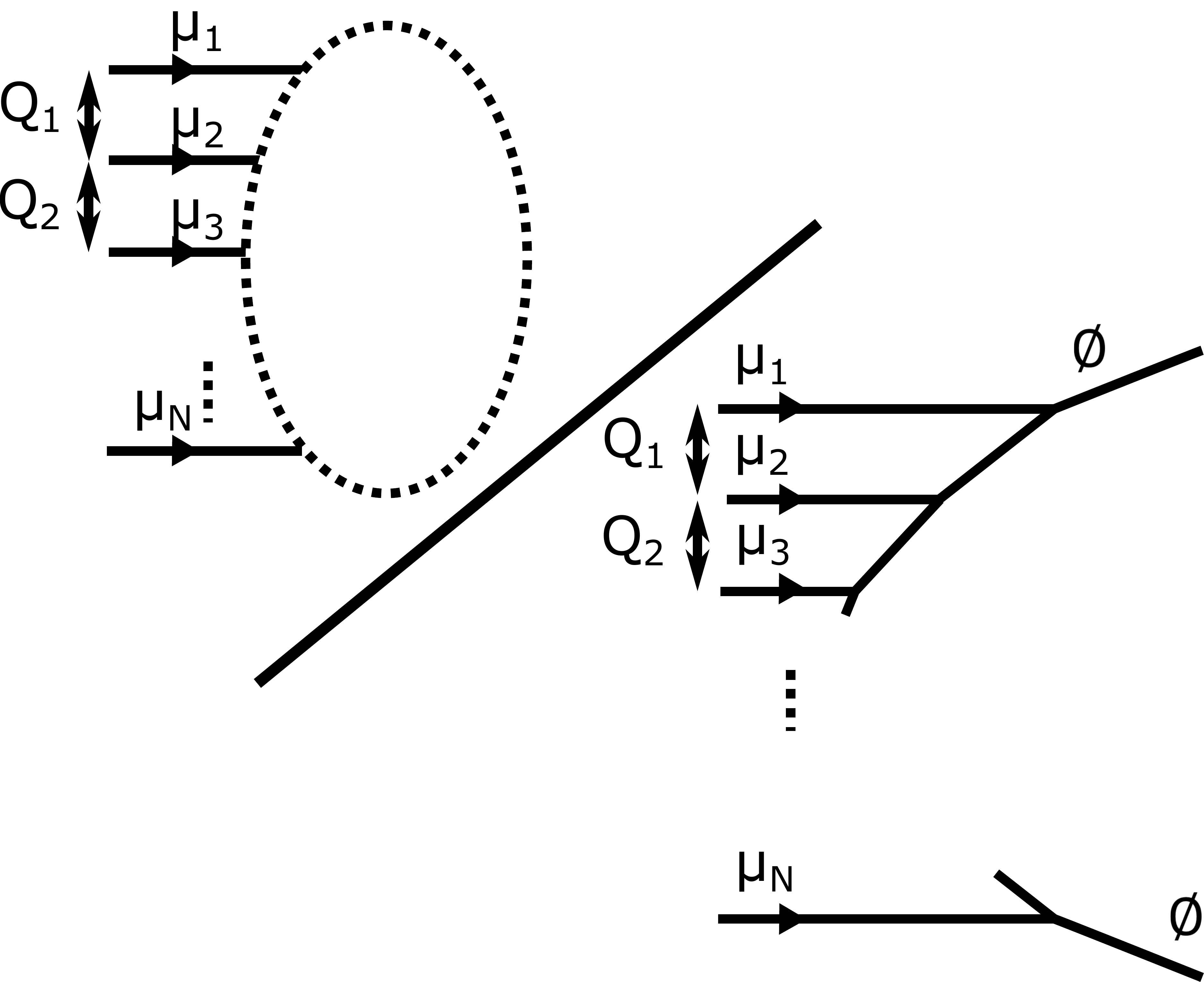}}
\hspace{2 cm}
\subfigure[]{\label{fig:left}
\includegraphics[width=4cm]{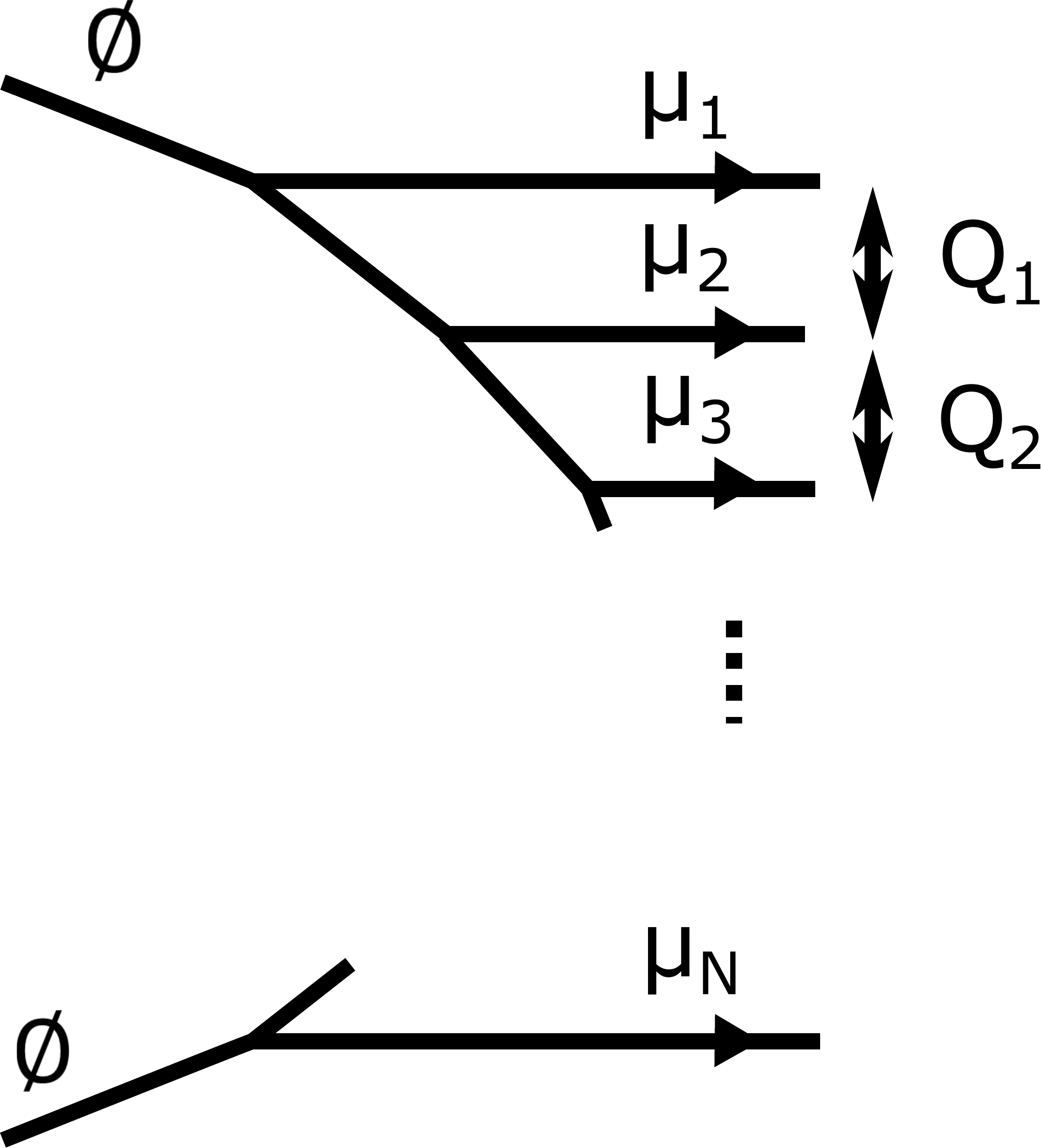}}
\caption{(a): The way of computing the contribution of each piece which we glue. (b): Another strip diagram. Combining this diagram with the diagram drawn in the denominator in Figure \ref{fig:each} yields a diagram for a pure $\SU(N)$ gauge theory.}
\label{fig:eachleft}
\end{figure}
For computing the partition function of a theory realized on a diagram with trivalent gluing or quadrivalent gluing, we can do in the following way. For each diagram which we will glue, we divide the partition function computed by the topological vertex by the contribution of a strip diagram with the same Young diagrams assigned to gluing legs as in Figure \ref{fig:each}. We also use the same K\"ahler parameters for the lengths between the gluing lines. The contribution of the strip diagram is roughly a half of the pure $\SU(N)$ Nekrasov partition function. Let $Z_{i.\{\mu_j\}}^{\text{top}}\; \left(\{\mu_j\} = \{\mu_1, \mu_2, \cdots, \mu_N\}\right)$ for $i=1, 2, 3 $ be the partition function computed by applying the topological vertex to each of the diagrams in Figure \ref{fig:gluing2}. Let $Z^{\SU(N), R}_{\{\mu_j\}}$ be the partition function of the strip diagram which is written in the denominator in Figure \ref{fig:each} and we also denote the partition function of another strip diagram in Figure \ref{fig:left} by $Z^{\SU(N), L}_{\{\mu_j\}}$. Then the partition function of the diagram in Figure \ref{fig:gluing2} is written as\footnote{This prescription may be rephrased by the $N^-_k$-vertex \cite{Hayashi:2019fsa}, which is a generalization of the $1^-_k$ vertex that is mirror of the standard topological vertex \cite{Aganagic:2004js}.}
\begin{align}
Z_{\text{trivalent}} = \sum_{\mu_1, \mu_2, \cdots, \mu_N}\left(\prod_{k=1}^N\left(-P_{k}\right)^{|\mu_k|}f_{\mu_k}^{n_k}(q)\right)Z^{\SU(N), L}_{\{\mu_j\}}Z^{\SU(N), R}_{\{\mu_j\}}\prod_{i=1}^3\frac{Z^{\text{top}}_{i, \{\mu_j\}}}{Z^{\SU(N), R}_{\{\mu_j\}}}.\label{trivalent}
\end{align}
$P_k$ is the K\"ahler parameter for the gluing line with the Young diagram $\mu_k$. The power $n_k$ of the framing factor can be determined from the local structure of the gluing. Namely it is possible to locally write down a web diagram around the part of the trivalent $\SU(N)$ gauging and the power of the framing factor is determined by the usual way reviewed in appendix \ref{sec:vertex}. Similarly for the diagram with the quadrivalent gauging, let $Z_{i.\{\mu_j\}}^{\text{top}}\; i=1, 2, 3, 4$ be the partition function computed by applying the topological vertex to each of the four diagrams in Figure \ref{fig:gluing3}. Then the partition function of the diagram in Figure \ref{fig:gluing3} is given by
\begin{align}
Z_{\text{quadrivalent}} = \sum_{\mu_1, \mu_2, \cdots, \mu_N}\left(\prod_{k=1}^N\left(-P_{k}\right)^{|\mu_k|}f_{\mu_k}^{n_k}(q)\right)Z^{\SU(N), L}_{\{\mu_j\}}Z^{\SU(N), R}_{\{\mu_j\}}\prod_{i=1}^4\frac{Z^{\text{top}}_{i, \{\mu_j\}}}{Z^{\SU(N), R}_{\{\mu_j\}}}.\label{quadrivalent}
\end{align} 
$P_k$ is the K\"ahler parameter for the gluing line and the power $n_k$ of the framing factor can be determined in the same way. 

\subsection{Nekrasov partition function}
\label{sec:Nek}
We close this appendix by remarking the Nekrasov partition funciton. The Nekrasov partition function consists of three factors,
\begin{align}
Z_{\text{Nek}} = Z_{\text{Cartan}}Z_{\text{roots}}Z_{\text{weights}}Z_{\text{inst}},
\end{align}
where $ Z_{\text{Cartan}}Z_{\text{roots}}Z_{\text{weights}}$ is the perturbative contribution and $Z_{\text{inst}}$ is the instanton contribution. The perturbative contribution has a univeral form. $Z_{\text{Cartan}}$ is the contribution from vector multiplets of the Cartan subalgebra and it is given by \eqref{cartan}. $Z_{\text{roots}}$ is the contribution from vector multiplets of the roots of a gauge group $G$. The explicit form is 
\begin{align}
Z_{\text{roots}}  
= \text{PE}\left[\frac{2q}{(1-q)^2}\sum_{\alpha \in \Delta_+}e^{-\alpha\cdot a}\right], \label{roots}
\end{align}  
where $\Delta_+$ is the set of the positive roots of the Lie algebra $\mathfrak{g}$ of the gauge group $G$ and $a = (a_1, a_2, \cdots, a_{\text{rank}(G)})$ is the Coulomb branch moduli in the Cartan subalgebra. Lastly $Z_{\text{weights}}$ is the contribution from hypermultiplets in a representation $R_f$ of $\mathfrak{g}$, which is given by
\begin{align}
Z_{\text{weights}} =\text{PE}\left[-\frac{q}{(1-q)^2}\sum_f\sum_{w_f \in R_f}e^{-|w_f\cdot a - m_f|}\right],
\end{align}
where $w_f$ is a weight of the representation $R_f$. $f$ parameterizes different hypermultiplets like the number of flavors. 

On the other hand the explicit form of the instanton part $\text{Z}_{\text{inst}}$ is more involved and does not have a universal formula in general. Since we will not use the instanton partition function in this paper, we do not present its explicit form here.

\bigskip

\bibliographystyle{JHEP}
\bibliography{refs}
\end{document}